\pdfoutput=1
\documentclass[a4paper,nobind]{ociamthesis} 
\usepackage{dsfont} 
\usepackage[utf8]{inputenc}
\usepackage[T1]{fontenc}
\usepackage{amsmath}
\usepackage{amssymb,enumerate}
\usepackage{hyperref}
\usepackage{xspace}
\usepackage{booktabs}
\usepackage{xcolor}
\usepackage[mathscr]{euscript}
\usepackage{bbm}
\usepackage{url}
\usepackage{epsfig}
\usepackage{wasysym} %For permil sign
\usepackage[capitalise,noabbrev]{cleveref}

\usepackage[square,comma,sort&compress,numbers]{natbib}
\usepackage{doi}
\usepackage{lmodern}
\usepackage{graphicx}
\usepackage{feynmf}
\usepackage{slashed}
\usepackage{tikz}
\usepackage{appendix}
\usepackage{amsthm}
\usepackage{rotating}
\usepackage{mathtools}
\usepackage{lipsum}
\newcommand{\bra}[1]{\mathinner{\langle{#1}|}}

\newcommand{\field}[1]{\mathbb{#1}} %For the field of real and complex numbers

\newcommand{\Imp}{\quad \Rightarrow \quad}

\title{At the Frontier of Precison QCD in the LHC~Era}
\author{Alexander Karlberg}
\college{Merton College}
\degree{Doctor of Philosophy}
\degreedate{Trinity 2016}

\newcommand{\as}{\alpha_s}
\newcommand{\alphas}{\alpha_s}

\newcommand{\GeV}{\;\mathrm{GeV}}
\newcommand{\TeV}{\;\mathrm{TeV}}

\newcommand{\LO}{\text{LO}}
\newcommand{\NLO}{\text{NLO}}
\newcommand{\NNLO}{\text{NNLO}}
\newcommand{\NNNLO}{\text{N$^3$LO}}
\newcommand{\POWHEGBOX}{\texttt{POWHEG-BOX}\xspace}
\newcommand{\POWHEG}{\texttt{POWHEG}\xspace}
\newcommand{\PYTHIA}{\texttt{PYTHIA}\xspace}

\definecolor{light-gray}{gray}{0.8}

\newcommand{\noun}[1]{{\texttt{#1}\xspace}}

\newcommand{\ZNLOPSpPYTHIA}{\noun{Zj-MiNLO}}
\newcommand{\ZNNLOPSpPYTHIA}{\noun{NNLOPS}}

\newcommand{\MINLO}{\noun{MiNLO}}

\newcommand{\ZJMINLO}{\noun{Zj-MiNLO}}
\newcommand{\WJMINLO}{\noun{Wj-MiNLO}}
\newcommand{\VJMINLO}{\noun{Vj-MiNLO}}

\newcommand{\DYNNLO}{\noun{DYNNLO}}
\newcommand{\FEWZ}{\noun{FEWZ}} 
\newcommand{\NLOPS}{\noun{NLOPS}}
\newcommand{\NNLOPS}{\noun{NNLOPS}}
\newcommand{\MCatNLO}{\noun{MC@NLO}} 

\newcommand{\DYQT}{\noun{DYqT}} 

\newcommand{\ZJ}{\noun{Zj}}
\newcommand{\WJ}{\noun{Wj}}
\newcommand{\VJ}{\noun{Vj}}
\newcommand{\Z}{\noun{Z}}
\newcommand{\V}{\noun{V}}

\newcommand{\JETVHETO}{\noun{JetVHeto}}
\newcommand{\FASTJET}{\noun{FastJet}}
\newcommand{\Kr}{K_{\scriptscriptstyle \mathrm{R}}}
\newcommand{\Kf}{K_{\scriptscriptstyle \mathrm{F}}}
\newcommand{\mur}{\mu_{\scriptscriptstyle \mathrm{R}}}
\newcommand{\muf}{\mu_{\scriptscriptstyle \mathrm{F}}}
\newcommand{\pt}{p_{\scriptscriptstyle \mathrm{T}}}

\newcommand{\ptz}{p_{\scriptscriptstyle \mathrm{T,Z}}}
\newcommand{\ptw}{p_{\scriptscriptstyle \mathrm{T,W}}}
\newcommand{\ptjone}{p_{\scriptscriptstyle
    \mathrm{T,j_{1}}}}
\newcommand{\kt}{k_{\scriptscriptstyle \mathrm{T}}}
\newcommand{\qt}{q_{\scriptscriptstyle \mathrm{T}}}
\newcommand{\mtw}{m_{\scriptscriptstyle \mathrm{T, W}}}

\newcommand{\hc}{\beta}
\newcommand{\hgam}{\gamma}

\renewcommand{\cal}{\mathcal}
\renewcommand{\rm}{\mathrm}
\renewcommand{\tt}{\texttt}

\newcommand\POWHEGBOXVT{{\tt{POWHEG~BOX V2}}}

\newcommand\HERWIG{{\tt{HERWIG}}}

\newcommand\VBFNLO{{\tt{VBFNLO}}}

\def\({\left(} 
\def\){\right)} 

\def\beq{\begin{equation}}
\def\beqn{\begin{eqnarray}}
\def\eeq{\end{equation}}
\def\eeqn{\end{eqnarray}}

\def\mr{\mathrm}

\def\vbfeemm{VBF $e^+e^-\mu^+\mu^- jj$\;}
\def\vbfww{VBF $W^+W^-jj$\;}

\def\llvv{\ell^+\ell^-\nu\bar\nu}
\def\llll{\ell^+\ell^-{\ell'}^+{\ell'}^-}
\def\llqq{\ell^+\ell^-\bar qq}
\def\eemm{e^+e^-\mu^+\mu^-}

\def\mc{\mathcal}
\def\muf{\mu_\mr{F}}
\def\mur{\mu_\mr{R}}

\usepackage{heppennames2}
\providecommand\HAWK{{\textsc{HAWK}}}
\providecommand\UGeV{\;\mathrm{GeV}}
\providecommand\UTeV{\;\mathrm{TeV}}
\providecommand\MV{M_{\scriptscriptstyle\mathrm{V}}}
\providecommand\MW{M_{\scriptscriptstyle\mathrm{W}}}
\providecommand\MZ{M_{\scriptscriptstyle\mathrm{Z}}}
\providecommand\MH{M_{\scriptscriptstyle\mathrm{H}}}
\providecommand\GF{G_{\scriptscriptstyle\mathrm{F}}}
\providecommand{\muR}{{\mu_{\mathrm{R}}}}
\providecommand{\muF}{{\mu_{\mathrm{F}}}}
\providecommand{\kT}{\ensuremath{k\sb{\scriptstyle\mathrm{T}}}}
\providecommand{\pT}{\ensuremath{p\sb{\scriptstyle\mathrm{T}}}}
\providecommand{\Bref}[1]{Reference \cite{#1}}
\providecommand{\Brefs}[1]{References \cite{#1}}
\providecommand\DIS{\mathrm{DIS}}
\providecommand\VBF{\mathrm{VBF}}
\providecommand\ELWK{\mathrm{EW}}
\providecommand\NNLO{\mathrm{NNLO}}
\providecommand\NLO{\mathrm{NLO}}
\providecommand\LO{\mathrm{LO}}
\providecommand\QCD{\mathrm{QCD}}
\providecommand\sw{\sin\theta_{\mathrm{W}}}
\providecommand\btilde{\tilde{\mathcal{B}}}
\providecommand\bbar{\bar{\mathcal{B}}}

\begin{document}

%%%%% CHOOSE YOUR LINE SPACING HERE
% This is the official option.  Use it for your submission copy and library copy:
\setlength{\textbaselineskip}{22pt plus2pt}
% This is closer spacing (about 1.5-spaced) that you might prefer for your personal copies:
%\setlength{\textbaselineskip}{18pt plus2pt minus1pt}

% You can set the spacing here for the roman-numbered pages (acknowledgements, table of contents, etc.)
\setlength{\frontmatterbaselineskip}{18pt plus1pt minus1pt}

% Leave this line alone; it gets things started for the real document.
\setlength{\baselineskip}{\textbaselineskip}

%%%%% CHOOSE YOUR SECTION NUMBERING DEPTH HERE
% You have two choices.  First, how far down are sections numbered?  (Below that, they're named but
% don't get numbers.)  Second, what level of section appears in the table of contents?  These don't have
% to match: you can have numbered sections that don't show up in the ToC, or unnumbered sections that
% do.  Throughout, 0 = chapter; 1 = section; 2 = subsection; 3 = subsubsection, 4 = paragraph...

% The level that gets a number:
\setcounter{secnumdepth}{3}
% The level that shows up in the ToC:
\setcounter{tocdepth}{2}

%%%%% ABSTRACT SEPARATE
% This is used to create the separate, one-page abstract that you are required to hand into the Exam
% Schools.  You can comment it out to generate a PDF for printing or whatnot.
%\begin{abstractseparate}
%	\input{abstract/abstract} % Create an abstract.tex file in the 'text' folder for your abstract.
%\end{abstractseparate}

% JEM: Pages are roman numbered from here, though page numbers are invisible until ToC.  This is in
% keeping with most typesetting conventions.
\begin{romanpages}

% Title page is created here
\maketitle

%%%%% DEDICATION -- If you'd like one, un-comment the following.
\begin{dedication}
\emph{Til mine forældre}
\end{dedication}

%%%%% ABSTRACT -- Nothing to do here except comment out if you don't want it.
\begin{abstract}
	This thesis discusses recent advances in precision calculations of quantum
chromodynamics and their application to the Large Hadron Collider (LHC)
physics program and beyond.

The first half of the thesis is dedicated to the study of vector boson fusion
Higgs (VBF) production; fully differential at the next-to-next-to-leading order
level (\NNLO{}), and inclusively at next-to-next-to-next-to-leading order
(\NNNLO{}). Both calculations are performed in the structure function
approximation, where the VBF process is treated as a double deep inelastic
scattering. For the differential calculation a new subtraction method,
``projection-to-Born'', is introduced and applied. We study VBF production in a
number of scenarios relevant for the LHC and for Future Circular Colliders
(FCC). We find \NNLO{} corrections after typical cuts of $5-6\%$ while
differential distributions show corrections of up to $10-12\%$ for some standard
observables. For the inclusive calculation we find \NNNLO{} corrections at the
order of $1-2\permil$.

The second half of the thesis presents recent results on the matching of fixed
order calculations with parton showers. We first present the POsitive Weight
Hardest Emission Generator (\POWHEG{}) method for matching next-to-leading order
(\NLO{}) calculations with parton showers. We then proceed to apply it to the
case of vector boson fusion $ZZjj$ production and discuss the results for
scenarios relevant for the LHC and a possible FCC. In order to present the
matching of a \NNLO{} calculation with a parton shower, we next discuss the
Multi-Scale Improved \NLO{} (\MINLO{}) procedure. By applying a reweighting
procedure to \MINLO{} improved Drell-Yan production, we obtain a generator which
is \NNLO{} accurate when integrated over all radiation while providing a fully
exclusive description of the final state phase space. We compare the calculation
to dedicated next-to-next-to-leading logarithm resummations and find very good
agreement. The generator is also found to be in good agreement with $7$ and
$8\TeV$ LHC data.
%\noindent
%\lipsum[1-3]

\end{abstract}

%%%%% ACKNOWLEDGEMENTS -- Nothing to do here except comment out if you don't want it.
\begin{acknowledgements}
 	There are many people without whom I would never have been able to complete this
thesis. First and foremost I am extremely grateful for the inspiration and
guidance my supervisor, Giulia Zanderighi, has provided me with over the
years. Everything I know about precision QCD and which is reflected in this
thesis I owe to her. The shortcomings of the thesis are my own.

I would also like to thank my collaborators Matteo Cacciari, Frédéric Dreyer,
Barbara Jäger, Emanuele Re and Gavin Salam for making my DPhil a very productive
one. All the work presented in this thesis has been carried out in collaboration
with them, and many aspects of QCD have become clearer to me only after
discussions with them. Additionally I have enjoyed fruitful interactions with
William Astill, Alan Barr, Wojciech Bizon, Stefan Dittmaier, Uli Haisch, Keith
Hamilton, Alexander Huss, David Kraljic, Gionata Luisoni, Michelangelo Mangano,
Pier Monni, Andy Powell (who also deserves special thanks for sharing an office
with me for four years), Kai Roehrig, James Scargill, James Scoville, Chuang
Sun, Jim Talbert, Ciaran Williams and Marco Zaro amongst others whom I have
shamefully forgotten. Mike Teper deserves my thanks for taking on the
co-supervisor role after Giulia moved to CERN. Poul Henrik Damgaard and Emil
Bjerrum-Bohr deserve credit for first suggesting Oxford for my DPhil.

I would like to extend my sincere thanks to James Buckee, without whose generous
financial support I would not have been able to take up my place as a student in
Oxford and at Merton College. Merton College has, in addition to being an
amazing social focus of my life over the past four years, awarded me with
several Research Grants to support my various conference activities. I have also
enjoyed generous support from Augustinus Fonden, Knud Højgaards Fond, Oticon
Fonden and Krista og Viggo Petersens Fond over the years.

Much of my work was carried out in the Theory Department of CERN. I am grateful
for the hospitality the department has shown me, and for the many people with
whom I interacted during my visits. These visits were financially supported
by the ERC Consolidator Grant HICCUP.

My family and friends (too many to list here) deserve huge thanks: those in
Oxford for making the last four years a treat, and those back home in Denmark
for not forgetting me. In particular I thank Anne, who turned out to be the real
reason to study at Oxford. 

Most importantly, I would like to thank my parents and brothers for their
unconditional love and support. They made all of this possible.

%Giulia/Mike
%Year group
%Collaborators
%Family/Anne
%Buckee/Merton
%\noindent
%\lipsum[1-3]

\end{acknowledgements}

%%%%% MINI TABLES
% This lays the groundwork for per-chapter, mini tables of contents.  Comment the following line
% (and remove \minitoc from the chapter files) if you don't want this.  Un-comment either of the
% next two lines if you want a per-chapter list of figures or tables.
\dominitoc % include a mini table of contents
%\dominilof  % include a mini list of figures
%\dominilot  % include a mini list of tables

% This aligns the bottom of the text of each page.  It generally makes things look better.
\flushbottom

% This is where the whole-document ToC appears:
\tableofcontents

%\listoffigures
%	\mtcaddchapter
% \mtcaddchapter is needed when adding a non-chapter (but chapter-like) entity to avoid confusing minitoc

% Uncomment to generate a list of tables:
%\listoftables
%	\mtcaddchapter

%%%%% LIST OF ABBREVIATIONS
% This example includes a list of abbreviations.  Look at text/abbreviations.tex to see how that file is
% formatted.  The template can handle any kind of list though, so this might be a good place for a
% glossary, etc.
%\include{text/abbreviations}

% The Roman pages, like the Roman Empire, must come to its inevitable close.
\chapter*{\label{ch:preface}Preface}
\addcontentsline{toc}{chapter}{Preface} The Standard Model of Particle Physics
is one of the greatest scientific triumphs of the $20^{\mathrm{th}}$
century. Since its conception more than $50$ years ago, experiments have
consolidated all of its numerous predictions, and with the discovery of the
Higgs Boson in $2012$, Nature finally revealed to us the last particle which
makes up the Standard Model.

One of the striking features of the Particle Physics program of the last several
decades was its guarantee to succeed. We knew that new physics had to be found
around the scale which we now associate with the weak vector bosons. We knew
that the top quark had to exist in order for the Standard Model to be anomaly
free. And we knew that the Higgs Boson, or something else, had to show up around
the$\TeV$-scale to save the Standard Model from breaking the fundamental
principle of unitarity. However, this guarantee has expired with the discovery
of the aforementioned Higgs Boson, as this last piece of the puzzle has rendered
the Standard Model self-consistent up to very large energy scales. Hence,
Particle Physics has transitioned from a phase of success into a phase of
unknowns.

We \emph{know} that there are phenomena the Standard Model cannot explain, like
Dark Matter, Dark Energy, neutrino masses, and Gravity, but for which we have
strong experimental evidence. We \emph{don't} know which of the innumerable
theories and models extending the Standard Model, if any, will prove to be the
correct answer(s). Until we either get direct experimental evidence of the
nature of Physics Beyond the Standard Model or stumble upon the correct model,
we may still learn much from studying the Standard Model in detail. Such studies
are currently under way at the Large Hadron Collider. Here protons are being
collided at a total centre-of-mass energy of $13 \TeV$ and the outcome of these
collisions measured by one of the four experiments ALICE, ATLAS, CMS, and LHCb.

On their own these measurements can tell us a lot about Nature, but they become
extremely powerful when compared with theoretical predictions. By looking for
deviations of data from Standard Model predictions, we may ultimately learn how
the Standard Model breaks down and what has to replace it. If the deviations are
small, the uncertainties on our theoretical predictions necessarily have to be
smaller.

%This thesis is about recent progress in making precise quantum
%chromodynamics (QCD) predictions for processes relevant to the Large Hadron Collider.

When I started my DPhil-studies in $2012$ precision QCD was at the end of a
revolution. For a long time it had been impossible to carry out
loop-calculations for more than the simplest processes and even tree-level
calculations with more than a few external legs were unfeasible. However, a few
unexpected developments quickly changed that, and within a few years most of the
processes the experimental community had requested computed to next-to-leading
order (\NLO{}) were available, and leading order (\LO{}) calculations had become
completely automated. In addition to that, methods for matching \NLO{}
calculations with parton showers had been developed and would also be fully
automated within a few years. Beyond \NLO{} only very few processes had been
computed and even fewer so fully differentially. 

Today, we are in a similar situation to the one experienced before the
\NLO{}-revolution. Next-to-next-to-leading order (\NNLO{}) calculations are more
than often needed to meet the experimental precision, but currently only
$2\rightarrow 2$ scattering processes can be computed at two-loops, effectively
providing the bottleneck for computing higher multiplicity processes at
\NNLO{}. However, more than $20$ processes have been computed differentially to
\NNLO{} and first steps have been taken towards breaking the ``$2\rightarrow
2$''-wall. The simplest of these \NNLO{} processes have been matched to a parton
shower and two processes have been computed inclusively to
next-to-next-to-next-to-leading order (\NNNLO).

Here I describe some of these recent results in precision QCD and the methods
used to obtain them. In particular I discuss in \cref{ch:incVBF}
inclusive vector boson fusion Higgs production at \NNNLO{} and in
\cref{ch:vbfnnlo} I discuss fully differential vector boson fusion Higgs
production at \NNLO{}. This work was first presented in
\Brefs{Cacciari:2015jma,Dreyer:2016oyx} and was done in collaboration with
Matteo Cacciari, Frédéric Dreyer, Gavin Salam, and Giulia Zanderighi, but has
been significantly expanded here due to the letter format of the two original
publications. In particular I discuss the structure function approach in some
detail and develop the ``projection-to-Born'' method. \cref{ch:vbfnnlo}
also includes results reported in \Brefs{Contino:2016spe,YR4}.

In \cref{ch:powheg} I give a brief introduction to the \POWHEG{} method
for matching \NLO{} calculations and parton showers, and apply it to
electroweak $ZZjj$ production. The latter work was done in collaboration with
Barbara Jäger and Giulia Zanderighi and was first published in
\Bref{Jager:2013iza}.

Following that, I introduce the \MINLO{} method in \cref{ch:minlo} and discuss
how a \MINLO{}-improved \VJ{} \POWHEG{} generator can be upgraded to an \NNLO{}
accurate $V$ generator through a reweighting procedure. This work was done in
collaboration with Emanuele Re and Giulia Zanderighi and was first presented in
\Bref{Karlberg:2014qua}. Contributions were subsequently made to
\Bref{Alioli:2016fum} but have not been included here. In \cref{ch:final} I sum
up the research presented in this thesis, and provide some final remarks. 
\newline

\hfill Alexander Karlberg

\hfill Oxford, 2016

%\lipsum[4-7]

\end{romanpages}

%%%%% CHAPTERS
% Add or remove any chapters you'd like here, by file name (excluding '.tex'):
\flushbottom

%\part{Setting the Stage}
%\include{QCD/qcd}
%\include{SM/sm}
\part{Vector Boson Fusion Higgs Production}
\chapter{Inclusive Vector Boson Fusion Higgs~Production}\label{ch:incVBF}
There have been few discoveries in high energy physics as greatly anticipated as
that of the Higgs boson in 2012~\cite{Aad:2012tfa,Chatrchyan:2012ufa}. It has
been known since before the commissioning of the LHC that it was guaranteed to
discover either the Higgs boson, or something else in its place, to save the
Standard Model from violating unitarity. As the LHC has now entered the phase of
Run II, we hope to precisely determine the boson's
properties~\cite{Khachatryan:2015bnx} and thereby discover the true nature of
electroweak symmetry breaking.

The most relevant production channels for the Higgs boson at the LHC are gluon
fusion (ggH), vector boson fusion (VBF), production in association with a
vector boson (VH) and with a top-quark pair (ttH)~\cite{Dittmaier:2011ti}.

\begin{figure}[ht]
  \centering
  \includegraphics[width=0.55\linewidth]{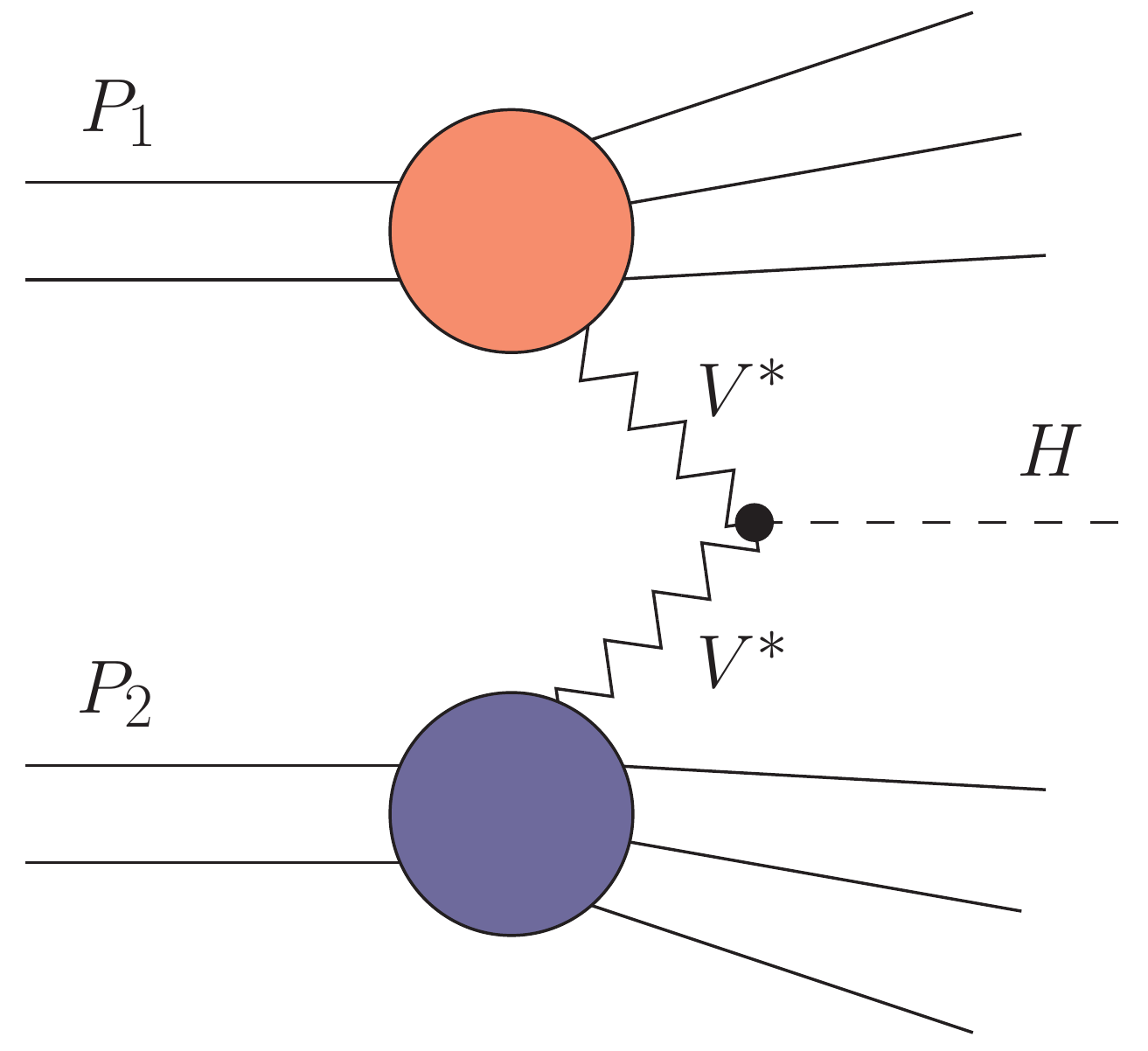}
  \caption{Illustration of Higgs production through vector boson fusion in the
    structure function approach. A vector boson is emitted from each of the two
    protons and fused into a Higgs boson. Each of the two protons are affected
    by separate but identical copies of QCD, here shown in orange and blue.}
  \label{fig:vbfh}
\end{figure}

Of these channels the cleanest one for studying the properties of the Higgs
Boson is the vector boson fusion channel~\cite{Jones:1979bq}, shown in
\cref{fig:vbfh}.
VBF is special for a number of
reasons~\cite{Dittmaier:2012vm,Heinemeyer:2013tqa,YR4}:
\begin{itemize}
\item it has the largest cross section of the processes that involves tree-level
  production of the Higgs boson (and is second largest among all processes);
\item it has a distinctive signature of two forward jets, which makes it
  possible to tag the events and so identify Higgs decays that normally have
  large backgrounds, e.g.\ $H \to \tau^+ \tau^-$;
\item the Higgs transverse momentum is non-zero even at lowest order, which
  facilitates searches for invisible decay
  modes~\cite{ATLAS:2015yda,CMS:2015dia};
\item and it also brings particular sensitivity to the charge-parity properties
  of the Higgs boson, and non-standard Higgs interactions, through the angular
  correlations of the forward jets~\cite{Plehn:2001nj}.
\end{itemize}

The forward jets are due to the t-channel topology of the process. The overall
energy of each jet is governed by the centre-of-mass energy of the collider
whereas the transverse momentum of the jets are set by the mass of the weak
vector bosons. For this reason VBF events also tend to have a large dijet
invariant mass. These features make it possible to separate the VBF signal from
the very large QCD background production through a set of cuts, which are
usually referred to as \emph{VBF cuts}. The VBF process therefore provides ideal
access for the intricate measurements of the Higgs
couplings~\cite{Zeppenfeld:2000td}.

Currently the VBF production signal strength has been measured with a
precision of about 24\%~\cite{Khachatryan:2016vau}, though significant
improvements can be expected during Run II and with the high luminosity
LHC.

The unique topology of the VBF process makes it not just experimentally very
accessible but also theoretically simple. One can view the VBF process as a
double Deep Inelastic Scattering (DIS) process, where a vector boson is radiated
independently from each proton after which the two vector bosons fuse into a
Higgs boson. In this picture the matrix element factorises into the contraction
of the two proton structure functions with the $HVV$ vertex, which is why it is
known as the \emph{structure function approach}~\cite{Han:1992hr}. The structure
function approach is exact to \NLO{} in the strong coupling constant and
receives only tiny corrections from non-factorisable contributions beyond this
order, which are both kinematically and colour suppressed. In fact, this
approach is exact in the limit in which one considers that there are two
identical copies of QCD associated with each of the two protons (shown orange
and blue in \cref{fig:vbfh}), whose interaction is mediated by the weak force.

Given the key role of VBF production at the LHC, it is of paramount
importance to have a precise prediction for its production. The total VBF rate
in the structure function approach was computed to \NNLO{} some years ago
~\cite{Bolzoni:2010xr,Bolzoni:2011cu,Zaro:2013twa}. This calculation found \NNLO{}
corrections of about $1\%$ and renormalisation and factorisation scale
uncertainties at the $5\permil$ level.

In this chapter we will first develop the structure function approach in some
detail and then proceed to compute the \NNNLO{} QCD corrections to the total VBF
cross section in this approximation. The calculation provides only the second
\NNNLO{} calculation for processes of relevance to the LHC physics program,
after a similar accuracy was recently achieved in the ggH
channel~\cite{Anastasiou:2015ema}. However, unlike the ggH calculation, our
calculation is fully differential in the Higgs kinematics. Since the \NNLO{}
corrections to VBF were already very small, the \NNNLO{} calculation is more of
theoretical interest than of phenomenological. As we will see, the \NNNLO{}
corrections are tiny and well within the scale uncertainty bands of the \NNLO{}
calculation. Hence our calculation shows very good convergence of perturbation
theory for the VBF process.

Since our calculation gives access to the \NNNLO{} structure functions we also
estimate missing higher order corrections to parton distribution functions,
which are currently only know to \NNLO{}. These corrections have not been
studied in much detail yet, but are likely to become interesting as more
processes become known at \NNNLO{}~\cite{Anastasiou:2016cez,Forte:2013mda}.

\section{The Structure Function Approach}
In the structure function approach, as discussed above, the VBF Higgs production
cross section is calculated as a double DIS process. Thus, it can be factorised
as the product of the hadronic tensors $\mathcal{W}^V_{\mu\nu}$ and the matrix
element for $V_1^* V_2^* \rightarrow H$, $\mathcal{M}^{\mu\nu}$. The cross
section can be expressed by~\cite{Han:1992hr}
\begin{align}
  \label{eq:vbfh-dsigma}
  d\sigma = &\frac{G_F^2}{s} M^2_{V_1}M^2_{V_2}
  \Delta_{V_1}^2(Q_1^2)
  \Delta_{V_2}^2(Q_2^2)
  %\notag
%  \\
%  &\times
  \mathcal{W}^{V_1}_{\mu\nu}(x_1,Q_1^2)\mathcal{M}^{\mu\rho} 
  \mathcal{M}^{*\nu\sigma} \mathcal{W}^{V_2}_{\rho\sigma}(x_2,Q_2^2)
  d\Omega_{\text{VBF}}.
\end{align}
Here $G_F$ is Fermi's constant, $M_V$ is the mass of the vector boson,
$\sqrt{s}$ is the collider centre-of-mass energy, $\Delta_V^2$ is the squared
vector boson propagator, $Q_i^2 = -q_i^2$ and $x_i = Q_i^2/(2P_i\cdot q_i)$ are the
usual DIS variables, and $d\Omega_{\text{VBF}}$ is the three-particle VBF phase
space given by
\begin{align}
  d\Omega_{\text{VBF}} = \frac{d^3 P_4}{(2\pi)^32E_4}\frac{d^3 P_5}{(2\pi)^32E_5}
  ds_4ds_5\frac{d^3 p_3}{(2\pi)^32E_3} (2\pi^4)
  \delta^4(P_1+P_2-p_3-P_4-P_5).
\end{align}
where $P_{1,2}$ are the incoming proton (not parton) momenta, $p_3$ is the
momentum of the Higgs, $P_{4,5}$ are the outgoing proton remnant momenta, and
$s_i = P_i^2 = (P_{i-3} +q_{i-3})^2$ are the invariant masses of the proton
remnants. From the knowledge of the vector boson momenta $q_i$, it is
straightforward to reconstruct the Higgs momentum. As such, the cross section
obtained using \cref{eq:vbfh-dsigma} is differential in the Higgs
kinematics. In the Standard Model, the $V_1^* V_2^* \rightarrow H$ matrix
element is given by
\begin{equation}
  \mathcal{M}^{\mu\nu} = 2 \sqrt{\sqrt{2}G_F}M_V^2g^{\mu\nu}.
  \label{eq:VVHME}
\end{equation}
The hadronic tensor $\mathcal{W}^V_{\mu\nu}$ can be expressed as
\begin{multline}
  \label{eq:hadr-tensor}
  \mathcal{W}^V_{\mu\nu}(x_i,Q_i^2) = 
  \Big(-g_{\mu\nu}+\frac{q_{i,\mu}q_{i,\nu}}{q_i^2}\Big) F_1^V(x_i,Q_i^2)
  \\
  + \frac{\hat{P}_{i,\mu}\hat{P}_{i,\nu}}{P_i\cdot q_i} F_2^V(x_i,Q_i^2)
  + i\epsilon_{\mu\nu\rho\sigma}\frac{P_i^\rho q_i^\sigma}{2 P_i\cdot q_i} 
  F_3^V(x_i,Q_i^2)\,,
\end{multline}
where we have defined $\hat{P}_{i,\mu} = P_{i,\mu} - \tfrac{P_i \cdot
  q_i}{q_i^2} q_{i,\mu}$, and the $F^V_i(x,Q^2)$ functions are the standard DIS
structure functions with $i=1,2,3$ and $V=Z,W^-,W^+$~\cite{Ellis:1991qj}.

Since the matrix element in \cref{eq:VVHME} is proportional to the flat
metric, it is obvious that the cross section must be proportional to the
hadronic tensor contracted with itself. This contraction is given by
\begin{align}
  & \mathcal{W}^{V_1}_{\mu\nu}(x_1,Q_1^2)\mathcal{W}^{V_2,\mu\nu}(x_2,Q_2^2)  = 
   F_1^{V_1}F_1^{V_2} \left[2 + \frac{(q_1\cdot q_2)^2}{q_1^2q_2^2}\right] \notag \\
  & + \frac{F_1^{V_1}F_2^{V_2}}{P_2\cdot q_2}\left[\frac{(P_2\cdot q_2)^2}{q_2^2} + \frac{1}{q_1^2}
    \left(P_2\cdot q_1 - \frac{(P_2\cdot q_2)(q_1\cdot q_2)}{q_1^2}\right)^2\right] \notag\\
  & + \frac{F_2^{V_1}F_1^{V_2}}{P_1\cdot q_1}\left[\frac{(P_1\cdot q_1)^2}{q_1^2} + \frac{1}{q_2^2}
    \left(P_1\cdot q_2 - \frac{(P_1\cdot q_1)(q_1\cdot q_2)}{q_2^2}\right)^2\right] \notag\\
   & + \frac{F_2^{V_1}F_2^{V_2}}{(P_1\cdot q_1)(P_2\cdot q_2)}\Bigg[P_1\cdot P_2 - \frac{(P_1\cdot q_1)(P_2\cdot q_1)}{q_1^2} \notag \\
     & \hspace{3.5cm} - \frac{(P_1\cdot q_2)(P_2\cdot q_2)}{q_2^2} + \frac{(P_1\cdot q_1)(P_2\cdot q_2)(q_1\cdot q_2)}{q_1^2 q_2^2}\Bigg]^2 \notag\\
   & + \frac{F_3^{V_1}F_3^{V_2}}{2(P_1\cdot q_1)(P_2\cdot q_2)}\left[(P_1\cdot P_2)(q_1\cdot q_2) - (P_1\cdot q_2)(P_2\cdot q_1)\right].
  \label{eq:hadrcontract}
\end{align}
where we have dropped the argument $(x_i,Q_i^2)$ from the structure functions to
ease notation. 

\begin{figure}[!tbh]
  \centering
  \includegraphics[width=0.6\textwidth]{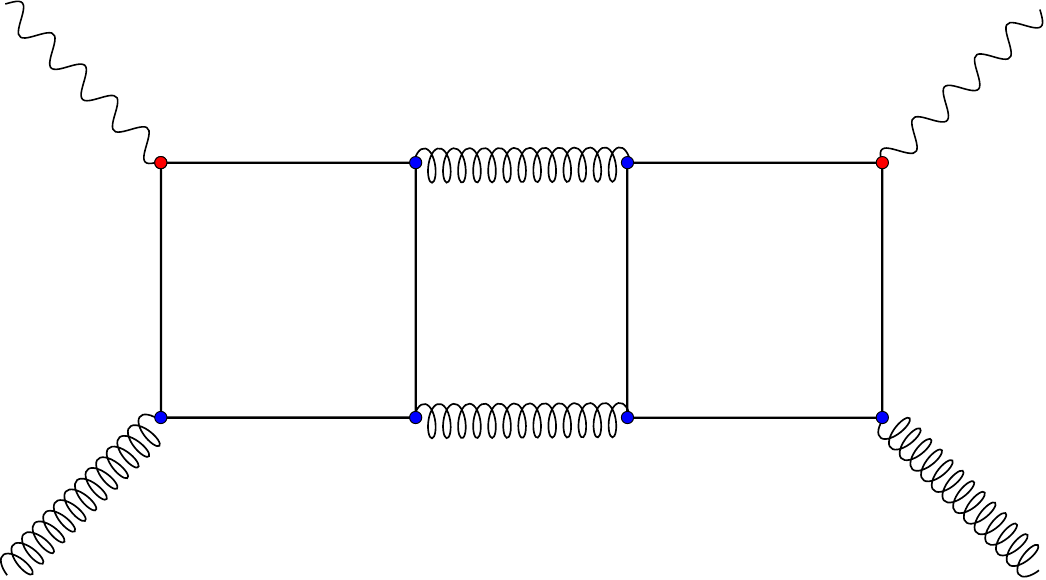}
  \caption{One of the new flavour structures appearing at \NNNLO{} for neutral
    current interactions but not charged current ones. Curly lines are gluons,
    wavy lines vector bosons and straight lines (anti)quarks.}
  \label{fig:fl11}
\end{figure}

In order to compute the N$^n$LO cross section, we require the structure
functions $F_i^V$ up to order $\mathcal{O}(\alpha_s^n)$ in the strong
coupling constant.
Using the QCD factorisation theorem, we may express the structure functions as
convolutions of the parton distribution functions (PDFs), $f_a$, with the short
distance Wilson coefficient functions, $C_i$
\begin{equation}
  \label{eq:conv-structf}
  F_i^V = \sum_{a=q,g} C_{i,a}^{V} \otimes f_a \,,\quad i=2,L,3\,, \quad V=Z,W^+,W^-\,,
\end{equation}
and
\begin{equation}
  F_L^V = F_2^V - 2xF_1^V.
\end{equation}
All the necessary coefficient functions are known up to third order in the
strong coupling constant
\footnote{The even-odd differences between charged-current coefficient functions
  were only known approximately when this work was carried
  out~\cite{Moch:2007rq}. However, the uncertainty associated with this
  approximation is less than $1\permil$ of the \NNNLO{} correction, and
  therefore completely negligible. Since then the exact result has been
  published~\cite{Davies:2016ruz} along with some approximate fourth-order
  results~\cite{Ruijl:2016pkm}.}.
To compute the \NNNLO{} VBF Higgs production cross section, we therefore
evaluate the convolution of the PDF with the appropriate coefficient functions
in \cref{eq:conv-structf}.
At \NNNLO{}, additional care is required due to the appearance of new
flavour topologies~\cite{Vermaseren:2005qc}, see \cref{fig:fl11}.
Therefore, contributions corresponding to interference of diagrams where the
vector boson attaches on different quark lines are to be set explicitly to zero
for charged boson exchanges.

To that end, it is useful to decompose the quark and anti-quark distributions,
$q_i(x,\mu)$ and $\bar{q}_i(x,\mu)$, into their pure-singlet contributions
\begin{equation}
  q_{\text{PS}} = \sum_{i=1}^{n_f}(q_i + \bar{q}_i)\,,
  \label{eq:qps}
\end{equation}
non-singlet valence contributions
\begin{equation}
  q_{\text{NS}}^{v} = \sum_{i=1}^{n_f}(q_i - \bar{q}_i)\,,
  \label{eq:qnsvalence}
\end{equation}
the flavour asymmetries\footnote{Note that this definition is different
  from what is being used in~\cite{Bolzoni:2011cu,Zaro:2013twa}. It leads to a
  slightly more intuitive definition of $F^Z_L$ and $F^Z_2$ and a slightly less
  intuitive definition of $F^Z_3$.}
\begin{equation}
  q_{\text{NS}}^{\pm} = (q_i \pm \bar{q}_i)\,,
  \label{eq:qns}
\end{equation}
and the asymmetry $\delta q_{\text{NS}^{\pm}}$, which parametrises the isotriplet
component of the proton
\begin{equation}
  \delta q_{\text{NS}}^{\pm} = \sum_{i\in u\text{-type}} (q_i \pm \bar{q}_i) -\sum_{i\in d-\text{type}} (q_i \pm \bar{q}_i)\,.
\end{equation}
We keep the gluon PDF, $g(x,Q^2)$, as is.

This requires us to decompose the quark coefficient functions in a similar
manner\footnote{This decomposition is completely analogous to what is typically
  done for splitting matrices.}
\begin{equation}
  C_{i,q} = C_{i,\text{NS}}^{+} + C_{i,\text{PS}}, \quad i=2,L
  \label{eq:Ciq}
\end{equation}
and define the valence coefficient functions
\begin{equation}
  C_{3,\text{NS}}^{v} = C_{3,\text{NS}}^{-} + C_{i,\text{NS}}^{s}.
\end{equation}
Here the superscript $s$ denotes the ``sea'' contribution to the valence
coefficient function. As it turns out, it is non-zero starting from third
order, and the pure-singlet piece of \cref{eq:Ciq} is non-zero starting from
second order~\cite{Bolzoni:2011cu}.

With these definitions the neutral current structure functions take the
form~\cite{Bolzoni:2011cu,Zaro:2013twa}
\begin{align}
  F_j^Z(x,Q^2) &= 2x\int_0^1 dz \int_0^1 dy\delta(x-yz)\sum_{i=1}^{n_f}\left[ v_i^2 + a_i^2\right] \notag \\
  &\quad \times \big[q_{\text{NS},i}^+ (y,\mu_F)C_{j,\text{NS}}^+(z,Q,\mu_R,\mu_F) \notag \\
    &\quad\qquad + q_{\text{PS}} (y,\mu_F)C_{j,\text{PS}}(z,Q,\mu_R,\mu_F)\notag \\
    &\quad\qquad + g(y,\mu_F)C_{j,g}(z,Q,\mu_R,\mu_F)\big], \quad j=2,L
  \label{eq:FiZ}
\end{align}
\begin{align}
  F_3^Z(x,Q^2) &= 2\int_0^1 dz \int_0^1 dy\delta(x-yz)\sum_{i=1}^{n_f} 2v_ia_i \notag \\
  &\quad \times \big[q_{\text{NS},i}^- (y,\mu_F)C_{3,\text{NS}}^-(z,Q,\mu_R,\mu_F) \notag \\
    &\quad\qquad + q_{\text{NS}}^v (y,\mu_F)C_{3,\text{NS}}^s(z,Q,\mu_R,\mu_F)\big].
  \label{eq:F3Z}
\end{align}
$v_i$ and $a_i$ are the vector and axial-vector couplings respectively. The
needed combinations are given by
\begin{equation}
  v_i^2 + a_i^2 =
  \begin{cases}
    \frac{1}{4}+\left(\frac{1}{2} -\frac{4}{3}\sin^2\theta_W\right)^2,& \text{if } i\in u\text{-type quark}\\
    \frac{1}{4}+\left(\frac{1}{2} -\frac{2}{3}\sin^2\theta_W\right)^2,& \text{if } i\in d\text{-type quark}    
\end{cases}
\end{equation}
and
\begin{equation}
  2v_ia_i =
  \begin{cases}
    \frac{1}{2} -\frac{4}{3}\sin^2\theta_W,& \text{if } i\in u\text{-type quark}\\
    \frac{1}{2} -\frac{2}{3}\sin^2\theta_W,& \text{if } i\in d\text{-type quark}\,.    
\end{cases}
\end{equation}
For the charged current case the structure functions are given by
\begin{align}
  F_j^{W^\pm}(x,Q^2) &= x\int_0^1 dz \int_0^1 dy\delta(x-yz)\frac{1}{n_f}\sum_{i=1}^{n_f}\left[ v_i^2 + a_i^2\right] \notag \\
  &\quad \times \big[\mp \delta q_{\text{NS}}^- (y,\mu_F)C_{j,\text{NS}}^-(z,Q,\mu_R,\mu_F) \notag \\
    &\quad\qquad + q_{\text{PS}} (y,\mu_F)C_{j,\text{q}}(z,Q,\mu_R,\mu_F)\notag \\
    &\quad\qquad + g(y,\mu_F)C_{j,g}(z,Q,\mu_R,\mu_F)\big], \quad j=2,L
  \label{eq:FiW}
\end{align}
\begin{align}
  F_3^{W^\pm}(x,Q^2) &= \int_0^1 dz \int_0^1 dy\delta(x-yz)\sum_{i=1}^{n_f} 2v_ia_i \notag \\
  &\quad \times \big[\mp \delta q_{\text{NS},i}^+ (y,\mu_F)C_{3,\text{NS}}^+(z,Q,\mu_R,\mu_F) \notag \\
    &\quad\qquad + q_{\text{NS}}^v (y,\mu_F)C_{3,\text{NS}}^v(z,Q,\mu_R,\mu_F)\big].
  \label{eq:F3W}
\end{align}
In this case the vector and axial-vector couplings are simply given by
\begin{equation}
  v_j=a_j =\frac{1}{\sqrt{2}}.
\end{equation}
These equations complete all the ingredients needed to evaluate the cross
section in \cref{eq:vbfh-dsigma}. As previously noted, all coefficient
functions are know to the precision of \NNNLO{}, whereas the PDFs themselves
have only been determined to \NNLO{}. 

\subsection{Scale Variation}
The coefficient functions appearing above are in the literature expressed in
terms of the vector boson momentum, $Q$. In general we are interested in
computing the cross section for a range of different factorisation and
renormalisation scales to asses the convergence of the perturbative series. In
order to compute the dependence of the cross section on the values of the
factorisation and renormalisation scales, we use renormalisation group
methods~\cite{Furmanski:1981cw,vanNeerven:2000uj,Buehler:2013fha} on the
structure functions
\begin{equation}
  \label{eq:conv-structf-suppl}
  F_i^V = \sum_a C_i^{V,a} \otimes f_a \,,\quad i=2,L,3\,.
\end{equation}
This requires us to compute the scale dependence to third order in the
coefficient functions as well as in the PDFs.

We start by evaluating the running coupling for $\as$ as an expansion in
$\as$. This is done by iteratively solving the renormalisation group equation
\begin{equation}
  \frac{d}{d\ln\mu^2}\as(\mu) = - \beta_0\as^2(\mu)  - \beta_1\as^3(\mu)  - \beta_2\as^4(\mu)  - \beta_3\as^5(\mu) + \mathcal{O}(\as^6)\,. 
\end{equation}
using
\begin{equation}
  \as(Q) = \as(\muR) - \int_0^{L_{RQ}} dL \frac{d}{dL}\as(\mu).
\end{equation}
Integrating yields
\begin{align}
  \label{eq:as-running}
  \as(Q) & = \as(\muR) + \as^2(\muR) \beta_0 L_{RQ} 
  + \as^3(\muR) (\beta_0^2 L_{RQ}^2 + \beta_1 L_{RQ}) \notag \\
  & + \as^4(\muR)\left(\beta_0^3 L_{RQ}^3 + \frac{5}{2}\beta_0 \beta_1 L_{RQ}^2 + \beta_2 L_{RQ} \right) + \mathcal{O}(\as^5(\muR))\,,
\end{align}
where we introduced the shorthand notation
\begin{equation}
  \label{eq:LRQ-notation}
  L_{RQ} = \ln\left(\frac{\muR^2}{Q^2}\right)\,,\quad
  L_{FQ} = \ln\left(\frac{\muF^2}{Q^2}\right)\,,\quad
  L = \ln\mu^2\,,
\end{equation}
as well as\footnote{Here defined in the $\overline{\mathrm{MS}}$
  scheme.}~\cite{Politzer:1973fx,Gross:1973id,Caswell:1974gg,Tarasov:1980au,Larin:1993tp}
\begin{align}
  4\pi\beta_0 &= 11 - \frac{2}{3} n_f \notag \\
  16\pi^2\beta_1 &= 102 - \frac{38}{3} n_f \notag \\
  64\pi^3\beta_2 &= \frac{2857}{2} - \frac{5033}{18}n_f + \frac{325}{54}n_f^2
  \label{eq:betacoeff}
\end{align}
where $n_f$ is the number of active flavours. We use the above to express the coefficient
functions as an expansion in $\as(\mu_R)$
\begin{align}
  \label{eq:coef-fct-expansion}
  & C_i = \sum_{k=0}  \left(\frac{\as(Q)}{2\pi}\right)^k C_i^{(k)} =  
  C_i^{(0)} + 
  \frac{\as(\muR)}{2\pi} C_i^{(1)} +\notag \\  
  & \left(\frac{\as(\muR)}{2\pi}\right)^2 \left(C_i^{(2)} + 2\pi \beta_0 C_i^{(1)} L_{RQ}\right) +\notag \\
  & \left(\frac{\as(\muR)}{2\pi}\right)^3 \bigg[
    C_i^{(3)} + 4\pi \beta_0 C_i^{(2)} L_{RQ}     
    + 4\pi^2 C_i^{(1)} L_{RQ} (\beta_1 + \beta_0^2 L_{RQ})
  \bigg] + \mathcal{O}(\as^4)\,.
\end{align}
To evaluate the dependence of the PDFs on the factorisation scale, $\muF$, we
integrate the
DGLAP~\cite{Lipatov:1974qm,Gribov:1972ri,Altarelli:1977zs,Dokshitzer:1977sg}
equation
\begin{equation}
  \frac{d}{d\ln\mu^2} f(x,\mu) = \frac{\as(\mu)}{2\pi} (P\otimes f)(x,\mu)\, .
\end{equation}
using
\begin{equation}
  \label{eq:pdf-integ}
  f(x,Q) = f(x,\muF) - \int_0^{L_{FQ}} dL \frac{d}{dL} f(x,\mu)\,. 
\end{equation}
Here both $P$ and $f$ are understood to be matrices expressed in terms of the
singlet and non-singlet parts as above. The splitting kernels, $P$, can be
expressed in terms of an expansion in $\as$
\begin{equation}
  \label{eq:split-fct-expansion}
  P(z,\as) = \sum_{i=0}{\left(\frac{\as}{2\pi}\right)^{i} P^{(i)}(z) }\,,
\end{equation}
where terms up to $P^{(2)}(z)$ are known~\cite{Vogt:2004mw,Moch:2004pa}.

It is then straightforward to express the PDF evaluated at $\muF$ in terms of an
expansion in $\as(\muR)$.
Evaluating, we obtain
\begin{align}
  \label{eq:pdf-expansion}
  f(x,Q) & = f(x, \muF) \Bigg( 1 - \frac{\as(\muR)}{2\pi} L_{FQ} P^{(0)} \notag \\
  & - \left(\frac{\as(\muR)}{2\pi}\right)^2 L_{FQ} \Big[
  P^{(1)} - \frac12 L_{FQ} (P^{(0)})^2
  -\pi\beta_0 P^{(0)} (L_{FQ} - 2 L_{RQ}) \Big]\notag  \\ 
  & - \left(\frac{\as(\muR)}{2\pi}\right)^3 L_{FQ} \Big[
  P^{(2)} - \frac12 L_{FQ} (P^{(0)} P^{(1)} + P^{(1)} P^{(0)}) \notag \\
  & \hspace{5cm} +  \pi \beta_0 (L_{FQ} - 2 L_{RQ}) (L_{FQ} (P^{(0)})^2 - 2 P^{(1)}) \notag  \\ 
  &\hspace{5cm}+ \frac16 L_{FQ}^2 (P^{(0)})^3 \notag \\
  &\hspace{5cm}+ 4\pi^2 \beta_0^2 P^{(0)} (L_{RQ}^2 - L_{FQ} L_{RQ} + \frac13
  L_{FQ}^2) \notag  \\ 
  &\hspace{5cm}- 2 \pi^2 \beta_1 P^{(0)} (L_{FQ} - 2 L_{RQ}) \Big] +\mathcal{O}(\as^4)\Bigg)\,.
\end{align}
where all the products are understood to be Mellin
transforms
\begin{equation}
  (f\otimes g)(x)=
  \int_x^1\frac{dy}{y}f(y)g(\frac{x}{y})\,.
\end{equation}

\cref{eq:coef-fct-expansion,eq:pdf-expansion}
allow us to evaluate the convolution in
\cref{eq:conv-structf-suppl} up to \NNNLO{} in perturbative QCD
for any choice of the renormalisation and factorisation scales.

\section{Impact of Higher Order PDFs}
There is one source of formally \NNNLO{} QCD corrections appearing in
\cref{eq:conv-structf} which is currently unknown, namely
missing higher order terms in the determination of the PDF.
Indeed, in order to truly claim \NNNLO{} accuracy of the cross section we must
use \NNNLO{} parton densities. However, only \NNLO{} PDF sets are available at
this time. 
These will be missing contributions from two main sources: from the
higher order corrections to the coefficient functions that relate
physical observables to PDFs; and from the higher order splitting
functions in the evolution of the PDFs.

To evaluate the impact of future \NNNLO{} PDF sets on the total cross
section, we consider two different approaches.
A first, more conservative estimate, is to derive the uncertainty
related to higher order PDF sets from the difference at lower orders,
as described in~\cite{Anastasiou:2016cez} (see
also~\cite{Forte:2013mda}).
We compute the \NNLO{} cross section using both the \NLO{} and the \NNLO{} PDF
set, and use their difference to extract the \NNNLO{} PDF uncertainty.
We find in this way that at $13 \TeV$ the uncertainty from missing higher
orders in the extractions of PDFs is
\begin{equation}
  \label{eq:pdf-uncert-schemeA}
  \delta_A^{\text{PDF}} = \frac12
  \left|\frac{\sigma^\text{NNLO}_\text{NNLO-PDF} - \sigma^\text{NNLO}_\text{NLO-PDF}}{\sigma^\text{NNLO}_\text{NNLO-PDF}}\right|
  = 1.1 \%\,.
\end{equation}
Because the convergence is greatly improved going from NNLO to \NNNLO{} compared
to one order lower, one might expect this to be rather conservative
even with the factor half in \cref{eq:pdf-uncert-schemeA}.
Therefore, we also provide an alternative estimate of the impact of
higher orders PDFs, using the known \NNNLO{} $F_2$ structure function.

We start by rescaling all the parton distributions using the $F_2$
structure function evaluated at a low scale $Q_0$
\begin{equation}
  \label{eq:n3lo-pdf-approx}
  f^{\NNNLO, \text{approx.}}(x, Q) = 
  f^{\NNLO}(x,Q) \frac{F_2^{\NNLO}(x, Q_0)}{F_2^{\NNNLO}(x, Q_0)}\,.
\end{equation}
In practice, we will use the $Z$ structure function.
We then re-evaluate the structure functions in
\cref{eq:conv-structf} using the approximate higher order
PDF given by \cref{eq:n3lo-pdf-approx}.
This yields
\begin{equation}
  \label{eq:pdf-uncert-schemeB}
  \delta_B^{\text{PDF}}(Q_0) = \left|\frac{\sigma^\NNNLO - \sigma^\NNNLO_\text{rescaled}(Q_0)}{\sigma^\NNNLO}\right|
  = 7.9 \text{\permil}\,,
\end{equation}
where in the last step, we used $Q_0=8\GeV$ and considered $13\TeV$
proton collisions.

By calculating a rescaled NLO PDF and evaluating the NNLO cross
section in this way, we can evaluate the ability of this method to
predict the corrections from NNLO PDFs.
We find that with $Q_0=8\GeV$, the uncertainty estimate obtained in
this way captures relatively well the impact of NNLO PDF sets.

The rescaled PDF sets obtained using
\cref{eq:n3lo-pdf-approx} will be missing \NNNLO{}
corrections from the evolution of the PDFs in energy.
We have checked the impact of these terms by varying the
renormalisation scale up and down by a factor two around the
factorisation scale in the splitting functions used for the PDF
evolution.
We find that the theoretical uncertainty associated with missing
higher order splitting functions is less than one permille of the
total cross section.
Comparing this with \cref{eq:pdf-uncert-schemeB}, it is
clear that these effects are numerically subleading, suggesting that a
practical alternative to full \NNNLO{} PDF sets could be obtained by
carrying out a fit of DIS data using the hard \NNNLO{} matrix element.

\begin{figure}
  \centering
  \includegraphics[width=0.98\linewidth]{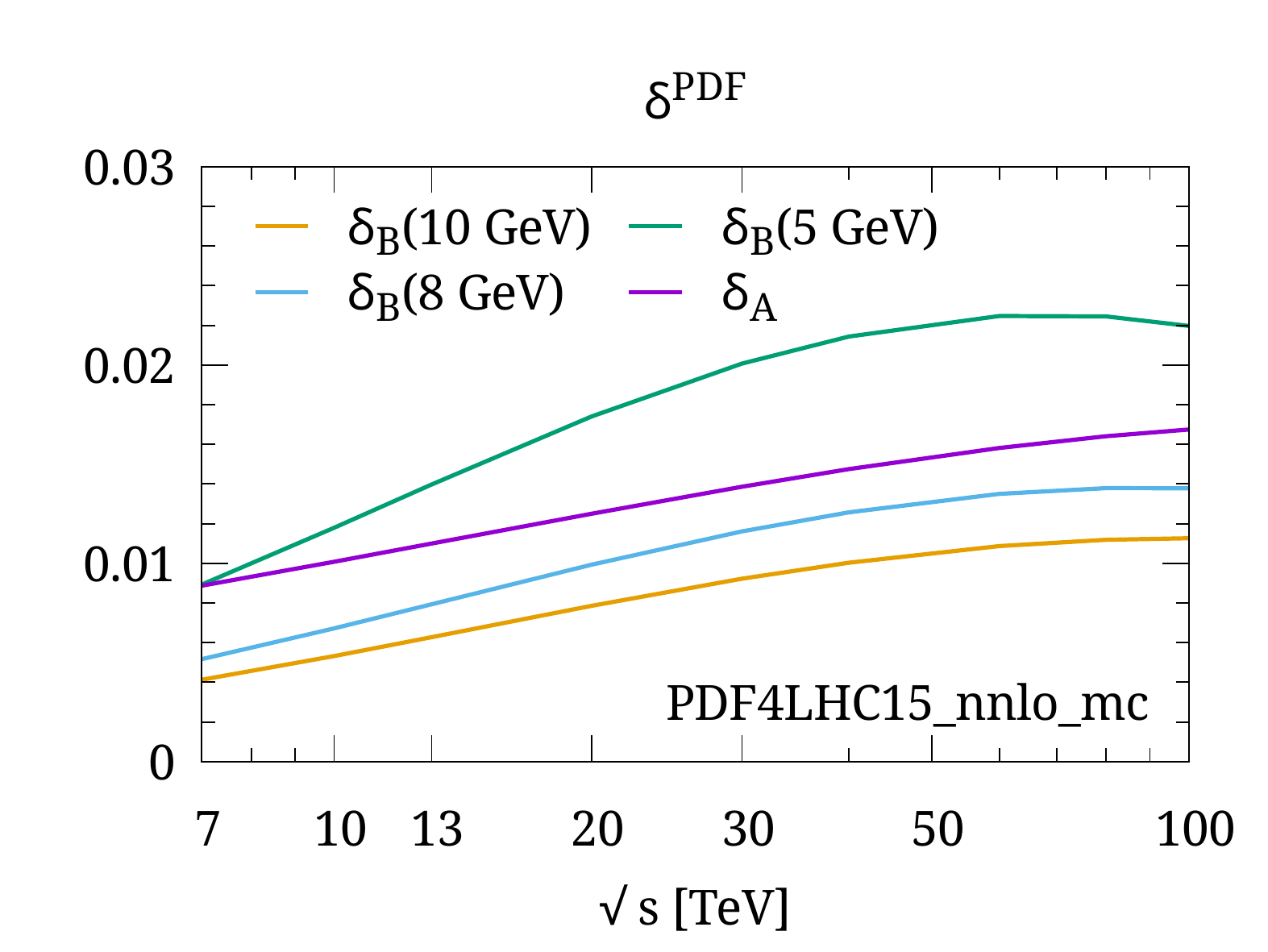}
  \caption{Estimate of the impact of missing higher orders corrections in PDFs,
    using \cref{eq:pdf-uncert-schemeA,eq:pdf-uncert-schemeB} with $Q_0=5$, $8$
    and $10\GeV$.
    \label{fig:n3lo-pdf}
  }
\end{figure}

The uncertainty estimates obtained with the two different methods described by
\cref{eq:pdf-uncert-schemeA,eq:pdf-uncert-schemeB} are shown in
\cref{fig:n3lo-pdf} as a function of centre-of-mass energy, and for a range of
$Q_0$ values.

One should note that the uncertainty estimates given in
\cref{eq:pdf-uncert-schemeA,eq:pdf-uncert-schemeB} do not include what is
usually referred to as PDF uncertainties.
While we are here calculating missing higher order uncertainties to
NNLO PDF sets, typical PDF uncertainties correspond to uncertainties
due to errors on the experimental data and limitations of the fitting
procedure.
These can be evaluated for example with the PDF4LHC15
prescription~\cite{Butterworth:2015oua}, and are of about $2\%$ at
$13\TeV$, which is larger than the corrections discussed above.
One can also combine them with $\as$ uncertainties, which for VBF are at the
$5\permil$ level. More detailed results on PDF and $\as$ uncertainties are given
in \cref{ch:vbfnnlo}.

\section{Phenomenological Results}\label{sec:vbfhn3lo-pheno}
Let us now discuss in detail the phenomenological consequences of the \NNNLO{}
corrections to VBF Higgs production.
We present results for a wide range of energies in proton-proton
collisions.
The central factorisation and renormalisation scales appearing in the structure
functions are set to the squared momentum of the corresponding vector boson.
To estimate missing higher-order uncertainties, we use a seven-point
scale variation, varying the scales by a factor two up and down
while keeping $0.5 < \mu_R/ \mu_F < 2$
\begin{equation}
  \label{eq:scale-choice}
  \mu_{R,i} = \xi_{\mu_{R}} Q_i\,,\quad
  \mu_{F,i} = \xi_{\mu_{F}} Q_i\,,
\end{equation}
where $\xi_{\mu_R},\xi_{\mu_F}\in\big\{\tfrac12,1,2\big\}$ and $i=1$, $2$
corresponds to the upper and lower hadronic sectors.

Our implementation uses the phase space from \POWHEG{}'s two-jet VBF Higgs
calculation~\cite{Nason:2009ai}.
The matrix element is derived from structure functions obtained with the
parametrised DIS coefficient
functions~\cite{SanchezGuillen:1990iq,vanNeerven:1991nn,Zijlstra:1992qd,Zijlstra:1992kj,vanNeerven:1999ca,vanNeerven:2000uj,Moch:2004xu,Vermaseren:2005qc,Vogt:2006bt,Moch:2007rq},
evaluated using \texttt{HOPPET} v1.2.0-devel~\cite{Salam:2008qg}.
We have tested our \NNLO{} implementation against the results of one of the
codes used in \Brefs{Bolzoni:2010xr,Bolzoni:2011cu} and found agreement,
both for the structure functions and the final cross sections.
We have also checked that switching to the exact DIS coefficient functions has a
negligible impact on both structure functions and total cross sections.
A further successful comparison of the evaluation of \NNLO{} structure functions was
made against \texttt{APFEL}~v.2.4.1\cite{Bertone:2013vaa}.

For our computational setup, we use a diagonal CKM matrix with five light
flavours ignoring top-quarks in the internal lines and final states. Full
Breit-Wigner propagators for the $W$, $Z$ and the narrow-width approximation for
the Higgs boson are applied.
We use the PDF4LHC15\_nnlo\_mc
PDF~\cite{Butterworth:2015oua,Dulat:2015mca,Harland-Lang:2014zoa,Ball:2014uwa}
and four-loop evolution of the strong coupling~\cite{vanRitbergen:1997va}, taking as our initial condition
$\as(M_Z) = 0.118$.
We set the Higgs mass to $M_H = 125.09\GeV$, in accordance with the
experimentally measured value~\cite{Aad:2015zhl}.  Electroweak
parameters are obtained from their PDG~\cite{Agashe:2014kda} values
and tree-level electroweak relations. As inputs we use
$M_W = 80.385\GeV$, $M_Z = 91.1876\GeV$ and
$G_F = 1.16637\times 10^{-5} \GeV^{-2}$. For the widths of the vector
bosons we use $\Gamma_W = 2.085 \GeV $ and $\Gamma_Z = 2.4952 \GeV$.

\begin{figure}
  \centering
  \includegraphics[page=1,width=0.98\linewidth]{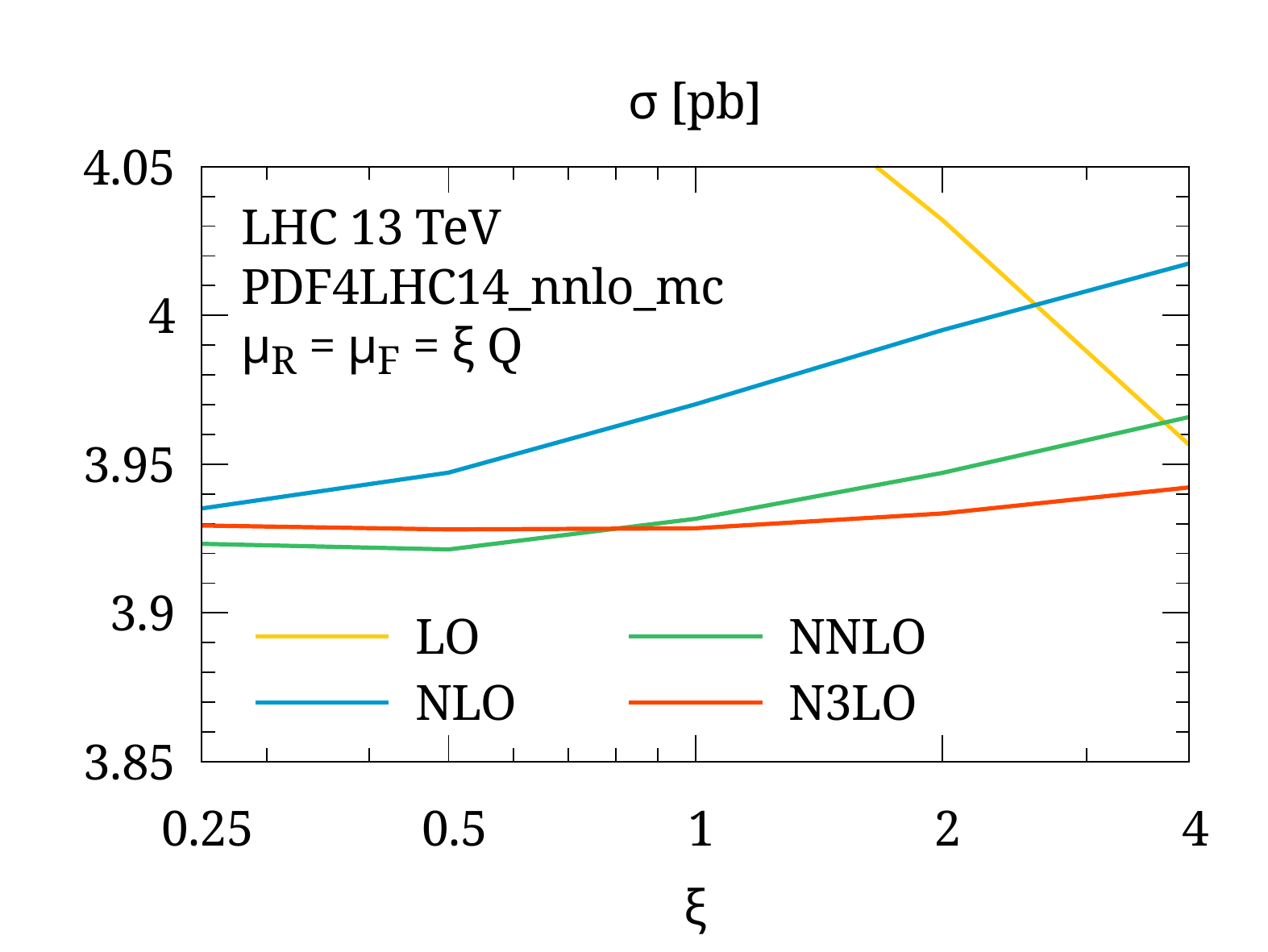}%
  \caption{Dependence of the cross section on the renormalisation 
    and factorisation scales for each order in perturbation theory.
    \label{fig:scale-var} }
\end{figure}

To study the convergence of the perturbative series, we show in
\cref{fig:scale-var} the inclusive cross section obtained at $13
\TeV$ with $\mu_R=\mu_F=\xi Q$ for $\xi\in [1/4,4]$.
Here we observe that at \NNNLO{} the scale dependence becomes
extremely flat over the full range of renormalisation and
factorisation scales.
We note that similarly to the results obtained in the ggH
channel~\cite{Anastasiou:2015ema}, the convergence improves
significantly at \NNNLO{}, with the \NNNLO{} prediction being well
inside of the NNLO uncertainty band, while at lower orders there is a
pattern of limited overlap of theoretical uncertainties.
\begin{figure}[!t]
  \centering
  \includegraphics[page=1,width=0.35\textwidth]{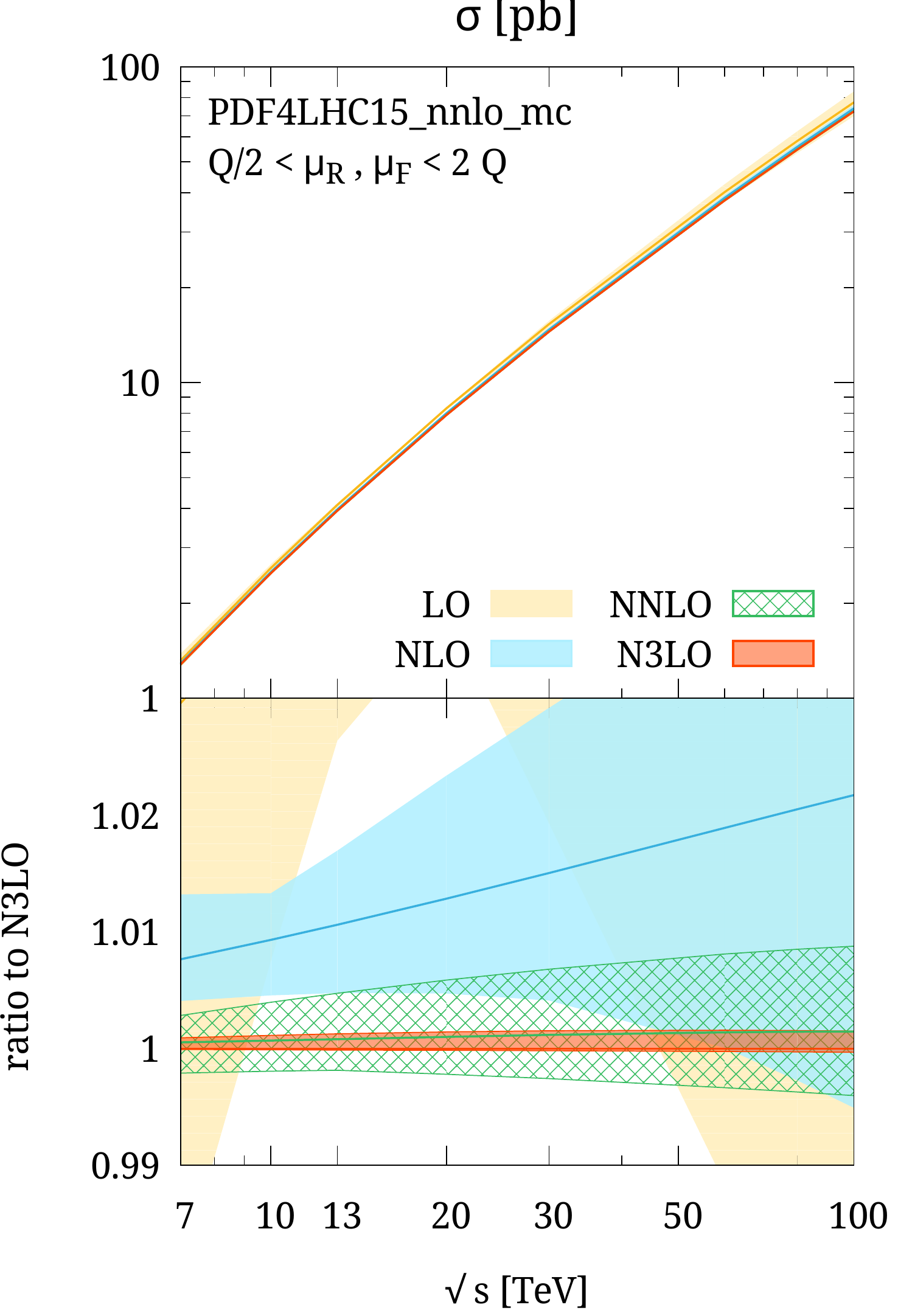}%
  \hspace{-5mm}%
  \hfill\includegraphics[page=1,width=0.35\textwidth]{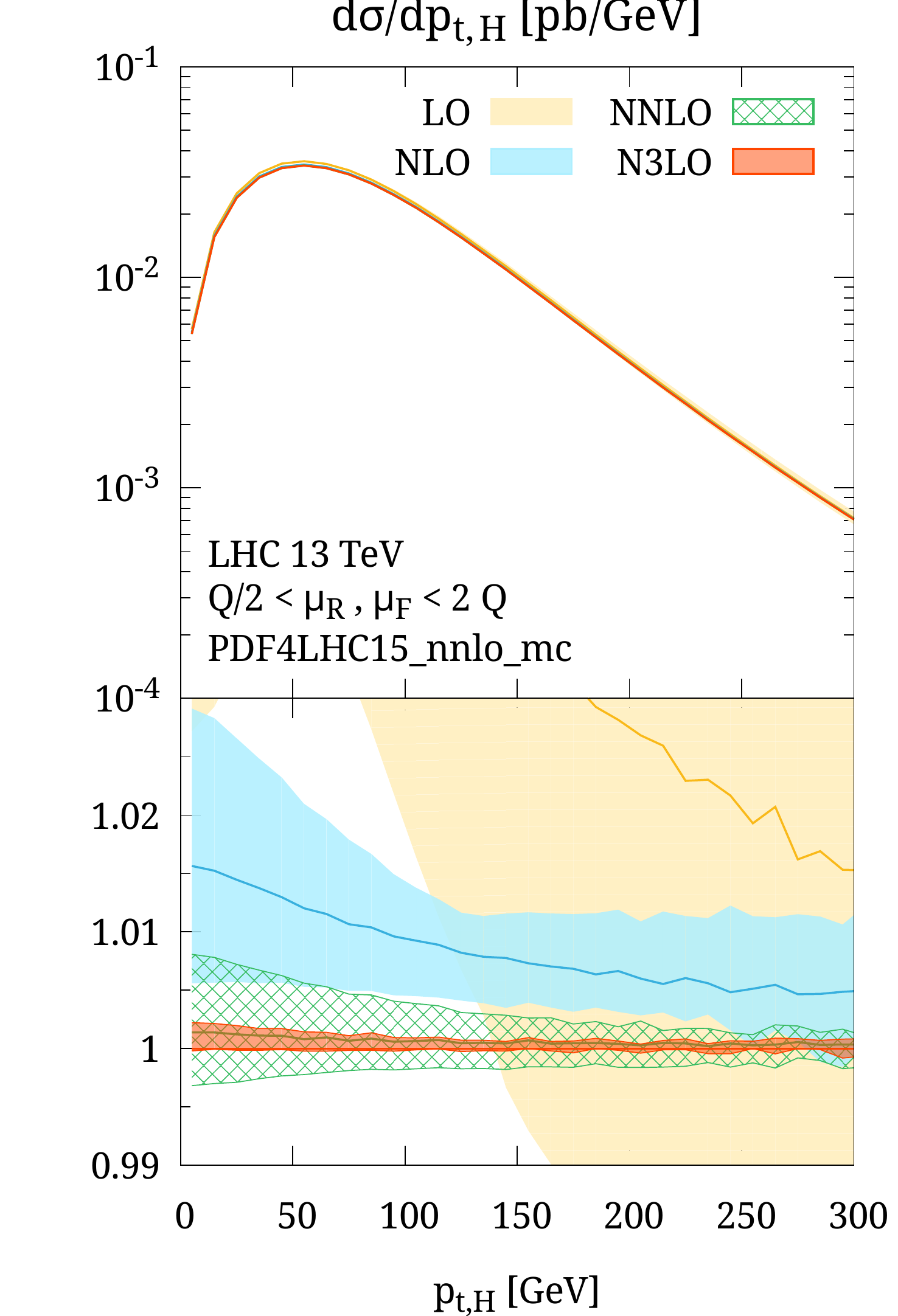}%
  \hspace{-5mm}%
  \hfill\includegraphics[page=2,width=0.35\textwidth]{VBFHN3LO/n3lo-hists}
  \caption{Cross section as a function of centre-of-mass energy
    (left), Higgs transverse momentum (centre) and Higgs
    rapidity (right).
    \label{fig:distributions} }
\end{figure}

In \cref{fig:distributions} (left), we give the cross section as
a function of centre-of-mass energy.
We see that at \NNNLO{} the convergence of the perturbative series
is very stable, with corrections of about $1\permil$ on the NNLO
result.
The scale uncertainty is dramatically reduced, going from
$7\permil$ at NNLO to $1.4\permil$ at \NNNLO{} at $13 \TeV$.
A detailed breakdown of the cross section and scale uncertainty
obtained at each order in QCD is given in
\cref{tab:cross-sections-n3lo} for $\sqrt{s}=13$, $14$ and $100\TeV$.

The centre and right plots of \cref{fig:distributions} show the Higgs transverse
momentum and rapidity distributions at each order in QCD, where we observe again
a large reduction of the theoretical uncertainty at \NNNLO{}. The \NNNLO{}
corrections are flat everywhere in phase space, except for at very high values
of the Higgs rapidity.

%----------------------------------------------------------------------
\begin{table}[t]
  \centering
  \phantom{x}\medskip
  \begin{tabular}{lcccccc}
    \toprule
    &&  $\sigma^{(13 \TeV)}$  [pb]  && $\sigma^{(14 \TeV)}$ [pb]  && $\sigma^{(100 \TeV)}$ [pb] \\
    \midrule
    LO       &&  $4.099\,^{+0.051}_{-0.067}$    &&  $4.647\,^{+0.037}_{-0.058}$   
             &&  $77.17\,^{+6.45}_{-7.29}$ \\[4pt]
    NLO      &&  $3.970\,^{+0.025}_{-0.023}$    &&  $4.497\,^{+0.032}_{-0.027}$   
             &&  $73.90\,^{+1.73}_{-1.94}$ \\[4pt]
    NNLO     &&  $3.932\,^{+0.015}_{-0.010}$    &&  $4.452\,^{+0.018}_{-0.012}$
             &&  $72.44\,^{+0.53}_{-0.40}$ \\[4pt]
    \NNNLO{} &&  $3.928\,^{+0.005}_{-0.001}$    &&  $4.448\,^{+0.006}_{-0.001}$ 
             &&  $72.34\,^{+0.11}_{-0.02}$ \\
    \bottomrule
  \end{tabular}
  \caption{Inclusive cross sections at LO, NLO, NNLO and \NNNLO{} for VBF Higgs production.
    The quoted uncertainties correspond to scale variations $Q/2 < \muR, \muF < 2 Q$,
    while statistical uncertainties are at the level of $0.2\permil$.
    \label{tab:cross-sections-n3lo}}
\end{table}

A comment is due on non-factorisable QCD corrections.
Indeed, for the results presented in this chapter, we have considered
VBF in the usual DIS picture, ignoring diagrams that are not of the
type shown in \cref{fig:vbfh}.
These effects neglected by the structure function approximation are
known to contribute less than $1\%$ to the total cross section at NNLO
\cite{Bolzoni:2011cu}.
The effects and their relative corrections are as follows:
\begin{itemize}
\item gluon exchanges between the upper and lower ha\-dro\-nic
  sectors, which appear at NNLO, but are kinematically and colour
  suppressed; 
  These contributions along with the heavy-quark loop induced
  contributions have been estimated to contribute at the permille
  level~\cite{Bolzoni:2011cu};
  
\item t-/u-channel interference which are known to contribute
  $\mathcal{O}(5\permil)$ at the fully inclusive level and
  $\mathcal{O}(0.5\permil)$ after VBF cuts have been applied
  \cite{Ciccolini:2007ec};
  
\item contributions from s-channel production, which have been calculated up to
  NLO~\cite{Ciccolini:2007ec}. At the inclusive level these contributions are
  sizeable but they are reduced to $\mathcal{O}(5\permil)$ after VBF cuts. The
  s-channel production is of course just associated Higgs production where the
  massive vector boson decays to a quark pair and hence it is usually considered
  a background process rather than an actual contribution to VBF;

\item single-quark line contributions, which contribute to the VBF
  cross section at NNLO. 
  At the fully inclusive level these amount to corrections of
  $\mathcal{O}(1\%)$ but are reduced to the permille level after VBF
  cuts have been applied~\cite{Harlander:2008xn};
  
\item loop induced interference between VBF and ggH
  production. 
  These contributions have been shown to be much below the permille
  level \cite{Andersen:2007mp}.
 
\end{itemize}

Furthermore, for phenomenological applications, one also needs to
consider NLO electroweak effects~\cite{Ciccolini:2007ec}, which amount
to $\mathcal{O}(5\%)$ of the total cross section.
In \cref{ch:vbfnnlo} we will study the impact of these electroweak
corrections in some detail, and also investigate how big of an impact PDF
uncertainties have on the total cross section.

\section{Conclusions}
In this chapter, we have presented the first \NNNLO{} calculation of a
$2\to 3$ hadron-collider process, made possible by the DIS-like
factorisation of the VBF process.
This brings the precision of VBF Higgs production to the same formal accuracy as
was recently achieved in the ggH channel in the heavy top mass
approximation~\cite{Anastasiou:2015ema}.
The \NNNLO{} corrections were found to be tiny, $1-2\permil$, and well
within previous theoretical uncertainties, but they provide a large
reduction of scale uncertainties, by a factor 5.
Thus, although the corrections are sub-leading to many known effects omitted in
the structure function approach, the calculation shows incredibly good
convergence of perturbative QCD.

We also studied the impact of missing higher order corrections to the PDFs. We
estimate that these corrections are at the $1\%$-level or below and that they
are dominated by the hard process coefficient functions rather than unknown
contributions to the splitting functions. Hence one could conceivably obtain
approximate \NNNLO{} PDFs from DIS data and the known \NNNLO{} DIS coefficient
functions.
Our calculation also provides the first element towards a
differential \NNNLO{} calculation for VBF Higgs production, which
could be achieved through the projection-to-Born
method (see \cref{ch:vbfnnlo}) using an \NNLO{} DIS 2+1 jet
calculation~\cite{Gehrmann:2009vu,Currie:2016ytq}.

%Remember to include results on pdf/errors etc which were never published.
\chapter{Fully differential NNLO Vector~Boson~Fusion Higgs~Production}\label{ch:vbfnnlo}
In the previous chapter we saw how the total cross section for VBF Higgs
production could be computed to \NNNLO{} using the structure function
approach. The calculation has the obvious disadvantage of not being differential
in the jet kinematics. The reason that the structure function approach does not
provide a fully differential cross section, is related to the fact that the DIS
coefficient functions used in the calculation implicitly integrate over hadronic
final states. Whereas the Higgs boson momentum can be reconstructed from the
knowledge of the momenta of the vector bosons emitted from the protons, only the
momenta of the outgoing proton remnants are known and not those of the
individual partons. In general it is therefore not possible to reconstruct the
full final state momenta\footnote{The structure function approach does reproduce
  the correct final state momenta at \LO{} where there can never be more than
  two jets in an event. In jet clustering algorithms with large clustering radii
  it will also be the case that the structure function approach often reproduces
  the correct final state momenta beyond \LO{} when there are only two jets
  present.}.

Given the smallness of the inclusive \NNLO{} and \NNNLO{} corrections we may ask
whether or not it is even relevant to study the differential corrections. In
addition to that, the differential \NLO{} corrections and their associated scale
uncertainties have been known for a long time to be
small~\cite{Figy:2003nv}. However, because of the use of transverse-momentum
cuts on the forward tagging jets, one might imagine that there are important
\NNLO{} corrections, associated with those jet cuts, that would not be seen in a
fully inclusive calculation. As we shall see later, that is indeed the case.

In this chapter we eliminate the limitation of the structure function approach
and present a fully differential \NNLO{} calculation for VBF Higgs
production. In order to do so, we will introduce the ``projection-to-Born''
method. An advantage of this approach is that it can be extended to any
perturbative order and that it therefore opens up for a fully differential
\NNNLO{} calculation as well. We proceed to present results relevant for the LHC
and discuss the inclusion of electroweak corrections. At the end of the chapter,
we discuss the prospects of studying VBF production at a $100\TeV$ proton-proton
collider.

\section{The ``Projection-to-Born'' Method}
Let us start by recalling that the cross section in the structure function
approach is expressed as a sum of terms involving products of structure
functions, e.g.  $F_2(x_1, Q_1^2) F_2(x_2, Q_2^2)$, where $Q_i^2 = -q_i^2 > 0$
is given in terms of the 4-momentum $q_i$ of the (outgoing) exchanged vector
boson $i$ (cf. \cref{eq:vbfh-dsigma,eq:hadrcontract}).
The $x_i$ values are fixed by the relation
\begin{equation}
  x_i = - \frac{Q_i^2}{(2P_i\cdot q_i)}\,,
\end{equation}
where $P_i$ is the momentum of proton $i$.
To obtain the total cross section, one integrates over all $q_1$,
$q_2$ that can lead to the production of a Higgs boson.
If the underlying upper (lower) scattering is Born-like, $\mathrm{quark}
\to \mathrm{quark} + V$, then it is straightforward to show that
knowledge of the vector boson momentum $q_1$ ($q_2$) uniquely
determines the momenta of both the incoming and outgoing (on-shell)
quarks,
\begin{equation}
  \label{eq:kinematics}
  p_{\mathrm{in},i} = x_i P_i,\qquad p_{\mathrm{out},i} = x_i P_i - q_i\,.
\end{equation}

We exploit this feature in order to assemble a full calculation from
two separate ingredients.
For the first one, the ``inclusive'' ingredient, we remain within the
structure function approach, and for each set of $q_1$ and $q_2$ use
\cref{eq:kinematics} to assign VBF Born-like kinematics to the
upper and lower sectors.
\begin{figure}[t]
  \centering
  \includegraphics[width=\textwidth]{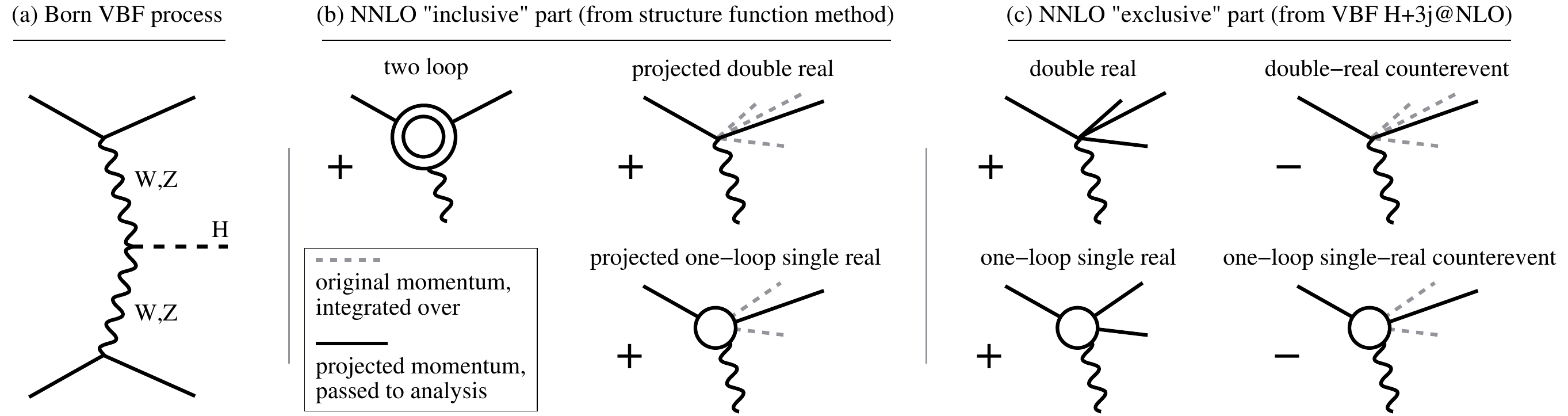}
  \caption{(a) Illustration of the Born VBF process. 
    (b) NNLO corrections to the upper sector of the VBF process, from the
    ``inclusive'' part of our calculation.
    (c) Corresponding ``exclusive'' part.
    The double-real and one-loop single-real counterevents in the
    exclusive part cancel the projected double-real and one-loop
    single-real contributions in the inclusive part.
    In the ``projected'' and ``counterevent'' contributions, the dashed
    lines corresponds to the full set of parton momenta that are
    integrated over (for the structure functions, this integral is
    implicit in the derivation of the coefficient functions), while the
    solid lines correspond to the partons that are left over after
    projection to Born-like kinematics and then passed to the analysis. 
    The projection does not change the direction of initial partons
    and so the corresponding incoming dashed lines are implicit.
  }
  \label{fig:ingredients}
\end{figure}

This is represented in \cref{fig:ingredients}b (showing just the upper
sector): for the two-loop contribution, the Born kinematics that we assign
corresponds to that of the actual diagrams;
for the tree-level double-real and one-loop single-real diagrams, it
corresponds to a projection from the true kinematics ($2\to H+n$ for
$n=3,4$) down to the Born kinematics ($2\to H+2$).
The projected momenta are used to obtain the ``inclusive'' contribution to
differential cross sections. It is important to understand that in the structure
function approach we are forced to construct the projected momenta rather than the
full real and double-real momenta. As previously mentioned, this is due to the
fact that the DIS coefficient functions are integrated over hadronic final state
momenta.

\begin{figure}[t]
  \centering
  \includegraphics[width=\textwidth]{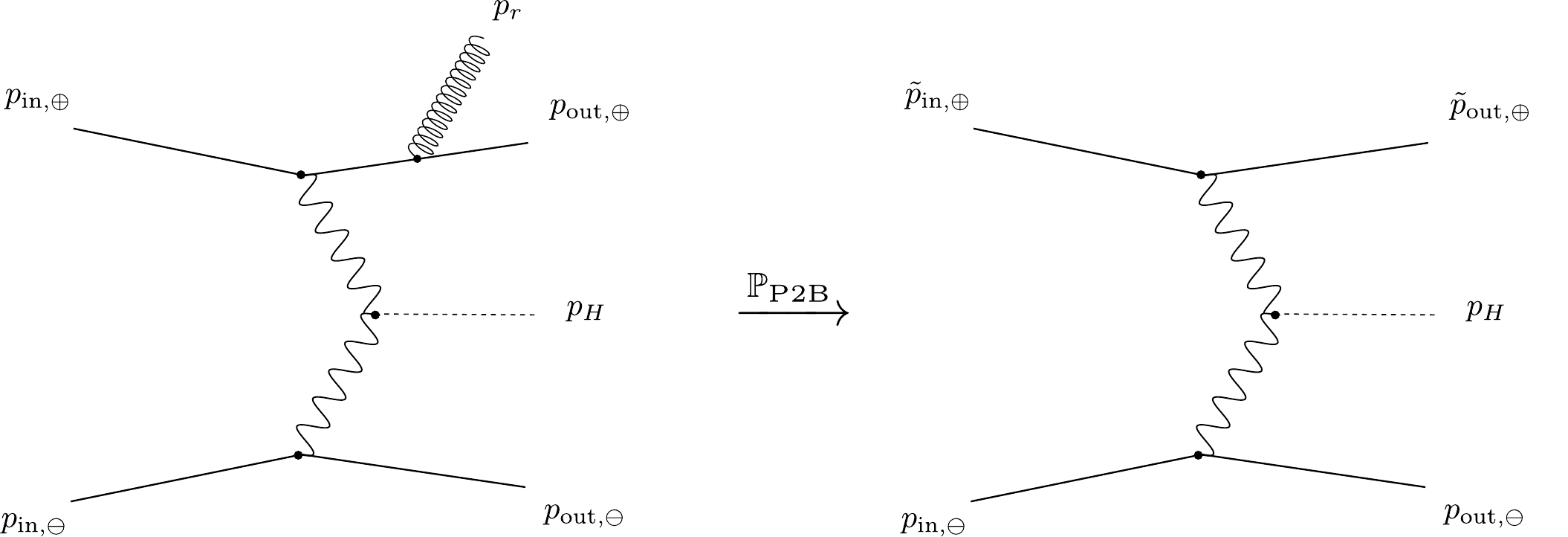}
  \caption{Illustration of the projection applied in the ``projection-to-Born''
    method when there is only one radiated parton on the upper line. All the
    momenta before the projection are assumed known. The momenta of the Higgs,
    $p_H$ and the momenta on the lower line, $p_{\mathrm{in},\ominus}$ and
    $p_{\mathrm{out},\ominus}$, are all left invariant under the projection. The
    projected momenta, $\tilde{p}_{\mathrm{in},\oplus}$ and $\tilde{p}_{\mathrm{out},\oplus}$,
    are given in \cref{eq:proj1,eq:proj2}. }
  \label{fig:projection}
\end{figure}

Here we aim to replace the projected real and double-real contributions with
their non-projected ones. We do so by adding a second, ``exclusive'', ingredient
to the ``inclusive'' one obtained from the structure function approach. This
ingredient will contain the full real and double-real contributions plus a set
of counterevents with projected kinematics. Let us first describe how to perform
the projection starting from the full kinematics. For simplicity we will assume
an event with only one real emission, see \cref{fig:projection}. Such an
event will have six external momenta, described by the vector
\begin{equation}
  \Omega_R = \left(p_1,p_2,p_3,p_4,p_5,p_6\right) = \left(p_{\mathrm{in},\oplus},p_{\mathrm{in},\ominus},p_H,p_{\mathrm{out},\oplus},p_{\mathrm{out},\ominus},p_r\right). 
\end{equation}
Here $\oplus$ ($\ominus$) refers to the upper (lower) VBF line and
$p_{\mathrm{r}}$ is the radiated gluon. We define the projected momenta by
\begin{equation}
  \field{P}_{\mathrm{P2B}} \Omega_R = \Omega_{\mathrm{P2B}} = \left(\tilde{p}_{\mathrm{in},\oplus},\tilde{p}_{\mathrm{in},\ominus},p_H,\tilde{p}_{\mathrm{out},\oplus},\tilde{p}_{\mathrm{out},\ominus}\right). 
\end{equation}
Let us now assume that the radiated particle is attached to the upper line of
the VBF diagram. In order to ease notation we therefore drop the
$\oplus$-subscript in the following. We then proceed to express the five momenta
$p_{\mathrm{in}},p_{\mathrm{out}},p_r,\tilde{p}_{\mathrm{in}},\tilde{p}_{\mathrm{out}}$
in lightcone coordinates
\begin{equation}
  p = (p^x,p^y,p^-,p^+)\,,
\end{equation}
where
\begin{equation}
  p^{\pm} = \frac{1}{\sqrt{2}} (p^E \pm p^z)\,,
\end{equation}
and hence
\begin{align}
  p_{\mathrm{in}} &= (0,0,0,p_{\mathrm{in}}^{+}) \\
  p_{\mathrm{out}} &= (p_{\mathrm{out}}^{x},p_{\mathrm{out}}^{y},p_{\mathrm{out}}^{-},p_{\mathrm{out}}^{+}) \\
  p_{r} &= (p_{r}^{x},p_{r}^{y},p_{r}^{-},p_{r}^{+}) \\
  \tilde{p}_{\mathrm{in}} &= (0,0,0,\tilde{p}_{\mathrm{in}}^{+}) \\
  \tilde{p}_{\mathrm{out}} &= (\tilde{p}_{\mathrm{out}}^{x},\tilde{p}_{\mathrm{out}}^{y},\tilde{p}_{\mathrm{out}}^{-},\tilde{p}_{\mathrm{out}}^{+})\,.
\end{align}
By momentum conservation we have
\begin{align}
  p_{\mathrm{in}} - p_{\mathrm{out}} - p_r &= \tilde{p}_{\mathrm{in}} - \tilde{p}_{\mathrm{out}} 
  \quad \Rightarrow \quad
  \begin{cases}
    \tilde{p}_{\mathrm{out}}^{x} = p_{\mathrm{out}}^{x} + p_{r}^{x} \\
    \tilde{p}_{\mathrm{out}}^{y} = p_{\mathrm{out}}^{y} + p_{r}^{y} \\
    \tilde{p}_{\mathrm{out}}^{-} = p_{\mathrm{out}}^{-} + p_{r}^{-} \\
    \tilde{p}_{\mathrm{in}}^{+} = p_{\mathrm{in}}^{+} - p_{\mathrm{out}}^{+} - p_{r}^{+} + \tilde{p}_{\mathrm{out}}^{+}\,.
  \end{cases}
  \label{eq:proj1}
\end{align}
In order to find $\tilde{p}_{\mathrm{out}}^{+}$ we impose that the projected
outgoing parton be massless
\begin{align}
  \left(\tilde{p}_{\mathrm{out}}\right)^2 & = 0 \notag \\
  & \Rightarrow  \quad \left(\tilde{p}_{\mathrm{out}}^{x}\right)^2 + \left(\tilde{p}_{\mathrm{out}}^{y}\right)^2 - 2 \tilde{p}_{\mathrm{out}}^{-} \tilde{p}_{\mathrm{out}}^{+} = 0 \notag \\
& \Rightarrow  \quad  \tilde{p}_{\mathrm{out}}^{+} = \frac{(p_{\mathrm{out}}^{x} + p_{r}^{x})^2 + (p_{\mathrm{out}}^{y} + p_{r}^{y})^2}{2(p_{\mathrm{out}}^{-} + p_{r}^{-})}\,,
  \label{eq:proj2}
\end{align}
which fixes all the momenta. In order to find the projection when the
radiated parton is on the lower line, we simply make the substitution $+
\leftrightarrow -$ everywhere. At \NNLO{} we will of course also have events
with two radiated partons. In this case they can either both be attached to the
same line or one on each line. In the former case we simply apply the projection
above with $p_r$ as the sum of the the two radiated parton momenta. In the
latter case we apply one projection to the upper line and one to the lower. Note
that the Higgs momentum is unaffected by the projection under all circumstances.

The ``exclusive'' ingredient starts from the NLO fully
differential calculation of vector boson fusion Higgs production with
three jets~\cite{Figy:2007kv,Jager:2014vna}, as obtained in a
factorised approximation, i.e. where there is no cross-talk between
upper and lower sectors.\footnote{The NLO calculation without this
  approximation is given in \Bref{Campanario:2013fsa}.}
Thus each parton can be uniquely assigned to one of the upper or lower
sectors and the two vector boson momenta can be unambiguously
determined.
For each event in a Monte Carlo integration over phase space, with weight $w$,
we add a counterevent, with weight $-w$, to which we assign projected Born VBF
kinematics as given in \cref{eq:proj1,eq:proj2} and illustrated
in \cref{fig:ingredients,fig:projection}.
From the original events, we thus obtain the full momentum structure
for tree-level double-real and one-loop single-real contributions.
Meanwhile, after integration over phase space, the counterevents
exactly cancel the projected tree-level double-real and one-loop
single-real contributions from the inclusive part of the calculation.
Thus the sum of the ``inclusive'' and ``exclusive'' parts gives the complete
differential NNLO VBF result.%

\section{Technical Implementation}
For the implementation of the ``inclusive'' part of the calculation we use the
implementation already described in \cref{ch:incVBF}.
as a starting point for the ``exclusive'' part of the calculation, we took the
NLO (i.e.\ fixed-order, but not parton-shower) part of the \POWHEG $H$+3-jet VBF
code~\cite{Jager:2014vna}, itself based on the calculation of
\Bref{Figy:2007kv}, with tree-level matrix elements from
MadGraph~4~\cite{Alwall:2007st}.
This code already uses a factorised approximation for the matrix
element, however for a given phase-space point it sums over
matrix-element weights for the assignments of partons to upper and
lower sectors.
We therefore re-engineered the code so that for each set of 4-momenta,
weights are decomposed into the contributions for each of the four
different possible sets of assignments of partons to the two sectors.
For every element of this decomposition it is then possible to
unambiguously obtain the vector boson momenta and so correctly
generate a counterevent.
The \POWHEGBOX's~\cite{Nason:2004rx,Alioli:2010xd} ``tagging''
facility was particularly useful in this respect, notably for the NLO
subtraction terms.

\begin{figure}[t]
  \centering
  \includegraphics[width=0.49\textwidth,page=1]{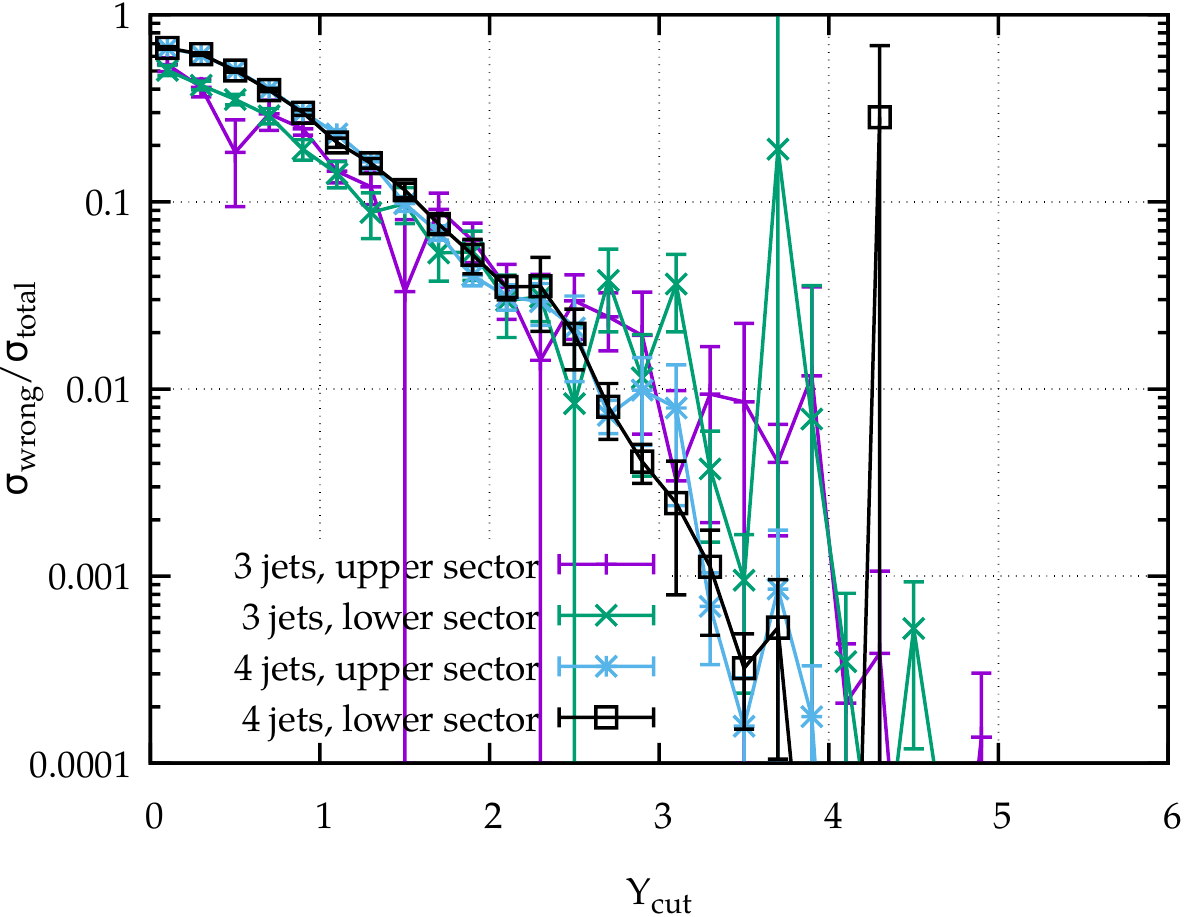}
  \includegraphics[width=0.49\textwidth,page=1]{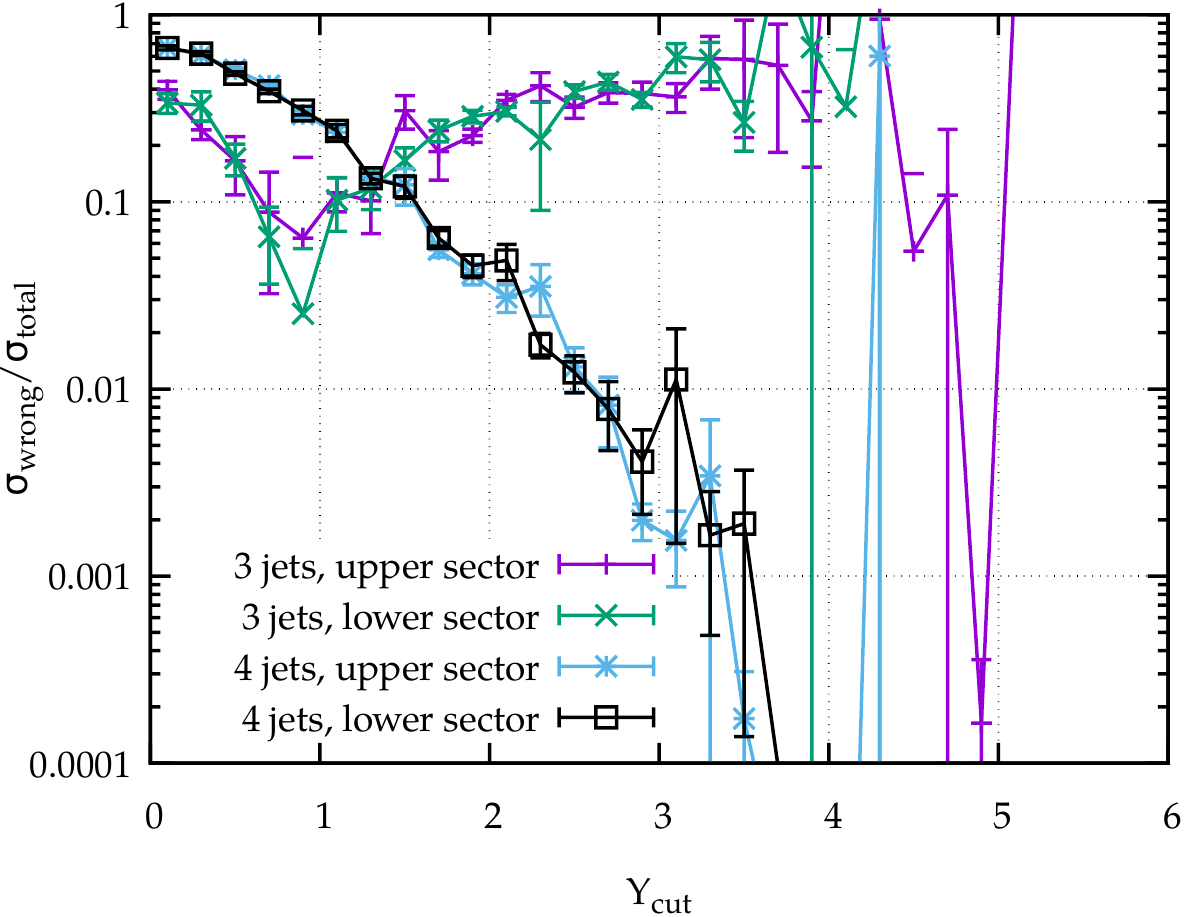}
  \caption{The relative fraction of the cross section for which partons assigned
    to the upper (lower) sector were found in the rapidity region associated
    with the lower (upper) leading jet as a function of the absolute rapidity
    separation of the two leading jets. We distinguish between events with 3 and
    4 jets (anti-$\kt$, $R=0.4$) and we separate between radiation on the upper
    and the lower line. \textbf{Left}: The distribution as found in our
    implementation. The relative fraction decreases exponentially with
    increasing rapidity separation. \textbf{Right}: Here we randomly re-assign
    the tags in the virtual correction, such that emissions sometimes end up in
    the wrong sector. The 3-jet distributions are clearly not decreasing
    anymore.}
  \label{fig:bug}
\end{figure}

To check the correctness of the assignment to sectors, we verified
that as the rapidity separation between the two leading jets
increases, there was a decreasing relative fraction of the cross
section for which partons assigned to the upper (lower) sector
were found in the rapidity region associated with the lower (upper)
leading jet, see \cref{fig:bug}.
This figure also shows a similar plot, after a small bug was introduced in the
program. The bug consisted of a random reassignment of tags, such that partons
belonging to the upper (lower) sector was identified with the lower (upper)
sector. We found that this bug gave visible results in the aforementioned
distribution even when only a few percent of the partons were given the wrong
tag. Furthermore, bugs of this type would ruin the internal \POWHEG{} check of
soft and collinear limits.
We also tested that the sum of inclusive and exclusive contributions
at NLO agrees with the \POWHEG NLO implementation of the VBF $H$+2-jet
process.

\section{Phenomenological Results}
To investigate the phenomenological consequences of the NNLO
corrections, we study $13 \TeV$ proton-proton collisions.
We use a diagonal CKM matrix, full Breit-Wigners for the $W$, $Z$ and
the narrow-width approximation for the Higgs boson.
We take NNPDF 3.0 parton distribution functions at NNLO with
$\alpha_s(M_Z) = 0.118$
(\texttt{NNPDF30\_nnlo\_as\_0118})~\cite{Ball:2014uwa}, also for our
LO and NLO results.
We have five light flavours and ignore contributions with top-quarks
in the final state or internal lines.
We set the Higgs mass to $M_H = 125\GeV$, compatible with the
experimentally measured value~\cite{Aad:2015zhl}.  Electroweak
parameters are set according to known experimental values and
tree-level electroweak relations. As inputs we use $M_W = 80.398\GeV$,
$M_Z = 91.1876\GeV$ and $G_F = 1.16637\times 10^{-5} \GeV^{-2}$. For
the widths of the vector bosons we use $\Gamma_W = 2.141 \GeV $ and
$\Gamma_Z = 2.4952 \GeV$.

\begin{figure}[t]
  \centering
  \includegraphics[width=0.8\textwidth,page=1]{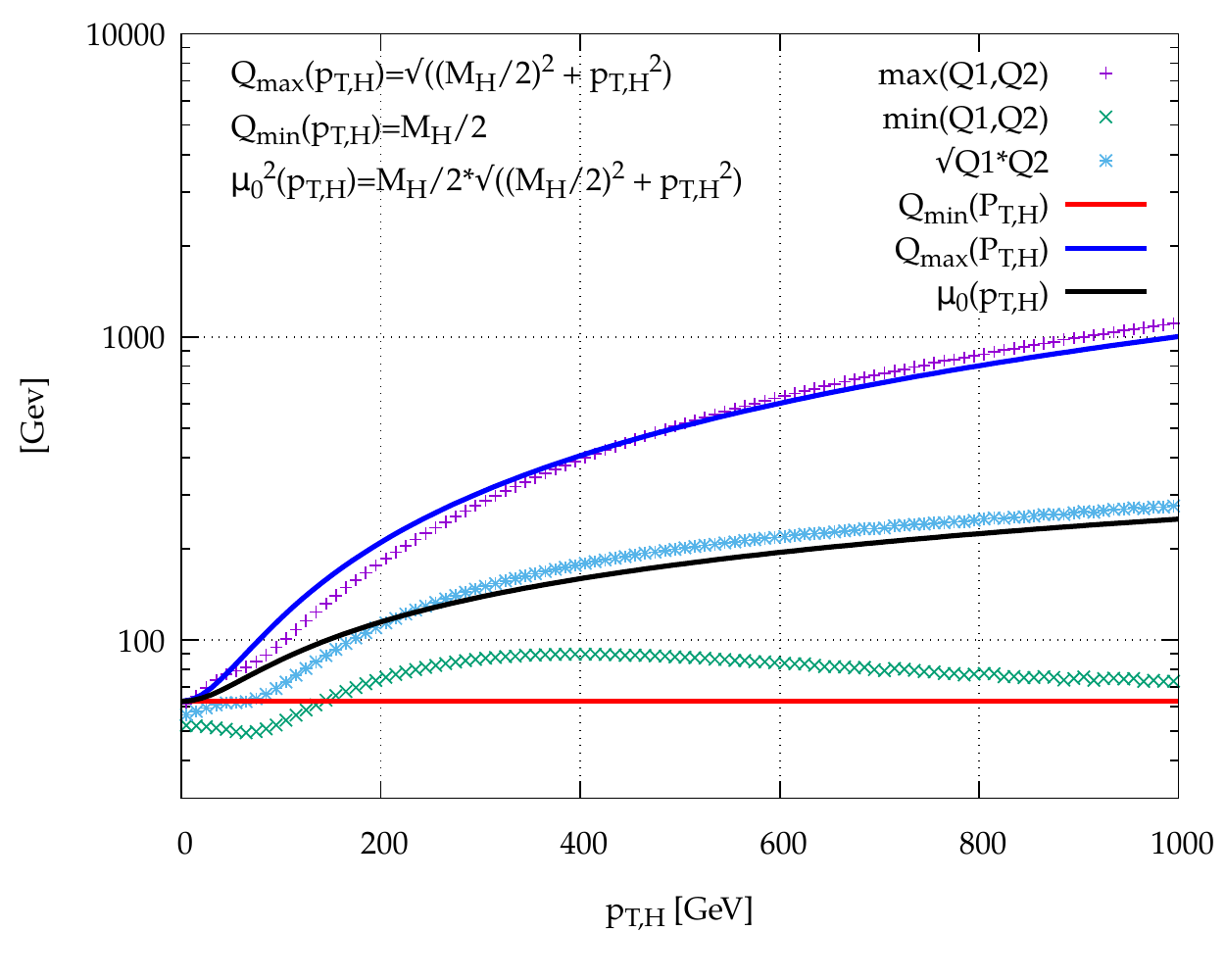}
  \caption{Our scale choice $\mu_0$ of \cref{eq:scale} (black) compared with
    $\sqrt{Q_1Q_2}$ (teal), $\mathrm{max}\{Q_1,Q_2\}$ (purple), and
    $\mathrm{min}\{Q_1,Q_2\}$ (green) as a function of the Higgs transverse
    momentum. Our scale choice is constructed as the product of two functions
    $Q_{\mathrm{max}}$ (blue) and $Q_{\mathrm{min}}$ (red) which approximate
    well the actual extrema.}
  \label{fig:scales}
\end{figure}

\begin{figure}[t]
  \centering
  \includegraphics[width=0.49\textwidth,page=1]{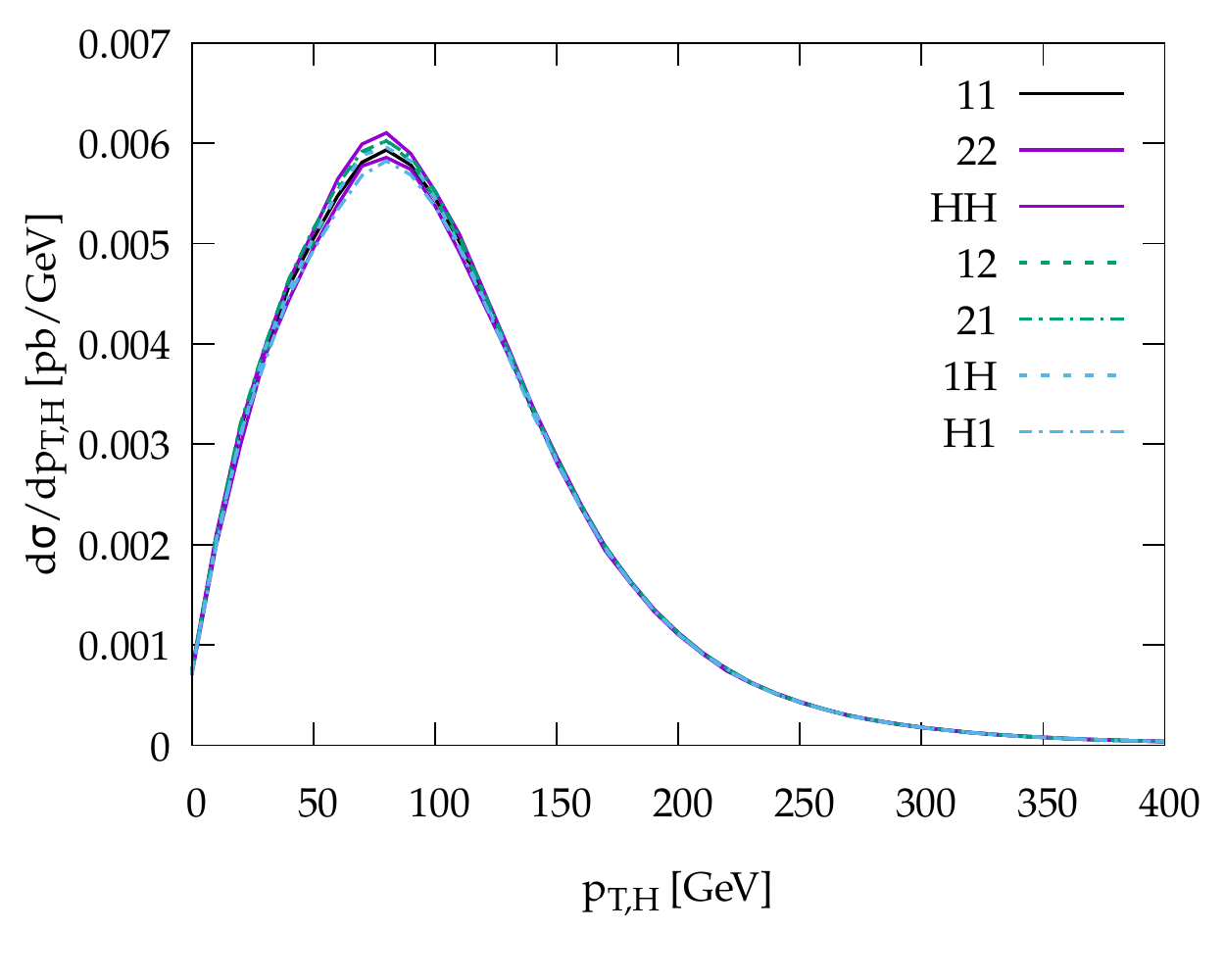}
  \includegraphics[width=0.49\textwidth,page=2]{VBFH/ptH}
  \caption{The transverse momentum of the Higgs boson evaluated at 7 different
    combinations of renormalisation, $\mu_R=K_R\mu_0$, and factorisation,
    $\mu_F=K_F\mu_0$, scales. The labels $XY$ refer to $K_R$ and $K_F$ in that
    order, where $H$ means \emph{half}, i.e. $1/2$. \textbf{Left}: The
    differential cross section. \textbf{Right}: The ratio to the central scale,
    $11$. The 7-point scale variation is almost completely contained within the
    3-point scale variation of $11,22$ and $HH$. If one takes the statistical
    fluctuations into account, the containment is perfect.}
  \label{fig:7point}
\end{figure}

Some care is needed with the renormalisation and factorisation scale
choice.
A natural option would be to use $Q_1$ and $Q_2$ as our central values
for the upper and lower sectors, as in \cref{ch:incVBF}.
While this is straightforward in the inclusive code, in the exclusive
code we had the limitation that the underlying \POWHEGBOX code can
presently only easily assign a single scale (or set of scales) to a
given event.
However, for each \POWHEG phase-space point, we have multiple
upper/lower classifications of the partons, leading to several
$\{Q_1,Q_2\}$ pairs for each event.
Thus the use of $Q_1$ and $Q_2$ would require some further degree of
modification of the \POWHEGBOX.
We instead choose a central scale that depends on the Higgs transverse
momentum $p_{\mathrm{T},H}$:
\begin{equation}
  \label{eq:scale}
  \mu_0^2(p_{\mathrm{T},H}) = \frac{M_H}{2} \sqrt{\left(\frac{M_H}{2}\right)^2 +
        p_{\mathrm{T},H}^2}\,.
\end{equation}
This choice of $\mu_0$ is usually close to
$\sqrt{Q_1 Q_2}$ as is seen in \cref{fig:scales}.
It represents a good compromise between satisfying the requirement of
a single scale for each event, while dynamically adapting to the
structure of the event.
In order to estimate missing higher-order uncertainties, we vary the
renormalisation and factorisation scales symmetrically (i.e. keeping
$\mu_R=\mu_F$) by a factor $2$ up and down around $\mu_0$. We verified that an
expanded scale variation, allowing $\mu_R \neq \mu_F$ with $\frac12 <
\mu_R/\mu_F < 2$, led only to very small changes in the NNLO scale uncertainties
for the VBF-cut cross section and the $p_{\mathrm{T},H}$ distribution, see
\cref{fig:7point}

To pass our VBF selection cuts, events should have at least two jets
with transverse momentum $\pt > 25\GeV$; the two hardest (i.e.\
highest $\pt$) jets should have absolute rapidity $|y|<4.5$, be
separated by a rapidity $\Delta y_{j_1,j_2} > 4.5$, have a dijet
invariant mass $m_{j_1,j_2} > 600\GeV$ and be in opposite hemispheres
($y_{j_1} y_{j_2} < 0$).
Jets are defined using the anti-$\kt$
algorithm~\cite{Cacciari:2008gp}, as implemented in \texttt{FastJet
  v3.1.2}~\cite{Cacciari:2011ma}, with radius parameter $R=0.4$.

%----------------------------------------------------------------------
\begin{table}[t] % why doesn't "t" work here?
  \centering
  \phantom{x}\medskip% add space to make table look nicer...
  \begin{tabular}{lcccc}
    \toprule
    &&  $\sigma^\mathrm{(no cuts)}$  [pb]  && $\sigma^\mathrm{(VBF cuts)}$ [pb] \\
    \midrule
    LO      &&  $4.032\,^{+0.057}_{-0.069}$    &&  $0.957\,^{+0.066}_{-0.059}$\\[4pt]
    NLO     &&  $3.929\,^{+0.024}_{-0.023}$    &&  $0.876\,^{+0.008}_{-0.018}$\\[4pt]
    NNLO    &&  $3.888\,^{+0.016}_{-0.012}$    &&  $0.826\,^{+0.013}_{-0.014}$\\
    \bottomrule
  \end{tabular}
  \caption{Cross sections at LO, NLO and NNLO for VBF Higgs production,
    fully inclusively and with VBF cuts.
    The quoted uncertainties correspond to scale dependence, while
    statistical errors at NNLO are about $0.1\%$ with VBF cuts and much smaller 
    without. }
\label{tab:cross-sections}
\end{table}
%----------------------------------------------------------------------

Results are shown in \cref{tab:cross-sections} for the fully
inclusive cross section and with our VBF cuts.
One sees that the NNLO corrections modify the fully inclusive cross
section only at the percent level, which is compatible with the
findings of \Bref{Bolzoni:2010xr}.
However, after VBF cuts, the \NNLO{} corrections are about 5 times
larger, reducing the cross section by $5-6\%$ relative to NLO.
The magnitude of the \NNLO{} effects after cuts imply that it will be
essential to take them into account for future precision studies.
Note that in both the inclusive and VBF-cut cases, the NNLO
contributions are larger than would be expected from NLO scale
variation.

\begin{figure}
  \centering
  \includegraphics[clip,height=0.7\textwidth,page=1,angle=0]{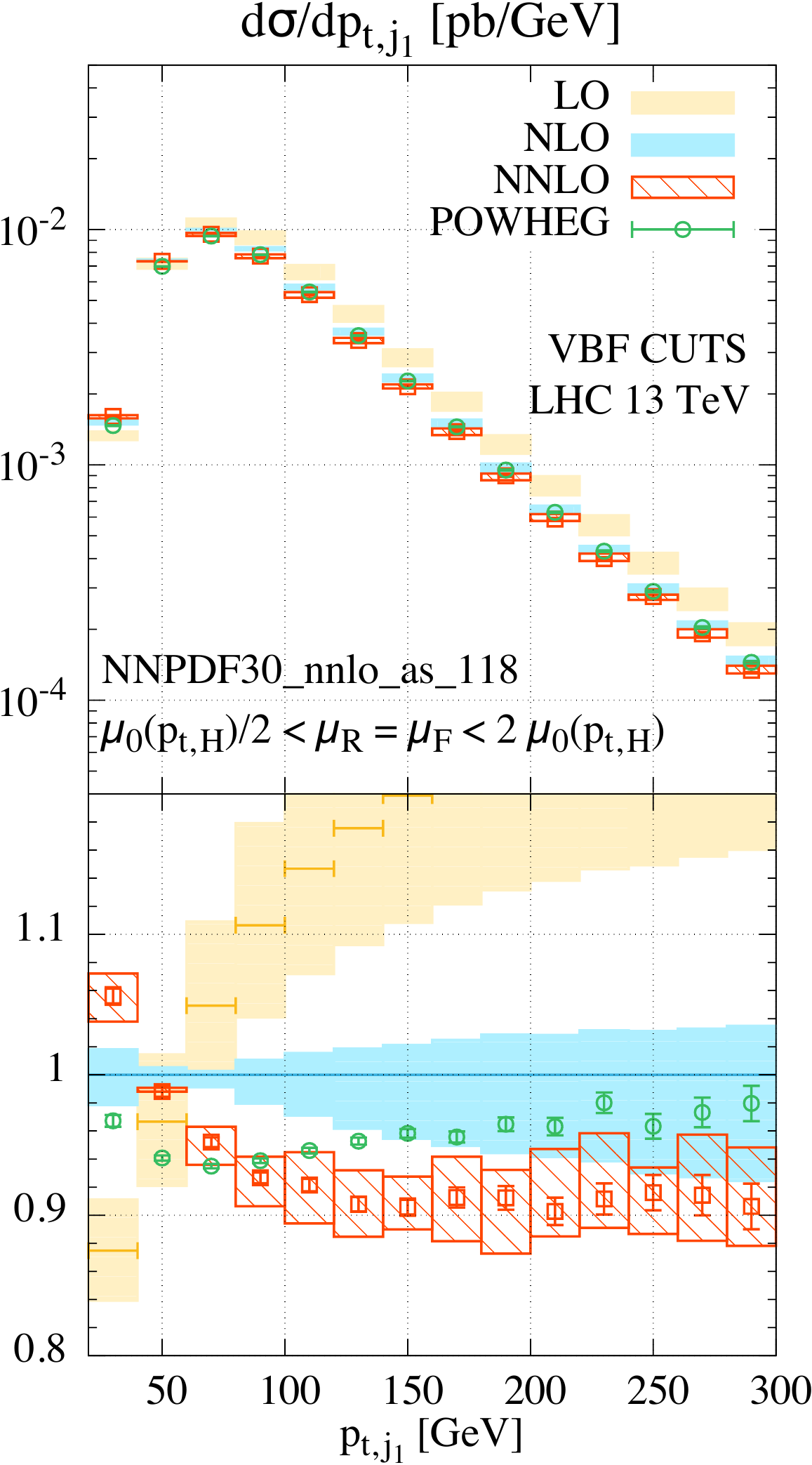}%
  \includegraphics[clip,height=0.7\textwidth,page=2,angle=0]{VBFH/NLO-NNLO-crop.pdf}%
    \caption{From left to right, differential cross sections for the transverse
      momentum distributions for the two leading jets, $p_{\mathrm{T},j_1}$ and
      $p_{\mathrm{T},j_2}$.}
    \label{fig:diff-cross-sections1}
\end{figure}

\begin{figure}
  \centering
  \includegraphics[clip,height=0.7\textwidth,page=3,angle=0]{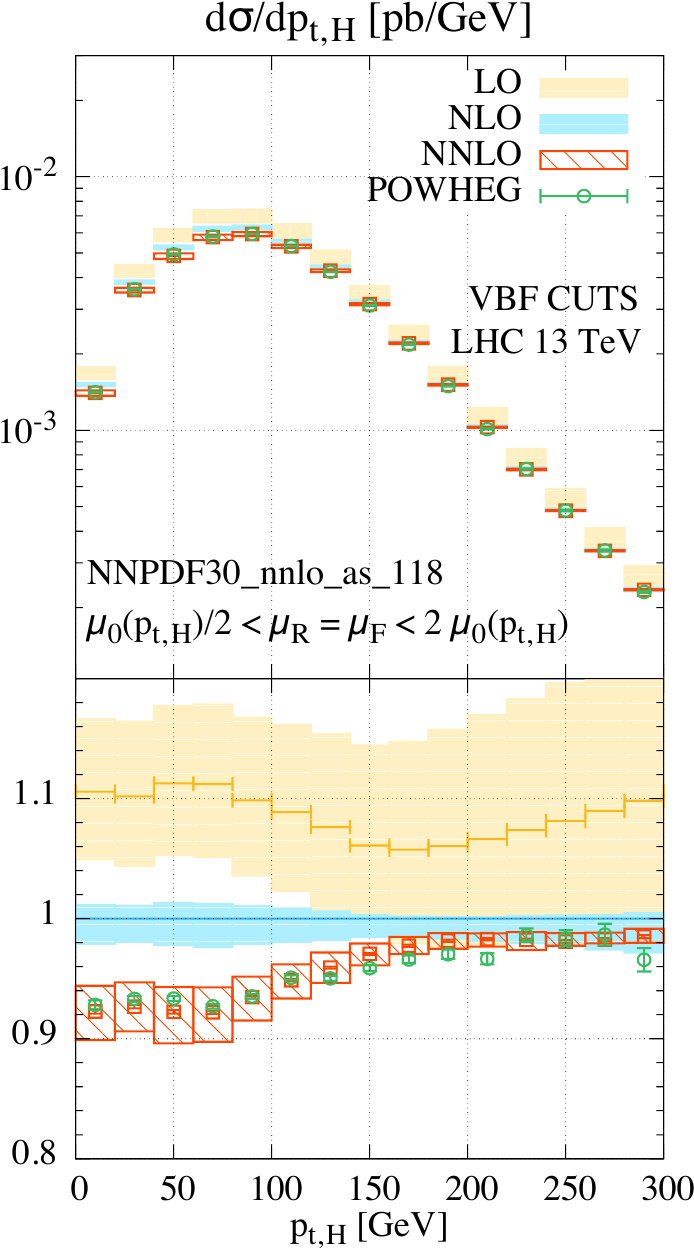}\hspace{0.8mm}%
    \includegraphics[clip,height=0.7\textwidth,page=4,angle=0]{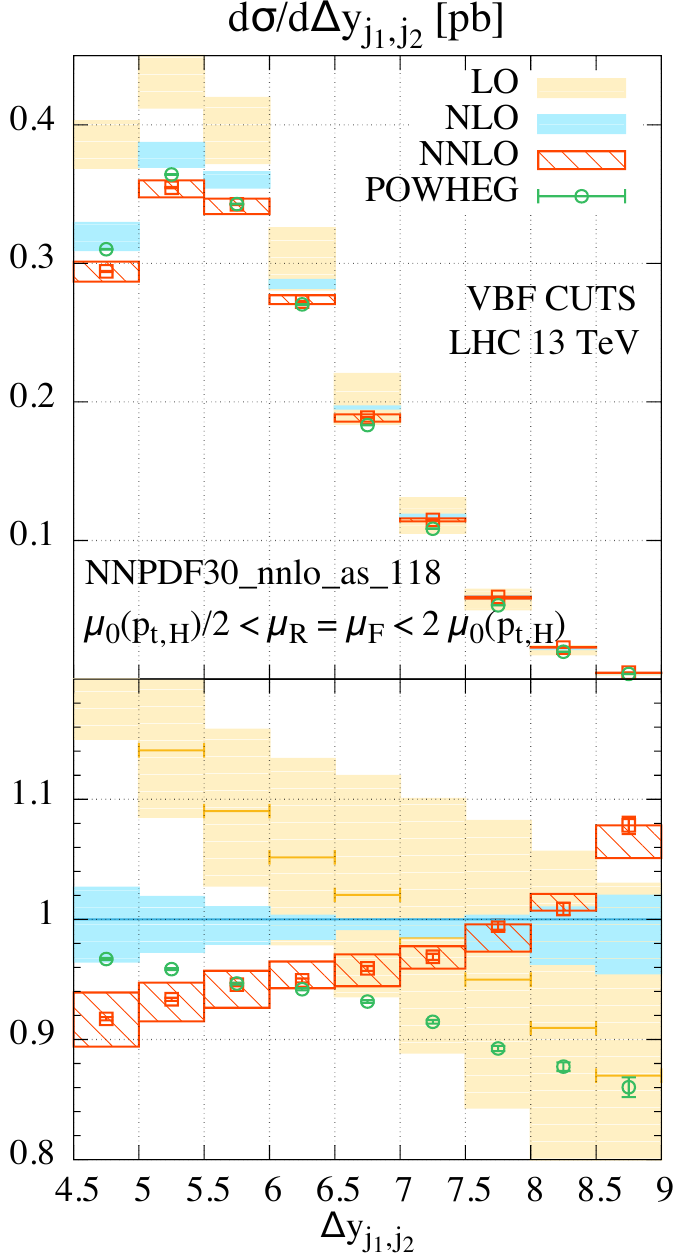}%
    \caption{From left to right, differential cross sections for the transverse
      momentum distributions for the Higgs boson, $p_{\mathrm{T},H}$, and the
      distribution for the rapidity separation between the two leading jets,
      $\Delta y_{j_1,j_2}$.  }
    \label{fig:diff-cross-sections2}
\end{figure}

Differential cross sections are shown in
\cref{fig:diff-cross-sections1,fig:diff-cross-sections2}, for events that pass
the VBF cuts.
We show in \cref{fig:diff-cross-sections1} distributions of the transverse
momentum of the two leading jets, $p_{\mathrm{T},j_1}$ and $p_{\mathrm{T},j_2}$,
and in \cref{fig:diff-cross-sections2} distributions of the Higgs boson
transverse momentum, $p_{\mathrm{T},H}$, and the rapidity separation between the
two leading jets, $\Delta y_{j_1,j_2}$.
The bands and the patterned boxes denote the scale uncertainties,
while the vertical error-bars denote the statistical uncertainty.
The effect of the NNLO corrections on the jets appears to be to reduce
their transverse momentum, leading to negative (positive) corrections
in regions of falling (rising) jet spectra.
One can see effects of up to $10-12\%$.
Turning to $p_{\mathrm{T},H}$, one might initially be surprised that such an
inclusive observable should also have substantial NNLO corrections, of
about $8\%$ for low and moderate $p_{\mathrm{T},H}$.
Our interpretation is that since NNLO effects redistribute jets from
higher to lower $\pt$'s (cf.\ the plots for $p_{\mathrm{T},j_1}$ and
$p_{\mathrm{T},j_2}$), they reduce the cross section for any observable defined
with VBF cuts.
As $p_{\mathrm{T},H}$ grows larger, the forward jets tend naturally to
get harder and so automatically pass the $\pt$ thresholds, reducing
the impact of NNLO terms. 

As observed above for the total cross section with VBF cuts, the NNLO
differential corrections are sizeable and often outside the
uncertainty band suggested by NLO scale variation.
One reason for this might be that NLO is the first order where the
non-inclusiveness of the jet definition matters, e.g.\ radiation
outside the cone modifies the cross section.
Thus NLO is, in effect, a leading-order calculation for the exclusive
corrections, with all associated limitations.
%
%----------------------------------------------------------------------
%----------------------------------------------------------------------

To further understand the size of the NNLO corrections, it is
instructive to examine a NLO plus parton shower (\NLOPS{}) calculation,
since the parton shower will include some approximation of the NNLO
corrections.
For this purpose we have used the \POWHEG VBF $H$+2-jet
calculation~\cite{Nason:2009ai}, showered with \PYTHIA version 6.428
with the Perugia 2012 tune~\cite{Skands:2010ak}.
The \POWHEG part of this \NLOPS{} calculation uses the same PDF, scale
choices and electroweak parameters as our full NNLO calculation. 
The \NLOPS{} results are included in
\cref{fig:diff-cross-sections1,fig:diff-cross-sections2}, at parton level, with
multi-parton interactions (MPI) switched off.
They differ from the NLO by an amount that is of a similar order of
magnitude to the NNLO effects.
This lends support to our interpretation that final (and
initial)-state radiation from the hard partons is responsible for a
substantial part of the NNLO corrections.
However, while the \NLOPS{} calculation reproduces the shape of the NNLO
corrections for some observables (especially $p_{\mathrm{T},H}$), there are
others for which this is not the case, the most
striking being perhaps $\Delta y_{j_1,j_2}$.
Parton shower effects were also studied in
\Bref{Frixione:2013mta}, using the MC@NLO
approach~\cite{Frixione:2002ik}.
Various parton showers differed there by up to about 10\%.
In addition to the NNLO contributions, precise phenomenological
studies require the inclusion of electroweak (EW) contributions and non-perturbative
hadronisation and MPI corrections.
The former are of the same order of magnitude as our NNLO
corrections~\cite{Ciccolini:2007ec}.
Using Pythia~6.428 and Pythia~8.185 we find that hadronisation
corrections are between $-2$ and $0\%$, while MPI brings up to $+5\%$
at low $\pt$'s.
The small hadronisation corrections appear to be due to a partial
cancellation between shifts in $\pt$ and rapidity.

\subsection{Precision Studies for the LHC}
\label{sec:VBF-XS}
Given that the effects mentioned above are of the same order as our \NNLO{}
corrections, it is useful to investigate their combined impact. In this
section\footnote{The results presented in this section are also reported
  in~\Bref{YR4}.}, we study in detail the combined effects of the \NNLO{} QCD
corrections presented here and EW contributions. In addition to that, we also
study the effect of PDF and $\as$ uncertainties. The numerical results presented
here have been computed using the values of the EW parameters given in
\cref{sec:vbfhn3lo-pheno}. The electromagnetic coupling is fixed in the $\GF$
scheme, \beq \alpha_{\GF} = \sqrt{2}\GF\MW^2(1-\MW^2/\MZ^2)/\pi, \eeq and the
weak mixing angle, $\theta_{\mathrm{w}}$, is defined in the on-shell scheme,
\begin{equation}
\sin^2\theta_{\mathrm{w}}=1-\MW^2/\MZ^2.
\end{equation}
The renormalisation and factorisation scales are set equal to the
$\PW$-boson mass,
\begin{equation}
\label{eq:VBF_ren_fac_scales}
\mu = \muR = \muF= \MW,
\end{equation}
and both scales are varied in the range $\MW/2 < \mu < 2\MW$ keeping
$\muF=\muR$, which catches the full scale uncertainty of integrated cross
sections (and of differential distributions in the essential regions). This
fixed scale choice is a consequence of limitations in the implementation of the
EW corrections.

The QCD corrections for inclusive cross sections and differential distributions
have been obtained as described earlier in this chapter. EW corrections have
been computed using \HAWK{}\footnote{The predictions from \HAWK{} were provided by
  Stefan Dittmaier.}~\cite{Denner:2014cla,HAWK}. \HAWK{} is a parton-level event
generator for Higgs production in vector boson fusion~\cite{Ciccolini:2007jr,
  Ciccolini:2007ec}, $\Pp\Pp\to\PH+2\,\mathrm{jets}$, and Higgs-strahlung
\cite{Denner:2011id}, $\Pp\Pp\to\PH\PW/\PZ\to \PH+2\,$leptons. It includes the
complete NLO-QCD and EW corrections and all weak-boson fusion and
quark--antiquark annihilation diagrams, i.e.~$t$-channel and $u$-channel
diagrams with VBF-like vector boson exchange and $s$-channel Higgs-strahlung
diagrams with hadronic weak-boson decay, as well as all interferences.  \HAWK{}
allows for an on-shell Higgs boson or for an off-shell Higgs boson (with
optional decay into a pair of gauge singlets).  The EW corrections include also
the contributions from photon-induced channels, but contributions from effective
Higgs--gluon couplings, which are part of the QCD corrections to Higgs
production via gluon fusion, are not taken into account.  External fermion
masses are neglected and the renormalisation and factorisation scales are set to
$\MW$ by default. Since version 2.0, \HAWK{} includes anomalous
Higgs-boson--vector boson couplings.  Further features of \HAWK{} are described
in \cite{Denner:2014cla} and on its web page~\cite{HAWK}.

In the calculation of the QCD-based cross sections, we have used the
PDF\-4LHC15\-\_nnlo\-\_100 PDFs~\cite{Butterworth:2015oua}, for the calculation
of the EW corrections we have employed the NNPDF2.3QED PDF
set~\cite{Ball:2013hta}, which includes a photon PDF.
%since it is the only up-to-date PDF set 
%with EW corrections and thus with a photon PDF in particular.
%
Note, however, that the relative EW correction factor, which is used in the following,
hardly depends on the PDF set, so that the uncertainty due to the
mismatch in the PDF selection is easily covered by the other remaining
theoretical uncertainties.

For the fiducial cross section and for differential distributions the following
reconstruction scheme and cuts have been applied. Jets are constructed
according to the anti-$\kT$ algorithm~\cite{Cacciari:2008gp} with $R=0.4$. Jets
are constructed from partons $j$ with
\begin{equation}
\label{eq:VBF_cuts1}
|\eta_j| < 5\,,
\end{equation}
where $\eta_j$ denotes the pseudo-rapidity.  Real photons, which
appear as part of the EW corrections, are an input to the jet
clustering in the same way as partons.  
Thus, in real photon radiation events, final states may consist of jets
only or jets plus a real identifiable photon, depending on whether 
the photon was merged into a jet or not, respectively. 
Both events with and without isolated photons are kept.

Jets are ordered according to their $\pT$ in decreasing
progression. The jet with highest $\pT$ is called leading jet $(j_1)$, the
one with next highest $\pT$ subleading jet $(j_2)$, and both are the tagging
jets.  Only events with at least two jets are kept.  They must satisfy
the additional constraints
\begin{equation}
\label{eq:VBF_cuts2}
{\pT}_j > 20\UGeV, \qquad 
|y_j| < 5, \qquad 
|y_{j_1} - y_{j_2}| > 3\,, \qquad M_{jj} > 130\UGeV,
\end{equation}
where $y_{j_{1,2}}$ are the rapidities of the two leading jets.
The cut on the 2-jet invariant mass $M_{jj}$ is sufficient to suppress
the contribution of $s$-channel diagrams to the VBF cross section
to the level of $1{-}2\%$, so that the DIS approximation of
taking into account only $t$- and $u$-channel contributions is justified.
In the cross sections given below, the $s$-channel contributions will be given
for reference, although they are not included in the final VBF cross sections
by default.

While the VBF cross sections in the DIS approximation are independent
of the CKM matrix, quark mixing has some effect on $s$-channel contributions.
For the calculation of the latter we employed a Cabbibo-like CKM matrix
(i.e.\ without mixing to the third quark generation) with Cabbibo angle,
$\theta_{\mathrm{C}}$, fixed by $\sin\theta_{\mathrm{C}}=0.225$.
Moreover, we note that we employ complex W- and Z-boson masses in the
calculation of $s$-channel and EW corrections in the standard \HAWK{} approach, 
as described in 
\cite{Ciccolini:2007jr, Ciccolini:2007ec}.

The Higgs boson is treated as on-shell particle in the following consistently,
since its finite-width and off-shell effects in the signal region are
suppressed in the SM.
\subsubsection{Integrated VBF Cross Sections}
\label{subsec:VBF-XS}
The final VBF cross section $\sigma^{\VBF}$ is calculated according to:
\begin{equation}
\sigma^{\VBF} = \sigma_{\NNLO \QCD}^{\DIS} (1+\delta_{\ELWK}) + \sigma_{\gamma},
\label{eq:sigmaVBF}
\end{equation}
where $\sigma_{\NNLO \QCD}^{\DIS}$ is the NNLO-QCD prediction for the VBF cross
section in DIS approximation with PDF4LHC15\_nnlo\_100 PDFs. The relative NLO
EW correction $\delta_{\ELWK}$ is calculated with \HAWK{}, but taking into
account only $t$- and $u$-channel diagrams corresponding to the DIS
approximation.  The contributions from photon-induced channels,
$\sigma_{\gamma}$, and from $s$-channel diagrams, $\sigma_{\mbox{\scriptsize
    $s$-channel}}$ are obtained from \HAWK{} as well, where the latter includes
NLO-QCD and EW corrections.  To obtain $\sigma^{\VBF}$, the photon-induced
contribution is added linearly, but $\sigma_{\mbox{\scriptsize $s$-channel}}$ is
left out and only shown for reference, since it is not of true VBF origin (like
other contributions such as H+2jet production via gluon fusion).

\cref{tab:vbf_XStot,tab:vbf_XSfiducial} summarise the total and
fiducial Standard Model VBF cross sections and the corresponding uncertainties
for the different proton--proton collision energies
for a Higgs-boson mass $\MH=125\UGeV$.
\begin{table}
\caption{Total VBF cross sections including QCD and EW corrections
and their uncertainties for different proton--proton collision energies
$\sqrt{s}$ for a Higgs-boson mass $\MH=125\UGeV$.}
\label{tab:vbf_XStot}
\begin{center}%
\begin{small}%
  \tabcolsep5pt
  \resizebox{\columnwidth}{!}{
\begin{tabular}{ccccccc|c}%
\hline
$\sqrt{s}$[GeV] & $\sigma^{\VBF}$[fb] & $\Delta_{\mathrm{scale}}$[\%] & 
$\Delta_{\mathrm{PDF}/\alphas/\mathrm{PDF\oplus\alphas}}$[\%] &
$\sigma_{\NNLO \QCD}^{\DIS}$[fb] & $\delta_{\ELWK}$[\%] & $\sigma_{\gamma}$[fb] & $\sigma_{\mbox{\scriptsize $s$-channel}}$[fb]
\\
\hline
$7$  & $1241.4(1)$ &$^{+0.19}_{-0.21}$ &$\pm 2.1/\pm 0.4/\pm2.2$ &$1281.1(1)$ & $-4.4$ & $17.1$ & $584.5(3)$
\\
$8$  & $1601.2(1)$ &$^{+0.25}_{-0.24}$ &$\pm 2.1/\pm 0.4/\pm2.2$ &$1655.8(1)$ & $-4.6$ & $22.1$ & $710.4(3)$
\\
$13$ & $3781.7(1)$ &$^{+0.43}_{-0.33}$ &$\pm 2.1/\pm 0.5/\pm2.1$ &$3939.2(1)$ & $-5.3$ & $51.9$ & $1378.1(6)$
\\
$14$ & $4277.7(2)$ &$^{+0.45}_{-0.34}$ &$\pm 2.1/\pm 0.5/\pm2.1$ &$4460.9(2)$ & $-5.4$ & $58.5$ & $1515.9(6)$
\\
\hline
\end{tabular}}
\end{small}%
\end{center}%
\vspace{2em}
\caption{Fiducial VBF cross sections including QCD and EW corrections
and their uncertainties for different proton--proton collision energies
$\sqrt{s}$ for a Higgs-boson mass $\MH=125\UGeV$.}
\label{tab:vbf_XSfiducial}
\begin{center}%
\begin{small}%
  \tabcolsep5pt
  \resizebox{\columnwidth}{!}{
\begin{tabular}{ccccccc|c}%
\hline
$\sqrt{s}$[GeV] & $\sigma^{\VBF}$[fb] & $\Delta_{\mathrm{scale}}$[\%] & 
$\Delta_{\mathrm{PDF}/\alphas/\mathrm{PDF\oplus\alphas}}$[\%] &
$\sigma_{\NNLO \QCD}^{\DIS}$[fb] & $\delta_{\ELWK}$[\%] & $\sigma_{\gamma}$[fb] & $\sigma_{\mbox{\scriptsize $s$-channel}}$[fb]
\\
\hline
$7$  & $602.4(5)$ &$^{+1.3}_{-1.6}$ &$\pm 2.3/\pm 0.3/\pm2.3$ & $630.8(5)$ & $-6.1$ &  $9.9$ & $8.2$
\\
$8$  & $795.9(6)$ &$^{+1.3}_{-1.5}$ &$\pm 2.3/\pm 0.3/\pm2.3$ & $834.8(7)$ & $-6.2$ & $13.1$ & $11.1$
\\
$13$ & $1975.4(9)$ &$^{+1.3}_{-1.2}$ &$\pm 2.1/\pm 0.4/\pm2.2$ & $2084.2(10)$ & $-6.8$ & $32.3$ & $29.0$
\\
$14$ & $2236.6(26)$ &$^{+1.5}_{-1.3}$ &$\pm 2.1/\pm 0.4/\pm2.1$ & $2362.2(28)$ & $-6.9$ & $36.7$ & $33.1$
\\
\hline
\end{tabular}}%
\end{small}%
\end{center}%
\end{table}
The scale uncertainty, $\Delta_{\mathrm{scale}}$, results from a variation
of the factorisation and renormalisation scales
in \cref{eq:VBF_ren_fac_scales} by a factor of $2$ keeping $\muF=\muR$,
as indicated above, and the combined PDF${\oplus}\alphas$ uncertainty
$\Delta_{\mathrm{PDF\oplus\alphas}}$ is obtained following the PDF4LHC
recipe~\cite{Butterworth:2015oua}. Both $\Delta_{\mathrm{scale}}$ and
$\Delta_{\mathrm{PDF\oplus\alphas}}$ are actually obtained from
$\sigma_{\NNLO \QCD}^{\DIS}$, but these QCD-driven uncertainties can be
taken over as uncertainty estimates for $\sigma^{\VBF}$ as well.  The
theoretical uncertainties of integrated cross sections originating
from unknown higher-order EW effects can be estimated by
\begin{equation}
\Delta_\ELWK = \max\{0.5\%,\delta_{\ELWK}^2,\sigma_\gamma/\sigma^{\VBF}\}.
\end{equation}
The first entry represents the generic size of NNLO-EW corrections, while the
second accounts for potential enhancement effects.  Note that the whole
photon-induced cross-section contribution $\sigma_\gamma$ is treated as
uncertainty here, because the PDF uncertainty of $\sigma_\gamma$ is estimated to
be $100\%$ with the NNPDF2.3QED PDF set. At present, this source, which is about
$1.5\%$, dominates the EW uncertainty of the integrated VBF cross
section\footnote{Very recently a more precise determination of the photon PDF has
  been proposed under the name of LUXqed~\cite{Manohar:2016nzj}. Although this
  fit does not have a lot of overlap with the NNPDF2.3QED PDF set, things are
  greatly improved in the more recent NNPDF3.0QED PDF. It would thus be of
  interest to investigate the photon contribution to VBF Higgs production from
  the LUXqed PDF set. The reported uncertainty of LUXqed is at the percent
  level.}.

\subsubsection{Differential VBF Cross Sections}

\begin{figure}
\includegraphics[width=.47\textwidth]{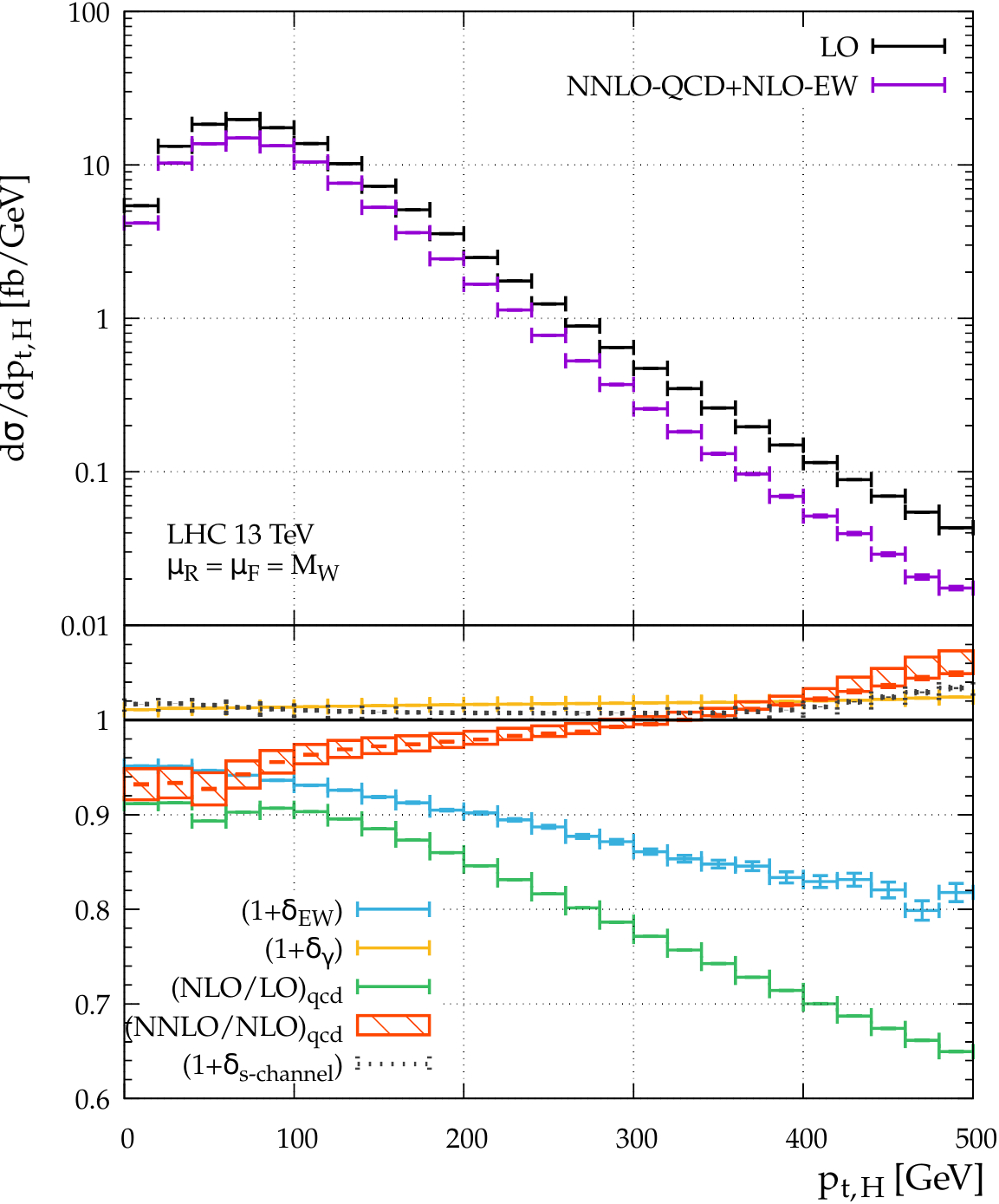}
\hfill
\includegraphics[width=.47\textwidth]{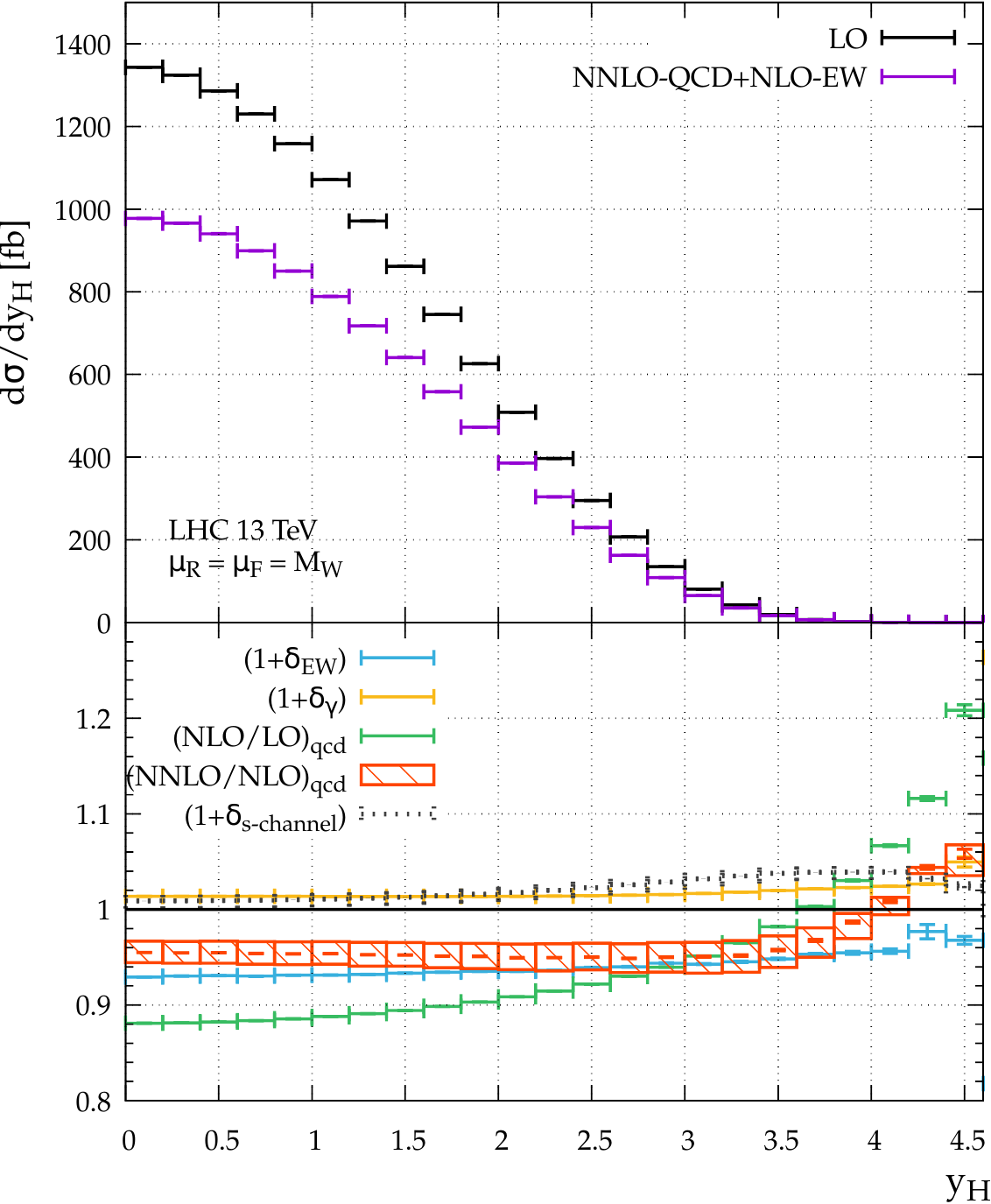}
\caption{Transverse-momentum and rapidity distributions of the Higgs boson in VBF
at LO and including NNLO-QCD and NLO-EW corrections (upper plots)
and various relative contributions (lower plots) for $\sqrt{s}=13\UTeV$ and $\MH=125\UGeV$.}
\label{fig:SM-VBF-ptH-yH}
\end{figure}

\begin{figure}
\includegraphics[width=.47\textwidth]{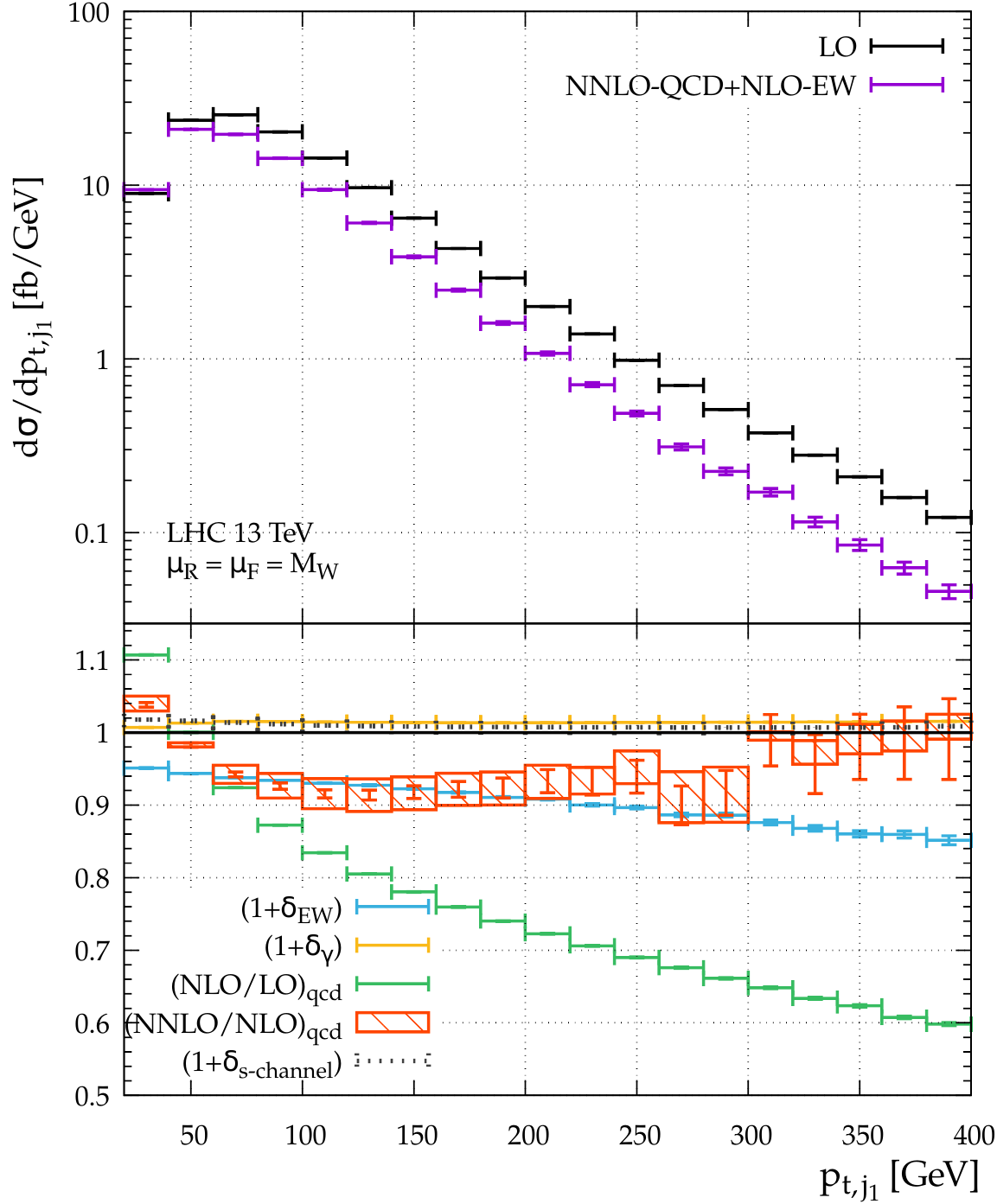}
\hfill
\includegraphics[width=.47\textwidth]{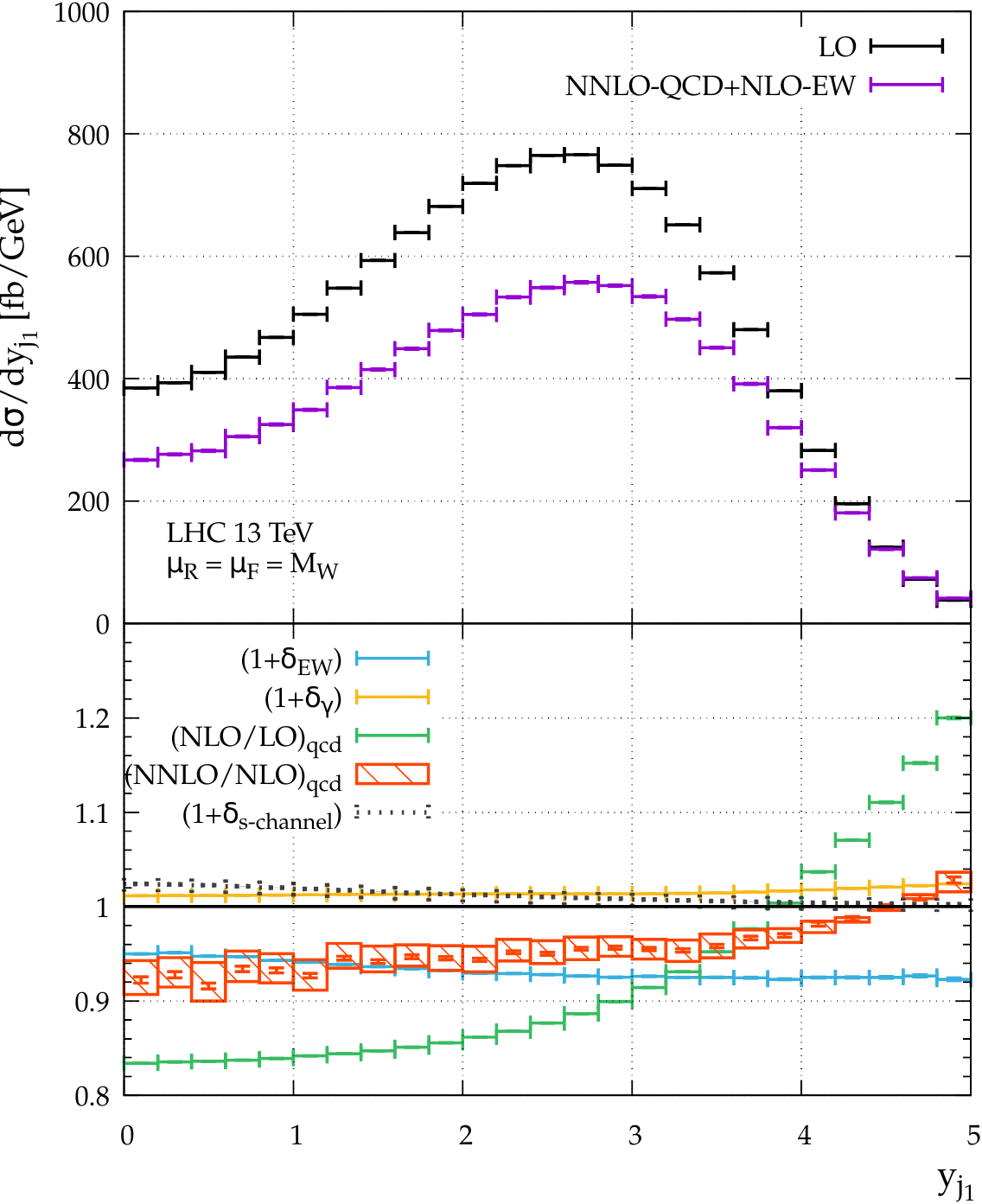}
\caption{Transverse-momentum and rapidity distributions of the leading jet in VBF
at LO and including NNLO-QCD and NLO-EW corrections (upper plots)
and various relative contributions (lower plots) for $\sqrt{s}=13\UTeV$ and $\MH=125\UGeV$.}
\label{fig:SM-VBF-ptj1-yj1}
\end{figure}

\begin{figure}
\includegraphics[width=.47\textwidth]{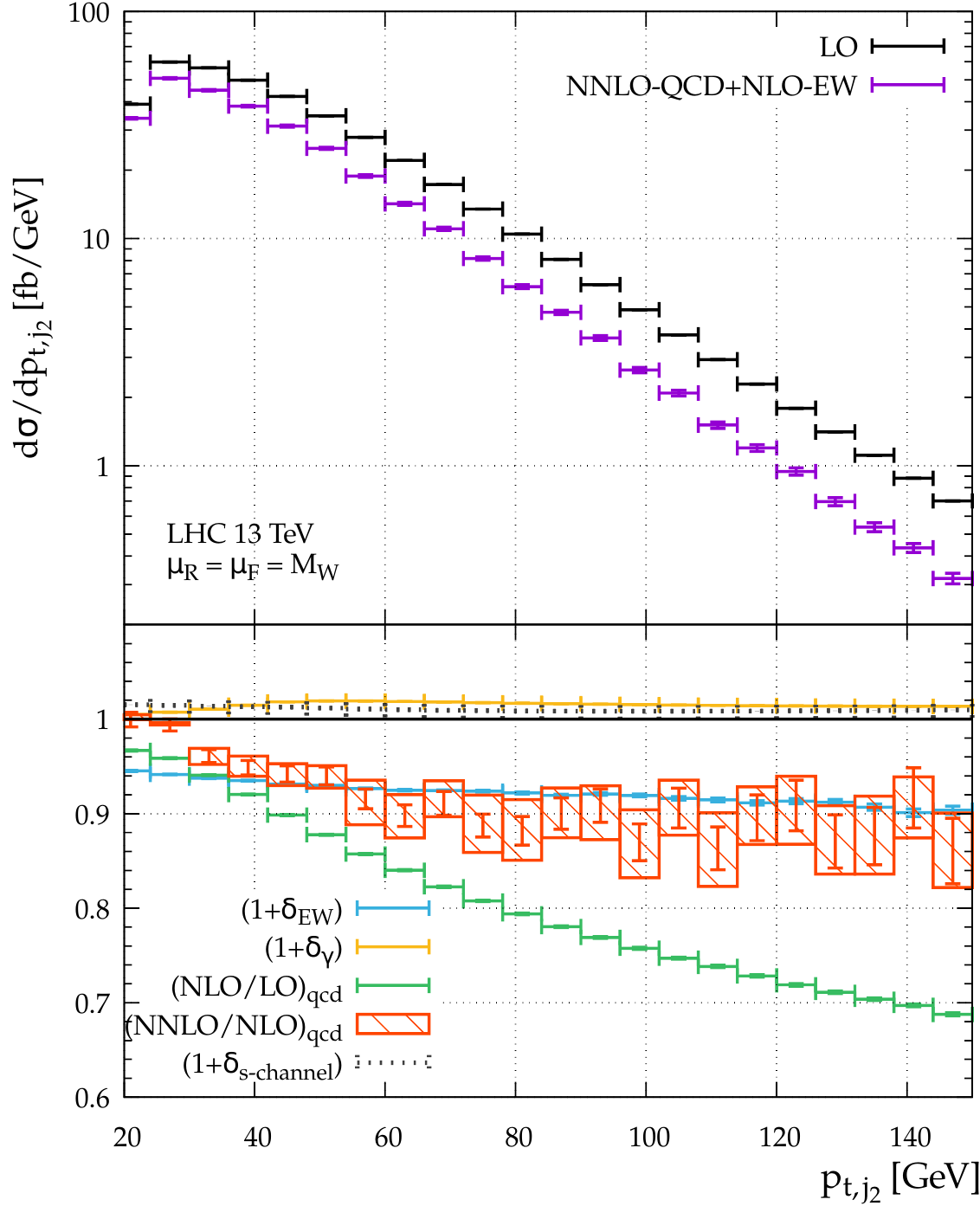}
\hfill
\includegraphics[width=.47\textwidth]{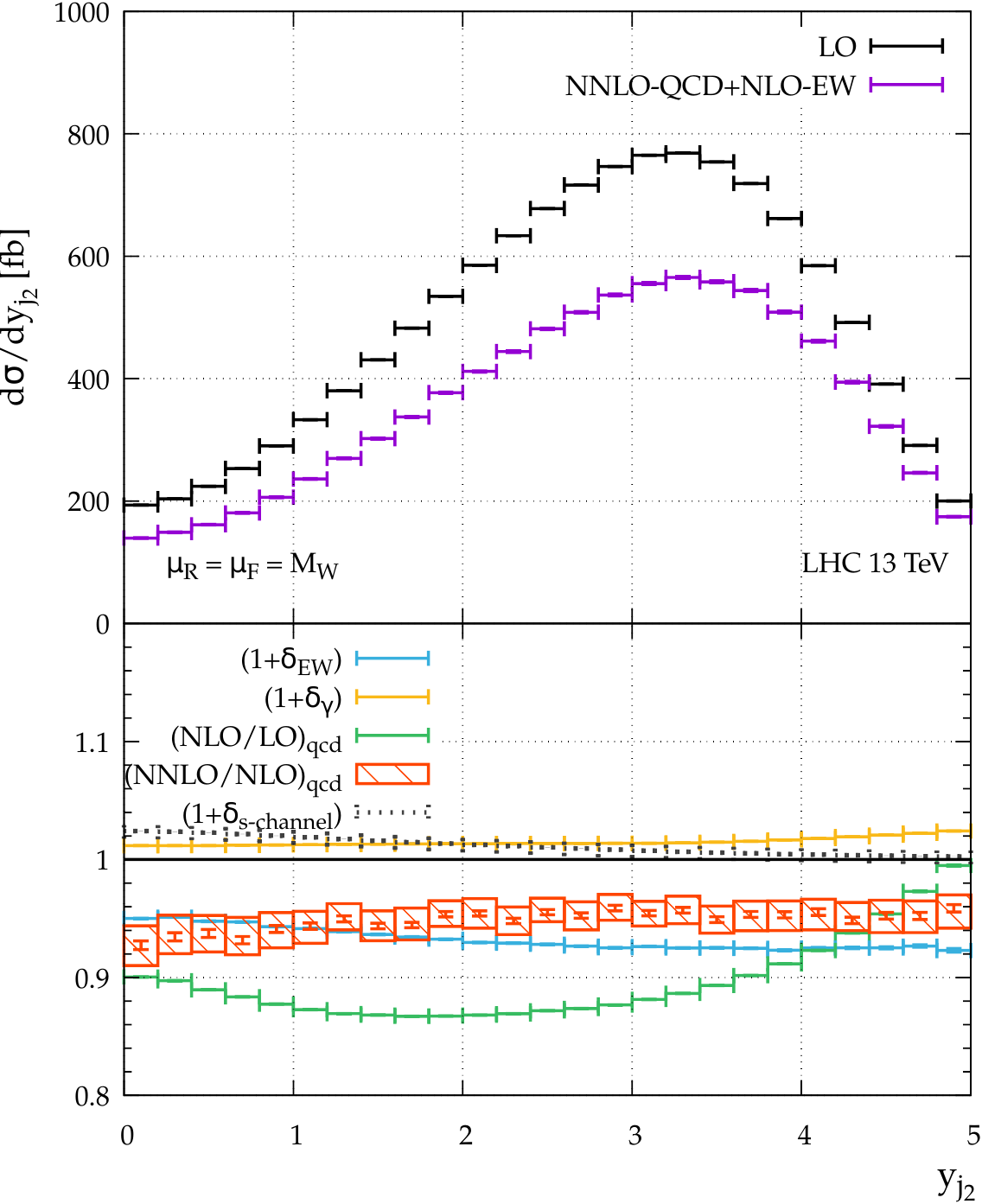}
\caption{Transverse-momentum and rapidity distributions of the subleading jet in VBF
at LO and including NNLO-QCD and NLO-EW corrections (upper plots)
and various relative contributions (lower plots) for $\sqrt{s}=13\UTeV$ and $\MH=125\UGeV$.}
\label{fig:SM-VBF-ptj2-yj2}
\end{figure}

\begin{figure}
\includegraphics[width=.47\textwidth]{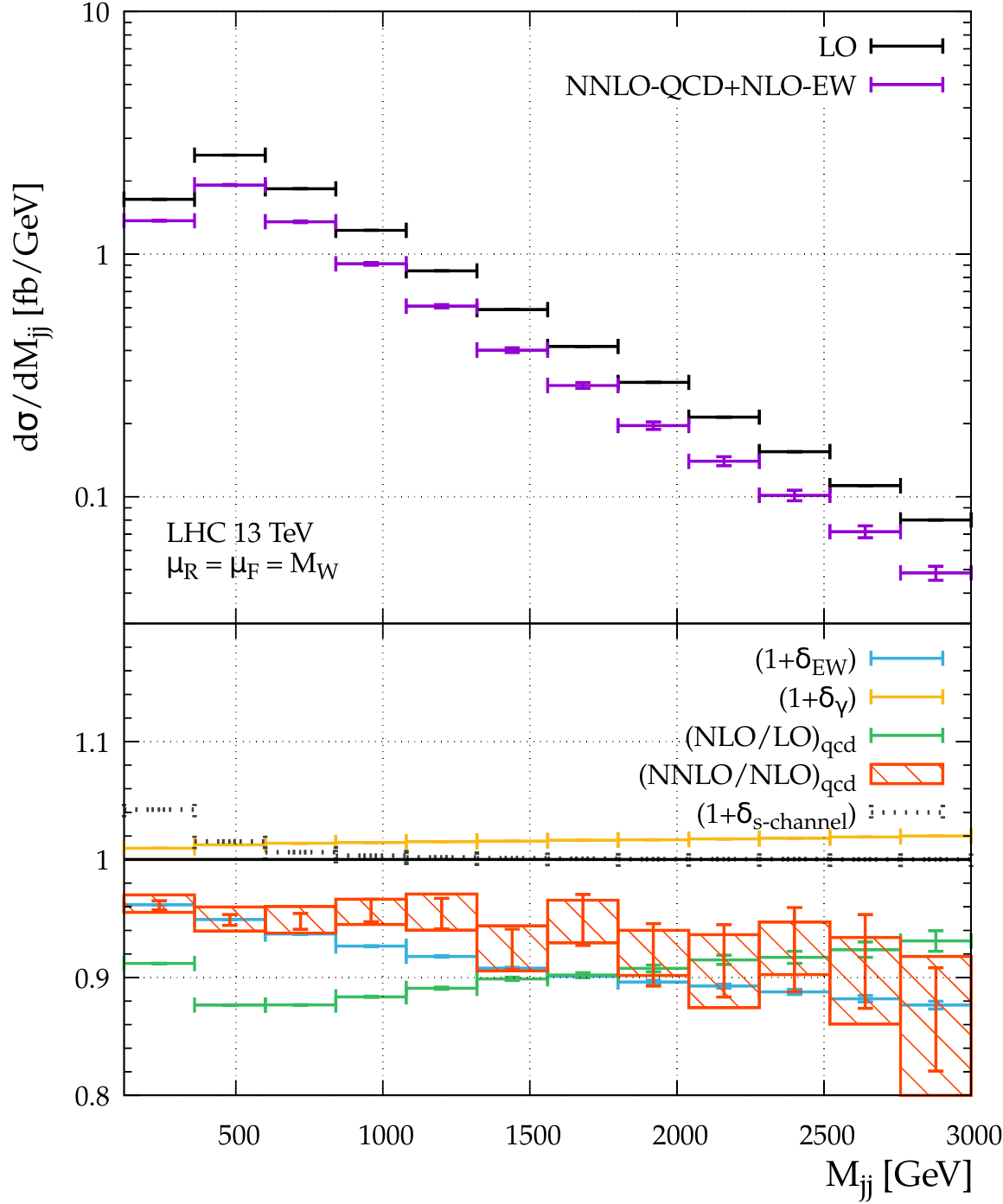}
\hfill
\includegraphics[width=.47\textwidth]{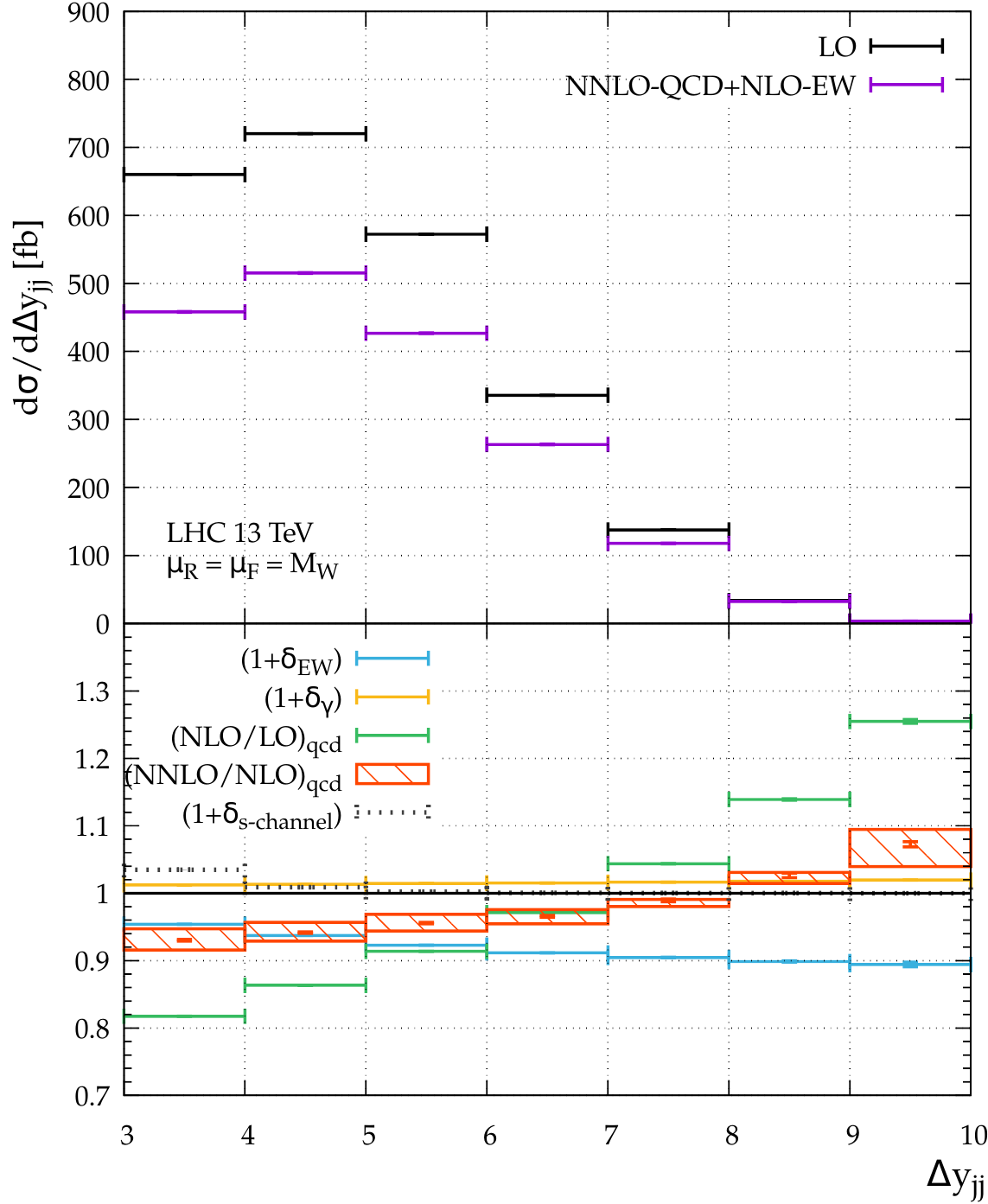}
\caption{Distributions in the invariant mass and in the rapidity difference of the first two
leading jets in VBF
at LO and including NNLO-QCD and NLO-EW corrections (upper plots)
and various relative contributions (lower plots) for $\sqrt{s}=13\UTeV$ and $\MH=125\UGeV$.}
\label{fig:SM-VBF-Mjj-yjj}
\end{figure}

\begin{figure}
\centerline{
\includegraphics[width=.47\textwidth]{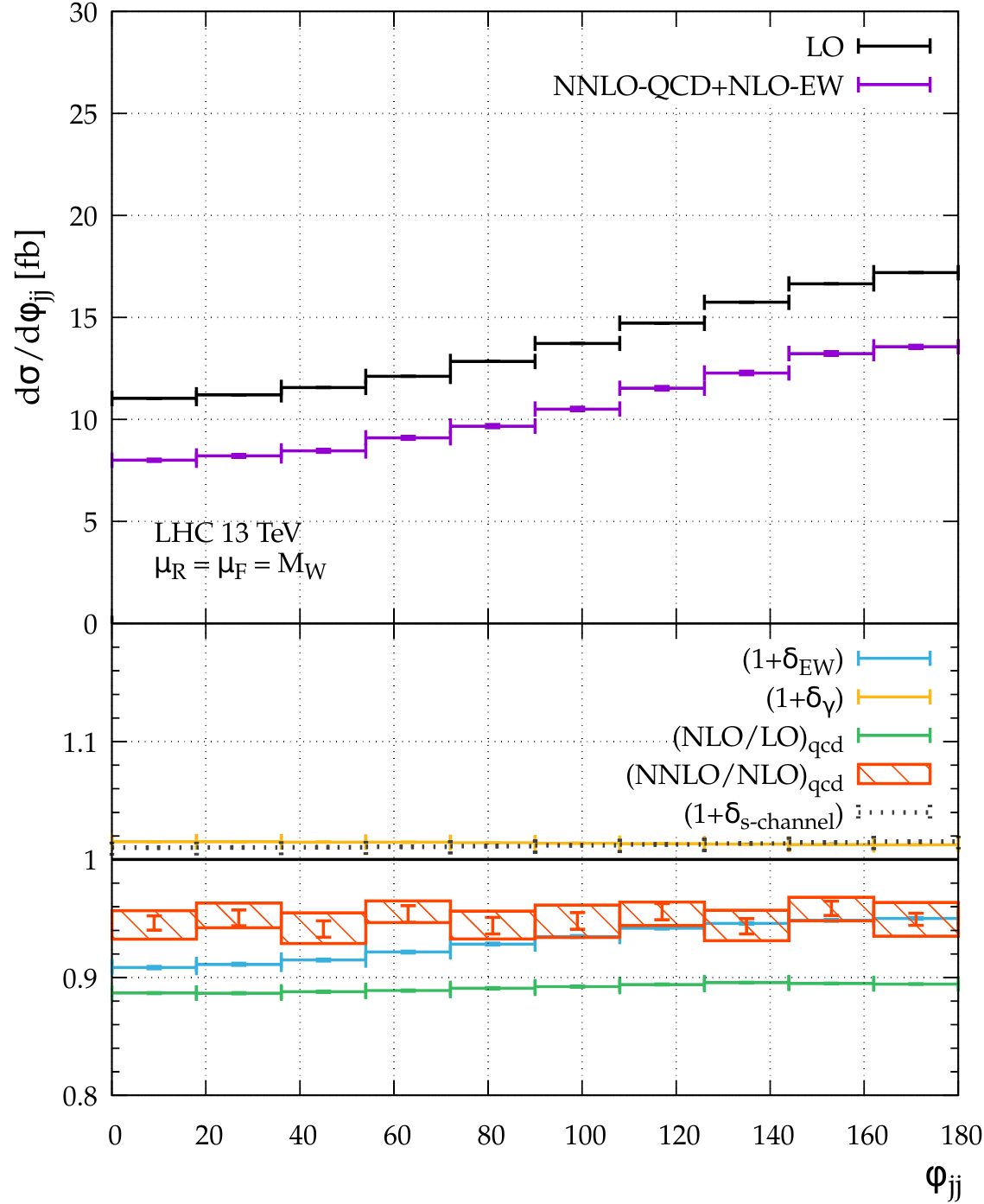}
}
\caption{Distribution in the azimuthal-angle difference of the first two
leading jets in VBF
at LO and including NNLO-QCD and NLO-EW corrections (upper plots)
and various relative contributions (lower plots) for $\sqrt{s}=13\UTeV$ and $\MH=125\UGeV$.}
\label{fig:SM-VBF-phijj}
\end{figure}

\cref{fig:SM-VBF-ptH-yH,fig:SM-VBF-ptj1-yj1,fig:SM-VBF-ptj2-yj2,fig:SM-VBF-Mjj-yjj,fig:SM-VBF-phijj}
show the most important differential cross sections for Higgs production via VBF
in the SM.
The upper panels show the LO cross section as well as the best fixed-order
prediction, based on the analogue of \cref{eq:sigmaVBF} for
differential cross sections.  The lower panels illustrate relative contributions
and the ratios $(\NLO/\LO)_{\mathrm{qcd}}$ and $(\NNLO/\NLO)_{\mathrm{qcd}}$ of
QCD predictions when going from LO to NLO-QCD to NNLO-QCD. Moreover, the
relative EW correction to the (anti)quark--(anti)quark channels
($\delta_{\ELWK}=\sigma_{\ELWK}/\sigma_{\LO}$) and the relative correction
induced by initial-state photons ($\delta_\gamma=\sigma_\gamma/\sigma_\LO$) are
shown.  Finally, the relative size of the $s$-channel contribution for
Higgs+2jet production ($\delta_{\mbox{\scriptsize
    $s$-channel}}=\sigma_{\mbox{\scriptsize $s$-channel}}/\sigma_\LO$) is
depicted as well, although it is not included in the definition of the VBF cross
section.  Integrating the differential cross sections shown in the following,
and all its individual contributions, results in the fiducial cross sections
discussed in the previous section.

The ratio $(\NLO/\LO)_{\mathrm{qcd}}$ shows a quite large impact of NLO-QCD
corrections, an effect that can be traced back to the scale choice $\mu=\MW$,
which is on the low side if mass scales such as $p_{\mathrm{T}}$ and $M_{jj}$
get large in some distributions. The moderate ratio
$(\NNLO/\NLO)_{\mathrm{qcd}}$, however, indicates nice convergence of
perturbation theory at NNLO-QCD. The band around the ratio
$(\NNLO/\NLO)_{\mathrm{qcd}}$ illustrates the scale uncertainty of the NNLO-QCD
cross section, which also applies to $\sigma^{\VBF}$.

The EW corrections $\delta_{\ELWK}$ to (pseudo)rapidity and angular
distributions are rather flat, resembling the correction to the integrated
(fiducial) cross section.  In the high-energy tails of the $p_{\mathrm{T}}$ and
$M_{jj}$ distributions, $\delta_{\ELWK}$ increases in size to $10$--$20\%$,
showing the onset of the well-known large negative EW corrections that are
enhanced by logarithms of the form $(\alpha/\sw^2)\ln^2(p_{\mathrm{T}}/\MW)$.
The impact of the photon-induced channels uniformly stays at the generic level
of $1$--$2\%$, i.e.\ they cannot be further suppressed by cuts acting on the
variables shown in the distributions.

The contribution of $s$-channel (i.e.\ VH-like) production uniformly shows the
relative size of about $1.5\%$ observed in the fiducial cross section, with the
exception of the $M_{jj}$ and $\Delta y_{jj}$ distributions, where this
contribution is enhanced at the lower ends of the spectra. Tightening the VBF
cuts at these ends, would further suppress the impact of
$\sigma_{s-\mathrm{channel}}$, but reduce the signal at the same time.  As an
alternative to decreasing $\sigma_{s-\mathrm{channel}}$, a veto on subleading
jet pairs with invariant masses around $\MW$ or $\MZ$ may be promising. Such a
veto, most likely, would reduce the photon-induced contribution $\delta_\gamma$,
and thus the corresponding uncertainty, as well.

The theoretical uncertainties of differential cross sections originating from unknown 
higher-order EW effects can be estimated by
\begin{equation}
\Delta_\ELWK = \max\{1\%,\delta_{\ELWK}^2,\sigma_\gamma/\sigma^{\VBF}\},
\end{equation}
i.e.\ $\Delta_\ELWK$ is taken somewhat more conservative than for integrated cross sections, accounting for possible
enhancements of higher-order effects due to a kinematical migration of events in distributions.
Note that $\delta_{\ELWK}^2$, in particular, covers the known effect of enhanced EW corrections at high
momentum transfer (EW Sudakov logarithms, etc.).
As discussed for integrated cross sections in the previous section, the large uncertainty of the
current photon PDF forces us to include the full contribution $\sigma_\gamma$ in the EW
uncertainties.

\subsection{Future Circular Collider Studies}
\label{sec:H_vbf}
In this section\footnote{The results presented in this section are also reported
  in~\Bref{Contino:2016spe}.} we study the production of a Standard Model Higgs
boson through vector boson fusion at a $100 \TeV$ proton-proton collider. As is
the case at $13 \TeV$, VBF has the second largest Higgs production cross section
and is interesting on its own for a multitude of reasons: 1) it is induced
already at tree-level; 2) the transverse momentum of the Higgs is non-zero at
lowest order which makes it suitable for searches for invisible decays; 3) it
can be distinguished from background processes due to a signature of two forward
jets. This last property is very important, as the inclusive VBF signal is
completely drowned in QCD background. One of the aims of this section is to
study how well typical VBF cuts suppress this background at a $100 \TeV$
proton-proton machine. In contrast to previous sections, we will here therefore
also study Higgs plus 2 jets production in the gluon fusion channel (QCD $Hjj$).

\subsubsection{Generators}
Fixed order QCD and EW predictions have been obtained in the same way as in
previous sections. NLO interfaced to a parton shower (\NLOPS{}) results have been
obtained using the
\POWHEGBOX{}~\cite{Nason:2009ai,Nason:2004rx,Frixione:2007vw,Alioli:2010xd}
together with version 6.428 of \textsc{PYTHIA}~\cite{Sjostrand:2006za} with the
Perugia Tune P12~\cite{Skands:2010ak}. LO QCD $Hjj$ results are obtained from the
\POWHEGBOX{} using the fixed-order part of~\cite{Campbell:2012am}. The \NLO{}
QCD $Hjj$ predictions have been obtained by using the setup developed for an
analogous analysis at $8$ and $13 \TeV$~\cite{Greiner:2015jha}, and is based on the
automated tools \textsc{GoSam}~\cite{Cullen:2011ac,Cullen:2014yla} and
\textsc{Sherpa}~\cite{Gleisberg:2008ta}, linked via the interface defined in the
Binoth Les Houches Accord~\cite{Binoth:2010xt,Alioli:2013nda}.

The one-loop amplitudes are generated with \textsc{GoSam}, and are based on an
algebraic generation of $d$-dimensional integrands using a Feynman diagrammatic
approach.  The expressions for the amplitudes are generated employing
\textsc{QGraf}~\cite{Nogueira:1991ex},
\textsc{Form}~\cite{Vermaseren:2000nd,Kuipers:2012rf} and
\textsc{Spinney}~\cite{Cullen:2010jv}. For the reduction of the tensor integrals
at running time, we used
\textsc{Ninja}~\cite{vanDeurzen:2013saa,Peraro:2014cba}, which is an automated
package carrying out the integrand reduction via Laurent
expansion~\cite{Mastrolia:2012bu}, and \textsc{OneLoop}~\cite{vanHameren:2010cp}
for the evaluation of the scalar integrals. Unstable phase space points are
detected automatically and reevaluated with the tensor integral library
\textsc{Golem95}~\cite{Heinrich:2010ax,Binoth:2008uq,Cullen:2011kv}.  The
tree-level matrix elements for the Born and real-emission contribution, and the
subtraction terms in the Catani-Seymour approach~\cite{Catani:1996vz} have been
evaluated within \textsc{Sherpa} using the matrix element generator
\textsc{Comix}~\cite{Gleisberg:2008fv}.

Using this framework we stored NLO events in the form of \textsc{Root}
Ntuples. Details about the format of the Ntuples generated by \textsc{Sherpa}
can be found in~\Bref{Bern:2013zja}.  The predictions presented in the following
were computed using Ntuples at $14$ and $100 \TeV$ with generation cuts specified by
\[
  p_{T,\,\mathrm{jet}}\;>\;25\:\mathrm{GeV}\qquad\mbox{and}\qquad
  |\eta_{\mathrm{jet}}|\;<\;10\,,
\]
and for which the Higgs boson mass $\MH$ and the Higgs vacuum expectation value
$v$ are set to $\MH=125\:\mathrm{GeV}$ and $v=246\:\mathrm{GeV}$,
respectively. To improve the efficiency in performing the VBF analysis using
the selection cuts described below, a separate set of Ntuples was
generated. This set includes an additional generation cut on the invariant mass
of the two leading transverse momentum jets. To generate large dijet masses
from scratch, we require $M_{jj}>1600\:\mathrm{GeV}$\footnote{The \NLO{} QCD
  Hjj predictions were provided by Gionata Luisoni. The text used here to
  describe the calculation is identical to the one found in section 3.3.1
  of~\Bref{Contino:2016spe}.}.

\subsubsection{Parameters}
The setup for this study is identical to the one used in the previous section
except that for VBF predictions we have used the
\textsc{MMHT2014nnlo68cl}~\cite{Harland-Lang:2014zoa} PDF set and for QCD $Hjj$
predictions we have used the \textsc{CT14nnlo}~\cite{Dulat:2015mca} PDF set as
implemented in \textsc{LHAPDF}~\cite{Buckley:2014ana}.

In order to estimate scale uncertainties we vary $\mu$ up and down a factor $2$
while keeping $\mu_R=\mu_F$. For QCD VBF results we use the scale choice of
\cref{eq:scale}, for EW predictions we use $\MW$ as central scale and
for QCD $Hjj$ predictions we use $\hat{H}'_\mathrm{T}/2$ as our central scale
defined as
\begin{equation}
  \frac{\hat{H}^\prime_T}{2}\;=\;
  \frac{1}{2}\left(\sqrt{\MH^2+p_{T,H}^2}+\sum_{i}|p_{T,\,j_i}|\right)\,.
\end{equation}
The sum runs over all partons accompanying the Higgs boson in the event.

\subsubsection{Inclusive VBF Production}
Due to the massive vector bosons exchanged in VBF production the cross section
is finite even when both jets become fully unresolved in fixed-order
calculations. In \cref{tab:vbf_XStot_fcc} we present the fully inclusive LO VBF
cross section and both NNLO-QCD and NLO-EW corrections at a $100 \TeV$
proton-proton collider.
\begin{table}[th]
  \caption{Total VBF cross section including QCD and EW corrections and their
    uncertainties for a $100 \TeV$ proton-proton
    collider. $\sigma^{\mathrm{VBF}}$ is obtained using \cref{eq:sigmaVBF} where
    $\sigma_{\mathrm{NNLO-QCD}}^{\mathrm{DIS}}$ is the total VBF cross section
    computed to NNLO accuracy in QCD, $\delta_{\mathrm{EW}}$ is the relative EW
    induced corrections and $\sigma_{\gamma}$ is the cross section induced by
    incoming photons. For comparison, the LO order cross section,
    $\sigma_{\mathrm{LO}}$, is also shown.}
  \label{tab:vbf_XStot_fcc}
  \begin{center}%
    \begin{small}%
      \tabcolsep5pt
      \begin{tabular}{cccccc}%
        \hline
        $\sigma^{\mathrm{VBF}}$[pb] & $\Delta_{\mathrm{scale}}$[\%] &$\sigma_{\mathrm{LO}}$[pb]  &$\sigma_{\mathrm{NNLO-QCD}}^{\mathrm{DIS}}$[pb] & $\delta_{\mathrm{EW}}$[\%] & $\sigma_{\gamma}$[pb]
        \\
        \hline
        $69.0$ &$^{+0.85}_{-0.46}$ & $80.6$ & $73.5$ & $-7.3$ & $0.81$ 
        \\
        \hline
      \end{tabular}%
    \end{small}%
  \end{center}%
\end{table}

In order to compute the VBF cross section we combine the NNLO-QCD and NLO-EW
corrections according to \cref{eq:sigmaVBF}.

The combined corrections to the LO cross section is about $14\%$ with QCD and EW
corrections contributing an almost equal amount. The scale uncertainty
$\Delta_{\mathrm{scale}}$ is due to varying $\mu$ by a factor $2$ up and down in
the QCD calculation alone keeping $\mu_F=\mu_R$. For comparison the total QCD
and EW corrections at $14 \TeV$ amount to about $7\%$ and the QCD induced scale
variations to about $0.4\%$, cf. \cref{tab:vbf_XStot}.

\subsubsection{VBF Cuts}
In order to separate the VBF signal from the main background of QCD $Hjj$
production we will extend typical VBF cuts used at the LHC to a $100 \TeV$
proton-proton collider. These cuts take advantage of the fact that VBF Higgs
production, and VBF production in general, has a very clear signature of two
forward jets clearly separated in rapidity. Examining the topology of a typical
VBF production diagram it becomes very clear that this is the case because the
two leading jets are essential remnants of the two colliding protons. Since the
$\pt$ of the jets will be governed by the mass scale of the weak vector bosons
and the energy by the PDFs the jets will typically be very energetic and in
opposite rapidity hemispheres.

%%%%%%%%%%%%%%%%%%%%%%%%%%
\begin{figure}[!th]
  \begin{minipage}{0.49\textwidth}
    \centering
    \includegraphics[width=1\textwidth,page=1]{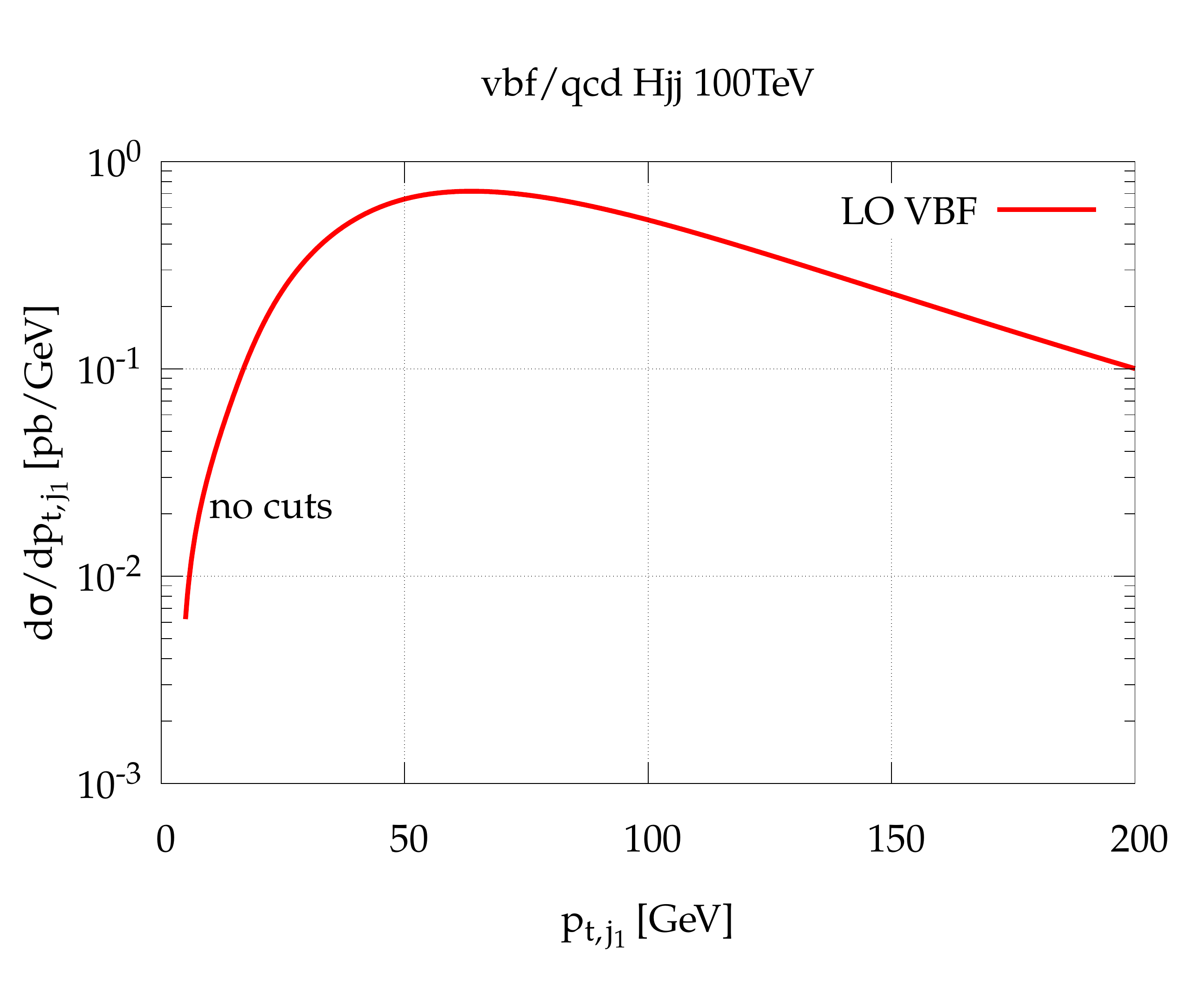}
  \end{minipage}
  \begin{minipage}{0.49\textwidth}
    \centering
    \includegraphics[width=1\textwidth,page=2]{VBFH/figs/VBF/vbf_100TeV.pdf}
  \end{minipage}
  \caption{\label{fig:VBFptj}Left panel: The $\pt$ of the hardest jet in VBF production at $100 \TeV$. We require at least two jets in the event but apply no other cuts; right panel: The $\pt$ of the second hardest jet in VBF and QCD $Hjj$ production at $100 \TeV$.}
\end{figure}
%%%%%%%%%%%%%%%%%%%%%%%%%%     

As is clear from \cref{fig:VBFptj} the hardest jet in VBF
production peaks at around $60 \GeV$. As discussed above, this value is set by the mass of the weak vector bosons and hence the $\pt$ spectra of the two hardest jets are very similar to what one finds at the LHC. From this point of view, and in order to maximise the VBF cross section, one should keep jets with $p_{\mathrm{T},cut} > 30 \GeV$. Here we present results for $p_{\mathrm{T},cut} = \{30,50,100\} \GeV$ to study the impact of the jet cut on both the VBF signal and QCD $Hjj$ background. We only impose the cut on the two hardest jets in the event.

\begin{figure}[!th]
  \begin{minipage}{0.49\textwidth}
    \centering
    \includegraphics[width=1\textwidth,page=15]{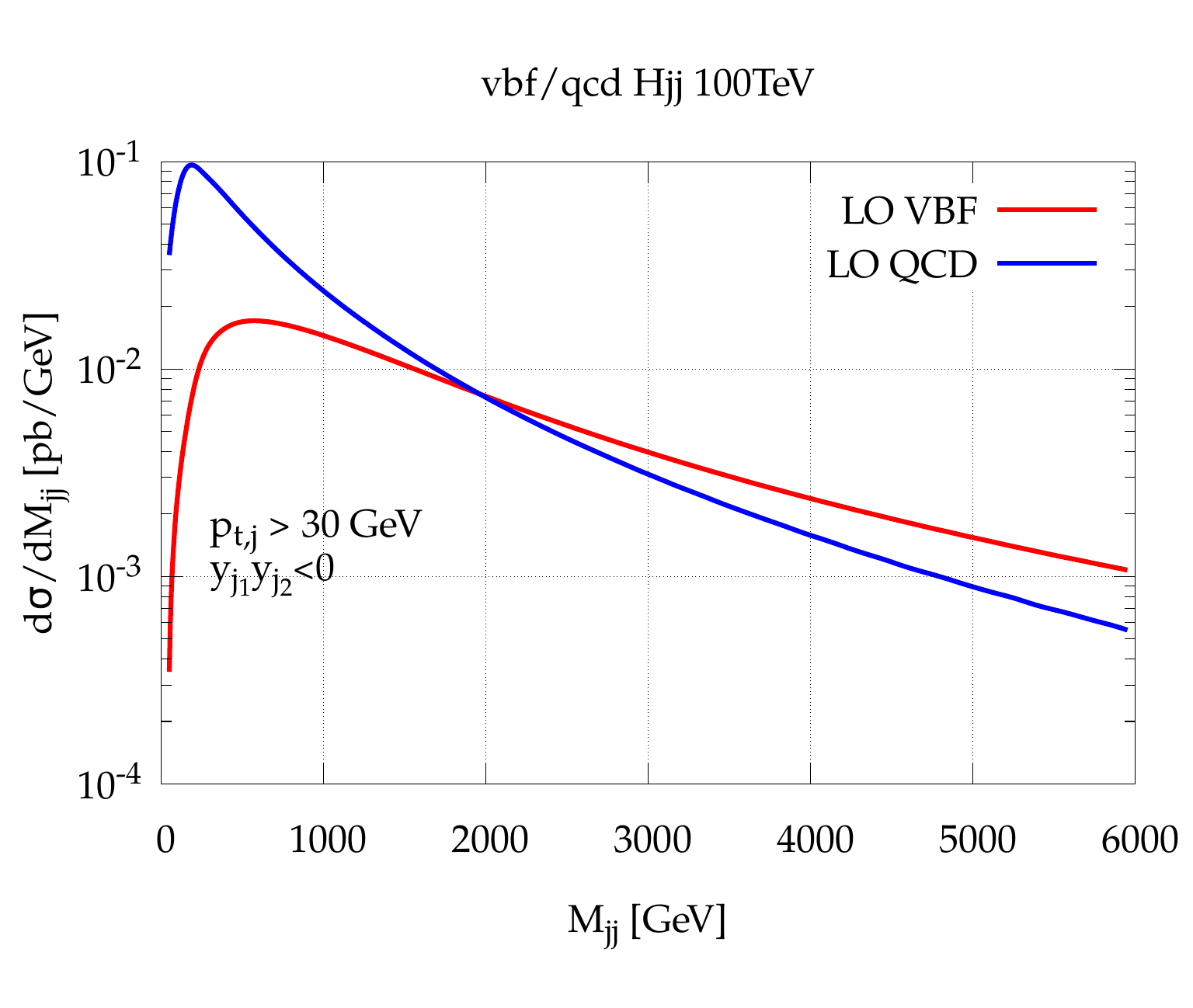}
  \end{minipage}
  \begin{minipage}{0.49\textwidth}
    \centering
    \includegraphics[width=1\textwidth,page=16]{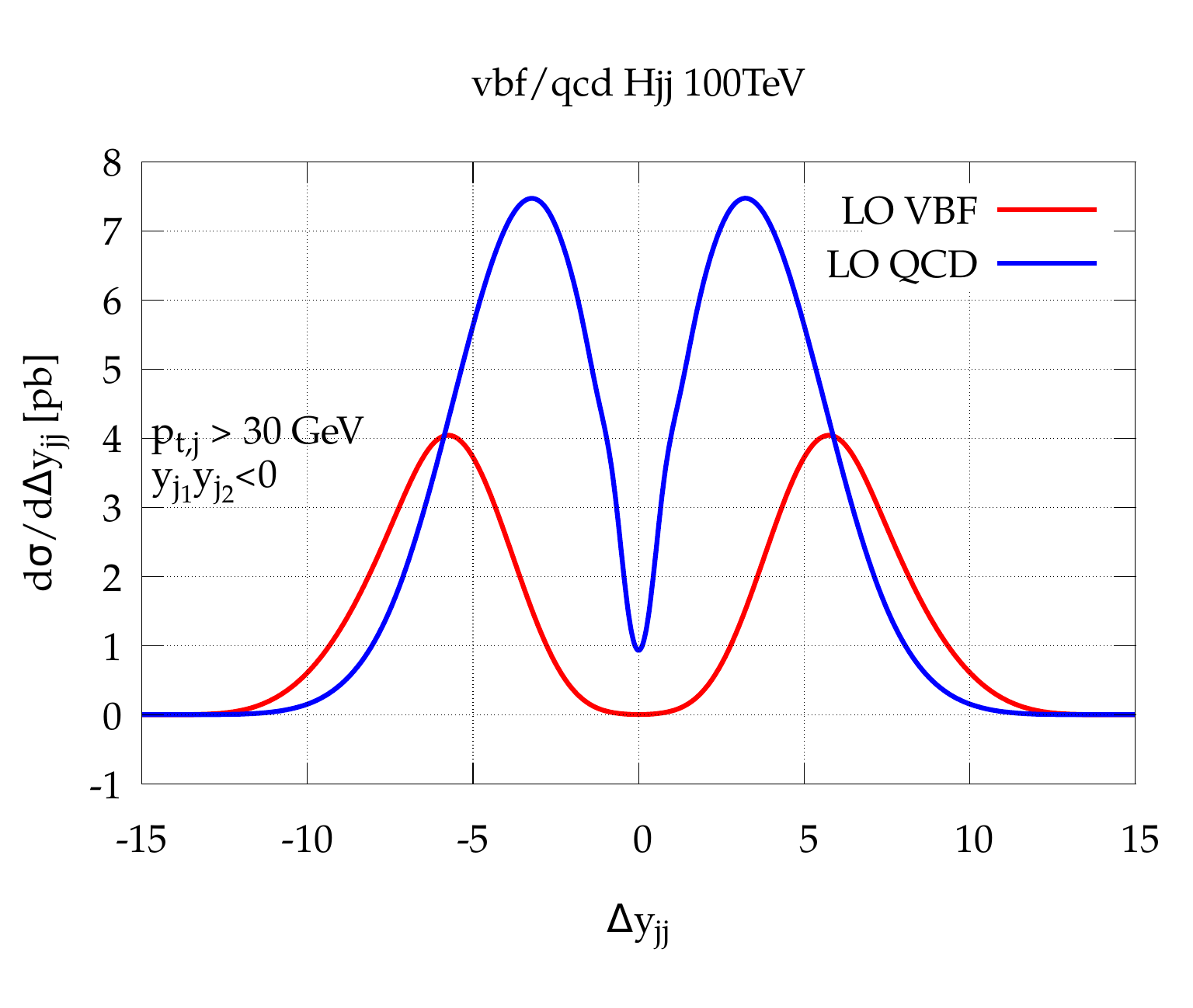}
  \end{minipage}
  \caption{\label{fig:VBFcuts1}\textbf{Left}: The invariant dijet mass $M_{jj}$ of the two hardest jets in VBF and QCD $Hjj$ production at $100 \TeV$. \textbf{Right}: The rapidity separation of the two hardest jets $\Delta y_{jj}$ in VBF and QCD $Hjj$ production at $100 \TeV$.}
\end{figure}

To establish VBF cuts at $100 \TeV$ we first study the variables which are typically used at the LHC. These are the dijet invariant mass, $M_{jj}$, the rapidity separation between the two leading jets, $\Delta y_{jj}$, the separation between the two leading jets in the rapidity-azimuthal angle plane, $\Delta R_{jj}$ and the azimuthal angle between the two leading jets $\phi_{jj}$. In \cref{fig:VBFcuts1} we show $M_{jj}$ and $\Delta y_{jj}$ after applying a cut on the two leading jets of $\pt > 30 \GeV$ and requiring that the two leading jets are in opposite detector hemispheres. This last cut removes around $60 \%$ of the background while retaining about $80 \%$ of the signal. 

%%%%%%%%%%%%%%%%%%%%%%%%%%                                                                               
%%%%%%%%%%%%%%%%%%%%%%%%%%   

\begin{figure}[!th]
  \begin{minipage}{0.49\textwidth}
    \centering
    \includegraphics[width=1\textwidth,page=33]{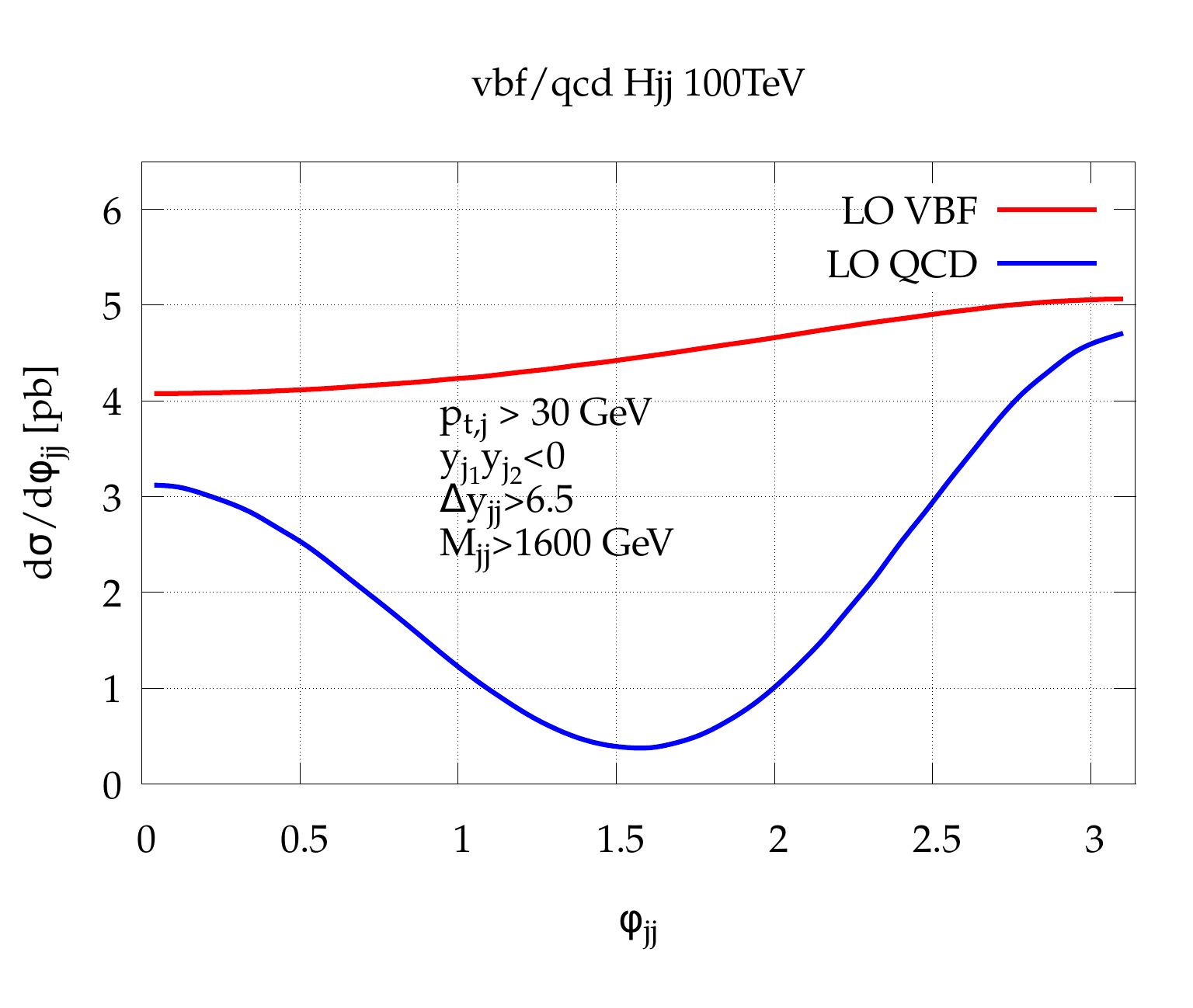}
  \end{minipage}
  \begin{minipage}{0.49\textwidth}
    \centering
    \includegraphics[width=1\textwidth,page=34]{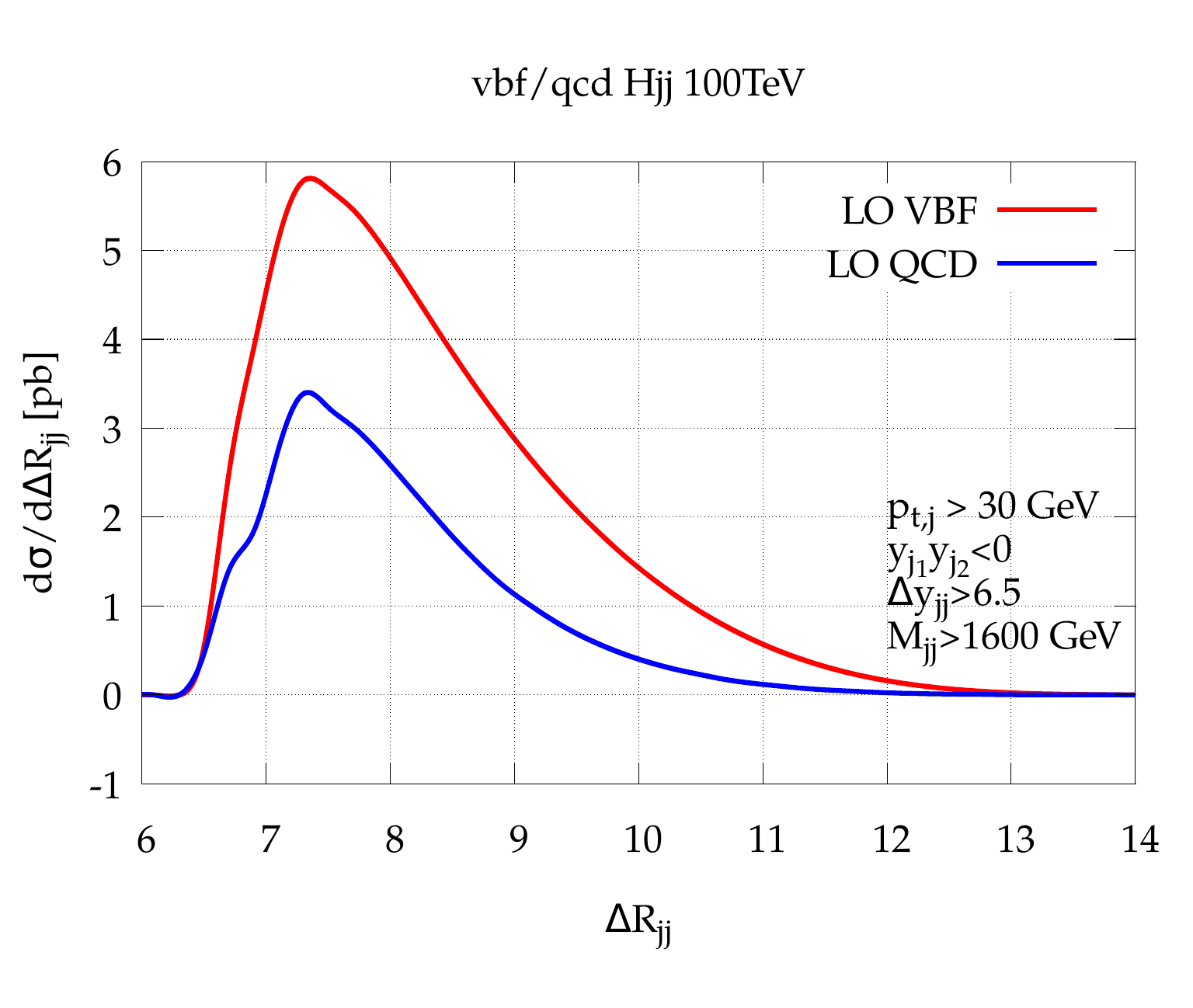}
  \end{minipage}
  \caption{\label{fig:VBFcuts2}\textbf{Left}: The azimuthal angle $\phi_{jj}$ between the two hardest jets in VBF and QCD $Hjj$ production at $100 \TeV$. \textbf{Right}: The rapidity-azimuthal angle separation of the two hardest jets $\Delta R_{jj}$ in VBF and QCD $Hjj$ production at $100 \TeV$.}
\end{figure}

In order to suppress the QCD background a cut of $\Delta y_{jj} > 6.5$ is
imposed. This cut also significantly reduces the QCD $M_{jj}$ peak and shifts
the VBF peak to about $2400 \GeV$. In order to further suppress the QCD
background we impose $M_{jj} > 1600 \GeV$. After these cuts have been applied,
and requiring $p_{\mathrm{T},j} > 30 \GeV$, the VBF signal to QCD background
ratio is roughly 3 with a total NNLO-QCD VBF cross section of about $12$
pb. From \cref{fig:VBFcuts2} it is clear that one could also impose a cut on
$\phi_{jj}$ to improve the suppression whereas a cut on $\Delta R_{jj}$ would
not help to achieve that. We hence state the VBF cuts that we will be using
throughout this section are

\begin{align}
  M_{j_1 j_2} > 1600 \GeV, \qquad \Delta y_{j_1 j_2} > 6.5, \qquad y_{j_1}y_{j_2} < 0.
  \label{eq:VBFcuts}
\end{align}

where $j_1$ is the hardest jet in the event and $j_2$ is the second hardest
jet. At a $13 \TeV$ machine the VBF cross section is $\mathcal{O}(1\;\mathrm{pb})$
  under typical VBF cuts and the QCD $Hjj$ background roughly a factor six
  smaller.

%%%%%%%%%%%%%%%%%%%%%%%%%%                                                                               
%%%%%%%%%%%%%%%%%%%%%%%%%%   

In \cref{tab:fidVBF_fcc} we show the fiducial cross section obtained after applying the VBF cuts of \cref{eq:VBFcuts} to VBF and QCD $Hjj$ production. The cross sections are reported at the three different jet $\pt$ cut values $\{30,50,100\} \GeV$. All numbers are computed at LO. It is clear from the table that requiring a somewhat higher jet $\pt$ cut than $30 \GeV$ leads to a lower $S/\sqrt{B}$ ratio. In going from $30 \GeV$ to $50 \GeV$ this reduction is however small. 

\begin{table}[!th]
  \caption{Fiducial VBF and QCD $Hjj$ cross sections for a $100 \TeV$ proton-proton collider at LO under the VBF cuts of \cref{eq:VBFcuts}. $S/\sqrt{B}$ is defined as the ratio between the VBF signal and the square root of the QCD background at an integrated luminosity of $20\mbox{ ab}^{-1}$.}
  \label{tab:fidVBF_fcc}
  \begin{center}%
    \begin{small}%
      \tabcolsep5pt
      \resizebox{\columnwidth}{!}{
      \begin{tabular}{l|ccc}%
        \hline
 & $\sigma(p_{\mathrm{T},j} > 30 \GeV)$ [pb] & $\sigma(p_{\mathrm{T},j} > 50 \GeV)$ [pb]& $\sigma(p_{\mathrm{T},j} > 100 \GeV)$ [pb] 
        \\
        \hline
        VBF & $14.1$ & $7.51$ & $1.08$ 
        \\
        QCD $Hjj$ & $5.04$ & $1.97$ & $0.331$ 
        \\
        \hline
        $S/\sqrt{B}@(20 \mbox{ ab}^{-1})$ & $28100$ & $24200$ & $8500$ 
        \\
        \hline
      \end{tabular}}%
    \end{small}%
  \end{center}%
\end{table}

In \cref{tab:totVBF_fcc} we show for comparison the cross sections obtained after only applying the three jet $\pt$ cuts. As expected the VBF signal is drowned in the QCD background. It is worth noticing that the $S/\sqrt{B}$ ratio is still very large when one assumes an integrated luminosity of $20\mbox{ ab}^{-1}$ and that it declines as the jet cut is increased. 

\begin{table}[!th]
  \caption{Total VBF and QCD $Hjj$ cross sections for a $100 \TeV$ proton-proton collider at LO with a cut on the two hardest jets. The numbers are obtained using the {\tt CT14nnlo} PDF. $S/\sqrt{B}$ is defined as the ratio between the VBF signal and the square root of the QCD background at an integrated luminosity of $20\mbox{ ab}^{-1}$.}
  \label{tab:totVBF_fcc}
  \begin{center}%
    \begin{small}%
      \tabcolsep5pt
      \resizebox{\columnwidth}{!}{
      \begin{tabular}{l|ccc}%
        \hline
 & $\sigma(p_{\mathrm{T},j} > 30 \GeV)$ [pb] & $\sigma(p_{\mathrm{T},j} > 50 \GeV)$ [pb]& $\sigma(p_{\mathrm{T},j} > 100 \GeV)$ [pb] 
        \\
        \hline
        VBF & $51.3$ & $28.5$ & $5.25$ 
        \\
        QCD $Hjj$ & $166$ & $78.6$ & $23.9$ 
        \\
        \hline
        $S/\sqrt{B}@(20 \mbox{ ab}^{-1})$ & $17900$ & $14300$ & $4900$ 
        \\
        \hline
      \end{tabular}}%
    \end{small}%
  \end{center}%
\end{table}

\subsubsection{Perturbative Corrections}
The results shown in the previous section were all computed at LO. Here we
briefly investigate the impact of NNLO-QCD, NLO-EW and parton shower corrections
to the VBF cross section computed with $p_{\mathrm{T},j} > 30 \GeV$ and under the
VBF cuts of \cref{eq:VBFcuts} at a $100 \TeV$ collider. We also compare
to the NLO-QCD predictions for QCD $Hjj$ production.

In \cref{tab:vbf_XStot_2} we show the best prediction for
$\sigma^{\mathrm{VBF}}$ as obtained by \cref{eq:sigmaVBF} and compare it to the
same cross section obtained by showering \POWHEG{} events with \textsc{PYTHIA6}
but including no effects beyond the parton shower itself. The NLO-EW and
NNLO-QCD corrections are found to be of roughly the same order, and amount to a
total negative correction of $\sim 23\%$. As was the case for the inclusive
cross section, the corrections are a factor two larger than at $14 \TeV$. Even
though the perturbative corrections to QCD $Hjj$ production are negative, the
effect of including higher order corrections to both VBF and QCD $Hjj$
production is that the $S/\sqrt{B}$ ratio at an integrated luminosity of
$20\mbox{ ab}^{-1}$ is decreased from $28100$ to $24300$.

\begin{figure}[!th]
  \begin{minipage}{0.49\textwidth}
    \centering
    \includegraphics[width=1\textwidth,page=1]{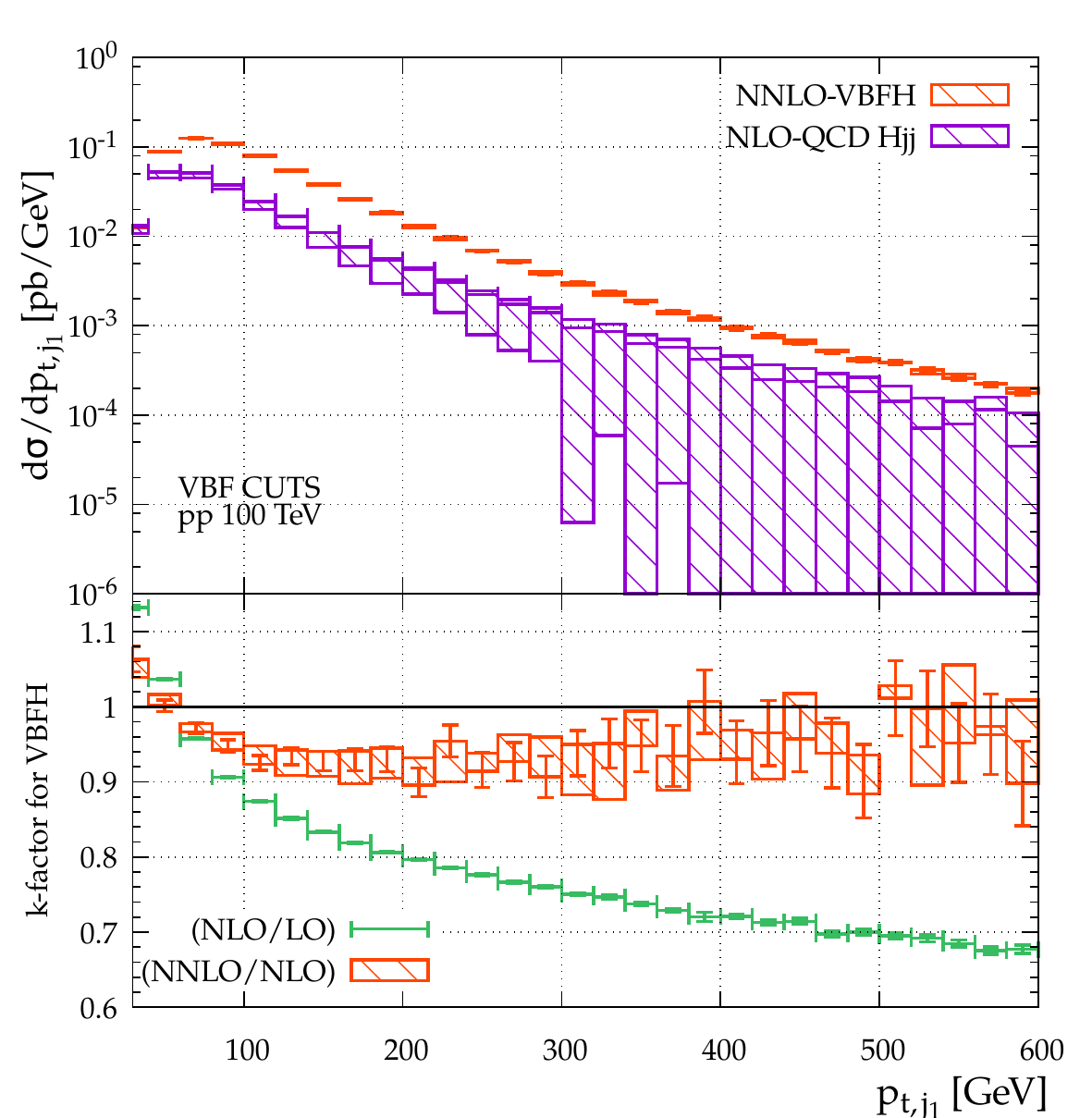}
  \end{minipage}
  \begin{minipage}{0.49\textwidth}
    \centering
    \includegraphics[width=1\textwidth,page=2]{VBFH/figs/VBF/diff-plots-100TeV.pdf}
  \end{minipage}
  \caption{\label{fig:VBFQCD1} Comparison between NNLO predictions for VBF
    production and NLO predictions for QCD $Hjj$ production under the VBF cuts
    of \cref{eq:VBFcuts}. The bands represent scale uncertainties obtained by
    varying $\mu_F=\mu_R$ by a factor two up and down. For the VBF production
    the statistical uncertainty is represented by the vertical line. No
    statistical uncertainties are shown for the QCD $Hjj$ result. The lower
    panel shows the k-factor for VBF production going from LO to NLO and NLO to
    NNLO. \textbf{Left}: Transverse momentum of the leading jet. \textbf{Right}:
    Transverse momentum of the subleading jet.}
\end{figure}

\begin{figure}[!th]
  \begin{minipage}{0.49\textwidth}
    \centering
    \includegraphics[width=1\textwidth,page=3]{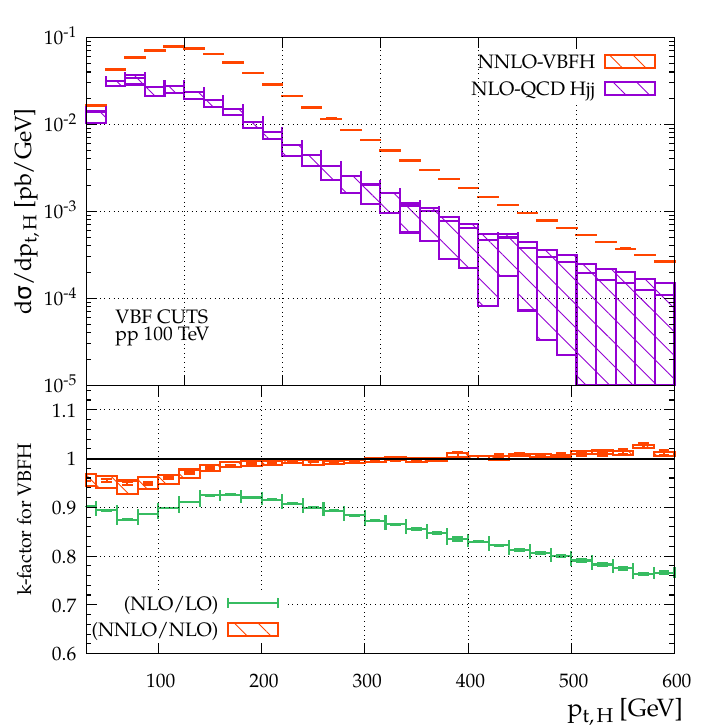}
  \end{minipage}
  \begin{minipage}{0.49\textwidth}
    \centering
    \includegraphics[width=1\textwidth,page=4]{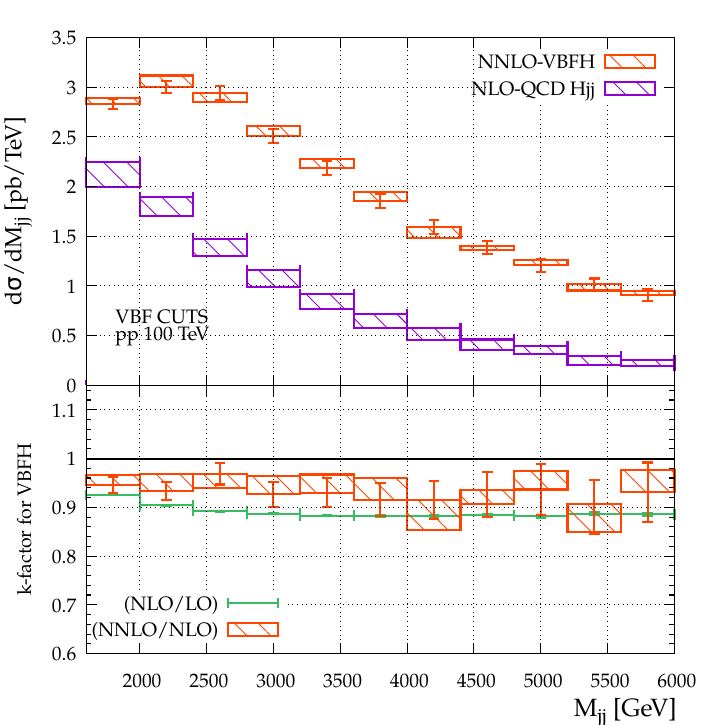}
  \end{minipage}
  \caption{\label{fig:VBFQCD2} Similar to \cref{fig:VBFQCD1}. \textbf{Left}:
  Transverse momentum of the Higgs Boson. \textbf{Right}: Invariant mass of the
  dijet pair.}
\end{figure}

\begin{figure}[!th]
  \begin{minipage}{0.49\textwidth}
    \centering
    \includegraphics[width=1\textwidth,page=5]{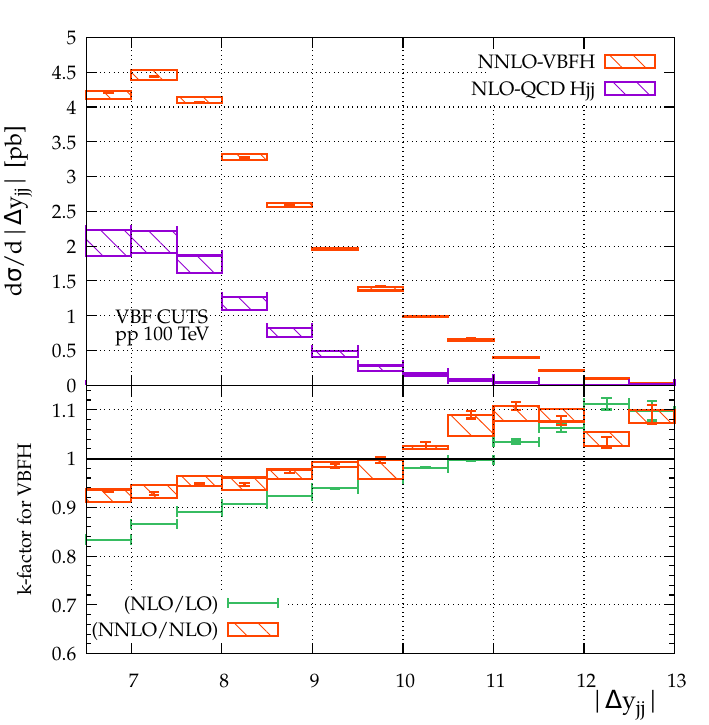}
  \end{minipage}
  \begin{minipage}{0.49\textwidth}
    \centering
    \includegraphics[width=1\textwidth,page=6]{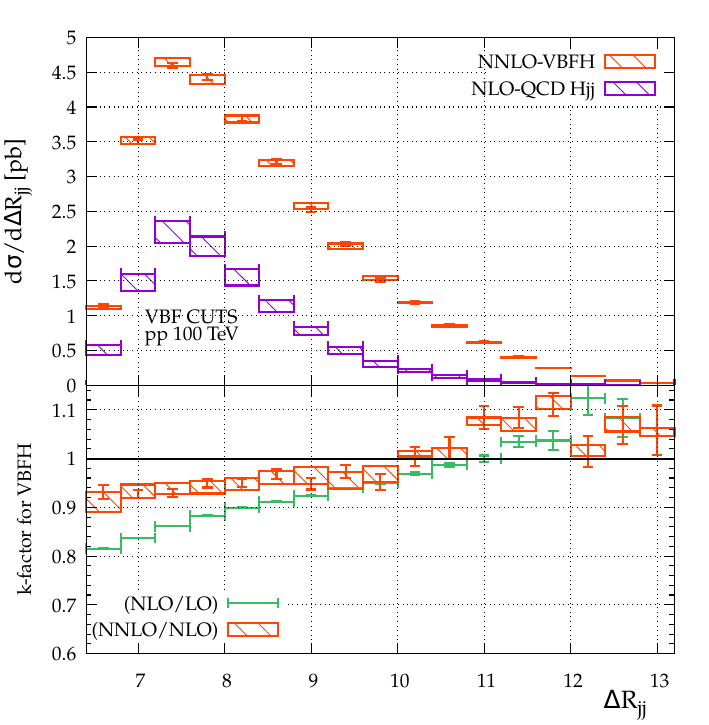}
  \end{minipage}
  \caption{\label{fig:VBFQCD3} Similar to \cref{fig:VBFQCD1}. \textbf{Left}:
  Absolute value of the rapidity separation between the two leading
  jets. \textbf{Right}: Distance between the two leading jets in the
  rapidity-azimuthal plane.}
\end{figure}

\begin{figure}[!th]
  \begin{center}
  \begin{minipage}{0.49\textwidth}
    \centering
    \includegraphics[width=1\textwidth,page=7]{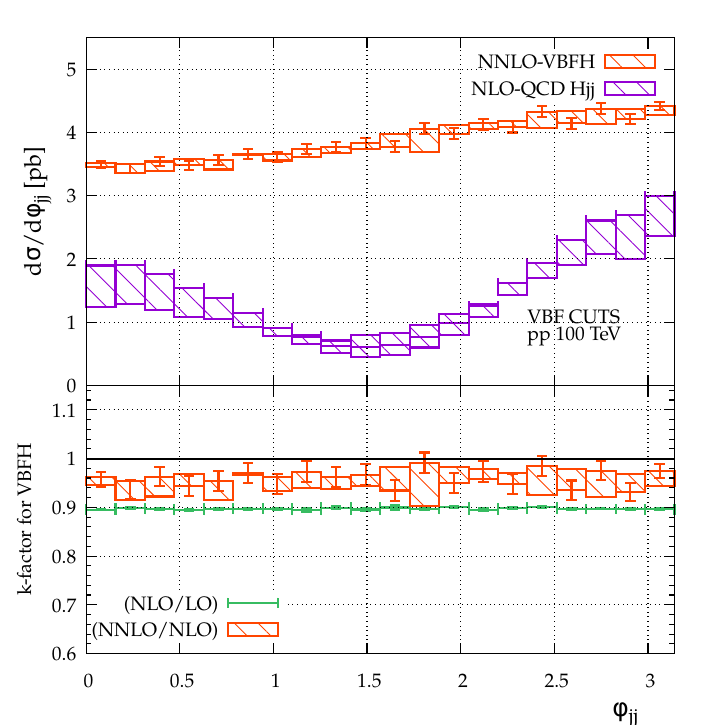}
  \end{minipage}
  \end{center}
  \caption{\label{fig:VBFQCD4} Similar to \cref{fig:VBFQCD1}. Shown here is the
  azimuthal angle between the two leading jets.}
\end{figure}

\begin{table}[!th]
  \caption{Fiducial VBF cross section including QCD and EW corrections
    and their uncertainties for a $100 \TeV$ proton-proton collider. For comparison the QCD induced Hjj cross section is also shown. At fixed-order QCD corrections are included at NNLO and EW corrections at NLO. }
  \label{tab:vbf_XStot_2}
  \begin{center}%
    \begin{small}%
      \tabcolsep5pt
      \begin{tabular}{l|ccccc}%
        \hline
 Process & $\sigma^{\mathrm{fid}}$[pb] & $\Delta_{\mathrm{scale}}$[\%] &$\sigma_{\mathrm{QCD}}$[pb] & $\delta_{\mathrm{EW}}$[\%] & $\sigma_{\gamma}$[pb]
        \\
        \hline
        VBF (NNLO-QCD/NLO-EW) & $10.8$ &$\pm 1.0$ & $12.1$ & $-12.6$ & $0.22$ 
        \\
        VBF (\NLOPS{}) & $11.9$ &$^{+0.56}_{-0.41}$ & $11.9$ & - & - 
        \\
        QCD $Hjj$ (NLO)& $4.79$ &$^{+0}_{-23}$ & $4.79$ & - & - 
        \\
        \hline
      \end{tabular}%
    \end{small}%
  \end{center}%
\end{table}
In \cref{fig:VBFQCD1,fig:VBFQCD2,fig:VBFQCD3,fig:VBFQCD4} we show comparisons between VBF and
QCD $Hjj$ production computed at NNLO and NLO in QCD respectively. We have applied
the VBF cuts of \cref{eq:VBFcuts}. Also shown is the k-factor for VBF
production going from LO to NLO and NLO to NNLO. Note that the QCD $Hjj$
predictions have been obtained in the effective theory where the top quark is
treated as infinitely heavy and hence the $\pt$ spectra should not be trusted
beyond $2 M_t$. As can be seen from the plots the VBF cuts have suppressed the
background QCD $Hjj$ production in all corners of phase space. One could still
imagine further optimising these cuts, for example by requiring $\phi_{jj}$ in
the vicinity of $\frac{\pi}{2}$ or a slightly larger invariant dijet mass. We
note in particular that requiring that the Higgs Boson has a transverse momentum
greater than $40 \GeV$ seems to favour the VBF signal. Since a cut on the
transverse momentum of the decay products of the Higgs would in any case have
to be imposed, this improves the efficiency of the VBF cuts in realistic
experimental setups.

%%%%%%%%%%%%%%%%%%%%%%%%%%                                                                             
%%%%%%%%%%%%%%%%%%%%%%%%%%

%%%%%%%%%%%%%%%%%%%%%%%%%%                                                                             
%%%%%%%%%%%%%%%%%%%%%%%%%%

%%%%%%%%%%%%%%%%%%%%%%%%%%                                                                             
%%%%%%%%%%%%%%%%%%%%%%%%%%

%%%%%%%%%%%%%%%%%%%%%%%%%%                                                                             
%%%%%%%%%%%%%%%%%%%%%%%%%%

\subsubsection{Differential Distributions}
\begin{figure}[!thb]
  \begin{minipage}{0.49\textwidth}
    \centering
    \includegraphics[width=1\textwidth,page=5]{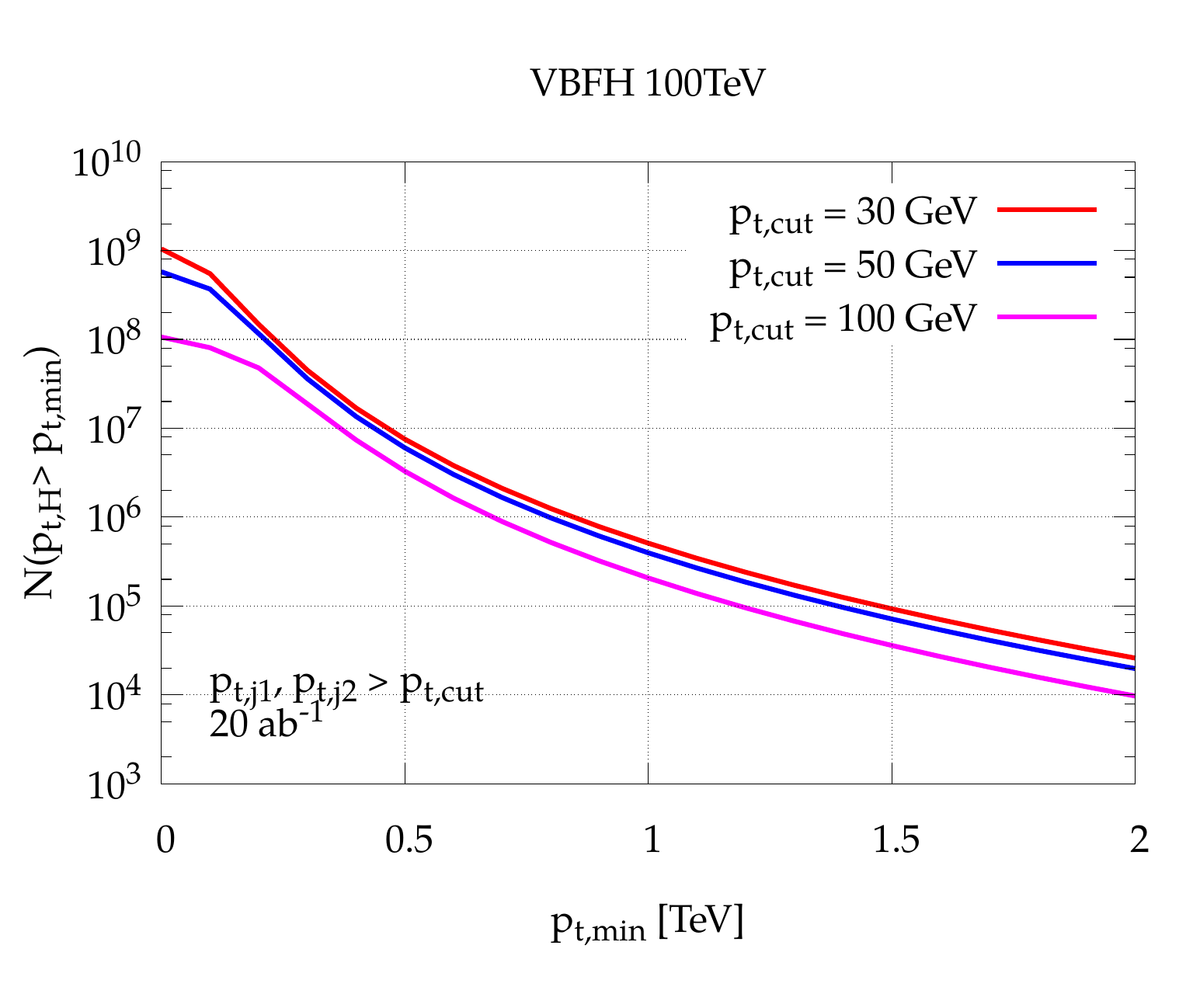}
  \end{minipage}
  \begin{minipage}{0.49\textwidth}
    \centering
    \includegraphics[width=1\textwidth,page=6]{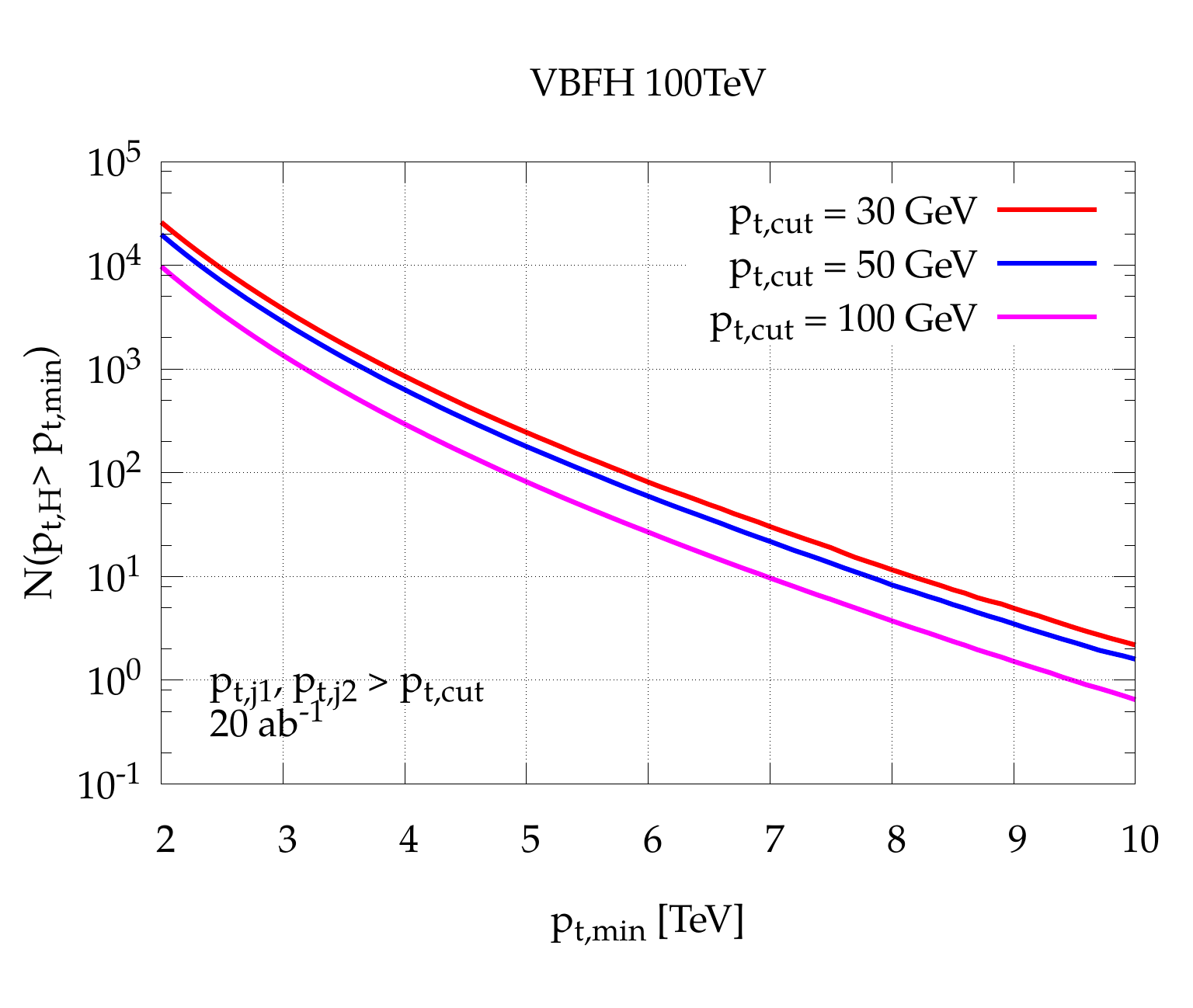}
  \end{minipage}
  \caption{\label{fig:VBFptHaccum} The total number of VBF events produced with
    $p_{{t,H}} > p_{{t,\mathrm{min}}}$ at a $100 \TeV$ collider with an integrated
    luminosity of $20\mbox{ ab}^{-1}$ under three different jet $p_{\mathrm{T}}$
    cuts. Left panel: $p_{\mathrm{T},H}$ in the range $0-2 \TeV$. Right panel:
    $p_{\mathrm{T},H}$ in the range $2-10 \TeV$.}
\end{figure}

\begin{figure}[!thb]
  \begin{minipage}{0.49\textwidth}
    \centering
    \includegraphics[width=1\textwidth,page=7]{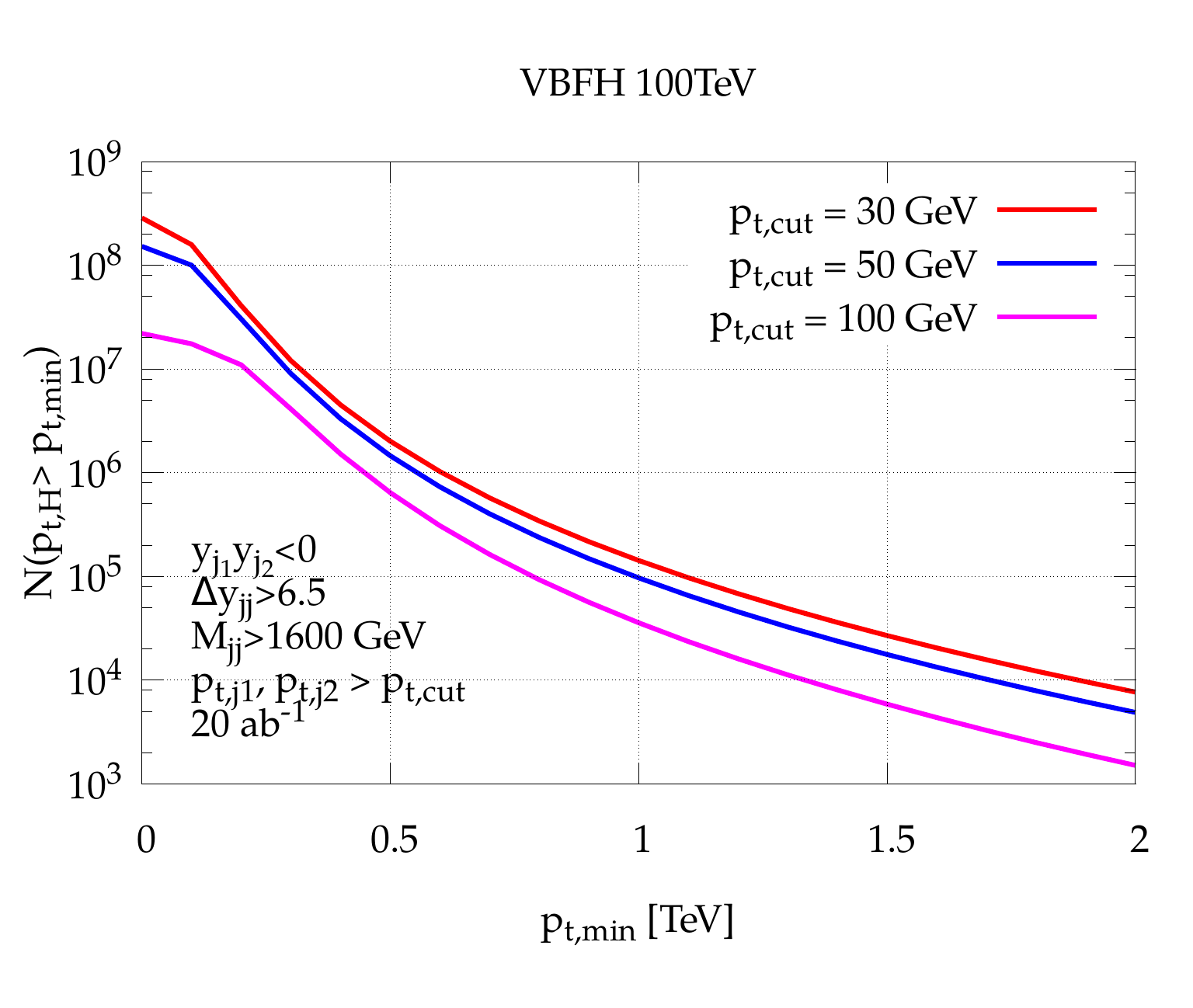}
  \end{minipage}
  \begin{minipage}{0.49\textwidth}
    \centering
    \includegraphics[width=1\textwidth,page=8]{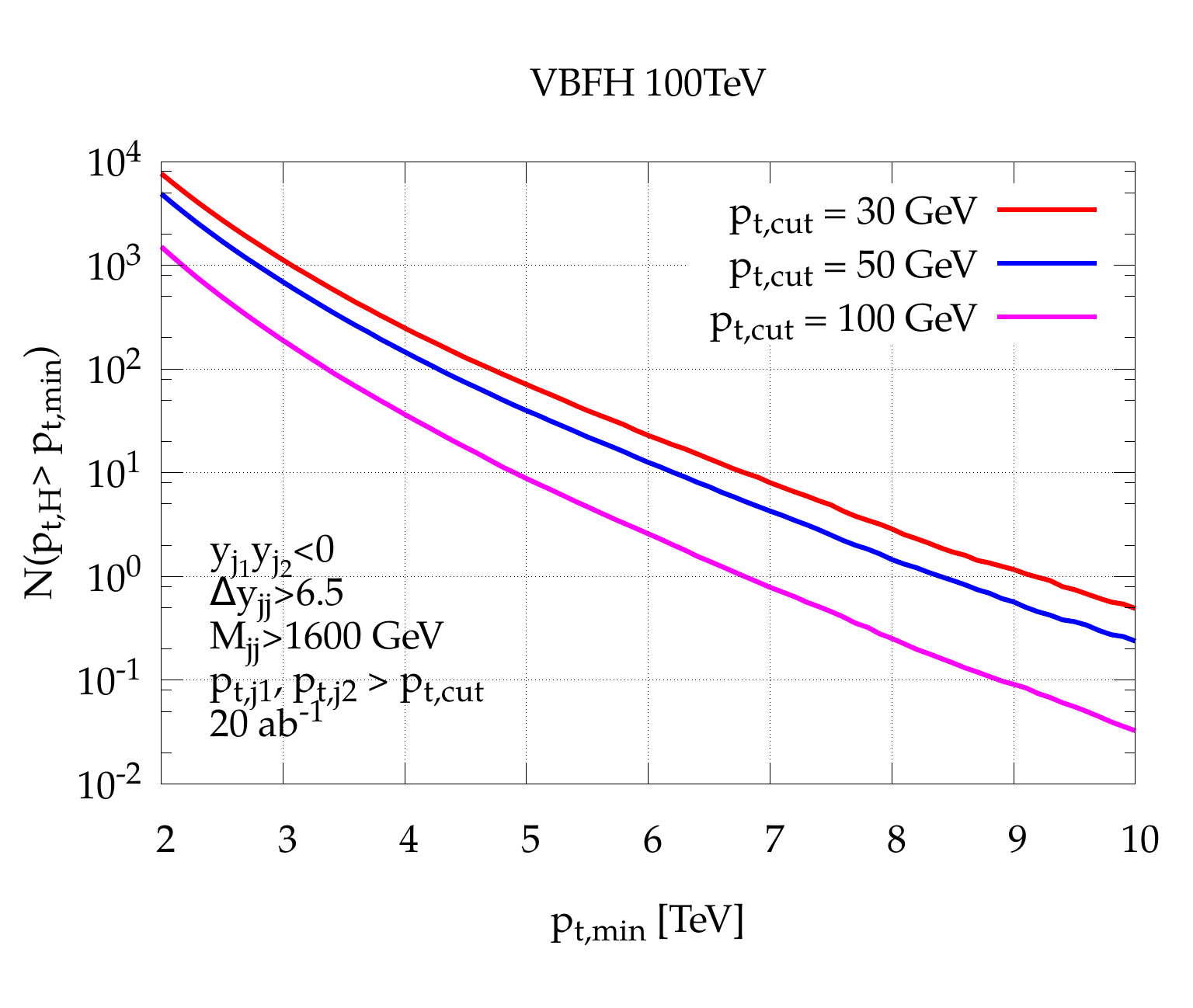}
  \end{minipage}
  \caption{\label{fig:VBFptHaccumVBF} The total number of VBF events produced with $p_{{t,H}} > p_{{t,\mathrm{min}}}$ at a $100 \TeV$ collider with an integrated luminosity of $20\mbox{ ab}^{-1}$under three different jet $p_{\mathrm{T}}$ cuts and with the VBF cuts of \cref{eq:VBFcuts} applied. Left panel: $p_{\mathrm{T},H}$ in the range $0-2 \TeV$. Right panel: $p_{\mathrm{T},H}$ in the range $2-10 \TeV$.}
\end{figure}

In addition to the distributions already presented in the previous section, we
here show a number of distributions to indicate the kinematical reach of the VBF
channel at $100 \TeV$. Assuming an integrated luminosity of $20$ ab$^{-1}$ we study
how many events will be produced with a Higgs whose transverse momentum exceeds
$p_{{t,\mathrm{min}}}$. In \cref{fig:VBFptHaccum,fig:VBFptHaccumVBF} we show
this distribution for various cut configurations. This variable is particularly
interesting in the context of anomalous couplings in the weak sector. It can be
seen that even under VBF cuts and requiring hard jets, a number of Higgs bosons
with transverse momentum of the order $6 \TeV$ will be produced in this
scenario.

\begin{figure}[!thb]
  \begin{minipage}{0.49\textwidth}
    \centering
    \includegraphics[width=1\textwidth,page=9]{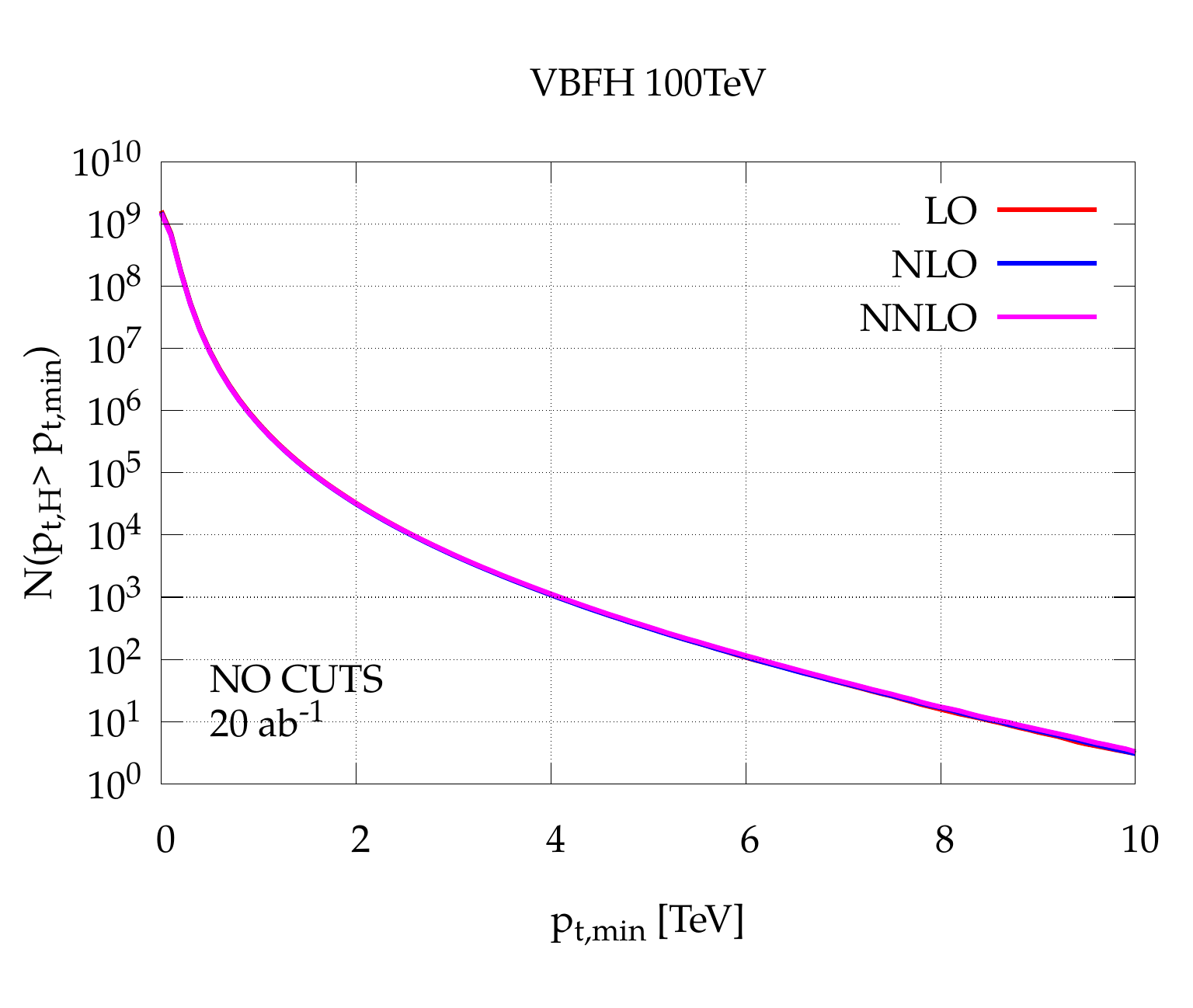}
  \end{minipage}
  \begin{minipage}{0.49\textwidth}
    \centering
    \includegraphics[width=1\textwidth,page=10]{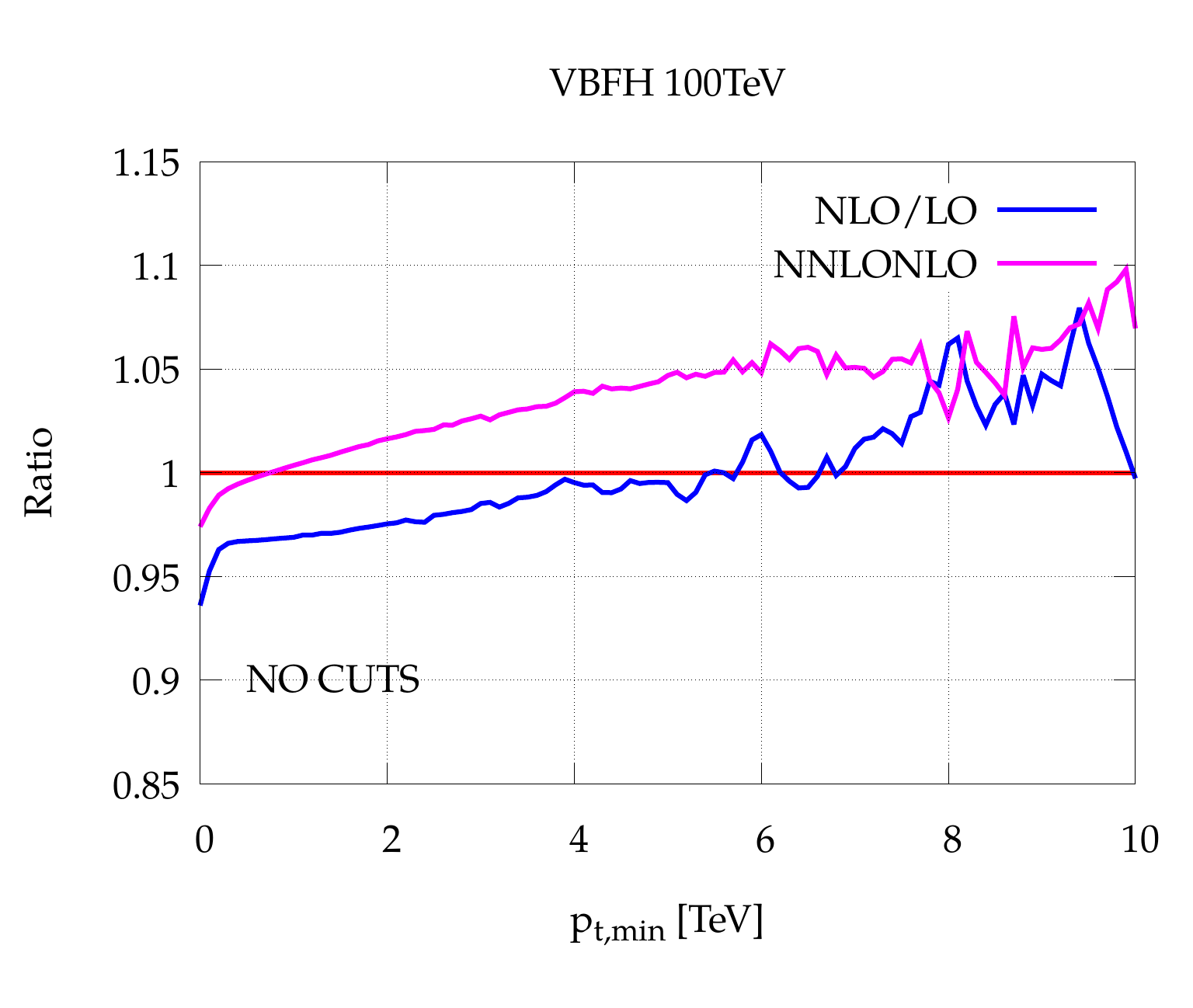}
  \end{minipage}
  \caption{\label{fig:VBFptHcorr} The total number of VBF events produced with $p_{{t,H}} > p_{{t,\mathrm{min}}}$ at a $100 \TeV$ collider with an integrated luminosity of $20\mbox{ ab}^{-1}$ with no cuts applied. Left panel: Spectrum computed at LO, NLO and NNLO in QCD. Due to the small corrections the difference between the three curves is hard to see by eye. Right panel: The k-factor going from LO to NLO and NLO to NNLO.}
\end{figure}

In \cref{fig:VBFptHcorr} we show the same distribution but fully
inclusively and at various perturbative orders. Also shown is the k-factor going
from LO to NLO and from NLO to NNLO. The perturbative corrections to this
variable are modest as it is not sensitive to real radiation at the inclusive
level. After applying VBF cuts and jet cuts the low $p_{{\mathrm{T},H}}$-spectrum
receives moderate corrections whereas the corrections at larger values of
$p_{{\mathrm{T},H}}$ can become very large as indicated in \cref{fig:VBFQCD2}.

\begin{figure}[!thb]
  \begin{minipage}{0.49\textwidth}
    \centering
    \includegraphics[width=1\textwidth,page=4]{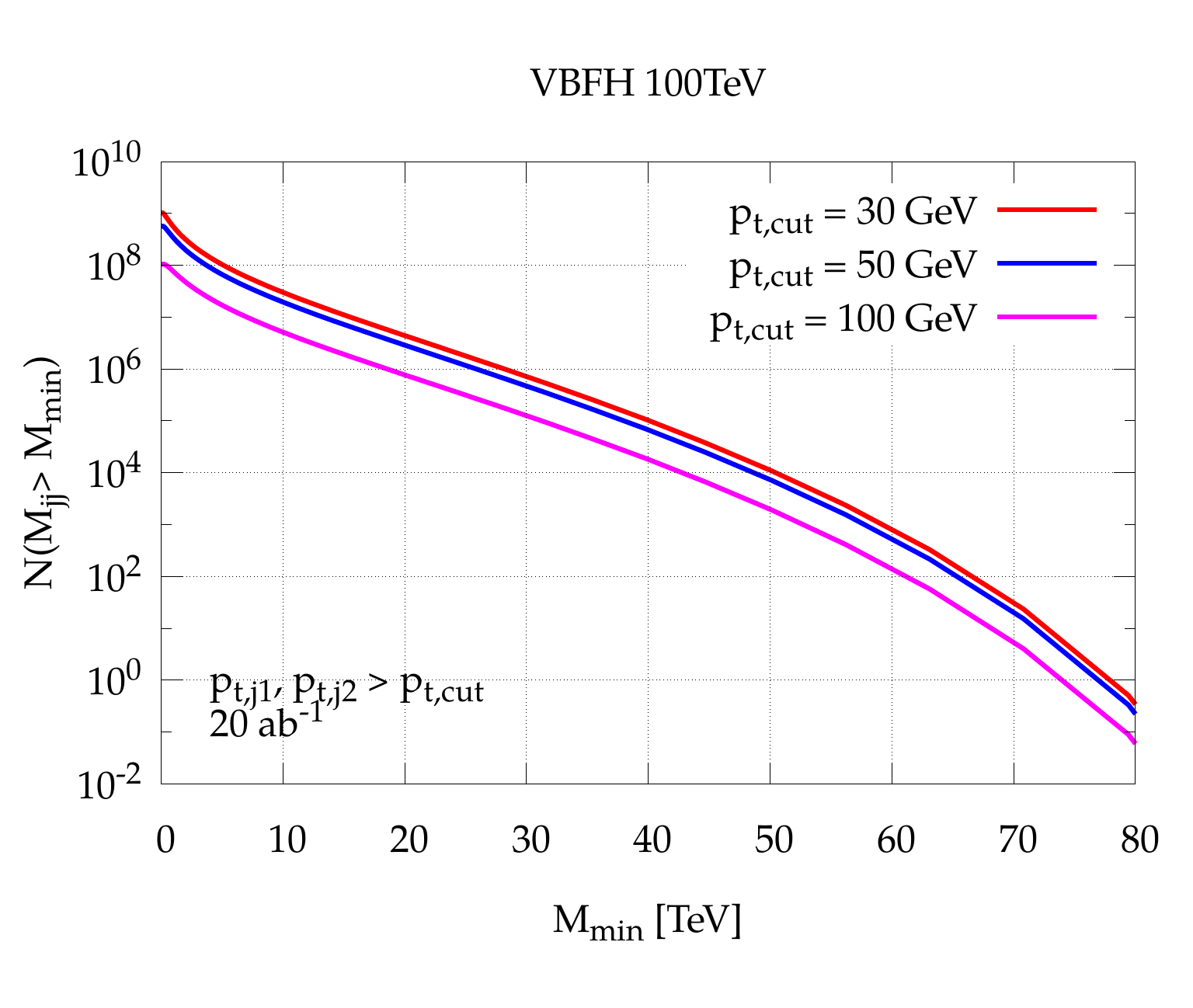}
  \end{minipage}
  \begin{minipage}{0.49\textwidth}
    \centering
    \includegraphics[width=1\textwidth,page=3]{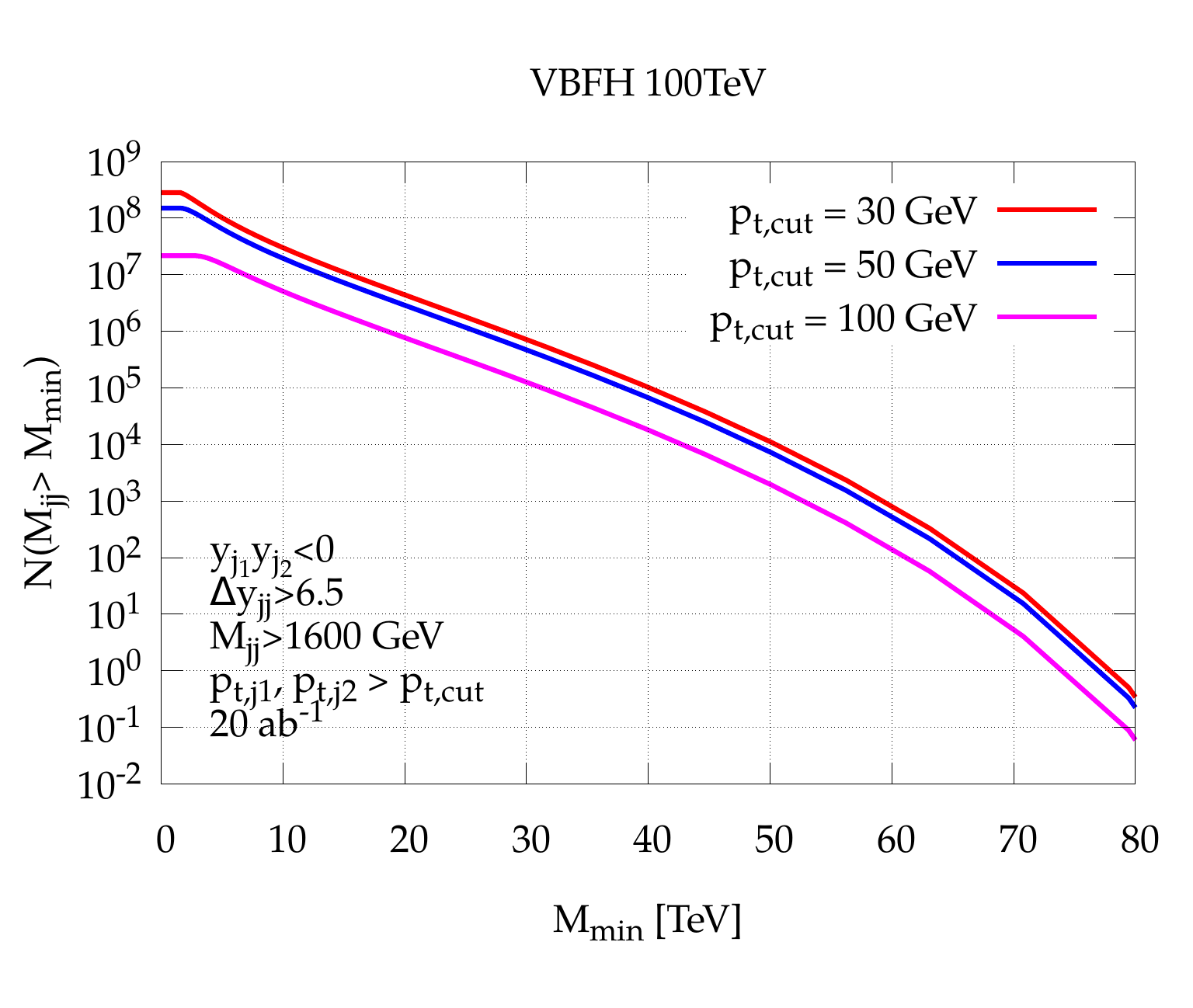}
  \end{minipage}
  \caption{\label{fig:VBFMjjaccum} The total number of VBF events produced with $M_{j_1 j_2} > M_{\mathrm{min}}$ at a $100 \TeV$ collider with an integrated luminosity of $20 \mbox{ ab}^{-1}$. Left panel: Three different jet $p_{\mathrm{T}}$ cuts applied but no VBF cuts applied. Right panel: VBF cuts of \cref{eq:VBFcuts} and three different jet $p_{\mathrm{T}}$ cuts applied.}
\end{figure}

In \cref{fig:VBFMjjaccum} we show how many events will be produced with a
dijet invariant mass exceeding $M_{\mathrm{min}}$ at various cut
configurations. Because the two hardest jets in the VBF event are typically the
proton remnants the invariant dijet mass can become very large. As can be seen
from the figure, even after applying VBF cuts and requiring very hard jets
hundreds of events with an invariant dijet mass larger than $60 \TeV$ is
expected. This is of interest when probing for BSM physics at the very highest
scales. It is also worth noticing that the tail of the distribution is almost
unaffected by the VBF cuts, as the VBF cuts are optimised to favour high
invariant dijet events.

%%%%%%%%%%%%%%%%%%%%%%%%%%                                                                             
%%%%%%%%%%%%%%%%%%%%%%%%%%

%%%%%%%%%%%%%%%%%%%%%%%%%%                                                                             
%%%%%%%%%%%%%%%%%%%%%%%%%%

%%%%%%%%%%%%%%%%%%%%%%%%%%                                                                             
%%%%%%%%%%%%%%%%%%%%%%%%%%

%%%%%%%%%%%%%%%%%%%%%%%%%%                                                                             
%%%%%%%%%%%%%%%%%%%%%%%%%%

\subsubsection{Detector Implications}
\begin{figure}[!thb]
  \begin{minipage}{0.49\textwidth}
    \centering
    \includegraphics[width=1\textwidth,page=1]{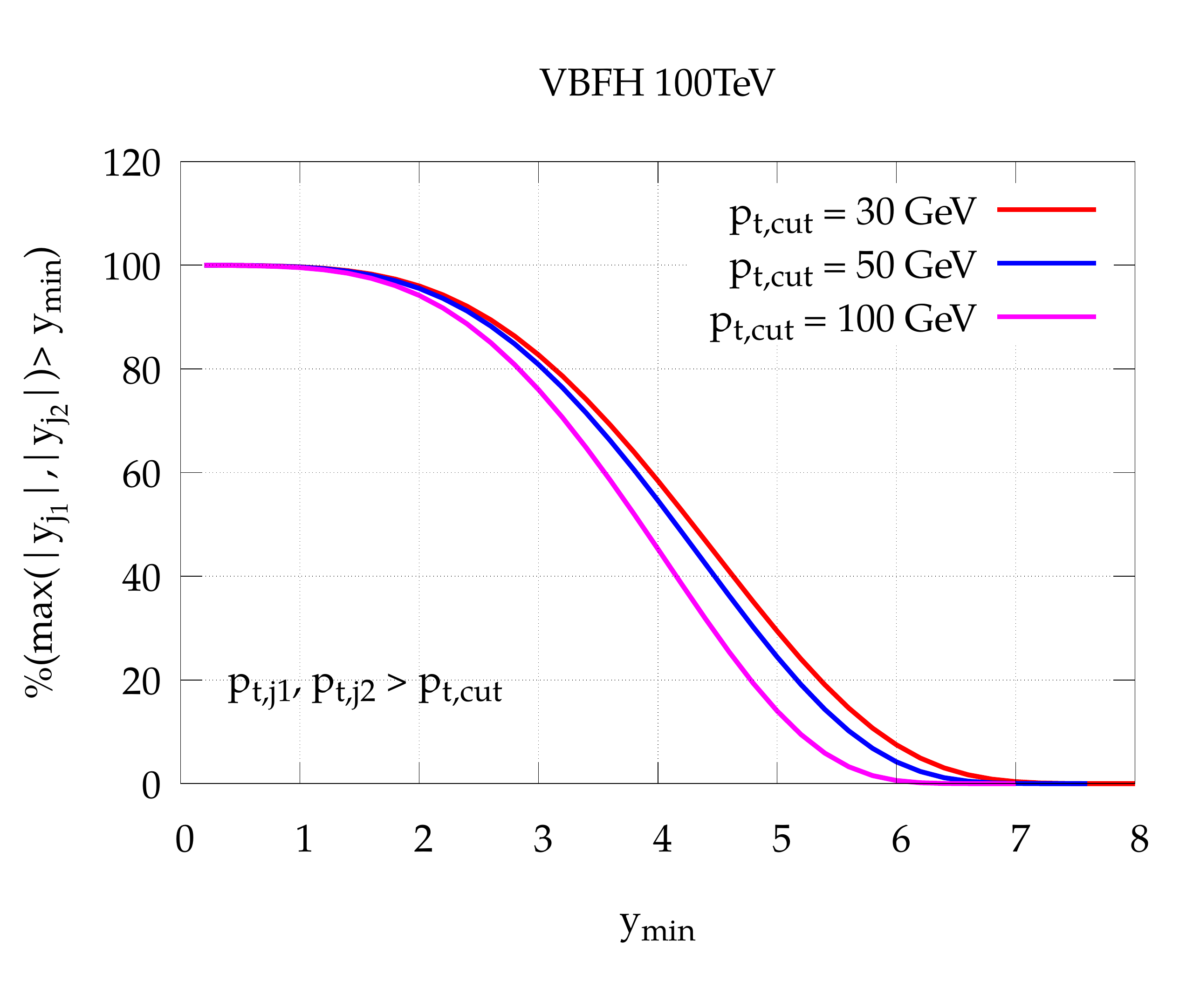}
  \end{minipage}
  \begin{minipage}{0.49\textwidth}
    \centering
    \includegraphics[width=1\textwidth,page=2]{VBFH/figs/VBF/vbf_100TeV_accumulated.pdf}
  \end{minipage}
  \caption{\label{fig:VBFmaxyj1yj2} The total fraction of events where $\mathrm{max}(|y_{j_1}|,|y_{j_2}|) > y_{\mathrm{min}}$ at a $100 \TeV$ collider.  Left panel: Three different jet $p_{\mathrm{T}}$ cuts applied but no VBF cuts applied. Right panel: VBF cuts of \cref{eq:VBFcuts} and three different jet $p_{\mathrm{T}}$ cuts applied.}
\end{figure}

The requirement that the two hardest jets are in opposite detector hemispheres
and are separated by at least $6.5$ units of rapidity, means that a symmetric
detector in the style of \textsc{ATLAS} or \textsc{CMS} must have a rapidity
reach well above $3.25$. In fact, looking at \cref{fig:VBFmaxyj1yj2},
which shows the fraction of events which satisfy
$\mathrm{max}(|y_{j_1}|,|y_{j_2}|) > y_{\mathrm{min}}$ for various cut
configurations, it becomes clear that a detector with a rapidity reach of $4.5$
would at best only retain $40\%$ of the VBF events after VBF cuts are applied.
Since a jet with $p_{\mathrm{T}}=30$~GeV can be produced at a rapidity of $\sim
8$ whereas a jet with $p_{\mathrm{T}}=100$~GeV can only be produced with
rapidities up to $\sim 6.8$, the required rapidity reach of the detector will
also depend on how well soft jets can be measured and controlled at $100 \TeV$. In
all cases a rapidity reach above 6 seems to be desirable.

%%%%%%%%%%%%%%%%%%%%%%%%%%                                                                             
%%%%%%%%%%%%%%%%%%%%%%%%%%   

\section{Conclusions}
With the calculation presented in this chapter, differential VBF Higgs
production has been brought to the same NNLO level of accuracy that
has been available for some time now for the
ggH~\cite{Catani:2007vq,Anastasiou:2004xq} and
VH~\cite{Ferrera:2011bk} production channels.
This constitutes the first fully differential NNLO $2\to 3$
hadron-collider calculation, an advance made possible thanks to the
factorisable nature of the process.
At both $13\TeV$ and $100\TeV$ the differential NNLO corrections are
non-negligible, $5$--$10\%$, i.e.\ an order of magnitude larger than the
corrections to the inclusive cross section. These corrections are found to be of
the same order of magnitude as the NLO-EW corrections.
Their size might even motivate a calculation one order higher, to N$^3$LO, which
we believe is within reach with the new ``projection-to-Born'' approach
introduced here.
It would also be of interest to obtain NNLO plus parton shower
predictions, again matching the accuracy achieved recently in
ggH~\cite{Hamilton:2013fea,Hoche:2014dla}.
Our studies of VBF at a $100\TeV$ collider showed that it is still possible to
suppress QCD backgrounds with suitable VBF cuts. However this suppression is
roughly a factor two worse than what can be obtained at a $14\TeV$ collider. In
addition to that, we found that experimental detectors with a rapidity reach in
excess of $6$ is needed to catch all of the VBF signal.

\part{Parton Shower Matching}
\chapter{The \POWHEG{} Method}\label{ch:powheg}
In this chapter we discuss how one might go about matching a fixed-order \NLO{}
calculation to a parton shower (\NLOPS{}). The reason such a matching is
desirable is obvious. Fixed-order calculations are very good at predicting
observables that are inclusive over QCD radiation, but can never make accurate
predictions for exclusive final states, due to the presence of soft and
collinear divergences in the fixed order results. On the other hand Shower Monte
Carlo programs can readily predict exclusive quantities through all order
resummations, currently done in the leading logarithmic approximation. The
showers are however limited to LO hard matrix elements at best, which do not
reproduce inclusive quantities with the desired accuracy. A matching of the two
approaches eliminate the deficiencies of both, retaining \NLO{} accuracy of
inclusive observables and exclusive final state generation.

Currently there exists two widely used methods for obtaining \NLOPS{} accurate
predictions: the \MCatNLO{} method~\cite{Frixione:2002ik} and the
\POWHEG{}\footnote{POsitive Weight Hardest Emission Generator}
method~\cite{Nason:2004rx,Frixione:2007vw}. Here we spend some time reviewing
the core results of the latter. In addition to that, a number of lesser used
prescriptions
exist~\cite{Dobbs:2001gb,Chen:2001nf,Kurihara:2002ne,Nagy:2005aa,Bauer:2006mk,Nagy:2007ty,Giele:2007di,Bauer:2008qh,Hoeche:2011fd},
some because they are relatively recent additions and others because development
halted after the initial ideas were presented. The \POWHEG{} method has been
implemented in two publicly available frameworks:
\noun{Sherpa}~\cite{Gleisberg:2008ta} and the \POWHEGBOX{}~\cite{Alioli:2010xd}.

\section{Matching NLO Calculations and Showers}\label{sec:POWHEG}
Deriving all the details of the \POWHEG{} method from first principles is a
rather involved affair. Here we will assume that the reader already has
familiarity with Shower Monte Carlos and only focus on the ingredients which are
particular to the \POWHEG{} method. The discussion below is therefore schematic
at best. For a thorough discussion we refer the reader
to~\cite{Marchesini:1983bm,Marchesini:1987cf,Nason:2004rx,Frixione:2007vw}.
\subsection{The Shower}
\begin{figure}[thb]
  \centering
  \includegraphics[width=0.7\textwidth]{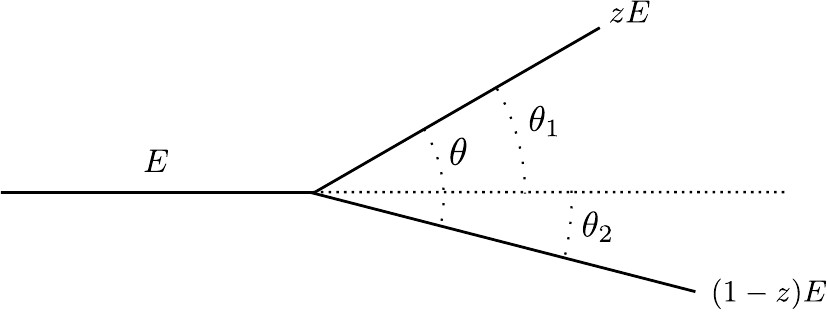}
  \caption{Illustration of a final state parton of energy $E$ splitting into two
    partons with energy fraction $z$ and $1-z$ respectively. }
  \label{fig:splitting}
\end{figure}
To set up the notation we start by briefly considering a splitting of an
incoming parton with energy $E$ into two partons with momentum fraction $z$ and
$1-z$ respectively as shown in \cref{fig:splitting}\footnote{In this
  chapter we will restrict ourselves to only discussing final state showers. The
  discussion generalises to initial state showers, cf. Section 6 of
  \Bref{Nason:2004rx}.}. By momentum conservation the transverse momentum of the
two outgoing partons with respect to the incoming parton will be
\begin{equation}
  \pt = \sin{\theta_1}zE = \sin{\theta_2}(1-z)E. 
\end{equation}
In the small angle approximation we have $\sin\theta \sim \theta$ from which it
follows that
\begin{equation}
  \theta = \frac{\theta_1}{1-z} = \frac{\theta_2}{z}.
\end{equation}
Defining $t=E^2\theta^2$ we have
\begin{equation}
  \pt = \sqrt{t}z(1-z),\label{eq:showerpt}
\end{equation}
and one can then show that the probability for a parton with ordering parameter
$t<t_{I}$ splitting into two partons with momentum fractions $z$ and $1-z$
is~\cite{Marchesini:1983bm}
\begin{equation}
  dP = F(z,t)dz\,dt\Delta_s(t_{I},t) = \frac{\as(\pt)}{2\pi}\hat{P}_{ab}(z)\frac{dt}{t}dz\Delta_s(t_{I},t).\label{eq:splitting}
\end{equation}
Here $\hat{P}_{ab}$ are the \emph{unregularised} Altarelli-Parisi splitting functions. They are intricately related to the splitting kernels of \cref{eq:split-fct-expansion}. To lowest order they are given by~\cite{Ellis:1991qj}
\begin{align}
  \hat{P}_{gg}(z) &= C_A \left[ \frac{1-z}{z}+\frac{z}{1-z} + z(z-1)\right]\,, \notag \\
  \hat{P}_{qg}(z) &= T_R \left[ z^2 + (1-z)^2\right]\,, \notag \\
  \hat{P}_{gq}(z) &= C_F \left[ \frac{1+(1-z)^2}{z}\right]\,, \notag \\
  \hat{P}_{qq}(z) &= C_F \left[ \frac{1+z^2}{1-z}\right]\,, 
  \label{eq:unreg-splitting}
\end{align}
where $C_F=4/3$, $C_A=3$ and $T_R = 1/2$. The Sudakov form factor, $\Delta_s$,
is the probability that no branching has taken place between $t_I$ and
$t$. Taking this as our ansatz we may now integrate \cref{eq:splitting} and
obtain
\begin{align}
  P(t,t_0) &= \int_{t_0}^{t}{F(z,t')dz\,dt'\Delta_s(t_I,t')} \Imp \notag \\
  & \Delta_s(t,t_0) = 1 - P(t,t_0) = 1 - \int_{t_0}^{t}{F(z,t')dz\,dt'\Delta_s(t_I,t')},
\end{align}
which has the solution
\begin{align}
\Delta_s(t,t_0) = e^{-\int_{t_0}^{t}{F(z,t')dz\,dt'}}.
\end{align}
$t_0$ is a lower cut-off to avoid the pole in $\as$. It is typically of the
order $1\GeV^2$. The angular ordered shower can now formally be expressed as the
following recursive equation\footnote{Following the notation
  of~\cite{Nason:2004rx} we have here dropped the azimuthal integration in the
  shower assuming that it is uniform in this variable.}
\begin{align}
  \field{S}(t_I) = \Delta_s(t_I,t_0)\bra{\field{I}} + \int_{t_0}^{t_I}{\Delta_s(t_I,t)F(z,t) \field{S}(z^2t)\field{S}((1-z)^2 t)dt\,dz},
  \label{eq:shower}
\end{align}
where $\bra{\field{I}}$ is the initial state parton. This equation can be solved
iteratively through ordinary Monte Carlo methods, i.e. one generates a random
number $r$ in the range $0$ to $1$ and solves $\Delta_s(t_I,t)=r$ to find
$t$. Then $z$ can be generated according to $F(z,t)$ and one repeats the
procedure until $t<t_0$, at which point there can be no more emissions and the
shower terminates. At leading-order it is straightforward to match the shower to
a matrix element. If $d\sigma_B=|\mathcal{M_B}|^2d\Phi_B$ is the Born-level
cross section, we simply act with $\field{S}$ on $d\sigma_B$
\begin{align}
  d\sigma_{SMC} = d\sigma_B \big\{ \Delta_s(t_I,t_0) &+ \Delta_s(t_I,t)F(z,t)dt\,dz \notag \\
  &\times\left\{ \Delta_s(t,t_0) + \Delta_s(t,t')F(z',t')dt'\,dz'\left\{ \dots \right\}\right\}\big\}.
\end{align}
In this case the starting scale of the shower, $t_I$, should be set to the scale
of the hard process, such that the parton shower correctly simulates soft and
collinear partons while the matrix element describes the hard partons.

In order to discuss the \POWHEG{} method we need to transform the above shower
into a $\pt$ ordered shower\footnote{\POWHEG{} also works with an angular
  ordered shower, although in this case one has to introduce a truncated shower
  to account for the fact that the first emission in the shower is not
  necessarily the hardest one.}, that is we want to make sure that the hardest
emission is generated first such that it can be described by the real matrix
element. It is clear from \cref{eq:showerpt} that if one generates first an
emission with $z\sim 1$ and then on the hardest emission line one generates an
emission with $z' \sim 1/2$ the $\pt$ of the second emission can easily be
larger than that of the first. We will state without proof than one can
implement a $\pt$ veto by modifying the Sudakov form factor in the obvious way
\begin{align}
  \ln\Delta_R(t,\pt) = -\int_{t_0}^{t}{dt'}\int{dz\,F(z,t')\theta(z(1-z)\sqrt{t'}-\pt)}.
  \label{eq:sudakovR}
\end{align}
Using \cref{eq:showerpt} we can immediately transform this into an integral in
$\pt$. For the first emission the Sudakov form factor is then given by
\begin{align}
  \ln\Delta_R(t_I,\pt) = -\int_{\pt}^{\sqrt{t_I}/4}{2\,dp'_{\scriptscriptstyle \mathrm{T}}}\int{dz'\,F(z',p'_{\scriptscriptstyle \mathrm{T}})\theta(t_I-t')}.
\end{align}
The differential cross section for the first emission can then be written
\begin{align}
  d\sigma = d\sigma_B \left\{ \Delta_R(t_I,\pt) + \Delta_R(t_I,\pt)F(z,t)\theta(t_I-t)dt\,dz\right\},
  \label{eq:firstemission}
\end{align}
which has the following $\mathcal{O}(\as)$ expansion
\begin{align}
  d\sigma = d\sigma_B \left\{ 1 - \int_{t_0}^{t_I}{dt'}\int{dz\,F(z,t')} + F(z,t)\theta(t_I-t)dt\,dz + \mathcal{O}(\as^2)\right\}.\label{eq:SMCNLO}
\end{align}
In order to turn this expression into something with which we can easily compare
later on, we introduce the ``+''-prescription to regulate the singularities in
the splitting kernel. We define it in the following way
\begin{equation}
  f(z)_+ = f(z) - \delta(1-z)\int_0^1 f(x) dx\,.
  \label{eq:plusprescription}
\end{equation}
Doing so the above equation reduces to
\begin{align}
  d\sigma = d\sigma_B \left\{ 1 + F(z,t)_+\theta(t_I-t)dt\,dz + \mathcal{O}(\as^2)\right\}.
  \label{eq:SMCNLO+}
\end{align}
In writing the cross section in this form we have isolated the approximate
Shower Monte Carlo \NLO{} contribution. In fact, the two non-trivial terms of
the bracket of \cref{eq:SMCNLO} correspond to the approximate virtual and
real contributions to the cross section. The ``+'' prescriptions simply
guarantees that the virtual and real singularities cancel. With this observation
it becomes clear why matching fixed-order \NLO{} calculations and partons
showers poses a problem. If we are not careful the shower can generate
configurations which are formally $\mathcal{O}(\as)$ spoiling the claimed \NLO{}
accuracy. We can avoid this double counting by generating the first emission
according to the \NLO{} matrix element and all subsequent emissions with the
shower. 

\subsection{\POWHEG{}}
Up until this point we have restricted our discussion to the effect of a parton
shower on one jet only. In general we are interested in studying $2 \,
\rightarrow\,n$ scattering in which $m$ of the $n$ final states are coloured. In
this case the above discussion is modified in the following way
\begin{itemize}
\item[1)] $F(z,t)\theta(t_I-t)dt\,dz$ is replaced by
  \begin{align}
    \sum_{i=1}^m F_i(z,t)\theta(t^i_I-t)dt\,dz,
  \end{align}
  and we define
  \begin{align}
    \Delta_R(\{t_I\},\pt) = \prod_{i=1}^m \Delta^i_R(t^i_I,\pt).
  \end{align}
  This in turn means that all lines, and not just the line emitting, will have
  $\Delta_R(t^i_I,\pt)$ factors associated to them. These factors are there to
  ensure consistency with the $\pt$ veto applied in the shower.
\item[2)] Then one generates the hardest $\pt$ according to $\Delta_R(\{t_I\},\pt)$
  and chooses values $i$ and $z$ according to $F_i(z,t)\theta(t^i_I-t)$. From
  these two values we can construct the two new partons from $i$ and let the
  full $m+1$ partons shower with a $\pt$ veto.
\end{itemize}

We now write the exact \NLO{} formula for the cross section in the following
schematic way
\begin{align}
  d\sigma^{\NLO{}} &= d\sigma^{\mathcal{B}} + d\sigma^{\mathcal{V}} + d\sigma^{\mathcal{R}} \notag\\
  &= \mathcal{B}(\Phi_{\mathcal{B}})d\Phi_{\mathcal{B}} + \mathcal{V}(\Phi_{\mathcal{B}})d\Phi_{\mathcal{B}} + \left[ \mathcal{R}(\Phi_{\mathcal{R}})d\Phi_{\mathcal{R}} - \mathcal{C}(\Phi_{\mathcal{R}})d\Phi_{\mathcal{R}}\field{M}\right]. 
\end{align}
For simplicity we are assuming that the real matrix element, $\mathcal{R}$, has
only one singular region, corresponding to the mapping $\field{M}$ from the
$m+1$-body kinematics to the $m$-body kinematics. Here we will assume the
simplest case, which is that $\field{M}\Phi_{\mathcal{R}} \rightarrow
\Phi_{\mathcal{B}}$ and we write
$d\Phi_{\mathcal{R}}=d\Phi_{\mathcal{B}}d\Phi_{r}$ where $d\Phi_{r}$ is the
three dimensional phase space associated with the emitted parton. In this way we
can rewrite the cross section in the following very suggestive way
\begin{align}
  d\sigma^{\NLO{}} &= \left[ \mathcal{V}(\Phi_{\mathcal{B}}) + \left( \mathcal{R}(\Phi_{\mathcal{R}}) - \mathcal{C}(\Phi_{\mathcal{R}})\right)d\Phi_{r}\field{M}\right]d\Phi_{\mathcal{B}} \notag \\
  &\qquad+ \left[\mathcal{B}(\Phi_{\mathcal{B}}) -\mathcal{R}(\Phi_{\mathcal{R}})\Phi_{r}\field{M} + \mathcal{R}(\Phi_{\mathcal{R}})\Phi_{r} \right] d\Phi_{\mathcal{B}} \notag \\
  &= \left[ \mathcal{V}(\Phi_{\mathcal{B}}) + \left( \mathcal{R}(\Phi_{\mathcal{R}}) - \mathcal{C}(\Phi_{\mathcal{R}})\right)d\Phi_{r}\field{M}\right]d\Phi_{\mathcal{B}} \notag \\
  &\qquad+ \mathcal{B}(\Phi_{\mathcal{B}})\left[1 + \frac{\mathcal{R}(\Phi_{\mathcal{R}})}{\mathcal{B}(\Phi_{\mathcal{B}})} (1-\field{M})d\Phi_{r} \right] d\Phi_{\mathcal{B}}\label{eq:exactNLO}
\end{align}
By comparing the last line of this expression to the approximate \NLO{}
expression of \cref{eq:SMCNLO+} we deduce how one would have to adjust
\cref{eq:firstemission} in order to obtain the exact \NLO{}
expression. We first define the \POWHEG{} Sudakov form factor by
\begin{align}
  \ln\bar{\Delta}(\Phi_{\mathcal{B}},\pt) = -\int{d\Phi_r \frac{\mathcal{R}(\Phi_{\mathcal{R}})}{\mathcal{B}(\Phi_{\mathcal{B}})} \theta(\kt(\Phi_{\mathcal{R}})-\pt)}\,.
\end{align}
The function $\kt(\Phi_{\mathcal{R}})$ is not unique, but has to correspond to
the transverse momentum of the emitted parton relative to the emitter close to
the singular limit. It is worth noticing, that in the collinear/soft limits this
Sudakov form factor corresponds to the one of \cref{eq:sudakovR} since in
this limit the real matrix element factorises and $\mathcal{R}\sim \mathcal{B}
\times F$. We then write the \NLO{} cross section as
\begin{align}
  d\sigma^{\NLO{}} 
  &= \left[ \mathcal{V}(\Phi_{\mathcal{B}}) + \left( \mathcal{R}(\Phi_{\mathcal{R}}) - \mathcal{C}(\Phi_{\mathcal{R}})\right)d\Phi_{r}\field{M}\right]d\Phi_{\mathcal{B}} \notag \\
  &\qquad+ \mathcal{B}(\Phi_{\mathcal{B}})d\Phi_{\mathcal{B}}\left[ \bar{\Delta}(\Phi_{\mathcal{B}},0)+ \bar{\Delta}(\Phi_{\mathcal{B}},\pt)\frac{\mathcal{R}(\Phi_{\mathcal{R}})}{\mathcal{B}(\Phi_{\mathcal{B}})} d\Phi_{r} \right].\label{NLOexactshower} 
\end{align}
By expanding $\bar{\Delta}$ in the above expression we exactly obtain
\cref{eq:exactNLO} analogously to the way \cref{eq:SMCNLO+} can be obtained from
\cref{eq:firstemission}. One could in principle stop here and implement
\cref{NLOexactshower} in a Shower Monte Carlo program. However, this expression
has the disadvantage of being able to generate negative weights, simply because
the first bracket is not required to be positive. A simple way of curing this
problem is by introducing the \emph{NLO weighted Born term}, $\bar{\mathcal{B}}$
\begin{align}
  \bar{\mathcal{B}}(\Phi_{\mathcal{B}}) = \mathcal{B}(\Phi_{\mathcal{B}}) + \mathcal{V}(\Phi_{\mathcal{B}}) + \int{\left[\mathcal{R}(\Phi_{\mathcal{R}}) - \mathcal{C}(\Phi_{\mathcal{R}})\right]d\Phi_r}\label{bbar}
\end{align}
and write the hardest emission cross section as
\begin{align}
  d\sigma^{\NLO{}} = \bar{\mathcal{B}}(\Phi_{\mathcal{B}})d\Phi_{\mathcal{B}}\left[ \bar{\Delta}(\Phi_{\mathcal{B}},\pt^{\mathrm{min}})+ \bar{\Delta}(\Phi_{\mathcal{B}},\kt(\Phi_{\mathcal{R}}))\frac{\mathcal{R}(\Phi_{\mathcal{R}})}{\mathcal{B}(\Phi_{\mathcal{B}})} d\Phi_{r} \right]\label{eq:powhegmaster}
\end{align}
which is equivalent to \cref{NLOexactshower} at $\mathcal{O}(\as)$. This
expression avoids the generation of negative events, since the
$\bar{\mathcal{B}}$ function is expected to be positive whenever perturbation
theory is valid. \cref{eq:powhegmaster} is in essence the master formula of
the \POWHEG{} method. For the purpose of our discussion of \MINLO{} later on, we
note that the $\bar{\mathcal{B}}$ function can receive corrections of
$\mathcal{O}(\as^2)$ without spoiling the accuracy of the cross section.

Before we continue and apply the \POWHEG{} method to VBF
$ZZjj$ production we will for completeness give the most general form of the
above equation when there are more than just one singular region.

\cref{eq:powhegmaster} takes the following
form~\cite{Frixione:2007vw}
\begin{align}
  d\sigma^{\POWHEG{}} &= \sum_{f_b}\bar{\mathcal{B}}^{f_b}(\Phi_{\mathcal{B}})d\Phi_{\mathcal{B}}\Big[ \bar{\Delta}^{f_b}(\Phi_{\mathcal{B}},\pt^{\mathrm{min}}) \notag \\
      & \qquad+ \sum_{\alpha_r\in\{\alpha_r|f_b\}}\frac{\big[\bar{\Delta}^{f_b}(\Phi_{\mathcal{B}},\kt(\Phi_{\mathcal{R}}))\theta(\kt(\Phi_{\mathcal{R}}-\pt^{\mathrm{min}}))\mathcal{R}(\Phi_{\mathcal{R}})d\Phi_{r}\big]^{\field{M}^{\alpha_r}\Phi_{\mathcal{R}}=\Phi_{\mathcal{B}}}_{\alpha_r}}{\mathcal{B}(\Phi_{\mathcal{B}})} \Big]\label{eq:powhegmasterfull}
\end{align}
where
\begin{equation}
 \ln\bar{\Delta}^{f_b}(\Phi_{\mathcal{B}},\pt) = -\sum_{\alpha_r\in\{\alpha_r|f_b\}}\int{\frac{\big[\theta(\kt(\Phi_{\mathcal{R}}-\pt))\mathcal{R}(\Phi_{\mathcal{R}})d\Phi_{r}\big]^{\field{M}^{\alpha_r}\Phi_{\mathcal{R}}=\Phi_{\mathcal{B}}}_{\alpha_r}}{\mathcal{B}(\Phi_{\mathcal{B}})}},
\end{equation}
and\footnote{For the ease of notation, we have omitted
  the two counter terms needed to cancel the initial state collinear
  singularities.}
\begin{align}
  \bbar^{f_b}(\Phi_{\mathcal{B}}) &= \left[\mathcal{B}(\Phi_{\mathcal{B}}) + \mathcal{V}(\Phi_{\mathcal{B}})\right]_{f_b} \notag\\
  & \qquad+ \sum_{\alpha_r\in\{\alpha_r|f_b\}}\int{\left[\left\{\mathcal{R}(\Phi_{\mathcal{R}}) - \mathcal{C}(\Phi_{\mathcal{R}})\right\}d\Phi_r\right]_{\alpha_r}^{\field{M}^{\alpha_r}\Phi_{\mathcal{R}}=\Phi_{\mathcal{B}}}}.
\end{align}

This expression looks complicated at first, but the various extra indices which
have been introduced are easily explained. The outer sum runs over all Born-like
flavour structures, $f_b$. To each such flavour structure exists a number of
real graphs which have the underlying flavour structure, $f_b$. We label these
real graphs by $\alpha_r$ according to their underlying flavour structure,
$\alpha_r\in\{\alpha_r|f_b\}$, and such that they each have one and only one
singular region. Hence
\begin{equation}
  \mathcal{R} = \sum_{\alpha_r}\mathcal{R}^{\alpha_r}.
\end{equation}
To each singular region corresponds a mapping,
$\field{M}^{\alpha_r}\Phi_{\mathcal{R}}\rightarrow\Phi_{\mathcal{B}}$ to a set
of underlying Born-like momenta. Hence we see that
\cref{eq:powhegmasterfull} simply sums over all possible Born-like flavour
states while including all real contributions with that underlying flavour
state.

It may so happen that certain real graphs have no singular regions and therefore
no underlying Born structure. This happens for instance in Higgs production
through gluon-fusion where the sub-process $q\bar{q} \rightarrow Hg$ can easily
be seen to be non-singular. It can be useful to split the real contributions
into the singular and non-singular ones to avoid the exponentiation of large
corrections. In this case $\mathcal{R}$ should be substituted by the singular
part of $\mathcal{R}$ and the non-singular part added outside of the
$\bar{\mathcal{B}}$-function.

To construct the $\bbar$ function it is convenient to introduce another
function, $\btilde$, given by
\begin{align}
  \btilde^{f_b}(\Phi_{\mathcal{B}}) &= \left[\mathcal{B}(\Phi_{\mathcal{B}}) + \mathcal{V}(\Phi_{\mathcal{B}})\right]_{f_b} \notag\\
  & \qquad+ \sum_{\alpha_r\in\{\alpha_r|f_b\}}\left[\left\{\mathcal{R}(\Phi_{\mathcal{R}}) - \mathcal{C}(\Phi_{\mathcal{R}})\right\}\left|\frac{\partial\Phi_r}{\partial X_r}\right|\right]_{\alpha_r}^{\field{M}^{\alpha_r}\Phi_{\mathcal{R}}=\Phi_{\mathcal{B}}},
\end{align}
such that
\begin{align}
  \bbar^{f_b}(\Phi_{\mathcal{B}}) = \int_0^1dX_r^{(1)}\int_0^1dX_r^{(2)}\int_0^1dX_r^{(3)}\btilde^{f_b}(\Phi_{\mathcal{B}}).
\end{align}
Here $X_r=\{X_r^{(1)},X_r^{(2)},X_r^{(3)}\}$ is simply a parametrisation of the
unit cube for the radiation variables.

Presently there exists a framework, the \POWHEGBOX{}~\cite{Alioli:2010xd}, where
a user may implement any fixed-order \NLO{} calculation to be matched with the
\POWHEG{} method. The \POWHEG{} method has been applied to a vast number of
Standard Model processes and several studies of Beyond the Standard Model
physics. The \POWHEGBOX{} requires the user to provide a list of Born, virtual,
and real processes along with their respective matrix elements. The user also
has to provide Born phase space mappings from the unit cube along with a
suitable Jacobian. From these ingredients the \POWHEGBOX{} automatically finds
the singular regions, constructs subtraction terms, and computes the
$\bar{\mathcal{B}}$ and $\tilde{\mathcal{B}}$ functions. An interface to
\HERWIG{} and \PYTHIA{} also exists, hence it is straightforward within this
framework to match existing \NLO{} fixed-order calculations to Shower Monte
Carlo programs. In the next section we study the implementation of VBF $ZZjj$
production in the \POWHEGBOX{}.

%Say something about regular contributions

\section{Electroweak $ZZjj$~Production} 
A primary goal of the CERN Large Hadron Collider (LHC) is an in-depth
understanding of the mechanism responsible for electroweak symmetry
breaking. Data collected and analysed by the ATLAS~\cite{Aad:2012tfa} and
CMS~\cite{Chatrchyan:2012ufa} collaborations have revealed the existence of a
scalar boson with a mass of about $125 \GeV$. Investigations on the properties of
this new particle consolidate its interpretation as the Higgs boson of the
Standard Model (SM). In particular, measurements of its spin and CP
properties~\cite{Aad:2013xqa,Chatrchyan:2012jja,Aad:2015rwa,Aad:2015mxa,Khachatryan:2014ira,Khachatryan:2014kca,Khachatryan:2014jba,Khachatryan:2016tnr}
as well as of its couplings to gauge bosons and fermions so far have disclosed
no deviation from the SM expectation of a spin-zero, CP-even particle. Should
physics beyond the Standard Model be realised in nature, its effects on
observables in the Higgs sector seem to be small, calling for high precision in
experiment as well as in theoretical predictions.

An ideal environment for the determination of the tensor structure and strengths
of the Higgs couplings to gauge bosons is provided by vector boson fusion
processes~\cite{Zeppenfeld:2000td,Duhrssen:2004cv,LHCHiggsCrossSectionWorkingGroup:2012nn}.
As we have already seen in previous chapters, because of their very pronounced
signature in phase space, featuring two well-separated jets in the forward
regions of the detector, VBF reactions can be separated well from QCD-induced
background reactions.

In this section we wish to present a tool for the simulation of $Z$-boson pair
production via vector boson fusion. The purely electroweak process $pp\to ZZjj$
predominantly proceeds via the scattering of two quarks by the exchange of weak
vector bosons in the $t$-channel with subsequent emission of two $Z$
bosons. Diagrams with a Higgs resonance contribute as well as weak boson
scattering graphs that are sensitive to triple and quartic gauge boson
couplings.

NLO-QCD corrections to this process, including
leptonic decays of the $Z$ bosons in the $\llll$ and $\llvv$ modes, have been
computed in \Bref{Jager:2006cp} and are publicly available in the computer
package \VBFNLO{}~\cite{Arnold:2008rz}. While that code allows the computation
of, in principle, arbitrary distributions within experimentally feasible
selection cuts, an interface to parton-shower Monte Carlo programs at NLO-QCD
accuracy is not yet available. We have therefore worked out a matching of the
NLO-QCD calculation with parton-shower programs in the framework of the
\POWHEG{} formalism described earlier in this chapter. To this end, we are
making use of the \POWHEGBOX{}~\cite{Alioli:2010xd}, a tool that provides all
the process-independent building blocks of the matching procedure, but requires
the user to implement process-specific ingredients in a specific format by
themselves.
Recently a Version 2 of the \POWHEGBOX{} has been released,
\POWHEGBOXVT. Version 2 includes a number of new features among which
are
\begin{itemize}
\item the possibility to produce grids in parallel and combine them;
\item the option to modify scales and parton distribution functions a
  posteriori, through a reweighting procedure of Les Houches events;
\item a faster calculation of upper bounds, and the possibility to
  store upper bounds and combine them; 
\item an improvement in the separation of regions for the real
  radiation~\cite{Campbell:2013vha}, which results in smoother
  distributions.
\end{itemize}
Given the complexity of electroweak $ZZjj$ production, we found it
useful to take full advantage of these features and therefore
implemented the process directly in Version 2 of the \POWHEGBOX{}.

In the following section we describe the technical details of our
implementation. In \cref{sec:pheno} we present phenomenological
results for some representative applications in the case of leptonic
final states, in the case of cuts suitable to study the continuum,
double-resonant production.  We also discuss the potential of this
process to constrain the size of dimension-six operators that arise in
effective field theory approaches to physics beyond the Standard
Model. In particular we study the capability of future Colliders to
constrain the couplings even further.
%
%%%%%%%%%%%%%%%%
%
\subsection{Technical Details of the Implementation}
\label{sec:tech}
Our implementation of electroweak $ZZjj$ production in the context of the
\POWHEG{}-\texttt{BOX}\xspace proceeds along the same lines as previous work
done for $Zjj$ \cite{Jager:2012xk}, $W^+W^+jj$ \cite{Jager:2011ms}, and
$W^+W^-jj$ production \cite{Jager:2013mu} via vector boson fusion.  We therefore
refrain from a detailed description of technical aspects that are common to all
vector boson fusion processes considered so far, but refer the interested reader
to the aforementioned references.

The first calculation of the NLO-QCD corrections to $ZZjj$ production via VBF in
the context of the Standard Model, including decays of the $Z$-boson pair into
four leptons or two leptons plus two neutrinos, has been presented in
\Bref{Jager:2006cp} and is publicly available in the context of the \VBFNLO{}
package~\cite{Arnold:2008rz}. We adapted the matrix elements of that calculation
to the format required by the \POWHEGBOX{}, and additionally computed the
scattering amplitudes for the semi-leptonic decay modes of the $Z$~bosons, i.e.
decay modes where one $Z$-boson decays leptonically and the other hadronically.

In addition to that we account for physics beyond the Standard Model in the weak
gauge boson sector by means of an effective field theory
approach~\cite{Degrande:2012wf} with operators of dimension six that affect
triple and quartic gauge boson vertices, but do not change the QCD structure of
the Standard Model. Details of the operators entering the Lagrangian are given
later. Notice that because decays are not affected by QCD corrections, it is
enough to have an LO implementation of the modified decay currents
even at NLO in QCD. We could therefore adapt the LO implementation of the
effective field theory in {\tt MadGraph~5}~\cite{Alwall:2011uj} for the modelling
of the modified electroweak building blocks needed for $pp\to ZZjj$.

\begin{figure}[t]
  \centering
  \includegraphics[angle=0,width=0.95\textwidth]{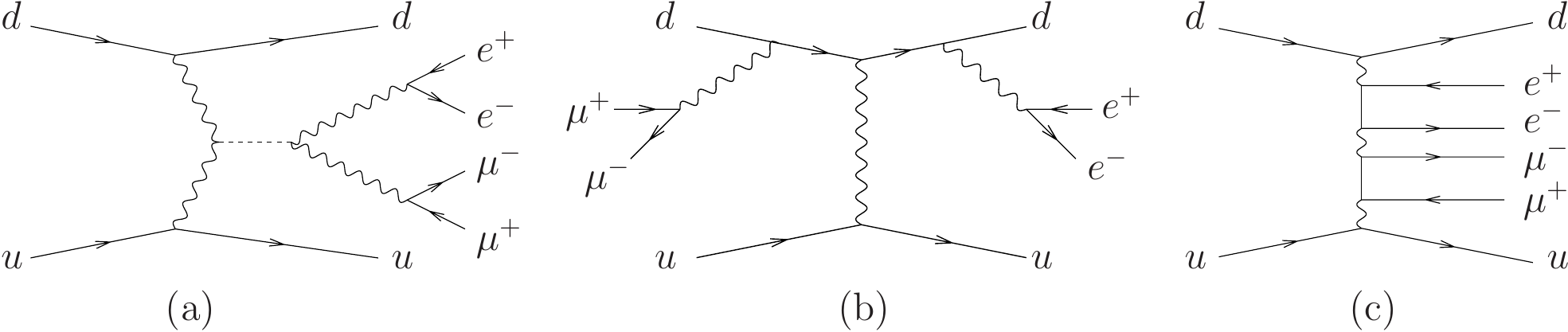}
  \caption{Representative Feynman diagrams for the partonic subprocess $u d \to \eemm u d$ at leading order. 
  }
\label{fig:feynman-graphs}
\end{figure}

In either model, at order $\mc{O}(\alpha^6)$ electroweak $ZZjj$ production
predominantly proceeds via the scattering of two (anti-)quarks mediated by
weak-boson exchange in the $t$-channel. The external $Z$~bosons that in turn
decay into a pair of leptons, neutrinos, or quarks can be emitted from either of
the two fermion lines, or stem from vector boson scattering sub-amplitudes of
the type $VV\to VV$ (with $V$ generically denoting a photon, a $W^\pm$, or a $Z$
boson). In order to maintain electroweak gauge invariance, contributions with
one or two photons instead of the $Z$~bosons and diagrams for single- and
non-resonant four-fermion production in association with two jets have to be
considered as well. A representative set of diagrams is depicted in
\cref{fig:feynman-graphs}.
%
%%%%%%%%%%%%%%%%%
%
%
%%%%%%%%%%%%%%%%%%
%
For partonic subprocesses with quarks of identical flavour, in addition to the
aforementioned $t$-channel exchange diagrams, $u$-channel diagrams arise that
are taken fully into account. However, the interference of $u$-channel with
$t$-channel contributions is neglected. We furthermore disregard contributions
induced by the exchange of a weak boson in the $s$-channel. This gauge-invariant
subset of diagrams is strongly suppressed in the phase-space regions that are
explored experimentally in vector boson fusion searches,
c.f.\ \Bref{Denner:2012dz} for a tree-level assessment of the numerical
impact these contributions have in a realistic setup for the related case of
electroweak $W^+W^+jj$ production at the LHC.

For $\ell^+\ell^-\nu_\ell\bar\nu_\ell$ final states we neglect the interference
with $W^+W^-jj$ production, when the $W$ bosons decay into the same final
state. In the case of QCD production, this interference has been shown to be
very small~\cite{Melia:2011tj}.  In the semi-leptonic decay modes, interference
effects between the scattering quarks and the decay quarks are neglected.

For simplicity, we will refer to the electroweak production processes
$pp\to\llll jj$, $pp\to\llvv jj$, and $pp\to\llqq jj$ within the aforementioned
approximations as $ZZjj$ production via VBF in the fully leptonic,
leptonic-invisible and semi-leptonic decay modes, respectively, even though we
always include contributions from off-resonant diagrams that do not arise from a
$ZZjj$ intermediate state.

In the case of semi-leptonic decay modes we do not explicitly take into account
QCD corrections to the hadronic decays of the $Z$~bosons, or QCD corrections
that connect the $ZZjj$ production with the $Z\to\bar q q$ decay
processes. While the latter corrections are expected to be very small,
corrections to the hadronic $Z$~decay are well-described by Monte-Carlo programs
that are interfaced to our NLO-QCD calculation.  In fact their decay machinery
is tuned to reproduce collider data.
 
We note that, similarly to the cases of electroweak $Zjj$ and $W^+W^-jj$
production, the \POWHEGBOX{} requires a prescription for dealing with
singularities emerging in the Born cross section for $pp\to ZZjj$ via VBF.  One
such type of singularities stems from collinear $q\to q\gamma$ configurations
that emerge when a photon of low virtuality is exchanged in the
$t$-channel. Phenomenologically, such contributions are irrelevant, as they are
entirely removed once VBF-specific selection cuts are applied on the $ZZjj$
cross section that require the two partons of the underlying Born configuration
to exhibit sufficient transverse momentum to be identified as tagging jets. We
therefore drop this type of contributions already at generation level, by
removing all events with an exchange boson in the $t$-channel with a virtuality
smaller than $Q_\mr{min}^2 = 4~\mr{GeV}^2$. To improve the efficiency of the
phase-space integration, we use a Born-suppression factor $F(\Phi_n)$ that
dampens the phase-space integrand whenever a singular configuration is
approached~\cite{Alioli:2010qp}. This is ensured by the choice
\beq
F(\Phi_n) = 
\left(\frac{p_{T,1}^2}{p_{T,1}^2+\Lambda^2}\right)^2
\left(\frac{p_{T,2}^2}{p_{T,2}^2+\Lambda^2}\right)^2\,,
\eeq
where the $p_{T,i}$ denote the transverse momenta of the two final-state partons
of the underlying Born configuration, and $\Lambda$ is a cutoff parameter that
we set to $10 \GeV$.

In VBF $ZZjj$ production processes, an additional type of singular
configurations is caused by diagrams with a quasi on-shell photon that decays
into a fermion pair, $\gamma^\star\to f\bar f$. Such contributions can easily be
identified by a small value of the invariant mass $m_{ff}$ of the decay
system. In our simulations, we remove all configurations with
$m_{ff}<m_{ff}^\mr{gen}$, where we set
\beq
m_{ff}^\mr{gen}=20~\mr{GeV}\,,
\eeq
unless explicitly stated otherwise. 

In the presence of a light Higgs boson, the VBF $ZZjj$ cross section receives
contributions from two regions of phase space with very different kinematic
properties. Therefore it is useful to split the phase space into two separate
regions, around and away from the Higgs resonance. The full result is then
obtained by adding the results of the two separate
contributions~\cite{Jager:2011ms}.

\subsection{Phenomenological Results}
We will concentrate in the following on the fully charged leptonic decay mode,
which has a smaller branching fraction than the semi-leptonic ($\llqq$) or the
leptonic-invisible ($\llvv$) decay modes, but is experimentally cleaner.
Because of the Higgs and $Z$ resonances, events tend to have either four leptons
with an invariant mass close to the Higgs mass, or two pairs of leptons with an
invariant mass close to the mass of the $Z$~boson each. Typically, according to
whether one is interested in studying Higgs production with subsequent $H\to
ZZ^{(\star)}$ decays or VBF $ZZjj$ production in the continuum one applies
different invariant mass cuts that suppress one of the two contributions, and
leave the other almost unchanged. Continuum VBF $ZZjj$ production is a rare SM
process that is well-suited to probe triple but also quartic gauge boson
couplings. In this section we present few sample results obtained with our
\POWHEGBOX{} implementation, both in the pure SM and involving anomalous
couplings that in our framework arise from an effective Lagrangian.

Let us stress here that the $\llqq$ mode, although plagued by large QCD
backgrounds, could in principle be studied with an analysis that uses boosted
techniques and jet-substructure (see e.g. \Bref{Altheimer:2013yza}), along
the lines of what was done in~\Bref{Jager:2013mu}. However, because of the
small production cross sections for VBF $ZZjj$, considering the boosted regime
where only a tiny part of the inclusive cross section survives is pointless at
the LHC.

\label{sec:pheno}
\subsubsection{Standard Model Results}
\label{sec:noanomalous}
With Run II already well under its way, and anticipating an energy upgrade of the
LHC, we consider proton-proton collisions at a centre-of-mass energy of
$\sqrt{s}=14 \TeV$. We use the NLO-QCD set of the MSTW2008
parametrisation~\cite{Martin:2009iq} for the parton distribution functions of
the proton, as implemented in the \tt{LHAPDF}
library~\cite{Whalley:2005nh}. Jets are defined according to the anti-$\kt$
algorithm~\cite{Cacciari:2005hq,Cacciari:2008gp} with $R=0.4$, making use of the
\tt{FASTJET} package~\cite{Cacciari:2011ma}. Electroweak input parameters are
set according to known experimental values and tree-level electroweak relations.
As input we use the mass of the $Z$ boson, $\MZ=91.188\GeV$, the mass of the
$W$~boson, $\MW=80.419\GeV$, and the Fermi constant, $G_F=1.16639\times
10^{-5}\GeV^{-2}$. For the widths of the weak bosons we use $\Gamma_Z =
2.51\GeV$ and $\Gamma_W=2.099\GeV$.  The width of the Higgs boson is set to
$\Gamma_H=0.00498\GeV$ which corresponds to $\MH=125\GeV$.  Factorisation and
renormalisation scales are set to $\muf=\mur=\MZ$ throughout, unless specified
otherwise.

Here, we present numerical results for VBF $ZZjj$ production at the LHC in the
fully leptonic decay mode.
Our analysis requires each lepton pair to have an invariant mass close to
$\MZ$. This completely excludes any contamination from a Higgs boson consistent
with the one observed by the \tt{ATLAS} and \tt{CMS} collaborations at
$\MH=125\GeV$~\cite{Aad:2012tfa,Chatrchyan:2012ufa}, which results in $H\to
ZZ^{(\star)}$ decays with at least one off-shell gauge boson.  Our
phenomenological study is inspired by~\cite{Aad:2012awa}.
In the following, we will always consider decays to
$e^+e^-\mu^+\mu^-$. Neglecting same-type lepton interference effects, the
cross-section for $Z$ bosons decaying to any combination of electrons and muons
is twice as large. In \cref{sec:noanomalous} all results are quoted for
the $pp \rightarrow e^+e^-\mu^+\mu^-jj$ decay mode only, whereas the results in
\cref{sec:anomalous} have been obtained for the $pp \rightarrow
e^+e^-\mu^+\mu^-jj$ decay mode and then multiplied by two to account for any
decay into electrons or muons, while neglecting same-type lepton interference
effects.

The VBF and invariant mass cuts that we apply in the following,
inspired by~\Brefs{Rainwater:1996ud,Rainwater:1999sd}, are very
effective in suppressing QCD-like processes with coloured objects in
the $t$-channel.
In particular, we require the presence of at least two jets with 
\begin{equation}
\label{eq:pttag-cuts}
p_{T,j}>25~\mr{GeV}\,, \quad
y_j<4.5\,.
\end{equation}
The two hardest jets satisfying these cuts are called ``tagging jets''
and are furthermore forced to be well separated %in rapidity by obeying 
by the VBF cuts 
\begin{equation}
  \label{eq:rap-cuts}
|y_{j_1}-y_{j_2}|>4.0\,, \quad
y_{j_1}\cdot y_{j_2}<0\,, \quad
m_{j_1 j_2}>600~\mr{GeV}\,.
\end{equation}
For the leptons we require
\begin{equation}
  \label{eq:lepton-cuts}
p_{T,\ell}>25~\mr{GeV}\,, \quad
y_\ell<2.4\,, \quad
R_{j\ell}>0.4\,.
\end{equation}
In addition to that, we request that the leptons fall in between the
two tagging jets
\begin{equation}
  \label{eq:leprap-cuts}
\mr{min}\{y_{j_1},y_{j_2}\}<y_\ell<\mr{max}\{y_{j_1},y_{j_2}\}\,.
\end{equation}
Furthermore  the two same-flavour opposite-charge
leptons have to be close to the on-shell mass of the $Z$ boson, 
\begin{equation}
\label{eq:mass-cut}
66~\mr{GeV}<m_{\ell\ell}<116~\mr{GeV}\,.
\end{equation}
%
%%%%%%%%%%%%%%%%%
%
\begin{figure}[t]
  \centering
  \includegraphics[angle=0,width=0.47\textwidth]{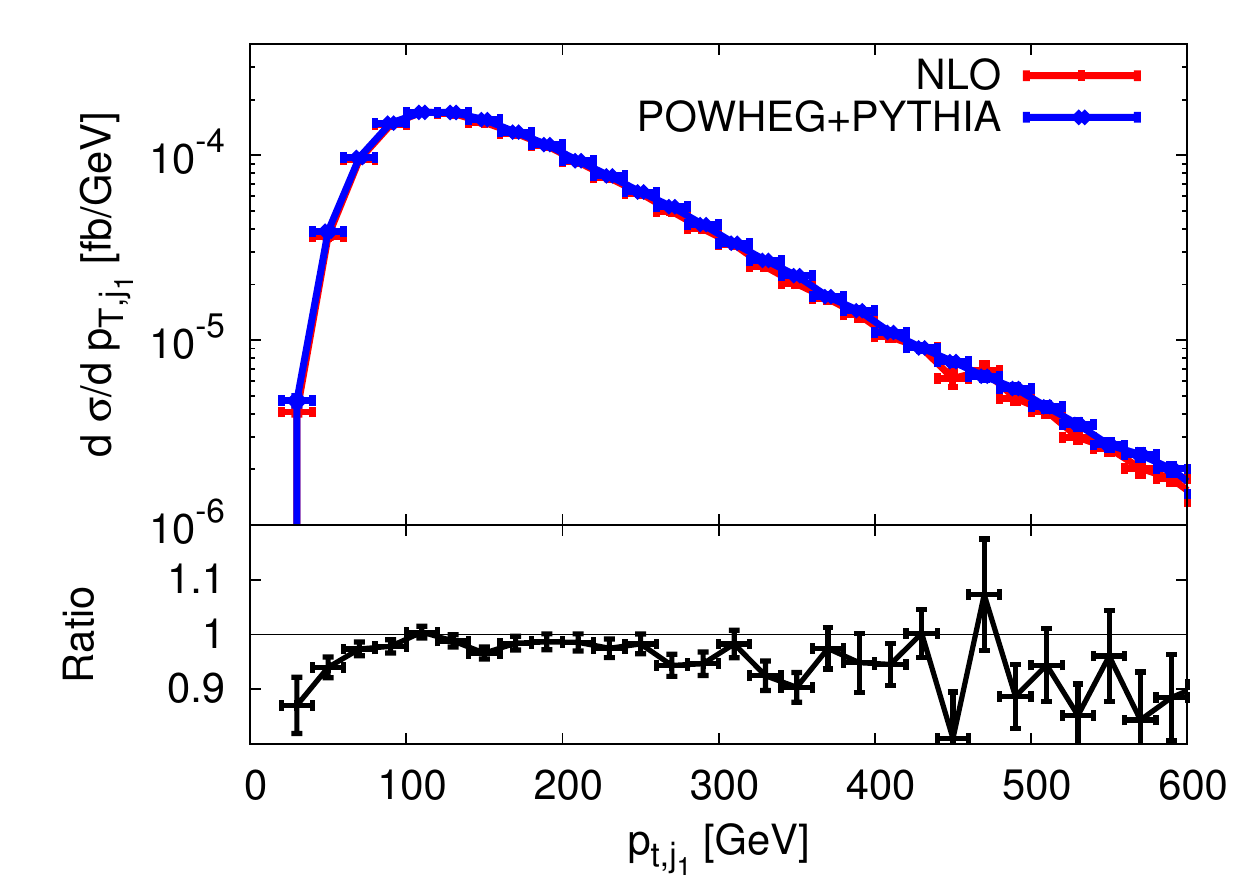}
  \includegraphics[angle=0,width=0.47\textwidth]{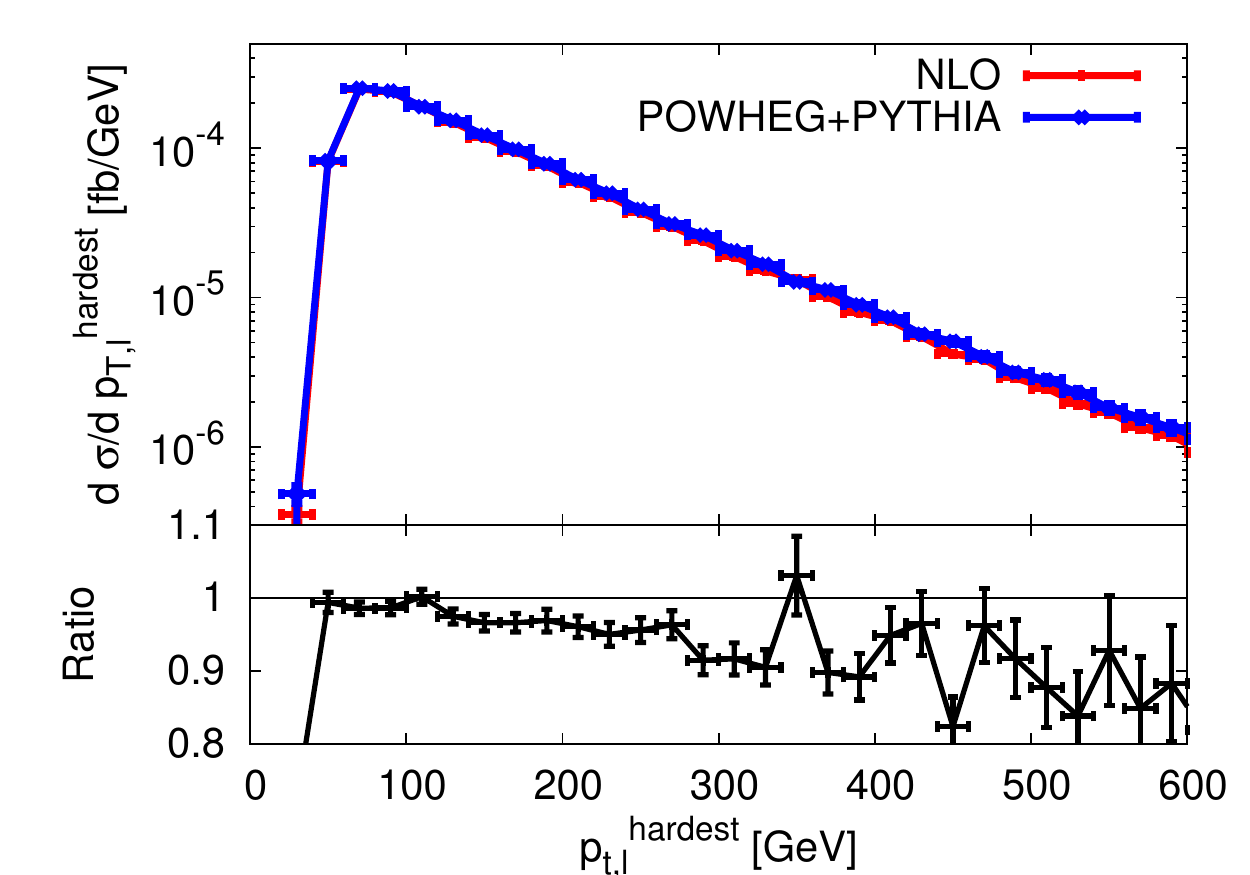}
  \caption{Transverse momentum distributions of the hardest jet (left) and the
    hardest lepton (right) for $pp\rightarrow e^+e^-\mu^+\mu^- jj$ at the LHC
    with $\sqrt{s}=14 ~\mr{TeV}$ within the cuts of
    \cref{eq:pttag-cuts,eq:rap-cuts,eq:lepton-cuts,eq:leprap-cuts,eq:mass-cut}
    at NLO~(red) and \NLOPS{}~(blue). The lower panel shows the ratio between
    NLO and \NLOPS{}.
    \label{fig:ptzz-vbfcuts}
  }
\end{figure}
%%%%%%%%%%%%%%%%%%
%
%
Because of the low mass of the Higgs boson and its very narrow width,
this last cut ensures that contributions with an intermediate Higgs
resonance are suppressed very strongly.

The inclusive cross section for \vbfeemm production after applying the cuts of
\cref{eq:pttag-cuts,eq:rap-cuts,eq:lepton-cuts,eq:leprap-cuts,eq:mass-cut} is
given by $\sigma_{ZZ}^\mr{VBF}=0.03003(7)~\mr{fb}$ at NLO in QCD and
$\sigma_{ZZ}^\mr{VBF}=0.03249(4)~\mr{fb}$ at LO, where the uncertainties quoted
are purely statistical.
We then match the NLO calculation with the parton-shower program \tt{PYTHIA}
6.4.25~\cite{Sjostrand:2006za} via \POWHEG~(\NLOPS{}). The parton shower is
run with the Perugia~0 tune, including hadronisation corrections, multi-parton
interactions and underlying events. We do not take QED radiation effects into
account.  At the \NLOPS{} level, we obtain an inclusive cross section of
$\sigma_{ZZ}^\mr{VBF}= 0.03084(7)~\mr{fb}$.  In order to estimate the
theoretical uncertainty of the calculation we have varied the renormalisation
and factorisation scales in the range $\MZ/2$ to $2 \MZ$, finding a change in
the \NLOPS{} cross section between $-0.0005~\mr{fb}$ and $+0.0001~\mr{fb}$
which is less than $3\%$.

\cref{fig:ptzz-vbfcuts} shows the transverse momentum
distributions of the hardest jet and the hardest lepton, respectively. The NLO and
the \NLOPS{} results agree very well for these two observables.
In general, distributions involving leptons or any of the two hardest
jets are only marginally distorted by the parton-shower. We notice
only a small increase in the VBF cross section by $3\%$ when going
from NLO to \NLOPS{}. This is comparable to the size of the scale
variation uncertainty.
%
%
%%%%%%%%%%%%%%%%%
%
\begin{figure}[t]
  \centering
  \includegraphics[angle=0,width=0.6\textwidth]{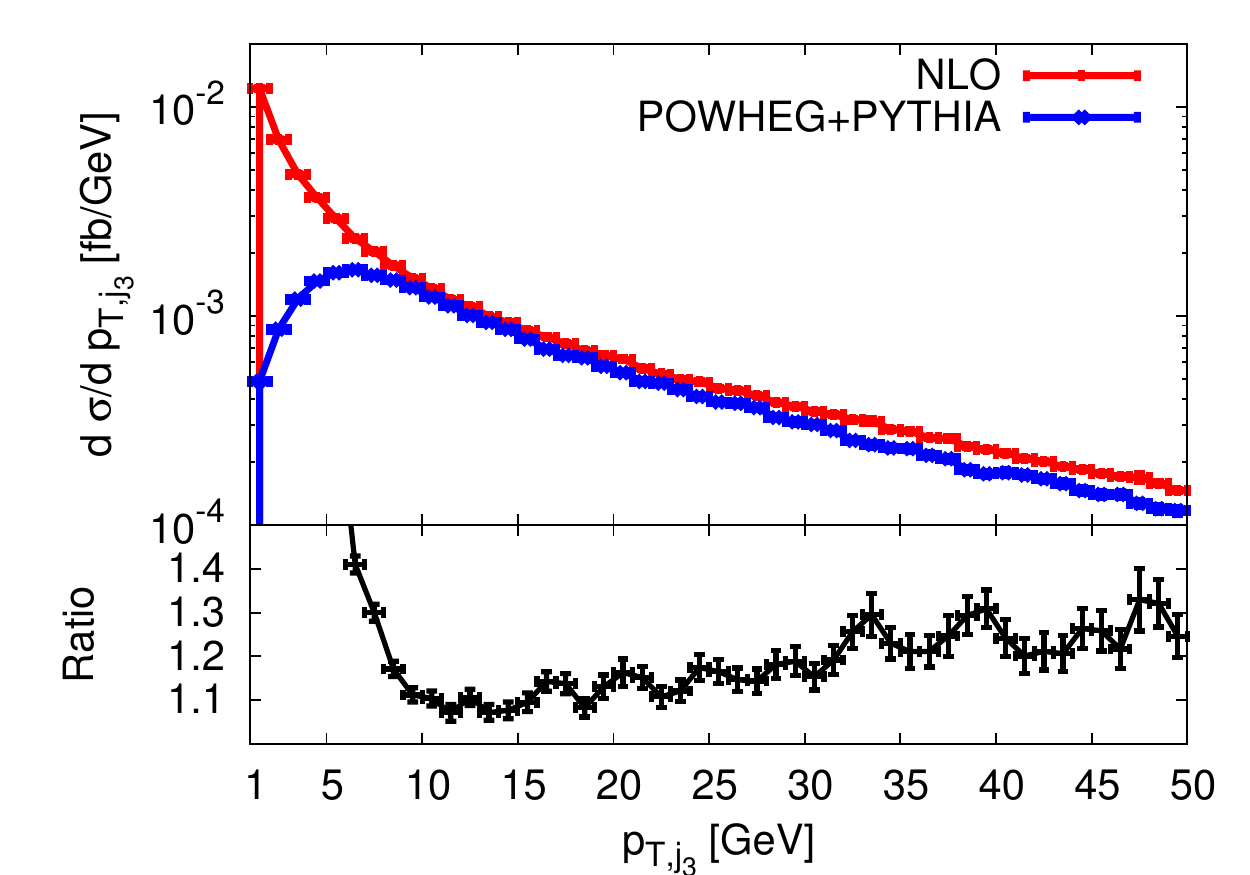}
  \caption{Transverse momentum distribution of the third jet for
  $pp\rightarrow e^+e^-\mu^+\mu^-jj$ at the LHC with $\sqrt{s}=14
  ~\mr{TeV}$ within the cuts of
  \cref{eq:pttag-cuts,eq:rap-cuts,eq:lepton-cuts,eq:leprap-cuts,eq:mass-cut}, at NLO (red) and at
  \NLOPS{} level (blue). The lower panel shows the ratio between NLO and \NLOPS{}.} 
\label{fig:ptj3_lep} 
\end{figure} 
%
%%%%%%%%%%%%%%%%%%
%
Illustrated in \cref{fig:ptj3_lep} is the effect of the
parton-shower on the transverse momentum of the third jet.  In the
NLO-QCD calculation for $pp\to ZZjj$ a third jet is described only at
lowest non-vanishing order, as it solely arises via the real-emission
contributions. When merged with the parton shower the soft-collinear
radiation is resummed at leading-logarithmic accuracy via the Sudakov
form factor, which results in the $p_{T,j_3}$ distribution being
damped at low transverse momentum.
At higher transverse momentum we observe that the parton shower tends
to slightly soften the spectrum of the third jet.
The parton shower also affects the rapidity of the third jet, giving
rise to an increased central jet activity.  This is expected since
soft QCD radiation tends to populate the central region. In
\cref{fig:yj3-lep} the rapidity of the third jet is shown with 
two different transverse-momentum cuts. Increasing the cut from
$10~\mr{GeV}$ to $20~\mr{GeV}$ decreases the central jet activity of
the parton shower without having any significant impact on the shape
of the distribution at fixed order.

Instead of considering the absolute position of the third jet it can
be useful to look at its relative position with respect to the two
hardest jets. This is usually measured through the $y^*$ quantity,
\begin{align}
y^*= y_{j_3} - \frac{y_{j_1}+y_{j_2}}{2}.
\label{eq:ystar}
\end{align}
\cref{fig:ystar-lep} shows that the parton shower populates the
region where $y^*$ is close to zero. This comes as no surprise, as we
require the two hardest jets to be in opposite hemispheres and with a
very large rapidity gap, and hence small values of $y^*$ will often
coincide with a very central third jet. If we increase the cut on the
transverse momentum of the third jet, we again see that the effect of
the parton shower is minimised.

In fact, and not surprisingly, the parton-shower has very much the
same impact on the distributions involving the third jet as was
reported in \cite{Jager:2013mu} for \vbfww.
%
%%%%%%%%%%%%%%%%%
%
\begin{figure}[t]
  \centering
\includegraphics[angle=0,width=0.47\textwidth]{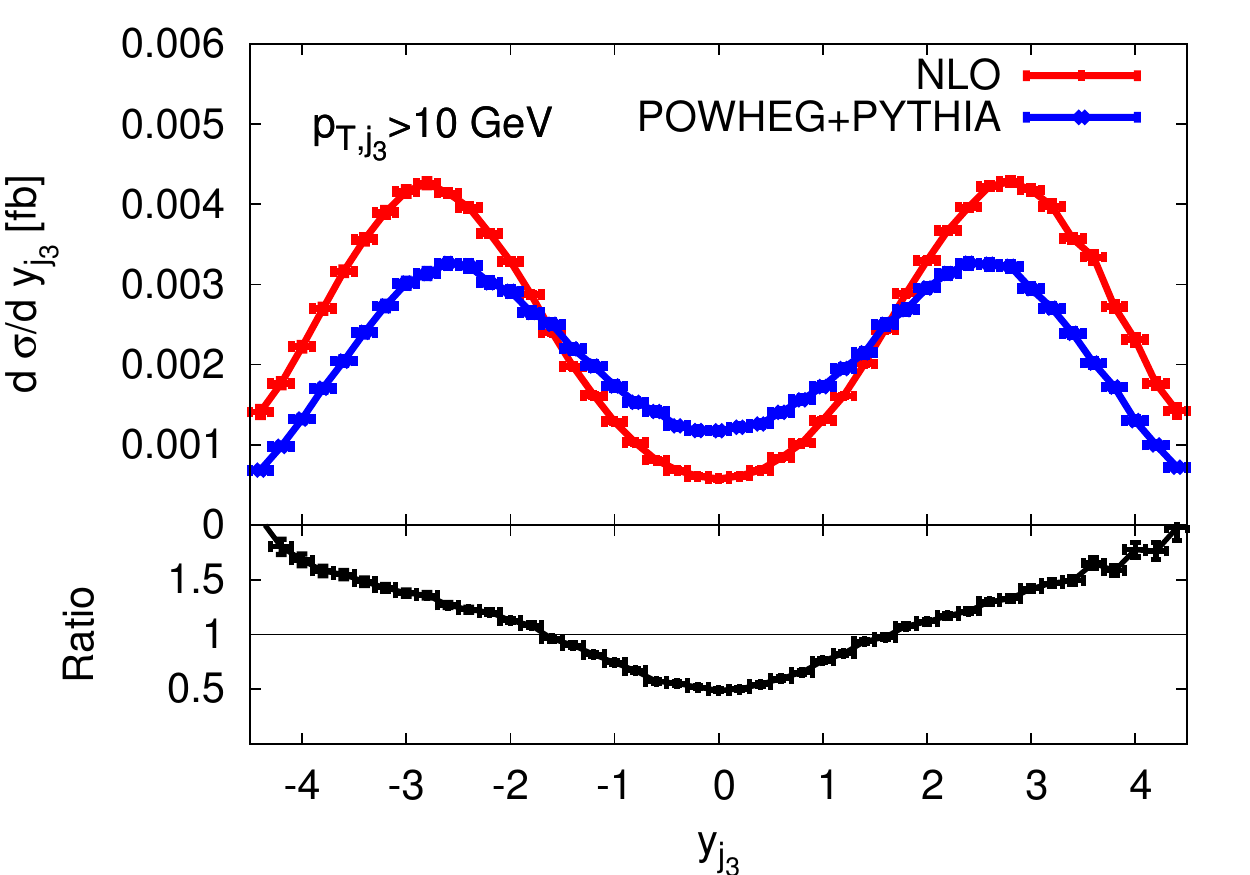}
\includegraphics[angle=0,width=0.47\textwidth]{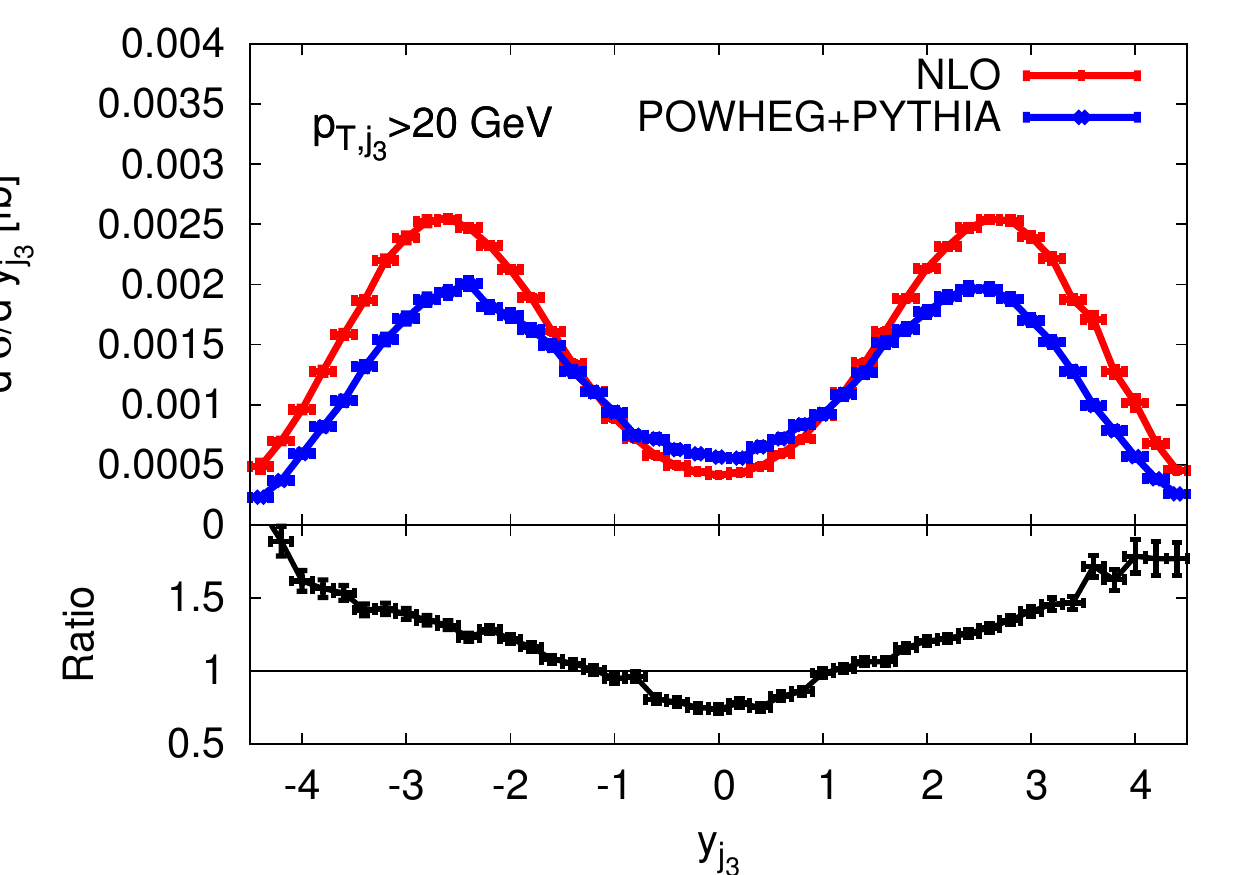}
\caption{Rapidity of the third jet for $pp\rightarrow e^+e^-\mu^+\mu^-
  jj$ at the LHC with $\sqrt{s}=14 ~\mr{TeV}$ within the cuts of
  \cref{eq:pttag-cuts,eq:rap-cuts,eq:lepton-cuts,eq:leprap-cuts,eq:mass-cut} and a transverse
  momentum cut on the third jet of $10~\mr{GeV}$~(left) and
  $20~\mr{GeV}$~(right), at NLO (red) and \NLOPS{} (blue). The lower panel shows the ratio between NLO and \NLOPS{}.}
\label{fig:yj3-lep}
\end{figure} 
%
%%%%%%%%%%%%%%%%%%
%
%
%%%%%%%%%%%%%%%%%
%
\begin{figure}[t]
  \centering
\includegraphics[angle=0,width=0.47\textwidth]{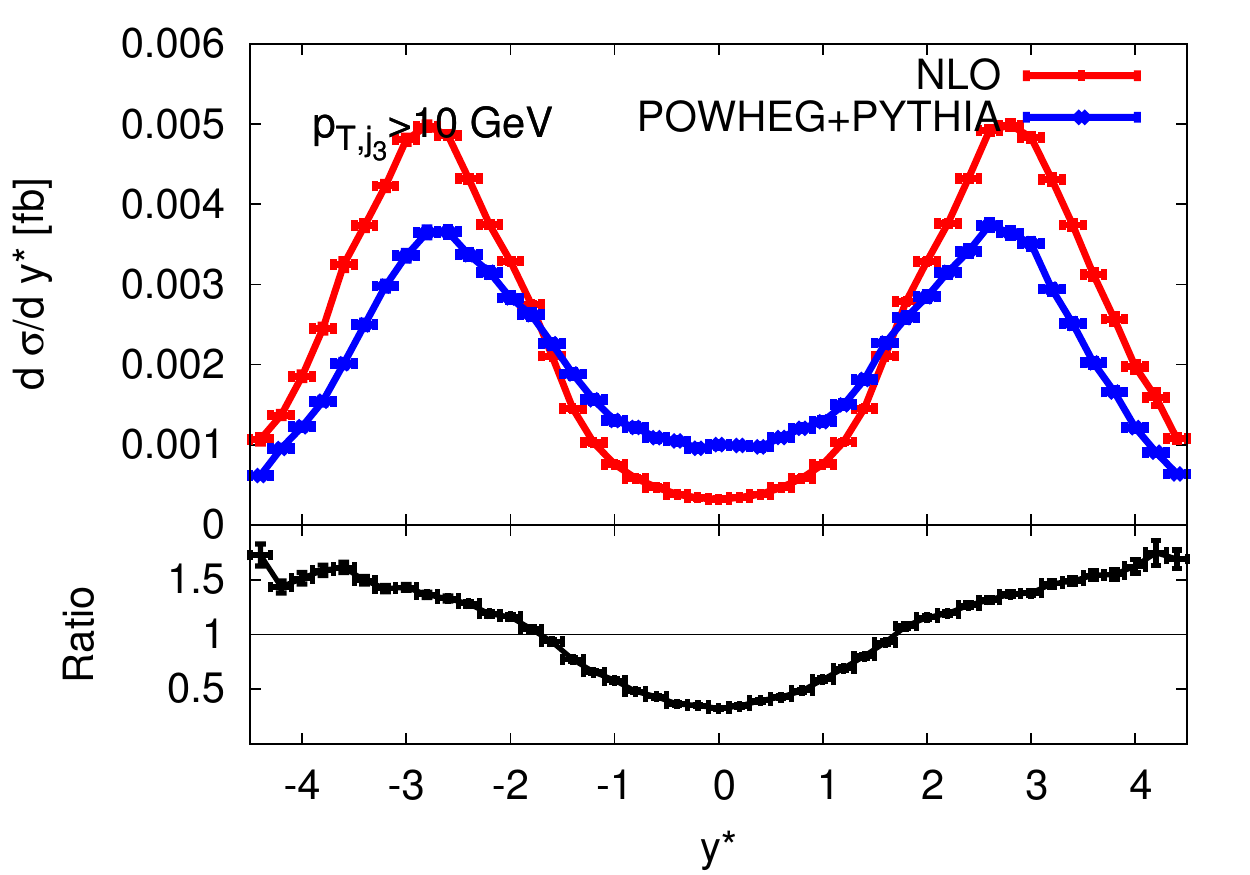}
\includegraphics[angle=0,width=0.47\textwidth]{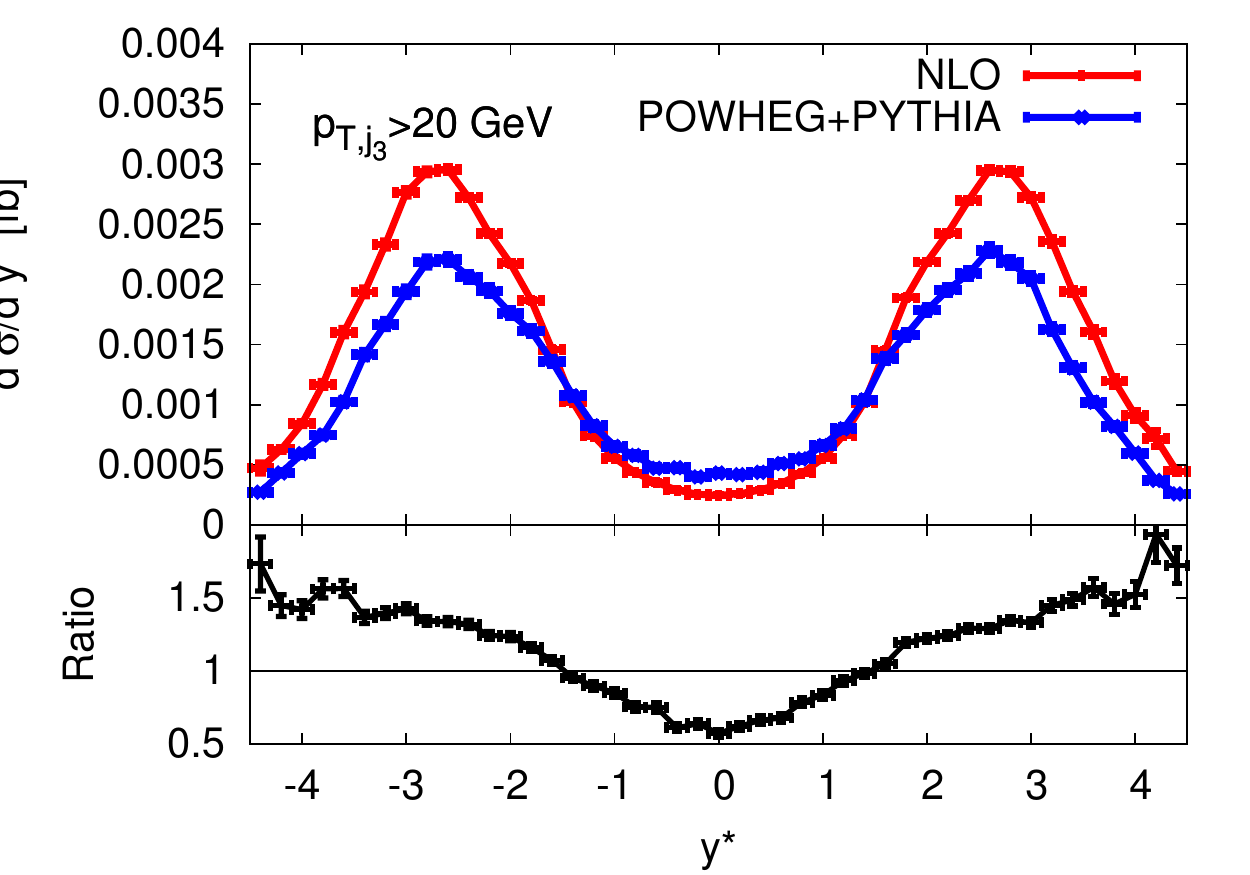}
\caption{$y^*$ as defined in \cref{eq:ystar} for $pp\rightarrow
  e^+e^-\mu^+\mu^- jj$ at the LHC with $\sqrt{s}=14 ~\mr{TeV}$ within
  the cuts of \cref{eq:pttag-cuts,eq:rap-cuts,eq:lepton-cuts,eq:leprap-cuts,eq:mass-cut} and a
  transverse momentum cut on the third jet of $10~\mr{GeV}$~(left) and
  $20~\mr{GeV}$~(right), at NLO (red) and \NLOPS{} (blue). The lower panel shows the ratio between NLO and \NLOPS{}.}
\label{fig:ystar-lep}
\end{figure} 
%
%%%%%%%%%%%%%%%%%%
%

\subsubsection{Effective Theory Results}
\label{sec:anomalous}
Vector boson scattering processes offer an excellent test bed for the
electroweak sector at the $\TeV$ scale. A convenient way to parametrise
deviations from the Standard Model is through anomalous couplings or,
alternatively, an effective field theory expansion.  Such an effective
theory is constructed as the low-energy approximation of a more
fundamental theory, and is valid up to an energy scale $\Lambda$.  The
explicit dependence of predictions on the scale $\Lambda$ can be used
to put limits on the scale of new physics itself.  For scales
$\Lambda$ much larger than the electroweak scale we can restrict
ourselves to the first correction to the SM contributions with
operators of dimension six.

The Lagrangian of the effective field theory can be written in the
form~\cite{Degrande:2012wf} 
\beq
\mathcal{L}_\mr{eff}
=\sum_{i,d}{\frac{c_i^{(d)}}{\Lambda^{d-4}}\,\mathcal{O}_i^{(d)}}
= \mathcal{L}_\mr{SM} 
+ \sum_{i}{\frac{c_i^{(6)}}{\Lambda^2}\,\mathcal{O}_i^{(6)}}+\ldots \,,
\eeq 
where $d$ is the dimension of the operators $\mathcal{O}_i^{(d)}$, and
the $c_i^{(d)}$ denote the coefficients of the expansion. For ease of
notation we therefore drop the superscript $d=6$ in the following.

For our analysis of VBF $ZZjj$ production we include the three
CP-conserving dimension-six operators
\cite{Hagiwara:1993ck,Degrande:2012wf,Degrande:2013rea},
\begin{align}
\label{eq:CP-con1}
\mathcal{O}_{WWW}&=\mr{Tr}[W_{\mu\nu}W^{\nu\rho}W_\rho^\mu]\,, \\
\label{eq:CP-con2}
\mathcal{O}_{W}&=(D_\mu \Phi)^\dagger W^{\mu\nu}(D_\nu\Phi)\,, \\
\label{eq:CP-con3}
\mathcal{O}_{B}&=(D_\mu \Phi)^\dagger B^{\mu\nu}(D_\nu\Phi)\,,
\end{align}
and the two CP-violating operators\,, 
\begin{align}
  \label{eq:CP-vio1}
\mathcal{O}_{\tilde{W}WW}&=\mr{Tr}[\tilde{W}_{\mu\nu}W^{\nu\rho}W_\rho^\mu]\,, \\
\mathcal{O}_{\tilde{W}}&=(D_\mu \Phi)^\dagger \tilde{W}^{\mu\nu}(D_\nu\Phi)\,,
\label{eq:CP-vio2}
\end{align}
where  
$\Phi$ is the Higgs doublet field, and 
\begin{align}
D_\mu&=\partial_\mu+\frac{i}{2}g\tau^I W_\mu^I+\frac{i}{2}g'B_\mu\,, \\
W_{\mu\nu}&=\frac{i}{2}g\tau^I(\partial_\mu W^I_\nu-\partial_\nu W^I_\mu +g\epsilon_{IJK}W^J_\mu W^K_\nu)\,, \\
B_{\mu\nu}&=\frac{i}{2}g'(\partial_\mu B_\nu-\partial_\nu B_\mu)\,. 
\end{align}
Here, the $B^\mu$ and $W^\mu$ denote the $U(1)$ and $SU(2)$ gauge fields with
the associated couplings $g'$ and $g$, respectively, and $\tau^I$ the weak
isospin matrices. The dual field strength tensor is defined as
\begin{align}
\tilde{W}_{\mu\nu}=\epsilon_{\alpha\beta\mu\nu}W^{\alpha\beta}.
\label{eq:Wtilde}
\end{align}  
It is worth noting that the five operators of
\cref{eq:CP-con1,eq:CP-con2,eq:CP-con3,eq:CP-vio1,eq:CP-vio2} form a minimal set
of operators affecting the triple and quartic gauge boson couplings.
%In
%principle one could also include an additional CP-violating operator
%\begin{align}
%\mathcal{O}_{\tilde{W}W}=\Phi^{\dagger}\Phi\tilde{W}_{\mu\nu}W^{\mu\nu},
%\end{align}
%%
%which, however, contributes to the dipole moment of the electron which
%is very constrained. 
%%
%We therefore choose not to include it
%here. 
For completeness we show which weak gauge boson vertices are affected
by the five operators defined above in \cref{table:EFTop}.

Our implementation of the effective Lagrangian approach for VBF $ZZjj$
production allows the user to specify the values of the $c_i/\Lambda^2$
coefficients in units of$\TeV^{-2}$ for each of the operators of
\cref{eq:CP-con1,eq:CP-con2,eq:CP-con3,eq:CP-vio1,eq:CP-vio2}. Here, we will
show the effect of the two operators $\mathcal{O}_{WWW}$ and
$\mathcal{O}_{\tilde{W}WW}$ for values of $c_{WWW}/\Lambda^2$ and
$c_{\tilde{W}WW}/\Lambda^2$ consistent with current limits on the anomalous
couplings $\lambda_Z$ and $\tilde{\lambda}_Z$.  These can be transformed into
limits on effective field theory parameters through a set of tree-level
relations. However, the relations between the anomalous couplings and the
effective field theory parameters are not exact, in the sense that they assume
no form factor dependence and disregard contributions from higher dimensional
operators.
%\begin{center}
\begin{table}[t]
  \centering
    \resizebox{\columnwidth}{!}
	{\begin{tabular} {l c c c c c | c c c c}
	\toprule 
	& $WWZ$ & $WW\gamma$ & $WWH$ & $ZZH$ & $\gamma ZH$ & $WWWW$ & $WWZZ$ & $WWZ\gamma$ & $WW\gamma \gamma$ \\ 
	\midrule
	$\mathcal{O}_{WWW}$ & x & x & - & - & - & x & x & x & x\\ 
	$\mathcal{O}_{W}$   & x & x & x & x & x & x & x & x & -  \\ 
	$\mathcal{O}_{B}$   & x & x & - & x & x & - & - & - & - \\
	\midrule 
	$\mathcal{O}_{\tilde{W}WW}$ & x & x & - & - & - & x & x & x & x\\ 
	$\mathcal{O}_{\tilde{W}}$   & x & x & x & x & x & - & - & - & -\\ 
	\bottomrule
    \end{tabular}}
  \caption{Crosses indicate triple and quartic weak boson vertices affected by
    the dimension-six operators of
    \cref{eq:CP-con1,eq:CP-con2,eq:CP-con3,eq:CP-vio1,eq:CP-vio2}. Taken from
    \Bref{Degrande:2013rea}.}
    \label{table:EFTop}
  \end{table}
%\end{center}
From \cite{Hagiwara:1993ck,Wudka:1994ny} we have
\begin{align}
\label{eq:coupcon}
\frac{c_{WWW}}{\Lambda^2}&=\frac{2}{3g^2\MW^2}\lambda_Z\,, \\
\frac{c_{\tilde{W}WW}}{\Lambda^2}&=\frac{1}{3g^2\MW^2}\tilde{\lambda}_Z\,,
\end{align}
which translates into 
\begin{align}
\label{eq:couplimits1}
-11.9~\mr{TeV}^{-2} < &\frac{c_{WWW}}{\Lambda^2} < -1.94~\mr{TeV}^{-2}\,, \\
\label{eq:couplimits2}
-19.4~\mr{TeV}^{-2} < &\frac{c_{\tilde{W}WW}}{\Lambda^2} < -2.42~\mr{TeV}^{-2}\,,
\end{align}
when using the current combined limits of \cite{Beringer:1900zz} at
the $68\%$ confidence level. The limits become compatible with the
Standard Model at the $95\%$ confidence level.

Our setup is identical to that of \cref{sec:noanomalous},
except we choose running factorisation and renormalisation scales,
\beq
\label{eq:runningscales}
\mu_R=\mu_F={\frac{\sqrt{\MZ^2+p_{T,Z_1}^2}+\sqrt{\MZ^2+p_{T,Z_2}^2}+\sum_i{p_{T,i}}}{2}}\,,
\eeq
where the $p_{T,i}$ are the transverse momenta of the (two or three)
final state partons and the $p_{T,Z_i}$ the transverse momenta of a
same-type lepton pair.
Such a dynamical scale is expected to optimally account for the
high-transverse momentum region where the effective operators have the
largest impact.

Our analysis within the effective field theory approach is done in analogy to
the SM analysis of \cref{sec:noanomalous}. We present results obtained for the
$\llll$ decay mode within the cuts of
\cref{eq:pttag-cuts,eq:rap-cuts,eq:lepton-cuts,eq:leprap-cuts,eq:mass-cut}. We
account for decays into any combination of electrons and muons by multiplying
results obtained for the $e^+e^-\mu^+\mu^-$ decay mode by a factor of two. As
mentioned earlier this procedure neglects any interference effects for the
leptons.
In order to illustrate the effect dimension-six operators can have on
observables we consider the operators $\mathcal{O}_{WWW}$ and
$\mathcal{O}_{\tilde{W}WW}$ independently by setting the associated expansion
coefficients to values compatible with the current experimental bounds of
\cref{eq:couplimits1,eq:couplimits2}, and all other to zero.  To this end, we
separately choose $c_{WWW}/\Lambda^2=-5~\mr{TeV}^{-2}$ and
$c_{\tilde{W}WW}/\Lambda^2=-5~\mr{TeV}^{-2}$. In diagrams where more than one
vertex could be affected by the effective operator, we only turn on the
effective coupling for one vertex at a time. This is consistent with only
considering dimension six operators, as diagrams suppressed by more than one
factor of $\Lambda^{-2}$ should also receive corrections from operators of
higher dimension.
Because of the explicit scale suppression of the effective operators,
it is expected that deviations from the Standard Model are most easily
seen in the tails of differential distributions that are sensitive to
the high-energy regime.  From our SM analysis of
\cref{sec:noanomalous} we may conclude that a parton shower has
very little effect on the distributions that do not involve the third
jet. In this section we therefore only discuss fixed-order results.
%
%
%%%%%%%%%%%%%%%%%
%
\begin{figure}[t]
  \centering
\includegraphics[angle=0,width=0.47\textwidth]{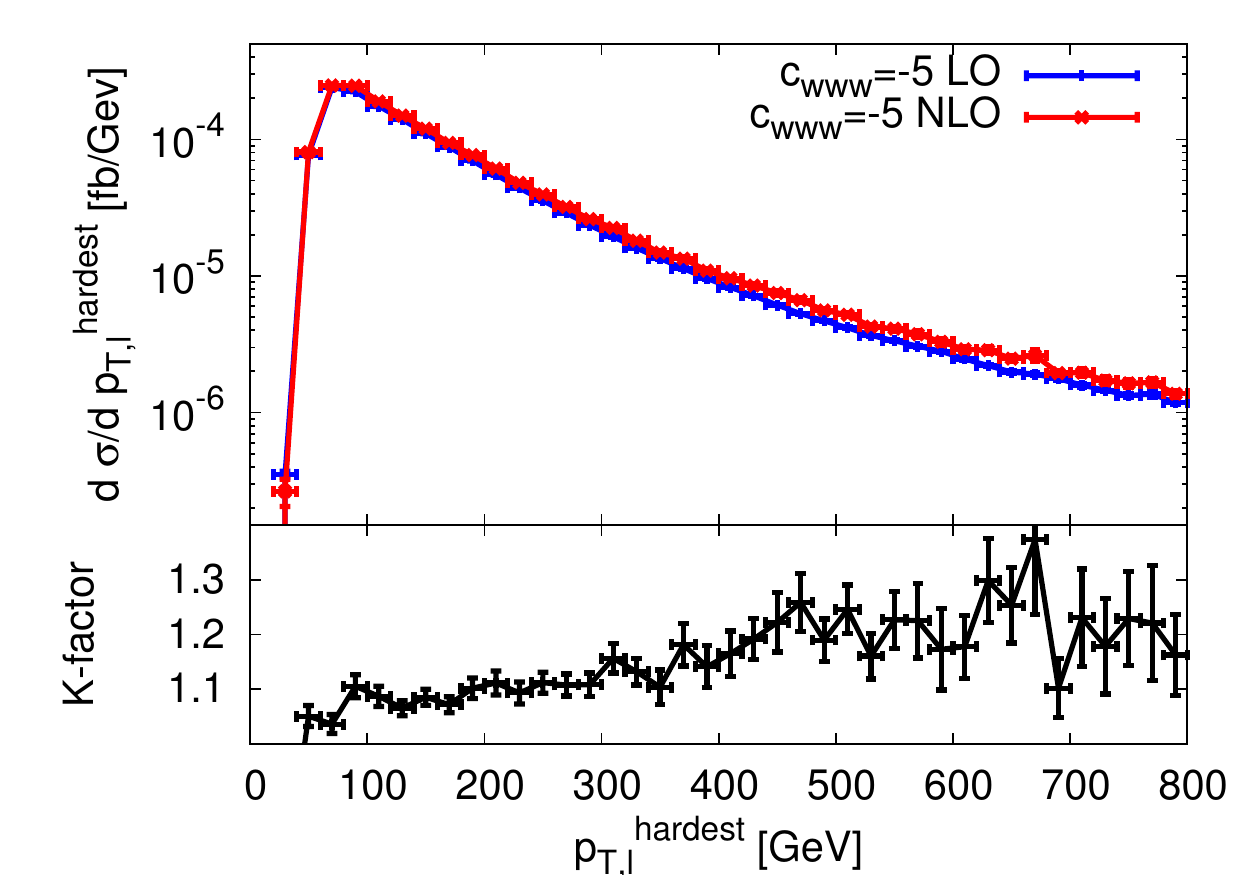}
\includegraphics[angle=0,width=0.47\textwidth]{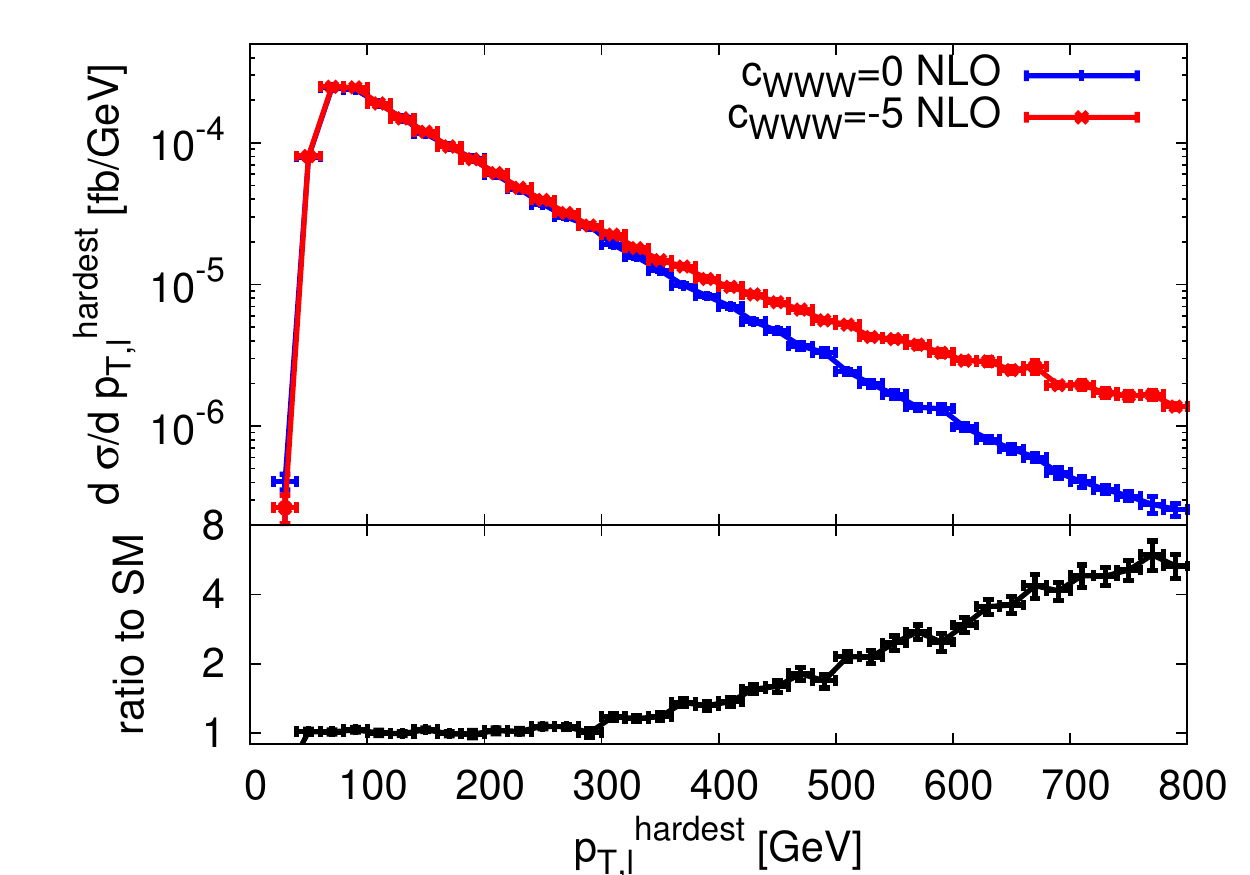}
\caption{
Transverse momentum distribution of the hardest lepton for $pp\rightarrow e^+e^-\mu^+\mu^- jj$ at the LHC
  with $\sqrt{s}=14 ~\mr{TeV}$ within the cuts of 
  \cref{eq:pttag-cuts,eq:rap-cuts,eq:lepton-cuts,eq:leprap-cuts,eq:mass-cut}, 
with $c_{WWW}/\Lambda^2=-5~\mr{TeV}^{-2}$ at LO and NLO (\textbf{left panel}), and at NLO with  $c_{WWW}/\Lambda^2=-5~\mr{TeV}^{-2}$ and $c_{WWW}/\Lambda^2=0$ (\textbf{right panel}).  The lower panels show the respective ratios. 
}
\label{fig:ptlhard-cwww}
\end{figure} 
%
%%%%%%%%%%%%%%%%%%
%
In \cref{fig:ptlhard-cwww} (left panel) we show the transverse
mass distribution of the hardest lepton at LO and NLO for
$c_{WWW}/\Lambda^2=-5~\mr{TeV}^{-2}$ together with the associated
dynamical $K$-factor, defined by
\begin{align}
K(x)=\frac{d \sigma_{NLO}/d x}{d \sigma_{LO}/ d x}\,,
\label{eq:kfactor}
\end{align}
and a comparison of the NLO prediction for
$c_{WWW}/\Lambda^2=-5~\mr{TeV}^{-2}$ with the SM result (right
panel). All other dimension-six operator coefficients are set to zero.
We note that the impact of the NLO-QCD corrections and the considered
dimension-six operator contributions are of the same order of magnitude
in the range of low to intermediate transverse momenta. For smaller
absolute values of $c_{WWW}/\Lambda^2$ the NLO corrections are
significant up to even higher transverse momenta. Hence, in that case,
in order to unambiguously distinguish new physics from higher-order perturbative effects in the Standard Model, full NLO-QCD results have to be considered.
%
%%%%%%%%%%%%%%%%%
%
\begin{figure}[t]
  \centering
\includegraphics[angle=-90,width=0.60\textwidth]{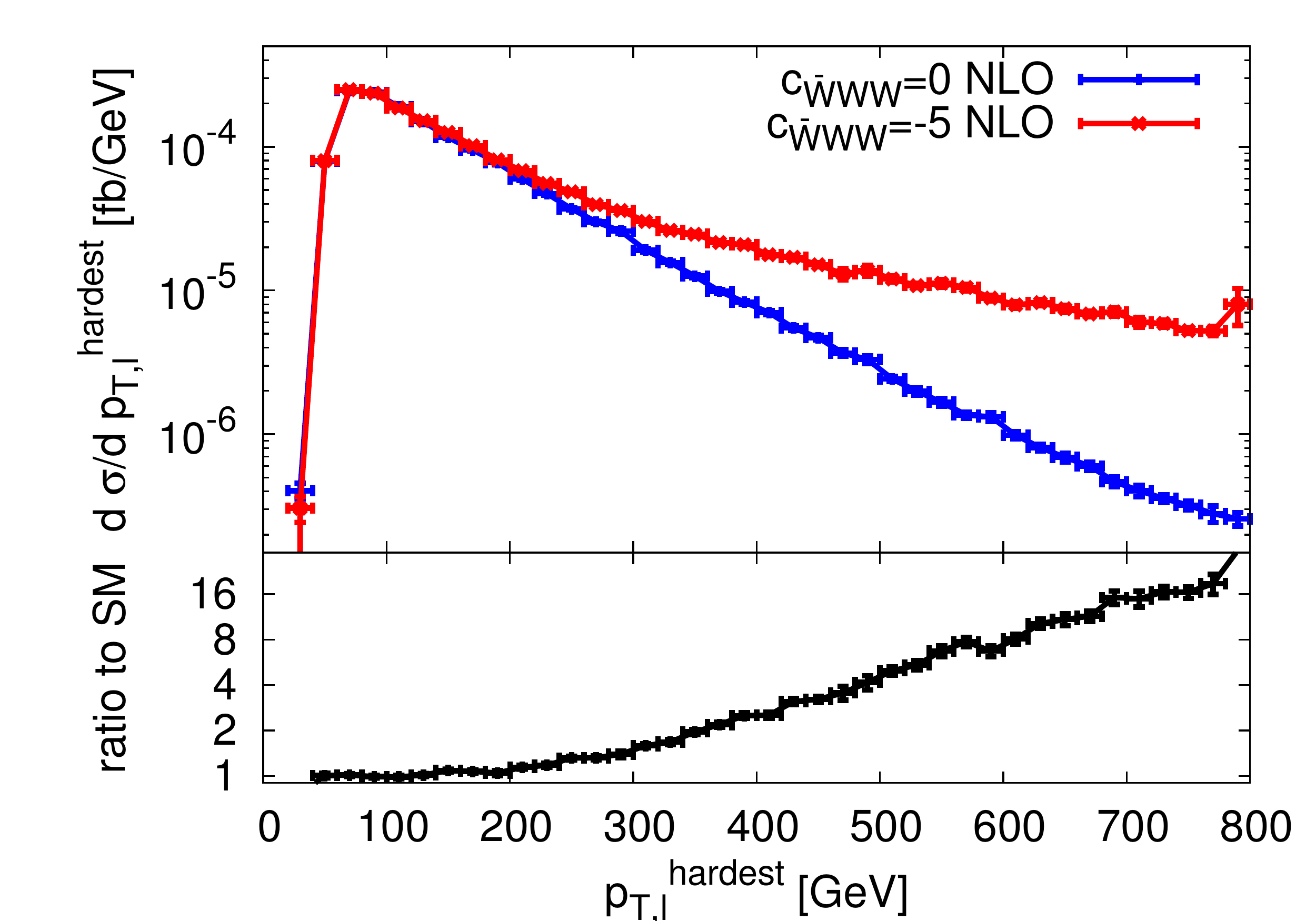}
\caption{Transverse momentum distribution of the hardest lepton for
  $pp\rightarrow e^+e^-\mu^+\mu^- jj$ at the LHC with $\sqrt{s}=14
  ~\mr{TeV}$ within the cuts of
  \cref{eq:pttag-cuts,eq:rap-cuts,eq:lepton-cuts,eq:leprap-cuts,eq:mass-cut}, at NLO with
  $c_{\tilde{W}WW}/\Lambda^2=-5~\mr{TeV}^{-2}$ and
  $c_{\tilde{W}WW}/\Lambda^2=0$, together with the respective ratio.}
\label{fig:ptlhard-cpwww}
\end{figure} 
%
%%%%%%%%%%%%%%%%%%
%
%
%%%%%%%%%%%%%%%%%
%
\begin{figure}[t]
  \centering
  \includegraphics[angle=0,width=0.47\textwidth]{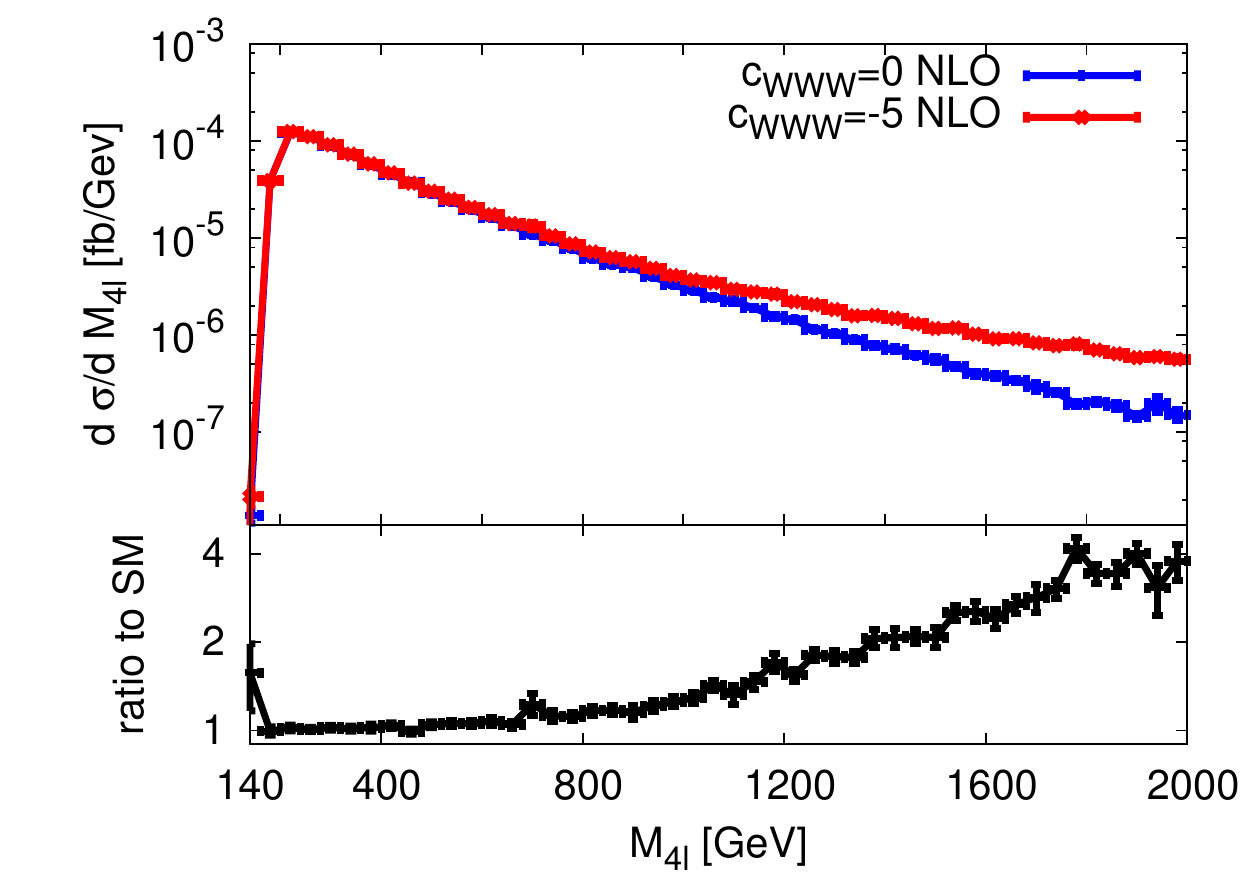}
  \includegraphics[angle=0,width=0.47\textwidth]{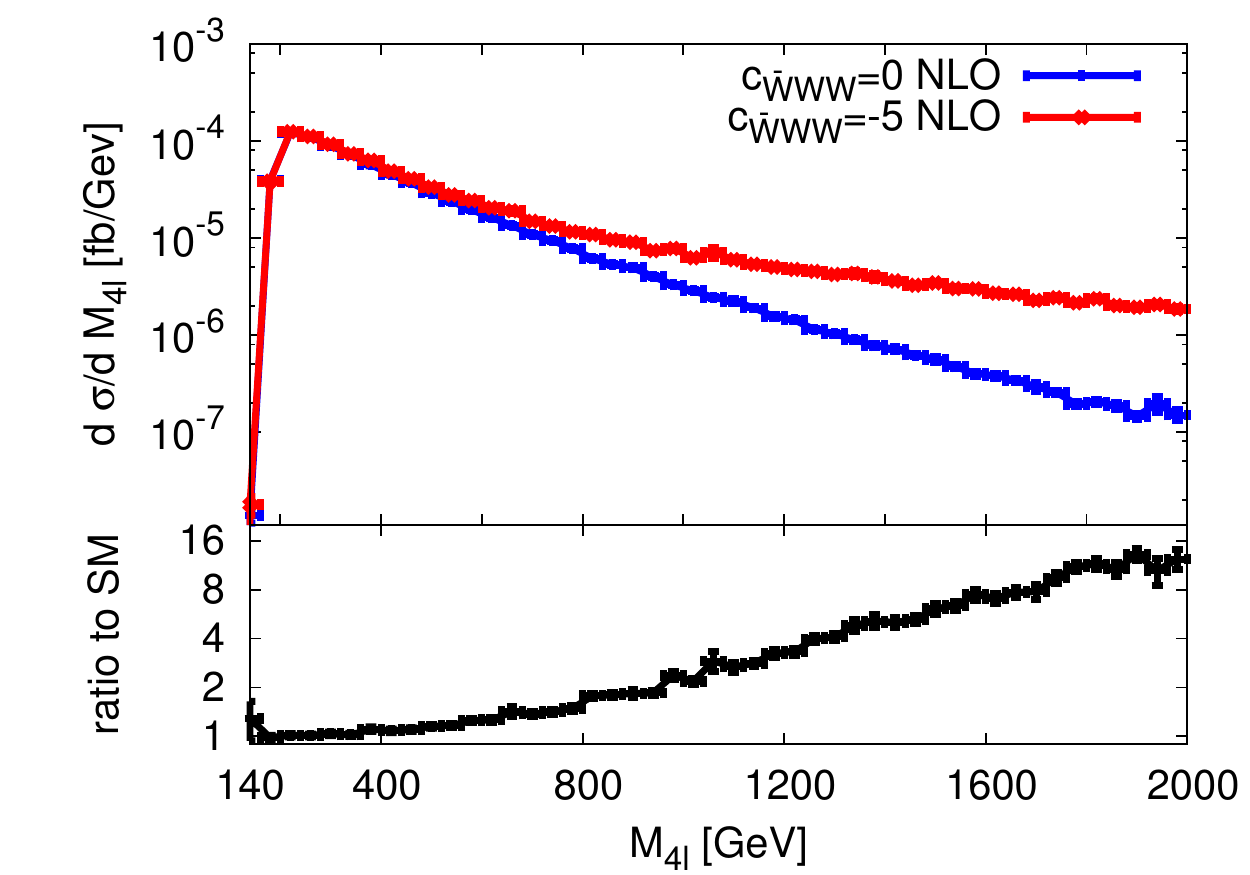}
  \caption{Invariant mass distribution of the four-lepton system in
    $pp\rightarrow e^+e^-\mu^+\mu^- jj$ at the LHC with $\sqrt{s}=14 ~\mr{TeV}$
    within the cuts of
    \cref{eq:pttag-cuts,eq:rap-cuts,eq:lepton-cuts,eq:leprap-cuts,eq:mass-cut},
    at NLO with $c_{WWW}/\Lambda^2=-5~\mr{TeV}^{-2}$ and $c_{WWW}/\Lambda^2=0$
    (\textbf{left panel}), $c_{\tilde{W}WW}/\Lambda^2=-5~\mr{TeV}^{-2}$ and
    $c_{\tilde{W}WW}/\Lambda^2=0$ (\textbf{right panel}), together with the
    respective ratio.  }
\label{fig:m4l}
\end{figure}

When fixing $c_{WWW} = c_{\tilde W WW}$ one notices that the
CP-violating operator $\mathcal{O}_{\tilde{W}WW}$ yields an
enhancement in the tails of the transverse momentum distribution of
the hardest lepton that is larger by roughly a factor two,
cf. \cref{fig:ptlhard-cpwww}. This is due to the normalisation
chosen in \cref{eq:Wtilde}, which would be more naturally defined
with a factor $1/2$ on the right-hand-side. 
Qualitatively similar results are obtained for the invariant mass of
the four-lepton system, as illustrated by \cref{fig:m4l}.

%
%
%%%%%%%%%%%%%%%%%
%
In order to estimate how sensitive the LHC is to the two couplings
$c_{WWW}/\Lambda^2$ and $c_{\tilde{W}WW}/\Lambda^2$ we have computed the number
of events in the tail of the transverse momentum distribution of the hardest
lepton, $p_{T,\ell}^\mr{hardest}>340~\mr{GeV}$, for the Standard Model, for
$c_{WWW}/\Lambda^2=-5~\mr{TeV}^{-2}$ and for
$c_{\tilde{W}WW}/\Lambda^2=-5~\mr{TeV}^{-2}$, respectively, with integrated
luminosities of $300~\mr{fb}^{-1}$ and $3000~\mr{fb}^{-1}$,
cf. \cref{tab:limits14NLO}. The cut-off value of $340~\mr{GeV}$ is chosen
upon inspection of the transverse-momentum distributions, where we start
observing a significant deviation from the Standard Model around this region,
cf. \cref{fig:ptlhard-cwww,fig:ptlhard-cpwww}. We note that for
different values of $c_{WWW}/\Lambda^2$ and $c_{\tilde{W}WW}/\Lambda^2$, we
would find different values for the cut-off. For consistency we will use
$p_{T,\ell}^\mr{hardest}=340~\mr{GeV}$ as cut-off for all values of
$c_{WWW}/\Lambda^2$ and $c_{\tilde{W}WW}/\Lambda^2$.
%
%%%%%%%%%%
%
\begin{table}[t]
  \begin{center}
    \resizebox{\columnwidth}{!}{
    \begin{tabular} {l | c c  | c c }
      & events @ $300~\mr{fb}^{-1}$
      & significance  & events @ $3000~\mr{fb}^{-1}$
      & significance \\
      \toprule 
      SM & 0.692 &  - & 6.92 & -  \\
	\midrule
      $\frac{c_{WWW}}{\Lambda^2}=-5~\mr{TeV}^{-2}$ & 1.49 & 0.96 &
      14.9 & 3.0 \\
      $\frac{c_{\tilde{W}WW}}{\Lambda^2}=-5 ~\mr{TeV}^{-2}$ & 3.76 &
      3.7 & 37.6 & 11.64 \\
      %	\midrule
      \bottomrule
    \end{tabular}}
    \caption{Number of events for $pp\rightarrow\llll jj$ at NLO-QCD
      at the LHC with $\sqrt{s}=14~\mr{TeV}$ within the cuts of
      \cref{eq:pttag-cuts,eq:rap-cuts,eq:lepton-cuts,eq:leprap-cuts,eq:mass-cut} and an
      additional cut of $p_{T,\ell}^\mr{hardest}>340~\mr{GeV}$, together
      with the significance of the signal defined in
      \cref{eq:deltasigma}.}
    \label{tab:limits14NLO}
  \end{center}
\end{table}
%
%%%%%%%%%%
%
%
%%%%%%%%%%
%
\begin{table}[t]
  \begin{center}
    \resizebox{\columnwidth}{!}{
    \begin{tabular} {l | c c  | c c }
      & events @ $300~\mr{fb}^{-1}$
      & significance  & events @ $3000~\mr{fb}^{-1}$
      & significance \\
      \toprule 
      SM & 0.599 &  - & 5.99 & -  \\
	\midrule
      $\frac{c_{WWW}}{\Lambda^2}=-5~\mr{TeV}^{-2}$ & 1.22 & 0.80 &
      12.2 & 2.5 \\
      $\frac{c_{\tilde{W}WW}}{\Lambda^2}=-5 ~\mr{TeV}^{-2}$ & 3.03 &
      3.1 & 30.3 & 9.9 \\
      %	\midrule
      \bottomrule
    \end{tabular}}
    \caption{Same as \cref{tab:limits14NLO}, but at LO.} 
    \label{tab:limits14LO}
  \end{center}
\end{table}

The significance of a non-Standard Model (nSM) signal is defined via
the number of events in the nSM and the SM scenarios as
\begin{align}
\frac{|\#\,\mr{events(nSM)}-\#\,\mr{events(SM)}|}{\sqrt{\#\,\mr{events(SM)}}}\,.
\label{eq:deltasigma}
\end{align}
Assuming that events are distributed according to a Gaussian
distribution, a one-, two- and three-sigma significance correspond to
the well-known 68\%, 95\% and 99.8\% probabilities. When the expected
number of SM events is low, one needs however to bear in mind that events
are distributed according to a Poisson distribution. In this case
these significances correspond to lower probabilities. However, when the
expected number of SM events is larger than five, we find that
these probabilities differ already by 1\%.

At $300~\mr{fb}^{-1}$ the dimension-six operators result in a
significant signal for the CP-violating coupling. However, in the
high-luminosity phase of the LHC with $3000~\mr{fb}^{-1}$,
CP-conserving operators with $c_{WWW}/\Lambda^2=-5~\mr{TeV}^{-2}$
become significant at the three sigma level and CP-violating operators
with $c_{\tilde{W}WW}/\Lambda^2=-5~\mr{TeV}^{-2}$ are significant to
more than five sigma.

It is worth noting that the significance decreases, if only LO results are
taken into account. Comparing \cref{tab:limits14NLO,tab:limits14LO} we observe that the significance increases by
$\sim20\%$, when NLO-QCD corrections are included. This strongly
favours including the NLO-QCD corrections, as a similar gain in
significance by technical means would require an increase in luminosity of $\sim44\%$.
%%%%%%%%%%
%
%
\begin{table}[t]
    \resizebox{\columnwidth}{!}{\begin{tabular} {r | c c  | c c | c c}
	$\frac{c_{WWW}}{\Lambda^2}$ & events @ $14~\mr{TeV}$
	& significance & events @ $33~\mr{TeV}$
      & significance & events @ $100~\mr{TeV}$
	& significance \\
	\toprule
	$0.0~\mr{TeV}^{-2}$ & 0.200 &  - & 3.26 & - & 32.1 & - \\
	\midrule

	$-2.0~\mr{TeV}^{-2}$ & 0.234 & 0.0765 & 4.47 & 0.671 & 74.6 & 7.51 \\

	$-4.0~\mr{TeV}^{-2}$ & 0.334 & 0.301 & 8.12 & 2.70 & 203 & 30.2 \\

	$-6.0~\mr{TeV}^{-2}$ & 0.496 & 0.663 & 14.3 & 6.10 & 419 & 68.3 \\

	$-8.0~\mr{TeV}^{-2}$ & 0.725 & 1.18 & 22.8 & 10.8 & 720 & 122 \\

	$-10.0~\mr{TeV}^{-2}$ & 1.01 & 1.82 & 33.7 & 16.9 & 1110 & 190 \\
	\bottomrule
    \end{tabular}}
    \caption{Number of events for $pp\rightarrow\llll jj$ for different collider
      energies with an integrated luminosity $100~\mr{fb}^{-1}$, within the cuts
      of
      \cref{eq:pttag-cuts,eq:rap-cuts,eq:lepton-cuts,eq:leprap-cuts,eq:mass-cut}
      and an additional cut of $p_{T,\ell}^{\mr{hardest}}>340~\mr{GeV}$ in the
      SM and including the effect of ${\mathcal O}_{WWW}$, together with the
      significance of the signal defined in \cref{eq:deltasigma}.}
    \label{tab:limitscwww}
\end{table} 
\begin{table}[t]
    \resizebox{\columnwidth}{!}{\begin{tabular} {r | c c  | c c | c c}
      $\frac{c_{\tilde{W}WW}}{\Lambda^2}$ & events @ $14~\mr{TeV}$
      & significance & events @ $33~\mr{TeV}$
      & significance & events @ $100~\mr{TeV}$
      & significance \\
      \toprule
      $0.0~\mr{TeV}^{-2}$ & 0.200 & - & 3.26 & - & 32.1 & - \\
      \midrule

      $-2.0~\mr{TeV}^{-2}$ & 0.331 & 0.293 & 8.12 & 2.70 & 205 & 30.3 \\

      $-4.0~\mr{TeV}^{-2}$ & 0.717 & 1.16 & 22.8 & 10.9 & 723 & 121 \\

      $-6.0~\mr{TeV}^{-2}$ & 1.36 & 2.60 & 47.3 & 24.4 & 1580 & 272 \\

      $-8.0~\mr{TeV}^{-2}$ & 2.27 & 4.64 & 81.7 & 43.5 & 2790 & 484 \\

      $-10.0~\mr{TeV}^{-2}$ & 3.43 & 7.23 & 125 & 67.7 & 4350 & 759 \\
      \bottomrule
    \end{tabular}}
    \caption{Same as \cref{tab:limitscwww}, but including the
      operator ${\mathcal O}_{\tilde WWW}$.}
    \label{tab:limitscpwww}
\end{table}

If we consider only contributions of one operator,
e.g. $\mathcal{O}_{WWW}$, the matrix element squared schematically
takes the form
\begin{align}
\label{eq:msq}
|\mathcal{M}|^2=|\mathcal{M}_{SM}|^2 +
 \frac{c^2_{WWW}}{\Lambda^4}|\tilde{\mathcal{M}}_{WWW}|^2
 +\frac{c_{WWW}}{\Lambda^2}(\tilde{\mathcal{M}}_{WWW}\mathcal{M}^*_{SM}
 +\mathcal{M}_{SM}\tilde{\mathcal{M}}^*_{WWW})\,.
\end{align}
By calculating the cross section for at least three different values
of the coupling $c_{WWW}/\Lambda^2$ (or $c_{\tilde WWW}/\Lambda^2$) it
is possible to interpolate the cross section for any value of the
coupling. Using this, we expect the following one sigma bounds for the
LHC at $300~\mr{fb}^{-1}$, 
\begin{align}
\label{eq:couplimits1a}
-4.98~\mr{TeV}^{-2} < &\frac{c_{WWW}}{\Lambda^2} < 5.12~\mr{TeV}^{-2}\,, \\
\label{eq:couplimits2a}
-2.54~\mr{TeV}^{-2} < &\frac{c_{\tilde{W}WW}}{\Lambda^2} < 2.54~\mr{TeV}^{-2}\,,
\end{align} 
and at $3000~\mr{fb}^{-1}$,    
\begin{align}
\label{eq:couplimits1b}
-2.77~\mr{TeV}^{-2} < &\frac{c_{WWW}}{\Lambda^2} < 2.91~\mr{TeV}^{-2}\,, \\
\label{eq:couplimits2b}
-1.43~\mr{TeV}^{-2} < &\frac{c_{\tilde{W}WW}}{\Lambda^2} < 1.43~\mr{TeV}^{-2}\,.
\end{align} 
We see that these limits are already tighter than  the current experimental limits
quoted in \cref{eq:couplimits1,eq:couplimits2}. Note
that the limits only improve by a factor $10^{1/4}\sim 1.8$ when the
luminosity is increased by a factor of $10$. This is related to the
fact that for large values of $c_{WWW}/\Lambda^2$ we are essentially
only probing $c^2_{WWW}/\Lambda^4$, see \cref{eq:msq}. 
The limits on the coupling improve then with a quartic root of the available luminosity. 
%
%
%%%%%%%%%%%%%%%%%
%
\begin{figure}[t]
  \resizebox{\columnwidth}{!}{
\begin{minipage}[c]{0.325\textwidth}
\begin{equation*}
  14~\mr{TeV}
\end{equation*}
\includegraphics[angle=0,width=1.0\textwidth]{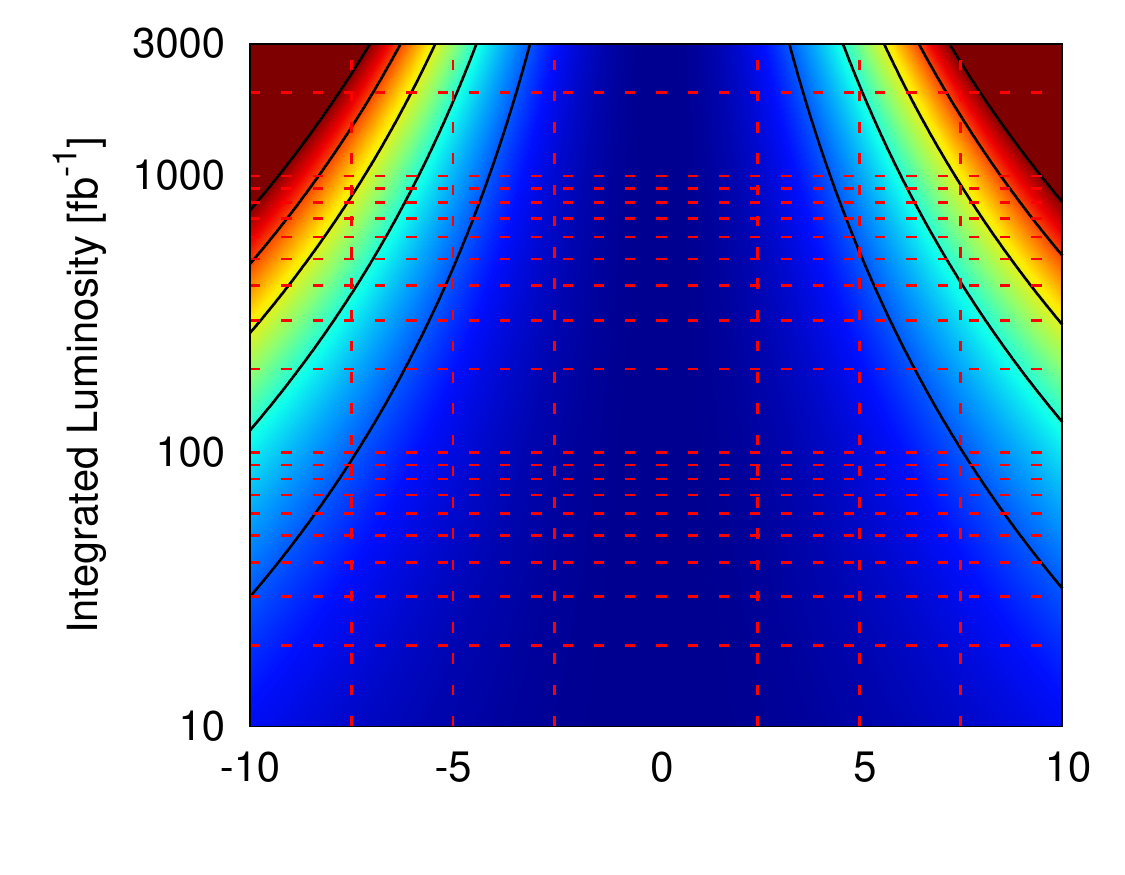}
\includegraphics[angle=0,width=1.0\textwidth]{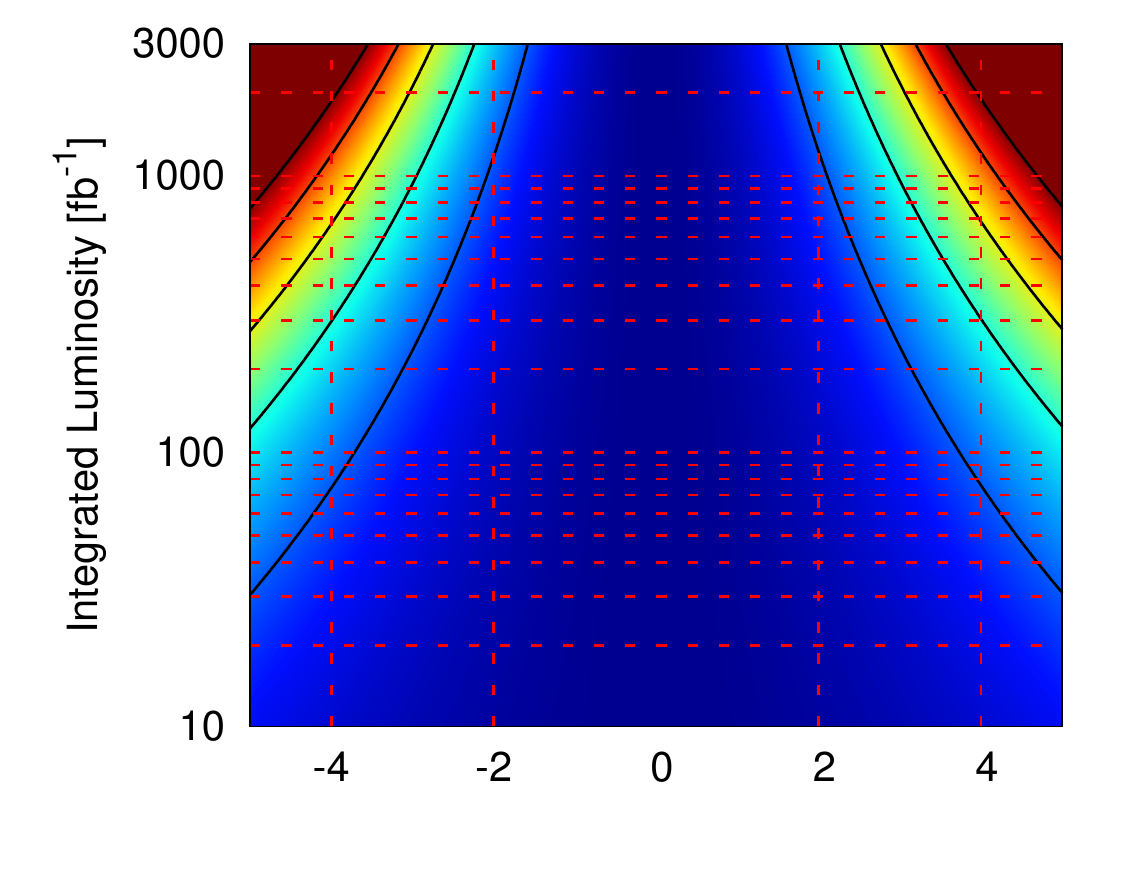}
\end{minipage}
\begin{minipage}[c]{0.325\textwidth}
\begin{equation*}
  33~\mr{TeV}
\end{equation*}
\includegraphics[angle=0,width=1.0\textwidth]{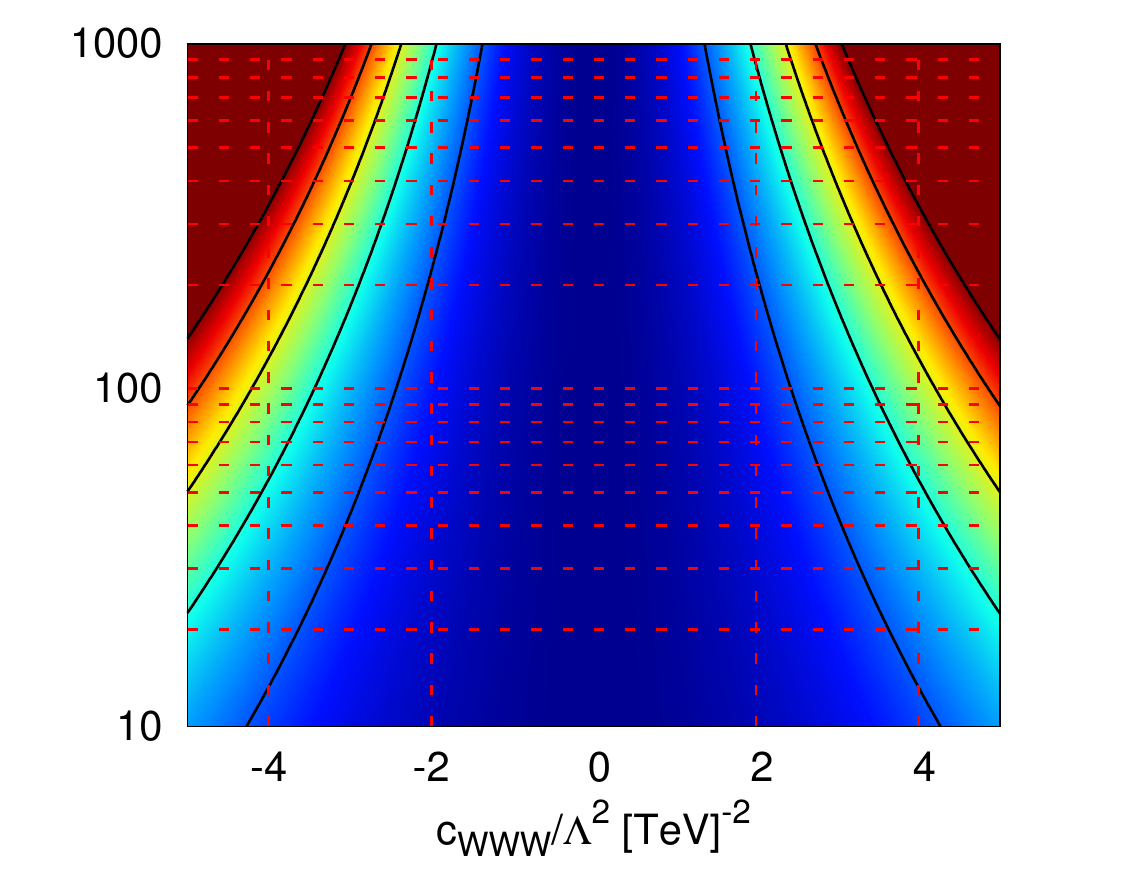}
\includegraphics[angle=0,width=1.0\textwidth]{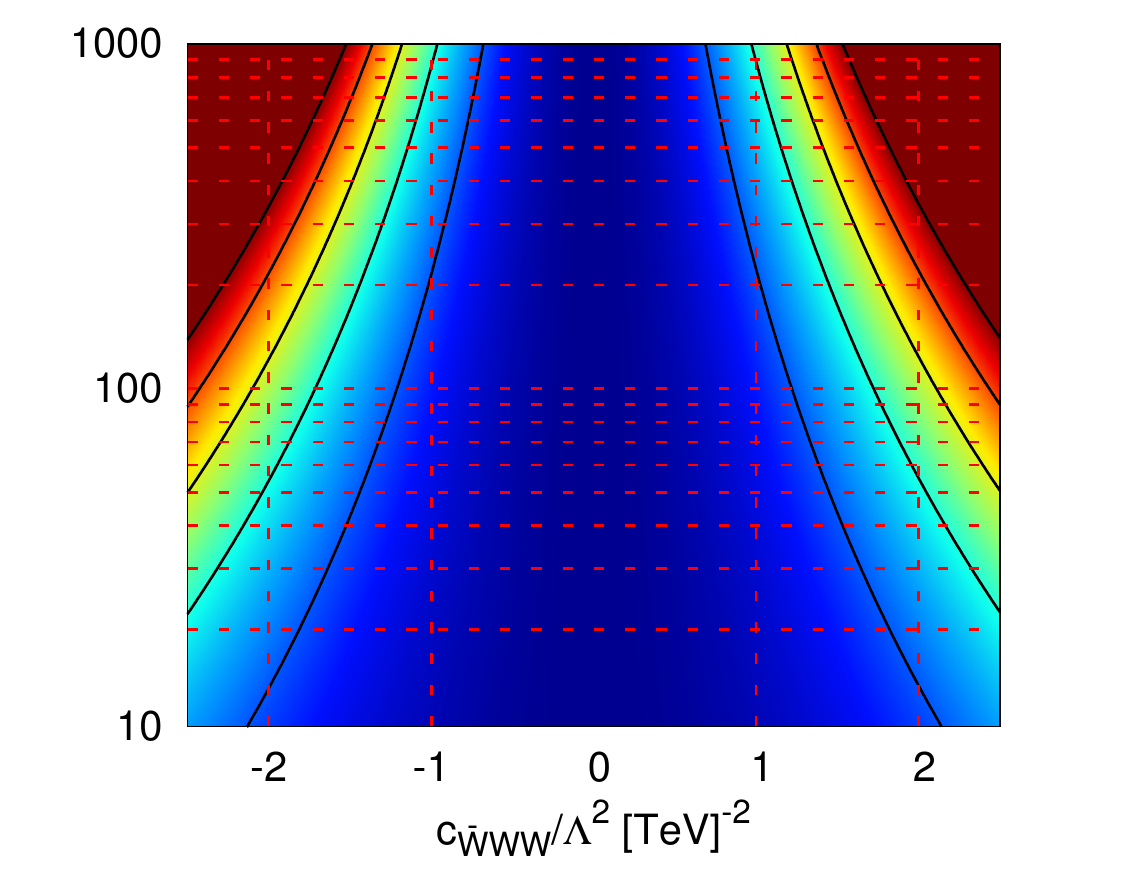}
\end{minipage}
\begin{minipage}[c]{0.325\textwidth}
\begin{equation*}
  100~\mr{TeV}
\end{equation*}
\includegraphics[angle=0,width=1.0\textwidth]{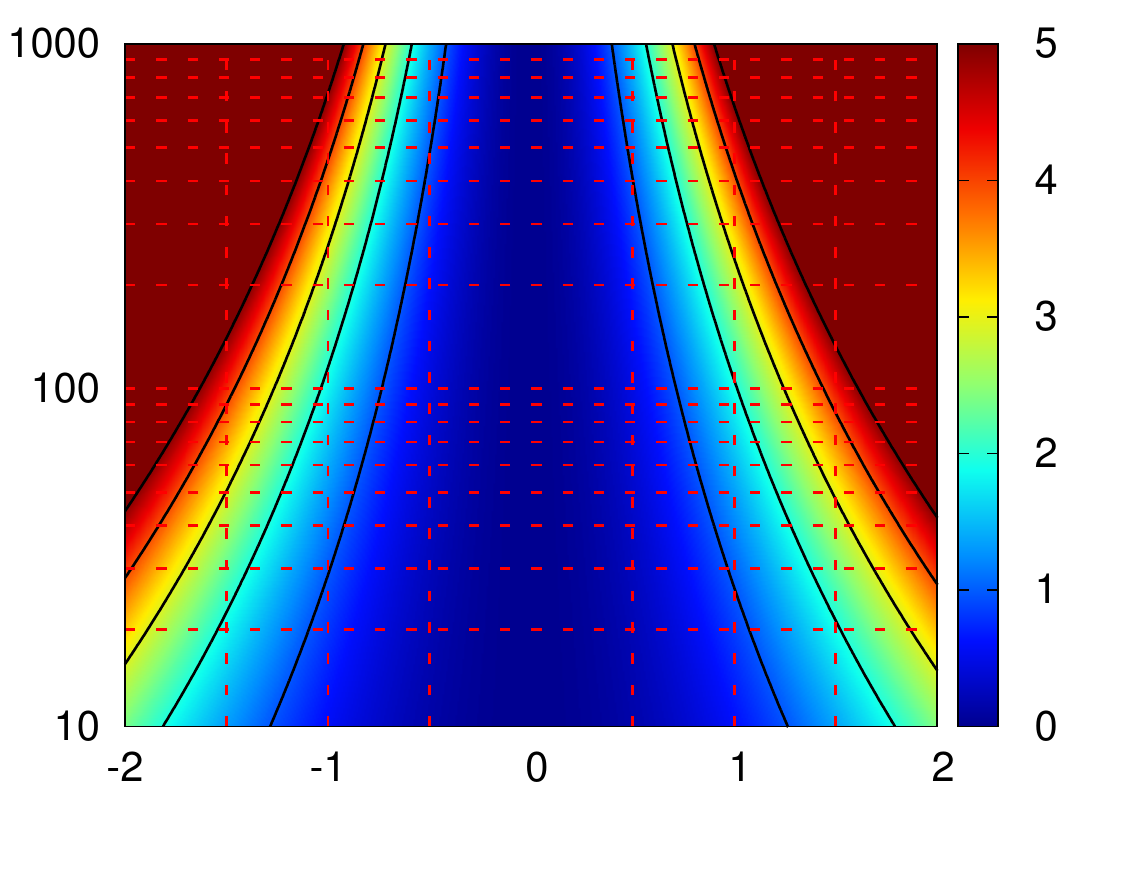}
\includegraphics[angle=0,width=1.0\textwidth]{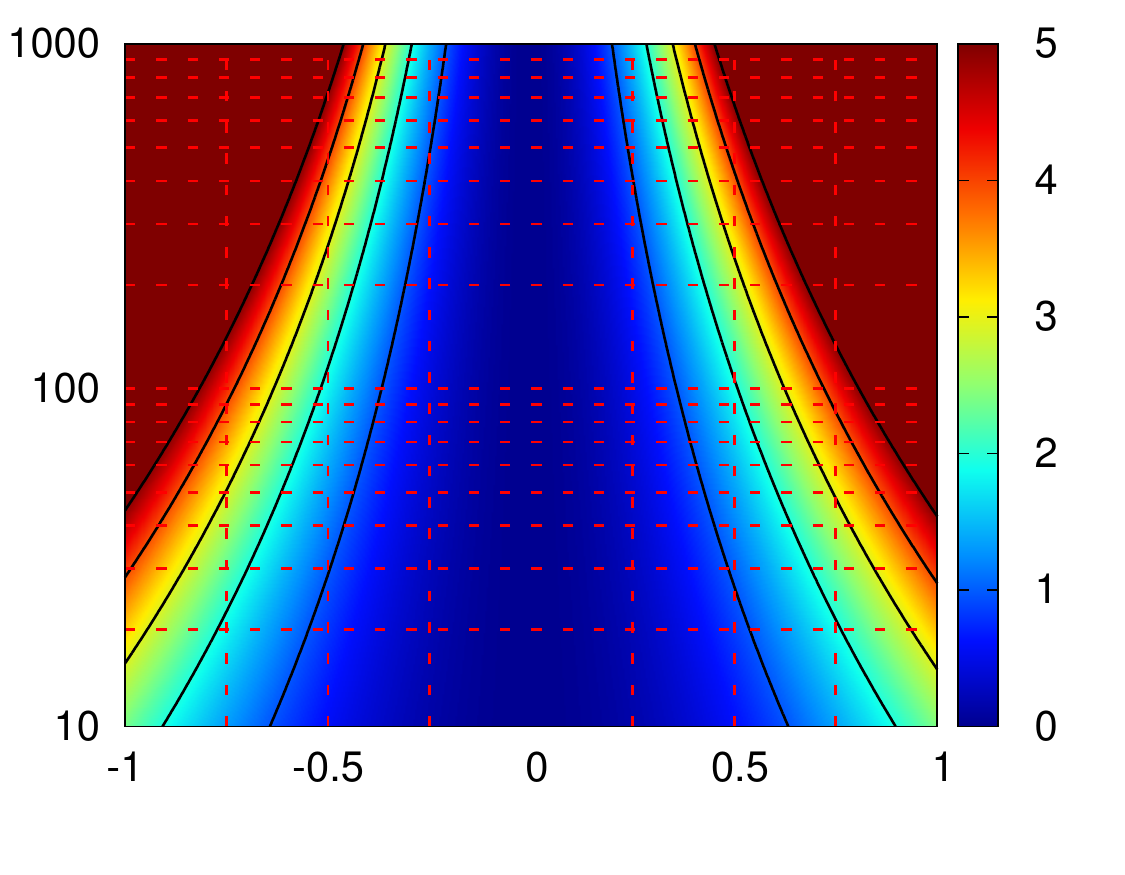}
\end{minipage}}
\caption{Significance of the two couplings $c_{WWW}/\Lambda^2$ and
  $c_{\tilde{W}WW}/\Lambda^2$ for the process $pp \rightarrow
  \ell^+\ell^-\ell'^+\ell'^- jj$ at $14~\mr{TeV}$, $33~\mr{TeV}$ and
  $100~\mr{TeV}$ within the cuts of
  \cref{eq:pttag-cuts,eq:rap-cuts,eq:lepton-cuts,eq:leprap-cuts,eq:mass-cut} and
  $p_{T,\ell}^\mr{hardest}>340~\mr{GeV}$, as a function of the integrated
  luminosity. The five black lines indicate one, two, three, four and five sigma
  significance defined in \cref{eq:deltasigma}, corresponding to the colour code
  indicated on the right-hand-side.  }
\label{fig:surface}
\end{figure} 
%
%%%%%%%%%%%%%%%%%%
%

Even better significances could be obtained with hadron colliders operating at
higher energies, such as the high-energy upgrade of the LHC (HE-LHC) with an
energy of $\sqrt{s}=33 \TeV$, or a FCC with an energy of up to
$\sqrt{s}=100 \TeV$. \cref{tab:limitscwww,tab:limitscpwww} show expected numbers
of events and associated significances for various scenarios at the LHC, HE-LHC
and FCC at LO. As reported above the significance is expected to increase at
NLO QCD.

Depending on the luminosity delivered, already for rather small
values of the operator coefficient $c_{WWW}/\Lambda^2$ an excess
over the SM values should be visible. Here we observe that the
significances grow faster than the collider energy squared.

To better illustrate the impact of increasing energy and integrated
luminosity, we have plotted the significance of a signal as a
function of the value of the coupling and the integrated luminosity
for each of the energies $14~\mr{TeV}$, $33~\mr{TeV}$, and
$100~\mr{TeV}$ in \cref{fig:surface}. It is obvious from these
plots that in order to improve the current limits on anomalous
couplings, higher energy is much more useful than higher luminosity.

It should be noted that we have disregarded the effects of various reducible and
irreducible background processes, e.g.\ QCD-induced $ZZjj$ production, which
would increase the SM contribution by about 50\% within our setup.  A realistic
assessment of the full significances would require to include these backgrounds,
as well as additional uncertainties, such as experimental efficiencies,
mis-identification issues, etc.  It is outside the scope of the present work to
systematically account for these effects.
%
%In addition to that, the uncertainties on the parton
%distribution functions are still considerable, in particular at high energies, such as $33$ and $100~\mr{TeV}$. However, uncertainties on parton distributions will be reduced with LHC run-2 data and with initial data of any high-energy hadron collider.  
%Studying such uncertainties at this time seems therefore premature. 

\section{Conclusions}
\label{sec:concl}
In this section we have reviewed the \POWHEG{} method for matching \NLO{}
calculations and parton showers, and presented an implementation of electroweak
$ZZjj$ production in the \POWHEGBOXVT{}. We take non-resonant contributions,
off-shell effects and spin correlations of the final-state particles into
account.
In the context of the Standard Model, we have considered the leptonic and
semi-leptonic decay modes of the $Z$~bosons. In addition, effects of new physics
in the gauge-boson sector that arise from an effective Lagrangian with operators
up to dimension six have been implemented. 

Here, we have discussed results for two specific scenarios: firstly, we have
performed a numerical analysis of Standard-Model $\eemm jj$ production at the
LHC with $\sqrt{s}=14\TeV$, in the regime where both $Z$~bosons are close to
on-shell. In this setup, the impact of the parton shower is small for most
observables related to the hard leptons and tagging jets, while larger effects
are observed in distributions related to an extra jet; secondly, we have
considered an effective field theory extension of the Standard Model with
operators of up to dimension six, and explored the impact of such operators on
observables in VBF $ZZjj$ processes. We found that tails of transverse-momentum
and invariant mass distributions of the hard jets and leptons are most sensitive
to such new-physics contributions.

Since the statistical significance of $ZZjj$ results at the LHC with an energy
of $\sqrt{s}=14\TeV$ and an integrated luminosity of about $300~\mr{fb}^{-1}$ is
limited, we additionally explored scenarios for high-energy proton colliders,
such as an HE-LHC and a FCC, with collider energies of $33\TeV$ and $100\TeV$,
respectively. We found that an increase in energy would help, much more than an
increase in luminosity, to substantially improve current limits on anomalous
couplings in the gauge boson sector.  For instance, an improvement in
significance by a factor of four can be obtained by increasing the energy by a
factor of (less than) two, or by increasing the luminosity by a factor of 16.
We also note that, for the process we have considered, relative NLO corrections
in the SM and in the effective field theory approach are of the same
size. However, since in the effective field theory scenario the number of events
in tails of transverse-momentum distributions is larger than in the SM, the NLO
corrections increase the significance by the square root of the $K$~factor, see
\cref{eq:deltasigma}. This, together with the well-known fact that uncertainties
are reduced significantly at NLO, strongly supports the use of NLO simulations
in the context of searches for new physics in the gauge-boson sector.

\chapter{\MINLO{} - A Path to \NNLO{} Matching}\label{ch:minlo}
After having studied the matching of fixed-order \NLO{} calculations with parton
showers, we now turn to the problem of matching fixed-order \NNLO{} calculations
with parton showers using \POWHEG{} and \MINLO{}\footnote{Multi-Scale Improved
  \NLO{}}~\cite{Hamilton:2012np,Hamilton:2012rf}. As it turns out, this is
related to the problem of merging samples of different jet multiplicities in a
consistent way. There exists a number of different merging procedures at both
\LO{} and \NLO{}, all of them having in common that one has to pick an
unphysical \emph{merging scale} to separate samples of different jet
multiplicities~\cite{Alioli:2011nr,Lavesson:2008ah,Hoeche:2012yf,Gehrmann:2012yg,Frederix:2012ps,Alioli:2012fc,Platzer:2012bs,Lonnblad:2012ix}.

Originally \MINLO{} was introduced as a procedure for assigning factorisation
and renormalisation scales to \NLOPS{} processes involving jets, in a
well-defined manner. The procedure involves assigning Sudakov form factors to
compensate for the scale choices, which in turn make the calculation finite even
in the presence of an unresolved jet. It was quickly realised by the authors
in~\cite{Hamilton:2012rf} that by a simple extension of the original procedure,
one could obtain a simulation of a colour singlet which was \NLO{} accurate in
both the 0 and 1-jet bin. By a reweighting procedure the simulation can be made
\NNLO{} accurate in the 0-jet bin and hence \NNLOPS{} accuracy can be
reached. One advantage of the \MINLO{} procedure compared with other merging
procedures is that one does not have to introduce an unphysical merging scale to
separate different event samples.

The method has already been applied to gluon-fusion Higgs
production~\cite{Hamilton:2013fea}, associate Higgs
production~\cite{Astill:2016hpa}, and to $WW$
production~\cite{Hamilton:2016bfu}. Here we will first discuss the details of
the \MINLO{} procedure and then apply it to Drell-Yan production. Currently
there exists two other schemes for matching \NNLO{} calculations and parton
showers based on \texttt{GENEVA}~\cite{Alioli:2013hqa} and
\texttt{UNLOPS}~\cite{Hoche:2014dla,Hoeche:2014aia} which we only mention here
for completeness.

\section{The Method}\label{sec:minlo}
When computing cross sections in hadron-hadron collisions we are always faced
with the problem of assigning \emph{unphysical} factorisation and
renormalisation scales, $\mu_F$ and $\mu_R$. We stress here that the scale
choice is unphysical, since the final cross section only depends on the scale
choice beyond the calculated accuracy. Hence if our calculation is accurate to
$\mathcal{O}(\as)$ we expect the dependence of the cross section on $\mu_F$ and
$\mu_R$ to be $\mathcal{O}(\as^2)$. Indeed, as we have seen in previous
chapters, this property of the cross section is used as a crude way of
estimating missing higher order corrections in a given calculation. In most
applications one therefore looks for a physical scale which seems to describe
the process well. In the simplest cases this can even be a fixed scale which
does not take the kinematics of individual events into account. However, this
simple picture breaks down as soon as one looks at processes with more than one
physical scale.

As an example take Drell-Yan production. At the fully inclusive level there is
really only one scale in the problem, namely the mass of the vector boson. Hence
the assignment of a factorisation scale is essentially unique. However if we
turn to vector boson production in association with one (or more) jets, the
situation changes. Now we are faced with two scales; the transverse momentum of
the hardest jet and the invariant mass of the vector boson. In this case we know
that the cross section has Sudakov double logarithms of the ratio of these two
scales and that they can become very large in the low $\pt$ range. These
logarithms can't be compensated by a smart choice of factorisation and
renormalisation scales but have to be dealt with through all-order
resummation. This can be done through the inclusion of Sudakov form factors.

At LO a procedure for uniquely assigning factorisation and
renormalisation scales and Sudakov form factors exists under the name of the
CKKW method~\cite{Catani:2001cc,Krauss:2002up,Mrenna:2003if}. \MINLO{} is
essentially an \NLO{} extension of the CKKW method, modified in such a way that
\NLO{} accuracy is not spoiled by the Sudakov form factors. In the CKKW method
one reconstructs the most likely branching history of an event and provides
matrix element rescaling in terms of the coupling constant and Sudakov form
factors. The Sudakov form factor in the CKKW procedure is given by
\begin{equation}
  \ln\Delta_f(Q_0,Q) = -\int_{Q_0}^{Q}\frac{dq^2}{q^2}\left[A^f(\as(q^2))\ln\left(\frac{Q^2}{q^2}\right)+B^f(\as(q^2))\right],
  \label{eq:minlosudakov}
\end{equation}
where $f$ is either a quark ($q$) or a gluon ($g$). The two functions $A$ and
$B$ have perturbative expansions given by
\begin{align}
  A^f(\as) = \sum_{i=1}A_i^f \as^i, \qquad B^f(\as) = \sum_{i=1}B_i^f \as^i, 
  \label{eq:ABexp}
\end{align}
where $A_1^f$, $A_2^f$ and $B_1^f$ are process independent and are given
by~\cite{Catani:1988vd,Kodaira:1981nh}
\begin{align}
  A_1^q &= \frac{C_F}{2\pi}, \qquad A_2^q = \frac{C_FK}{4\pi^2}, \qquad B_1^q = -\frac{3C_F}{4\pi}, \\ 
  A_1^g &= \frac{C_A}{2\pi}, \qquad A_2^g = \frac{C_AK}{4\pi^2}, \qquad B_1^g = -\beta_0. 
\end{align}
Here $\beta_0$ is the lowest order coefficient in the running of $\as$ as
defined in \cref{eq:betacoeff} , $C_F=4/3$, $C_A=3$, and the two-loop cusp
anomalous dimension is given by
\begin{equation}
  K = \left( \frac{67}{18} - \frac{\pi^2}{6}\right)C_A - \frac{5}{9}n_f,
\end{equation}
where $n_f$ is the number of active flavours. For the purpose of our discussion
we shall also need the $B_2^q$ coefficient in Drell-Yan production which is
given by~\cite{Davies:1984sp,Davies:1984hs}
\begin{align}
  B_2^q = \frac{1}{2\pi^2}&\Big[\left(\frac{\pi^2}{4}-\frac{3}{16}-3\zeta_3\right)C_F^2 + \left(\frac{11}{36}\pi^2 - \frac{193}{48}+\frac{3}{2}\zeta_3\right)C_FC_A \notag\\
    & \quad + \left(\frac{17}{24} - \frac{\pi^2}{18}\right)C_Fn_f\Big] + 4 \zeta_3(A_1^q)^2,
  \label{eq:B2DY}
\end{align}
where $\zeta_3 = 1.202056903\dots$ is the Riemann zeta function evaluated at
$3$. The Sudakov form factor itself also has an expansion in terms of $\as$
given by
\begin{equation}
  \Delta_f(Q_0,Q) = 1 + \sum_{i=1}\as^i\Delta_f^{(i)}(Q_0,Q).\label{eq:Sudakovexp}
\end{equation}
The first order result can be found by expanding the exponent in $\as$ and
then taking the first functional derivative. It is given by
\begin{equation}
  \Delta_f^{(1)}(Q_0,Q) = -\frac{1}{2}A_1^f\ln^2\frac{Q_0^2}{Q^2}+B_1^f\ln\frac{Q_0^2}{Q^2}.
\end{equation}
The scale at which the factors of $\as$ in \cref{eq:Sudakovexp} should be
evaluated will be specified a little later. This expansion makes it explicit
that providing a matrix element with one or more Sudakov form factors
effectively introduces an \NLO{} correction. In the CKKW method, which is
formulated at leading order, this doesn't spoil the accuracy. However in order
to generalise the procedure to \NLO{} it is clear that we need to compensate for
the Sudakov form factors by subtracting the $\mathcal{O}(\as)$ term in the right
way. In that sense the \MINLO{} procedure extends the CKKW procedure by
subtracting these terms and by assigning a scale to the extra factor of $\as$
which arises at \NLO{} and therefore isn't present in the CKKW procedure.

\subsection{\MINLO{}}
In order to formulate \MINLO{}, we start by writing the $\bar{\mathcal{B}}$ term
of the \POWHEG{} formula of \cref{bbar} in this very condensed way
\begin{align}
  \bar{\mathcal{B}}_{\POWHEG{}} = \as^N(\mu)\left[\mathcal{B} + \as(\mu) \left(\mathcal{V}(\mu) + \int d\Phi_r \mathcal{R}\right)\right],\label{eq:bbarinit} 
\end{align}
where $\mu$ is some arbitrary fixed scale. The aim is to construct
$\bar{\mathcal{B}}_{\MINLO{}}$ such that it is equal to
$\bar{\mathcal{B}}_{\POWHEG{}}$ up to relative $\mathcal{O}(\as^2)$
corrections. In practice we will do so by acting with a rescaling factor,
$\mathcal{W}_{\MINLO{}}$, on $\bar{\mathcal{B}}_{\POWHEG{}}$ while subtracting
spurious terms which spoil the \NLO{} accuracy.

The prescription given here is not exactly identical to the one found
in~\cite{Hamilton:2012np}. Here we adopt the convention that $\int d\Phi_r
\mathcal{R}$ is given exactly the same rescaling factor as the underlying Born
process whereas in the original proposal the real momenta were to be clustered
separately discarding the lowest splitting scale. This makes for a much simpler
but formally identical procedure.

Given a list of Born-level momenta and flavours, the \MINLO{} procedure is as follows:

\begin{itemize}
  \item[1.] We cluster all partons with the $\kt$-clustering algorithm in order
    to determine the most probable branching history of the event. The
    clustering is done such that only partons which are compatible in flavour
    are clustered together, i.e. gluons clustered with gluons, yielding gluon
    pseudopartons, gluons with quarks yielding quark pseudopartons of the same
    flavour, and quarks with opposite flavour anti-quarks, yielding gluon
    pseudpartons. We keep doing this until there are no partons left to
    cluster. From the clustering we obtain $n \le N$ clustering scales
    $q_1<\dots <q_n$. In general the clustering will stop when there are no more
    coloured particles left\footnote{This is however not the case if we are
      studying for instance dijet or VBF production, in which case one has to
      stop the clustering when two partons are left.}. We then define the
    resolution scale by $Q_0=q_1$ which will be the low scale entering all
    Sudakov form factors.
  \item[2.] We proceed to determine the scale $Q$ of the \emph{primary system},
    which is simply the system leftover from the clustering. In general we will
    take $Q$ to be the invariant mass of the primary system, but for complicated
    primary systems that might not be the best choice. In the event that $Q<q_n$
    we set $Q=q_n$. This scale will be used to evaluate the remaining $m=N-n$
    factors of $\as$ not associated with a branching. For ease of notation we
    will define $q_{n+1}=Q$.
  \item[3.] Compute $\mu_Q = K_R Q$ and $\mu_i = K_R q_i$ where $K_R$ is the
    renormalisation scale factor.
  \item[4.] Set the renormalisation scale explicitly appearing in the virtual
    corrections equal to $\mu_R = \left((\mu_Q)^m\cdot
    \prod_{i=1}^{n}\mu_i\right)^{\frac{1}{N}}$ and the factorisation scale
    explicitly appearing in the PDFs and collinear subtraction remnants equal
    to $\mu_F = K_F Q_0$ where $K_F$ is the factorisation scale factor. In
    practice we set the renormalisation scale by adding a term
    $\mathcal{B}\beta_0 \ln(\mu_R^2/\mu^2)$ to the virtual contribution.
  \item[5.] To each external line, $l$, we attach a Sudakov form factor
    $\Delta_{f_l}(Q_0,q_{k_l})$ where $f_l$ is the flavour of the external line
    and $k_l$ is the node to which the external line attaches. Note that all
    external lines attaching to the first node $q_1$ have
    $\Delta_{f_l}(Q_0,q_{1})=1$. To skeleton lines joining two nodes $i$ and
    $j$, we include form factors of the form
    $\Delta_{f_{ij}}(Q_0,q_i)/\Delta_{f_{ij}}(Q_0,q_j)$, where $q_i>q_j$ and
    $f_{ij}$ is the flavour of the line joining the two nodes. We define the
    overall Sudakov factor by
    \begin{equation}
      \mathcal{F}(Q_0;q_2,...,q_{n+1})=\prod_{l}\Delta_{f_l}(Q_0,q_{k_l})\prod_{ij}\frac{\Delta_{f_{ij}}(Q_0,q_i)}{\Delta_{f_{ij}}(Q_0,q_j)}\label{eq:totsudakov}
    \end{equation}
    where the two products extend over all external and internal legs as
    described above.
    \item[6.] To compensate for the $\mathcal{O}(\as)$ term in
      \cref{eq:totsudakov} we rescale the Born-term, $\mathcal{B}$, in
      \cref{eq:bbarinit} such that
      \begin{align}
        B \rightarrow B \times \left[ 1 - \as' \mathcal{F}^{(1)}(Q_0;q_2,...,q_{n+1}) \right],
      \end{align}
      where
      \begin{equation}
        \mathcal{F}^{(1)}(Q_0;q_2,...,q_{n+1}) = \sum_{l}\Delta^{(1)}_{f_l}(Q_0,q_{k_l}) + \sum_{ij}\left[\Delta^{(1)}_{f_{ij}}(Q_0,q_i) - \Delta^{(1)}_{f_{ij}}(Q_0,q_j)\right].
      \end{equation}
      The value of $\as'$ to be used in this expression is set in
      \cref{eq:alphasprime}.
    \item[7.] Finally we rescale the factor of $\as(\mu)$ appearing in front of
      the virtual and real corrections, such that it is set equal to
      \begin{equation}
        \as' = \frac{1}{N}\left[m\as(\mu_Q) + \sum_{i=1}^{n}\as(\mu_i)\right]. \label{eq:alphasprime}
      \end{equation}
\end{itemize}
In the end we therefore write
\begin{align}
  \bar{\mathcal{B}}_{\MINLO{}} &= {\as^m(\mu_Q)\prod_{i=1}^n\as(\mu_i)} \mathcal{F}(Q_0;q_2,...,q_{n+1})\bigg[\mathcal{B} + \notag \\
    & \quad \as' \left(\mathcal{V}(\mu) + \int d\Phi_r \mathcal{R} + \mathcal{B}\left(\beta_0 \ln\frac{\mu_R^2}{\mu^2} - \mathcal{F}^{(1)}(Q_0;q_2,...,q_{n+1})\right)\right)\bigg]. \label{eq:bbarminlo}
\end{align}
This formula has the property that it is finite when integrated over radiation
arising beyond the primary system, and it retains \NLO{} accuracy due to the
subtraction of the lowest order expansion of $\mathcal{F}$. Events generated
according to the above $\bar{\mathcal{B}}$ function with the \POWHEG{} method
can be interfaced to a parton shower in exactly the same way as normal \POWHEG{}
events.

\section{Merging Without a Merging Scale}
\begin{figure}[thb]
  \centering
  \includegraphics[width=0.8\textwidth]{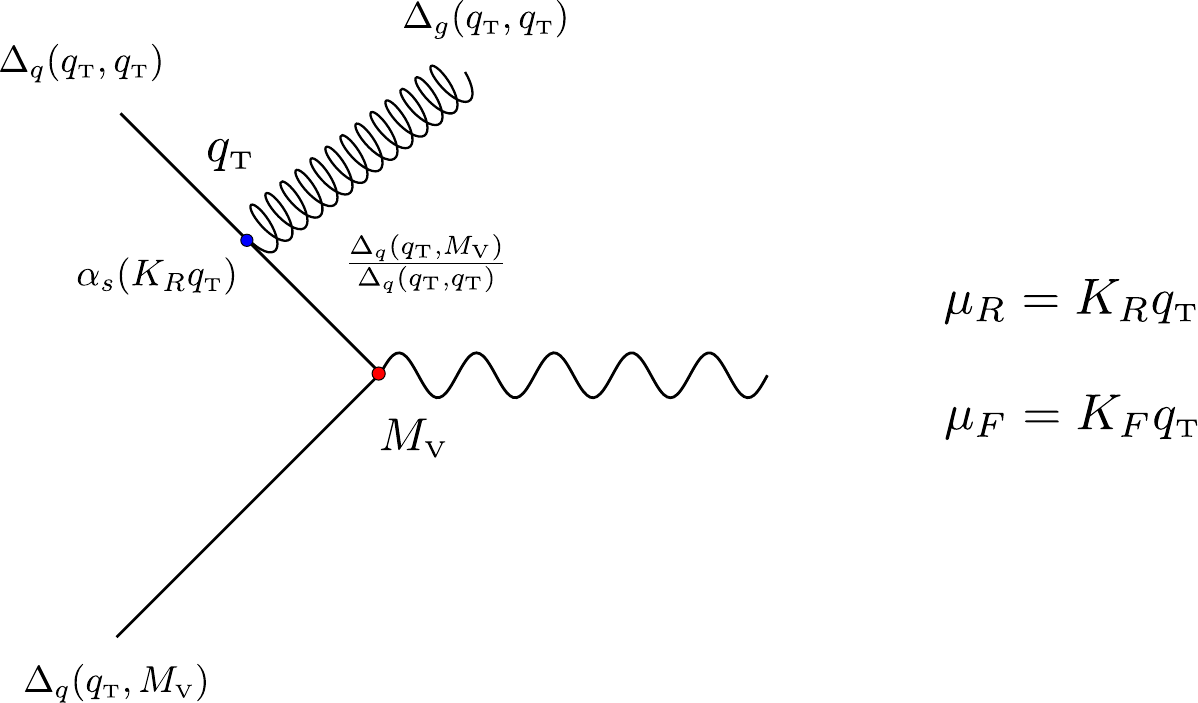}
  \caption{Drell-Yan production in association with one jet. $\MV$ is the
    invariant mass of the vector boson, $\qt$ is the transverse momentum of the
    jet (or the vector boson). We have attached Sudakov form factors,
    $\Delta_s$, according to the \MINLO{} procedure as outlined above. The
    strong coupling associated with the emitted gluon is evaluated in $K_R
    \qt$. The renormalisation and factorisation scales are set to $\mu_R = K_R
    \qt$ and $\mu_F = K_F \qt$ respectively.}
  \label{fig:DYdiagram}
\end{figure}In order to further study the \MINLO{} modified $\bar{\mathcal{B}}$ function we
turn to the production of a massive vector boson associated with one
jet, \VJ{} as shown in \cref{fig:DYdiagram}. In this case we have $m=0$ and $n=1$. We
cluster with the $\kt$-algorithm setting the radius parameter $R=1$. Since
there is only one branching in this process at Born-level, the lowest clustering
scale, $Q_0$, is simply the transverse momentum of the emitted parton, which is
equal to the transverse momentum of the vector boson, $\qt$. The primary scale,
$Q$, is taken to be the invariant mass of the vector boson, $\MV$. Using the
procedure described above, we find
\begin{equation}
  \mathcal{F}(\qt;\MV) = \Delta_q(\qt,\qt)\Delta_g(\qt,\qt)\Delta_q(\qt,\MV)\frac{\Delta_q(\qt,\MV)}{\Delta_q(\qt,\qt)} = \Delta^2_q(\qt,\MV),
\end{equation}
and hence
\begin{equation}
  \mathcal{F}^{(1)}(\qt;\MV) = 2\Delta^{(1)}_q(\qt,\MV).
\end{equation}
The \MINLO{} modified $\bar{\mathcal{B}}$ function is then given by
\begin{align}
  \bar{\mathcal{B}}_{\MINLO{}} &= \as(K_R \qt) \Delta^2_q(\qt,\MV) \bigg[B + \notag \\
    & \as(K_R\qt) \left(V(\mu) + \int d\Phi_r R + B\left(\beta_0 \ln\frac{K_R^2\qt^2}{\mu^2} - 2\Delta^{(1)}_q(\qt,\MV)\right)\right)\bigg], \label{eq:bbarvjminlo}
\end{align}
where $\mu$ is some fixed scale which we will take to be the on-shell mass of
the vector boson in our implementation. As already discussed above, this formula
will lead to \NLO{} accurate \VJ{} production, but it is not entirely clear how
accurate our expression is, when the hardest jet becomes unresolved. In the
original \MINLO{} paper~\cite{Hamilton:2012np} this question was not addressed
formally but very good behaviour of the predictions in the Sudakov peak regions
were observed even for observables which should not obviously benefit from the
\MINLO{} procedure. It was later shown that by modifying the \MINLO{} procedure
slightly, the \VJ{} generator would retain \NLO{} accuracy even in the Sudakov
peak region. Hence the \VJMINLO{} generator would effectively have merged an
\NLO{} \V{} sample with an \NLO{} \VJ{} sample. We say that the \VJMINLO{}
generator is both $\NLO^{(0)}$ and $\NLO^{(1)}$ accurate where $0,1$ refers to
the number of resolved jets.

The modification which is needed to the \MINLO{} procedure is very simple as we
will see shortly. However to show that the \VJMINLO{} generator is $\NLO^{(0)}$
accurate is fairly involved and we shall not reproduce the proof
of~\cite{Hamilton:2012rf} here. We will however arrive at the same conclusion as
the proof through naive power counting. We start by defining $L=\ln Q_0^2/Q^2$
and then rewrite \cref{eq:Sudakovexp} as
\begin{equation}
  \Delta_f(Q_0,Q) = 1 + \as\left( -\frac{1}{2}A_1 L^2 + B_1 L\right)  + \mathcal{O}(\as^2).
\end{equation}
As an expansion in $\as$ the $A_1$ and $B_1$ terms are equally important, but
near the Sudakov peak $L$ becomes large, and the $A_1$ term starts
dominating. We will take the Sudakov peak to mean $\as L^2 \sim 1$. In this
limit the above expansion is not a good one as higher order terms of the form
$(\as L^2)^n$ become as important as the leading order term. Of course the full
Sudakov form factor correctly takes care of these higher order terms, but here
we will use the expansion to quantify the precision of the calculation in the
Sudakov peak region.

Observe first that
\begin{equation}
  \as L^2 \sim 1 \quad \Rightarrow \quad L \sim \as^{-1/2},
\end{equation}
which means that the term proportional to $B_1$ goes as $\as L \sim
\as^{1/2}$. A necessary condition for the \VJMINLO{} generator to be
$\NLO^{(0)}$ accurate is that we are only neglecting terms of
$\mathcal{O}(\as^2)$. Further expanding the Sudakov form factor we get
\begin{align}
  &\ln\Delta_f(Q_0,Q) = \notag \\ & \qquad \as\left( -\frac{1}{2}[A_1 + \as A_2 +\as^2 A_3]L^2 + [B_1 + \as B_2 +\as^2 B_3] L\right)  + \dots
\end{align}
from which we easily read off that the $A_2$ term is $\mathcal{O}(\as)$, the
$B_2$ term is $\mathcal{O}(\as^{3/2})$, the $A_3$ term is $\mathcal{O}(\as^2)$,
and the $B_3$ term is $\mathcal{O}(\as^{5/2})$. Hence by including the $B_2$
term in the Sudakov form factor in addition to the $A_1$, $A_2$, and $B_1$
terms, we fulfil the condition that only terms of $\mathcal{O}(\as^2)$ are
neglected.

In addition to including the $B_2$ term in our Sudakov form factor, we also find
it useful to specify how the Sudakov form factor behaves when the
renormalisation scale is varied up and down around the central scale. This
variation can be absorbed into a redefinition of the $A$ and $B$ coefficients
given by~\cite{Hamilton:2012rf}
\begin{align}
  \tilde{A}_1 &= A_1 \\
  \tilde{A}_2 &= A_2 + 2A_1\beta_0 \ln K_R \\
  \tilde{B}_1 &= B_1 + 2A_1 \ln K_R \\
  \tilde{B}_2 &= B_2 + n\beta_0^2\ln K_R + 2A_2 \ln K_R + 2 A_1\beta_0 \ln^2K_R. 
\end{align}

\section{{\NNLOPS{} Accurate Drell-Yan Production}}
During Run I of the LHC, at $7$ and then $8 \TeV$ centre of mass energy, both
the ATLAS and CMS collaboration collected almost $30\;\mathrm{fb}^{-1}$ of
data. Because of their very large cross-sections and the very small systematic
uncertainties, Drell-Yan production through $W$ and $Z$ exchange are standard
candles at the LHC. Given the high-statistics reached, kinematic distributions
can be studied over many orders of magnitude. With Run II at $13 \TeV$ even more
$W$ and $Z$ events have become available~\cite{Aad:2016naf}, and in particular
it will be possible to study distributions over an even larger kinematic
range. These distributions provide important input to constrain parton
distribution functions. For example the rapidity distribution and the dilepton
invariant mass of both neutral and charged Drell-Yan data have been used
recently to constrain the photon content of the proton~\cite{Ball:2013hta}.

Higher-order corrections are indispensable for these studies, and Drell-Yan
production is to date the theoretically best described process at the LHC. The
cross-section is known through next-to-next-to-leading order (NNLO) in QCD
including the decay of $W$ and $Z$ bosons to
leptons~\cite{Anastasiou:2003yy,Melnikov:2006kv,Catani:2009sm}.
Two public codes exist (\DYNNLO{}~\cite{Catani:2009sm} and
\FEWZ{}~\cite{Gavin:2010az}) that implement QCD NNLO corrections to
the hadronic $W$ and $Z$ production.
Furthermore electroweak corrections have been the subject of intensive
studies~\cite{Baur:1997wa,Baur:2001ze,Dittmaier:2001ay,CarloniCalame:2007cd,Arbuzov:2007db,Dittmaier:2009cr}.
Electroweak Sudakov effects become more important at large invariant
mass~\cite{Denner:2000jv,Kuhn:2001hz,CarloniCalame:2007cd,Campbell:2013qaa}, a
region that was already interesting at Run I and which is being explored even
more during the current Run II of the LHC. Version 3 of
\FEWZ{}~\cite{Li:2012wna} implements also NLO EW corrections and the leading
photon initiated processes.
Recently, a framework for the calculation of the mixed QCD-electroweak ${\cal
  O}(\alpha_s\alpha)$ corrections to Drell-Yan processes in the resonance region
has been developed~\cite{Dittmaier:2014qza,Dittmaier:2015rxo}. The impact of
non-factorising (initial-final state) corrections is shown to be very
small. Factorisable ${\cal O}(\alpha_s\alpha_{ew})$ corrections in the pole
approximation have been computed. While QCD corrections are predominantly
initial state corrections, and EW contributions are predominantly final-state
corrections, because of kinematic effects there are sizeable differences between
the result of \Bref{Dittmaier:2014qza} and the naive product of NLO QCD and NLO
EW corrections.
 
Having constructed a Drell-Yan generator which is $\NLO^{(0)}$, $\NLO^{(1)}$,
and $\LO^{(2)}$ accurate it is obvious to ask if it is possible to reweight the
\VJMINLO{} events such that they exhibit $\NNLO^{(0)}$ accuracy without spoiling
the $\NLO^{(1)}$ and $\LO^{(2)}$ accuracy. We will here show that it is indeed
possible. The argument was first presented in~\cite{Hamilton:2012rf} and then
refined in~\cite{Hamilton:2013fea} for gluon fusion Higgs production. Here we
discuss how the procedure works for Drell-Yan production.

Our implementation relies on the following inputs:
\begin{itemize}
\item Les Houches events for the $Z$+one jet or $W$+one jet
  process~\cite{Alioli:2010qp} (respectively \ZJ{} and \WJ{} from now on), as
  implemented in \POWHEG{}, upgraded with the improved \MINLO{} procedure
  described in \cref{sec:minlo} in such a way that NLO accuracy is
  guaranteed for inclusive distributions without any jet cut;
\item NNLO accurate distributions, differential in the Born kinematics
  of the leptons, as obtained from \DYNNLO{}~\cite{Catani:2009sm};
\item a local reweighting procedure, described in details in
  \cref{sec:Theoretical-framework} below.
\end{itemize}

Compared with the standard \NLOPS{} implementation of $W$ and $Z$ in \POWHEG{} we
find considerably reduced theoretical uncertainties for inclusive
distributions. This is expected since our predictions have full NNLO accuracy.
Furthermore the 1-jet region is described at NLO accuracy in our
framework, even at small transverse momentum.

Throughout this work we pay particular attention to the issue of assigning a
theory uncertainty to our predictions. We describe our procedure in
\cref{subsec:Estimating-uncertainties}. Unlike the standard \POWHEG{}
approach (without a separation between singular and finite real contributions),
that is known to underestimate the theoretical uncertainty for the $W$, $Z$ or
Higgs boson transverse momentum~\cite{Nason:2012pr}, we believe that our
uncertainties are more reliable.  Compared with pure NNLO predictions, we find in
general a better description of observables sensitive to multiple emissions,
such as the boson transverse momentum, $\phi^*$, and jet-resolution variables
$d_{i}$ (which just vanish at NNLO starting from $i=2$). This is both because of
the underlying \MINLO{} procedure, and because of the \POWHEG{} framework.

\subsection{\NNLOPS{} Reweighting}\label{sec:Theoretical-framework}
We denote by $d\sigma^{{\scriptscriptstyle \mathrm{MINLO}}}/d\Phi$ the
cross-section obtained from the \VJMINLO{} event generator, fully differential
in the final state phase space, $\Phi$, at the level of the hardest emission
events, i.e. before parton shower.
Because of the properties of \MINLO{}, upon integration, this
distribution reproduces the next-to-leading order accurate,
$\mathcal{O}\left(\as\right)$, leptonic distributions inclusive in all
QCD radiation.
After integration over all QCD radiation only the leptonic system is left. This
can be characterised by three independent variables, for instance one can choose
the invariant mass of the lepton pair, $m_{\rm {ll}}$, the rapidity of the boson
before decay $y_V$, and the angle of the negatively (or positively) charged
lepton with respect to the beam, $\theta_l$. In the following we will use
$\Phi_B$ to denote the ensemble of these three Born variables.

We denote schematically the fixed order NNLO cross-section
differential over $\Phi_B$ by $d\sigma^{{\scriptscriptstyle
    \mathrm{NNLO}}}/d\Phi_B$ and the cross-section obtained from
\VJMINLO{} by $d\sigma^{{\scriptscriptstyle \mathrm{MINLO}}}/d\Phi_B$.
Since these distributions are identical up to terms of
$\mathcal{O}\left(\as\right)$, their ratio
is equal to one up to $\mathcal{O}\left(\alpha_{{\scriptscriptstyle
    \mathrm{S}}}^{2}\right)$ terms:
\begin{equation} 
  \label{eq:W}
  \mathcal{W}(\Phi_B) = \frac{\frac{d\sigma^{\scriptscriptstyle
        \mathrm{NNLO}}}{d\Phi_B}}{
    \frac{d\sigma^{\scriptscriptstyle \mathrm{MINLO}}}{d\Phi_B}} =
  \frac{c_{0}+c_{1}\alpha_{{\scriptscriptstyle
        \mathrm{S}}}+c_{2}\alpha_{{\scriptscriptstyle
        \mathrm{S}}}^{2}\phantom{+\ldots}}{c_{0}+c_{1}\alpha_{{\scriptscriptstyle
        \mathrm{S}}}+c_{2}^{\prime}\alpha_{{\scriptscriptstyle
        \mathrm{S}}}^{2}+\ldots} =
  1+\frac{c_{2}-c_{2}^{\prime}}{c_{0}}\,\alpha_{{\scriptscriptstyle
      \mathrm{S}}}^{2}+\ldots\,,
\end{equation} 
where the $c_{i}$ are $\mathcal{O}\left(1\right)$ coefficients.  Since the
\VJMINLO{} generator reproduces the inclusive fixed-order result up to and
including NLO terms, the NLO accuracy of the cross-section in the presence of
one jet (that starts at order $\as$) is maintained if the cross-section is
reweighed by the factor in \cref{eq:W}. This follows from the simple fact
that the reweighting factor combined with this cross-section yields spurious
terms of $\mathcal{O}(\as^{3})$ and higher.

It is also obvious that by reweighting \VJMINLO{} distributions with
this ratio, any of the three $\Phi_B$ distributions acquire NNLO
accuracy, and in fact coincide with the NNLO distributions.
We will now argue that the \VJMINLO{} generator reweighed with the
procedure of \cref{eq:W} maintains the original NNLO accuracy of
the fixed-order program, used to obtain the
$d\sigma^{\scriptscriptstyle \mathrm{NNLO}}/d\Phi_B$ distribution, for
all observables. The proof trivially extends the proof given in
\Bref{Hamilton:2013fea} by replacing the Higgs rapidity with the
chosen set of three variables $\Phi_B$ associated with the Born phase
space. 

The claim is based on the following theorem:
\begin{quote} 
\textit{A parton level V boson production generator that is accurate
  at ${\cal O}(\alpha_{{\scriptscriptstyle \mathrm{S}}}^{2})$ for all
  IR safe observables that vanish with the transverse momenta of all
  light partons, and that also reaches ${\cal
    O}(\alpha_{{\scriptscriptstyle \mathrm{S}}}^{2})$ accuracy for the
  three Born variables $\Phi_B$ achieves the same level of precision
  for all IR safe observables, i.e. it is fully NNLO accurate.}
\end{quote} 
To this end, we consider a generic observable $F$, including cuts,
that is an infrared safe function of the final state kinematics. $F$
could for instance be a bin of some distribution. Its value will be
given by
\begin{equation} 
  \label{eq:NNLOPS-F-i}
  \langle F\rangle=\int d\Phi\,\frac{d\sigma}{d\Phi}\,
  F(\Phi)
\end{equation} 
with a sum over final state multiplicities being implicit in the phase
space integral. Infrared safety ensures that $F$ has a smooth limit
when the transverse momenta of the light partons vanish.

Since the Born kinematics are fully specified by the Born variables $\Phi_B$,
such a limit may only depend upon the value of Born kinematic variables
$\Phi_B$. We generically denote such a limit by $F_{\Phi_B}$.  The value of
$\langle F\rangle$ can be considered as the sum of two terms: $\left\langle
F-F_{\Phi_B}\right\rangle + \left\langle F_{\phi_B}\right\rangle $.  Since
$F-F_{\Phi_B}$ tends to zero with the transverse momenta of all the light
partons, by hypothesis its value is given with ${\cal
  O}(\alpha_{{\scriptscriptstyle \mathrm{S}}}^{2})$ accuracy by the parton level
generator.  We may also write
\begin{equation} \left\langle F_{\Phi_B}\right\rangle =\int d
\Phi_B^{\prime}\,\frac{d\sigma}{d \Phi_B^{\prime}}\,
F_{\Phi_B}\left(\Phi_B^{\prime}\right)\label{eq:NNLOPS-F-ii}\,,
\end{equation} which is exact at the ${\cal
O}(\alpha_{{\scriptscriptstyle \mathrm{S}}}^{2})$ level by hypothesis. Thus,
$\langle F\rangle=\left\langle F-F_{\Phi_B}\right\rangle +\left\langle
F_{\Phi_B}\right\rangle $ is accurate at the ${\cal
  O}(\alpha_{{\scriptscriptstyle \mathrm{S}}}^{2})$ level.

The \VJMINLO{} parton level generator (in fact even just \VJ) fulfils the first
condition of the theorem, since it predicts any IR safe observable that vanishes,
when the transverse momenta of the light partons vanish, with
$\mathcal{O}(\as^{2})$ accuracy.  The second hypothesis of the theorem,
regarding NNLO accuracy of the Born variables, is simply realised by augmenting
the \VJMINLO{} generator by the reweighting procedure described above. We note
that \MINLO{} is crucial to preserve the first property after rescaling.  The
proof of $\mathcal{O}(\as^{2})$ accuracy for these observables thus corresponds
to the general proof of NLO accuracy of the \POWHEG{} procedure, given in
\Brefs{Nason:2004rx,Frixione:2007vw}.

Note that for observables of the type $\left\langle F-F_{\Phi_B}\right\rangle$,
adding the full shower development does not alter the $\mathcal{O}(\as^{2})$
accuracy of the algorithm, for the same reasons as in the case of the regular
\POWHEG{} method.  The only remaining worry one can have, concerns the
possibility that the inclusive Born variable distributions are modified by the
parton shower evolution at the level of $\mathcal{O}(\as^{2})$ terms. However,
our algorithm already controls the two hardest emissions with the required
$\as^{2}$ accuracy. A further emission from the shower is thus bound to lead to
corrections of higher order in $\as$. This concludes our proof.

\subsubsection{Variant Schemes}
\label{subsec:Simple-variations}
As discussed in detail in \Bref{Hamilton:2013fea} the reweighting in
\cref{eq:W} treats low and high transverse momentum distributions equally,
i.e. it distributes the virtual correction over the full transverse momentum range
considered. On the other hand, if one considers the transverse momentum
distribution of the vector boson $V$ or of the leading jet, in the high
transverse-momentum region \VJMINLO{} and the NNLO calculation have formally the
same (NLO) accuracy. Hence, while it is not wrong to do so, there is no need to
``correct'' the \VJMINLO{} result in that region. It is therefore natural to
introduce a function that determines how the two-loop virtual correction is
distributed over the whole transverse momentum region,
\begin{equation} h(\pt)=\frac{(\hc\, M_{{\scriptscriptstyle
        \mathrm{V}}})^{\hgam}}{(\hc\, M_{{\scriptscriptstyle
        \mathrm{V}}})^{\hgam}+\pt^{\hgam}},\label{eq:hfact}
\end{equation} where $\hc$ and $\hgam$ are constant parameters, and $M_{{\scriptscriptstyle
    \mathrm{V}}}$ is the mass of the vector boson. This function has the
property that when $\pt$ goes to zero it tends to one, while
when $\pt$ becomes very large, it vanishes.

The function $h(\pt)$ can therefore be used to split the cross-section
according to
\begin{eqnarray} 
d\sigma_{\phantom{0}} & = &
d\sigma_{A}+d\sigma_{B}\,,\label{eq:NNLOPS-dsig-eq-dsig0-plus-dsig1}
\\ d\sigma_{A} & = & d\sigma\, h\left(\pt\right)\,,\\ d\sigma_{B} & =
& d\sigma\,\left(1-h\left(\pt\right)\right)\,. \label{eq:NNLOPS-dsig1}
\end{eqnarray} 
We then reweight the \VJMINLO{} prediction using the following factor

\begin{eqnarray} 
  \mathcal{W}\left(\Phi_B,\, p_{{\scriptscriptstyle
      \mathrm{T}}}\right)&=&h\left(\pt\right)\,\frac{\smallint
    d\sigma^{{\scriptscriptstyle
        \mathrm{NNLO\phantom{i}}}}\,\delta\left(\Phi_B-\Phi_B\left(\Phi\right)\right)-\smallint
    d\sigma_{B}^{{\scriptscriptstyle
        \mathrm{MINLO}}}\,\delta\left(\Phi_B-\Phi_B\left(\Phi\right)\right)}{\smallint
    d\sigma_{A}^{{\scriptscriptstyle
        \mathrm{MINLO}}}\,\delta\left(\Phi_B-\Phi_B\left(\Phi\right)\right)}\nonumber\\ &+&\left(1-h\left(\pt\right)\right)\,,\label{eq:NNLOPS-overall-rwgt-factor-1}
\end{eqnarray} 
which preserves the exact value of the NNLO cross-section
\begin{eqnarray}
  \left(\frac{d\sigma}{d\Phi_B}\right)^{{\scriptscriptstyle
      \mathrm{NNLOPS}}} & = &
  \left(\frac{d\sigma}{d\Phi_B}\right)^{{\scriptscriptstyle
      \mathrm{NNLO}}}\,.\label{eq:NNLOPS-NNLOPS-eq-NNLO_0+MINLO_1-1}
\end{eqnarray}

The proof of NNLO accuracy for this rescaling scheme is completely
analogous to the proof given above, so we omit it here.

\subsection{Phenomenological Analysis}
\label{sec:Results}
In this section, in order to validate our results, we make a number of
comparisons with fixed-order calculations and dedicated resummations. In
particular, we find good agreement between our \NNLOPS{} calculations and
\DYQT{}~\cite{Bozzi:2010xn}\footnote{We thank Giancarlo Ferrera and Massimiliano
  Grazzini for providing us with a preliminary version of this code.} for the $Z$
transverse momentum distribution.
We also compare our results for the $\phi^*$ distribution to the
NLO+NNLL resummation of \Bref{Banfi:2011dx}, and the jet-veto
efficiency with the NNLL+NNLO results of \Bref{Banfi:2012jm}.

An implementation of Drell-Yan lepton pair production at NNLO accuracy
including parton shower effects using the UN$^2$LOPS algorithm in the
event generator \textsc{Sherpa} was recently presented in
\Bref{Hoeche:2014aia}. An important difference of this
implementation with respect to our approach, is that in
\Bref{Hoeche:2014aia} the pure NNLO correction sits in the bin
where the $Z$ boson has zero transverse momentum. These events do not
undergo any parton showering, populating only the zero-$\pt$
bin. Thus, in the approach of \Bref{Hoeche:2014aia}, the $Z$
boson transverse momentum is insensitive to the NNLO correction.

A general approach to matching NNLO computations with parton showers was also
proposed in \Bref{Alioli:2013hqa}. The \MINLO{} procedure that we use in this
work was discussed by the authors of \Bref{Alioli:2013hqa} in the context of the
general formulation of their method. Given that an implementation of
\Bref{Alioli:2013hqa} now exists for Drell-Yan production
\cite{Alioli:2015toa}, it will be interesting to see how the three different
matching schemes compare\footnote{First steps towards such a comparison can be
  found in \Bref{Alioli:2016fum}.}.

Before showing our validation results, we define our procedure for estimating
theoretical uncertainties.

\subsubsection{Estimating Uncertainties}
\label{subsec:Estimating-uncertainties}
As is standard, the uncertainties in the \VJMINLO{} generator are obtained by
varying the renormalisation scales appearing in the \MINLO{} procedure by $\Kr$
and the factorisation scale by $\Kf$, by a factor 2 up and down keeping $1/2 \le
\Kr/\Kf \le 2$. This leads to 7 different scale choices given by
\begin{equation} (K_{{\scriptscriptstyle
      \mathrm{R}}},K_{{\scriptscriptstyle
      \mathrm{F}}})=(0.5,0.5),(1,0.5),(0.5,1),(1,1),(2,1),(1,2),(2,2)\,.\label{eq:Errors-KR-KF-list}
\end{equation} 
We will consider the variation in our results induced by the above
procedure. The seven scale variation combinations have been obtained
by using the reweighting feature of the \POWHEGBOX{}.

For the pure NNLO results, the uncertainty band is the envelope of the
7-scale variation obtained by varying the renormalisation and
factorisation scale by a factor 2 around the central value $M_V$
keeping $1/2 \le \Kr/\Kf \le 2$.

For the \NNLOPS{} results, we have first generated a single of \ZJMINLO{}
event file with all the weights needed to compute the integrals
$d\sigma_{A/B}^{\scriptscriptstyle \mathrm{MINLO}}/d\Phi_B$, in
\cref{eq:NNLOPS-overall-rwgt-factor-1}, for all 7 scale choices.
The differential cross-section $d\sigma^{\scriptscriptstyle
  \mathrm{NNLO}}/d\Phi$ was tabulated for each of the three scale
variation points corresponding to $\Kr'=\Kf'$. 
The analysis is then performed by processing the
\MINLO{} event for given values of $(\Kr,\Kf)$, and multiplying its
weight with the factor
\begin{equation} h\left(\pt\right)\times\,\frac{\smallint
d\sigma_{(\Kr',\Kf')}^{{\scriptscriptstyle
\mathrm{NNLO\phantom{i}}}}\,\delta\left(\Phi_B-\Phi_B(\Phi)\right)-\smallint
d\sigma_{B,(\Kr,\Kf)}^{{\scriptscriptstyle
\mathrm{MINLO}}}\,\delta\left(\Phi_B-\Phi_B(\Phi)\right)}{\smallint
d\sigma_{A,(\Kr,\Kf)}^{{\scriptscriptstyle
\mathrm{MINLO}}}\,\delta\left(\Phi_B-\Phi_B(\Phi)\right)}+\left(1-h\left(\pt\right)\right)\,.\label{eq:master}
\end{equation} 
The central value is obtained by setting $(\Kr,\Kf)$ and $(\Kr',\Kf')$
equal to one, while to obtain the uncertainty band we apply this
formula for all the seven $(K_{{\scriptscriptstyle
    \mathrm{R}}},K_{{\scriptscriptstyle \mathrm{F}}})$ and three
$(K_{{\scriptscriptstyle \mathrm{R}}}^{\prime},K_{{\scriptscriptstyle
    \mathrm{F}}}^{\prime})$ choices. This yields 21 variations at
\NNLOPS{} level.

The reasoning behind varying scales in the NNLO and \ZJMINLO{} results
independently is that we regard uncertainties in the overall
normalisation of distributions, as being independent of the respective
uncertainties in the shapes. This is consistent with the recently
introduced efficiency method~\cite{Banfi:2012yh}, used for estimating
errors on cross-sections in the presence of cuts.

\subsubsection{Practical Implementation}
\label{subsec:practical}
There is some degree of freedom in the way the reweighting procedure
described above is carried out in practice. 
In particular, one is free to choose the three Born variables with
respect to which one performs the reweighting, as well as the form of
the damping factor $h$ in \cref{eq:hfact}. Our choice of the Born
variables is driven by the fact that one wants to populate all bins in
the three-dimensional histograms sufficiently well. To produce the
results presented in the following we used the rapidity of the $Z$
boson, $y_Z$, a variable directly related to the dilepton invariant
mass, $a_\rm{mll} = \rm{\arctan}((m_{\rm{ll}}^2 - \MZ^2)/(\MZ
\Gamma_Z))$ and $\theta^*_l$, where the latter is the angle between
the beam and a charged lepton in the frame where the boson has no
longitudinal momentum.\footnote{For the $W$ boson we always pick the
  charged lepton, for the $Z$ boson we always pick the negatively
  charged one.}

In the following we will use the values of $\hc = 1$ and $\hgam = 2$
in \cref{eq:hfact}.  The choice for $\hc$ is motivated by the
fact that typical resummation scale for Drell-Yan production is set to
the boson mass. The second choice is partly driven by the fact that in
the case of Higgs production, with this choice the \NNLOPS{} Monte
Carlo agrees well with the Higgs transverse momentum and leading jet
NNLL+NNLO resumed results.
We stress however that while $\hc$ and $\hgam$ are arbitrary, the
dependence on $\hc$ and $\hgam$ is formally ${\cal O}(\as^3)$ (or
exactly zero in the case of inclusive quantities).  $\pt$ in
\cref{eq:hfact} denotes the transverse momentum that is used to
decide how to distribute the virtual corrections. One could for
instance choose the $V$ transverse momentum, the leading jet
transverse momentum, or the total transverse momentum of the event. In
the following, we will adopt the choice of
\Bref{Hamilton:2013fea}, namely to use the leading jet transverse
momentum when clustering events according to the inclusive
$\kt$-algorithm with $R=0.7$~\cite{Catani:1993hr}.  This choice
ensures that $h$ goes to one when no radiation is present, since in
that case the leading jet transverse momentum vanishes. On the
contrary, the $V$ transverse momentum can vanish also in the presence
of radiation.

Finally we note that for the reweighting we used 25 bins per variable,
meaning that our 3-dimensional distributions involve a total of
$15625$ equal bins.
Results presented in the following are based on generating 20 million
events.
Even if we perform high-statistics runs, there might be bins that are
not well-populated. This can give rise to a rescaling factor that
is unphysically large just because too few events ended up in one
bin, and this in turn can give rise to spikes in kinematical
distributions. To avoid these occurrences, instead of using the local
reweighting factor, we use the global $K_\rm{NNLO}/K_\rm{NLO}$
factor to perform the reweighting whenever the local reweighting
factor (which is formally $1+{\cal O}(\alpha_s)^2$) exceeds 5. This
happens rarely, in about 0.3\% of the points. We checked that this
procedure has no visible systematic effect on distributions, other
than that of removing unphysical spikes.  Other workarounds could of
course also be adopted.

Before showing validation plots, we list here the settings used for
the results obtained in this Chapter.  We used the code
\DYNNLO{}~\cite{Catani:2009sm} to obtain NNLO
predictions.\footnote{Even with high-statistics runs with
  \FEWZ{}~\cite{Gavin:2010az}, we did not obtain high-quality triple
  differential distributions, as required here.}
Throughout this work we consider
the MSTW2008NNLO parton distribution functions~\cite{Martin:2009iq}
and set $\MZ = 91.1876 \GeV$, $\Gamma_Z = 2.49595 \GeV$, $\MW = 80.398 $
GeV and $\Gamma_W = 2.008872 \GeV$. We choose to use $\alpha_\rm{em} =
1/128.94$ and $\sin^2\theta_W = 0.22264585$.
Jets have been constructed using
\FASTJET{}~\cite{Cacciari:2005hq,Cacciari:2011ma}. To compute the $h(\pt)$
factor we use the $\kt$- algorithm~\cite{Catani:1993hr,Ellis:1993tq} with
$R=0.7$.
To shower partonic events, we have used both
\PYTHIA{8}~\cite{Sjostrand:2007gs} (version 8.185) with the ``Monash
2013''~\cite{Skands:2014pea} tune and \PYTHIA{6} (version 6.4.28) with
the MSTW2008LO variation of the ``Perugia'' tune (``Perugia
P12-M8LO'', tune 378). 
We have generated $Z$ and $W$ events with decays into electrons and
positrons, and always switched off QED radiation off leptons and quarks
in the showering stage.
Moreover, in the following, in order to define
the leptons from the boson decays we will always use the Monte Carlo
truth, \emph{i.e.}~we disregard complications due to the fact that
there might be other leptons in the event.

To obtain the results shown in the following, we have switched on the
``doublefsr'' option introduced in \Bref{Nason:2013uba} and used
a standard driver for \PYTHIA{6} and the \PYTHIA{8} driver suggested
by the \PYTHIA{} authors for showering events generated with
\POWHEG{}.  Although we have also explored the effect of using the
alternative prescription first introduced in Section 4 of
\Bref{Nason:2013uba} to compute the scale used by the parton
shower to veto hard emissions (which is also available as an option in
the \PYTHIA{8} driver)\footnote{A factor 2 is missing in equation~(4) of
  \Bref{Nason:2013uba}, but not in the practical implementation.},
the plots shown throughout the chapter have been obtained keeping the
veto scale equal to the default \POWHEG{} prescription, both for
\PYTHIA{6} and \PYTHIA{8}.

\subsubsection{Validation Plots}
\label{subsec:val}

In this section we consider the case of inclusive $Z\to e^+e^-$ production at 14
TeV, with no cuts on the final state other than requiring the dilepton invariant
mass to be in the range $66 \GeV \le m_\rm{ll} \le 116 \GeV$. In order to
validate our results, we will show comparisons to NNLO predictions and to
\MINLO{}-improved \ZJ{}-\POWHEG{} results (henceforth denoted as
\ZNLOPSpPYTHIA{}). Since we compare with \DYNNLO{}, for the \ZNLOPSpPYTHIA{} and
\NNLOPS{} results it is useful to consider here pure parton level results
before hadronisation (with underlying event and multiparton interaction switched
off). Unless otherwise stated, for \DYNNLO{} we have set $\muf=\mur=\MZ$, and
the associated uncertainty bands are obtained from a 7-points scale variation
envelope. Also unless stated otherwise, we will shower events with \PYTHIA{8}.

\begin{figure}[!tbh]
  \begin{centering}
    \includegraphics[clip,width=0.8\textwidth]{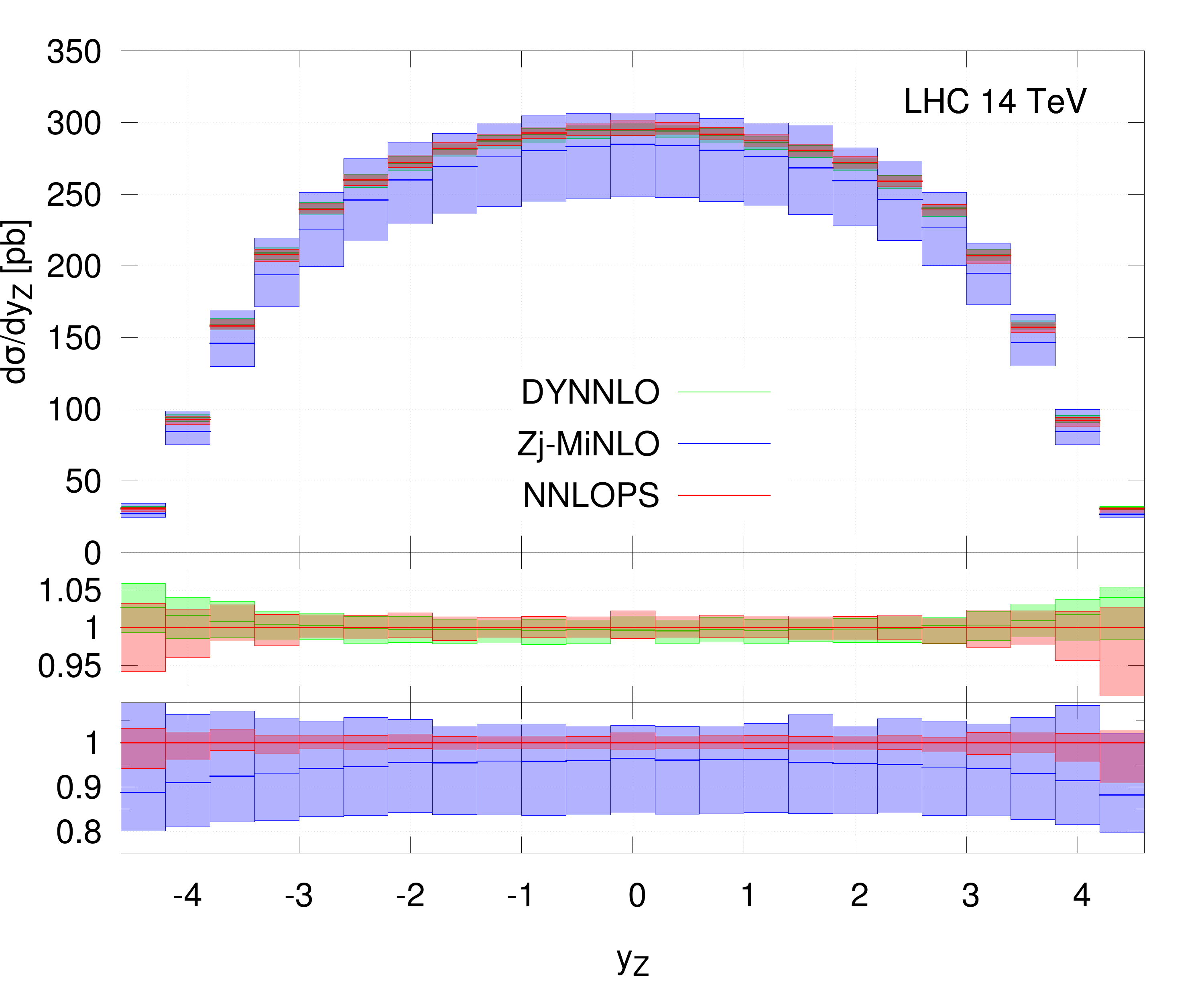}
    \par\end{centering}
  \caption{Comparison of the \ZNNLOPSpPYTHIA{} (red), \ZNLOPSpPYTHIA{}
    (blue) and \DYNNLO{} (green) results for the $Z$ boson fully
    inclusive rapidity distribution at the LHC running at 14 TeV.  The
    \DYNNLO{} central scale is $\muf=\mur=\MZ$, and its error band is
    the 7-point scale variation envelope.  For \ZNNLOPSpPYTHIA{} and
    \ZNLOPSpPYTHIA{} the procedure to define the scale uncertainty is
    described in detail in \cref{subsec:Estimating-uncertainties}.
    The two lower panels show the ratio of \DYNNLO{} and
    \ZNLOPSpPYTHIA{} predictions with respect to \ZNNLOPSpPYTHIA{}
    obtained with its central scale choice.}
    \label{fig:val-yz}
\end{figure}
In \cref{fig:val-yz} we show the $Z$ boson rapidity distribution $y_Z$ as
predicted at NNLO (green), with \ZNLOPSpPYTHIA{} (blue), and at
\ZNNLOPSpPYTHIA{} (red). As expected \ZNNLOPSpPYTHIA{} agrees very well with
\DYNNLO{} over the whole rapidity range, both for the central value and the
uncertainty band, defined as detailed in \cref{subsec:Estimating-uncertainties}.
We also note that as expected, the uncertainty band of the \ZNNLOPSpPYTHIA{}
result is considerably reduced compared with the one of \ZNLOPSpPYTHIA{}, which is
NLO accurate. In the central region the uncertainty decreases from about
$(+5:-15) \%$ to about $(+2:-2) \%$. We finally note that because of the
positive NNLO corrections, the central value of \ZNNLOPSpPYTHIA{} lies about 5
\% above the one of \ZNLOPSpPYTHIA{}, while no considerably difference in shape
is observed in the central rapidity region. Moderate but slightly more
pronounced shape differences can be seen at large rapidity.

\begin{figure}[!tbh]
  \begin{centering}
    \includegraphics[clip,width=0.49\textwidth]{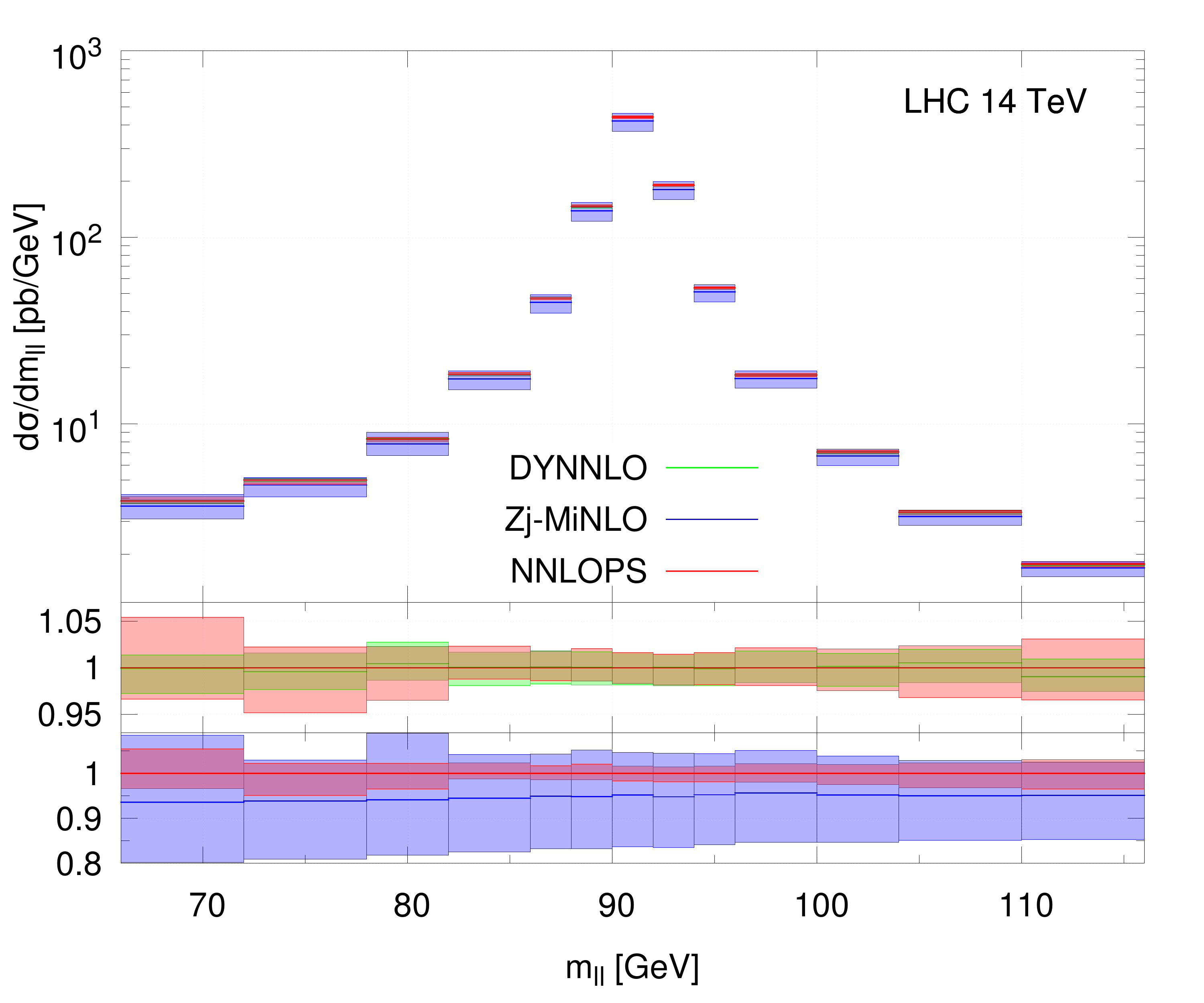}
    \hfill{}\includegraphics[clip,width=0.49\textwidth]{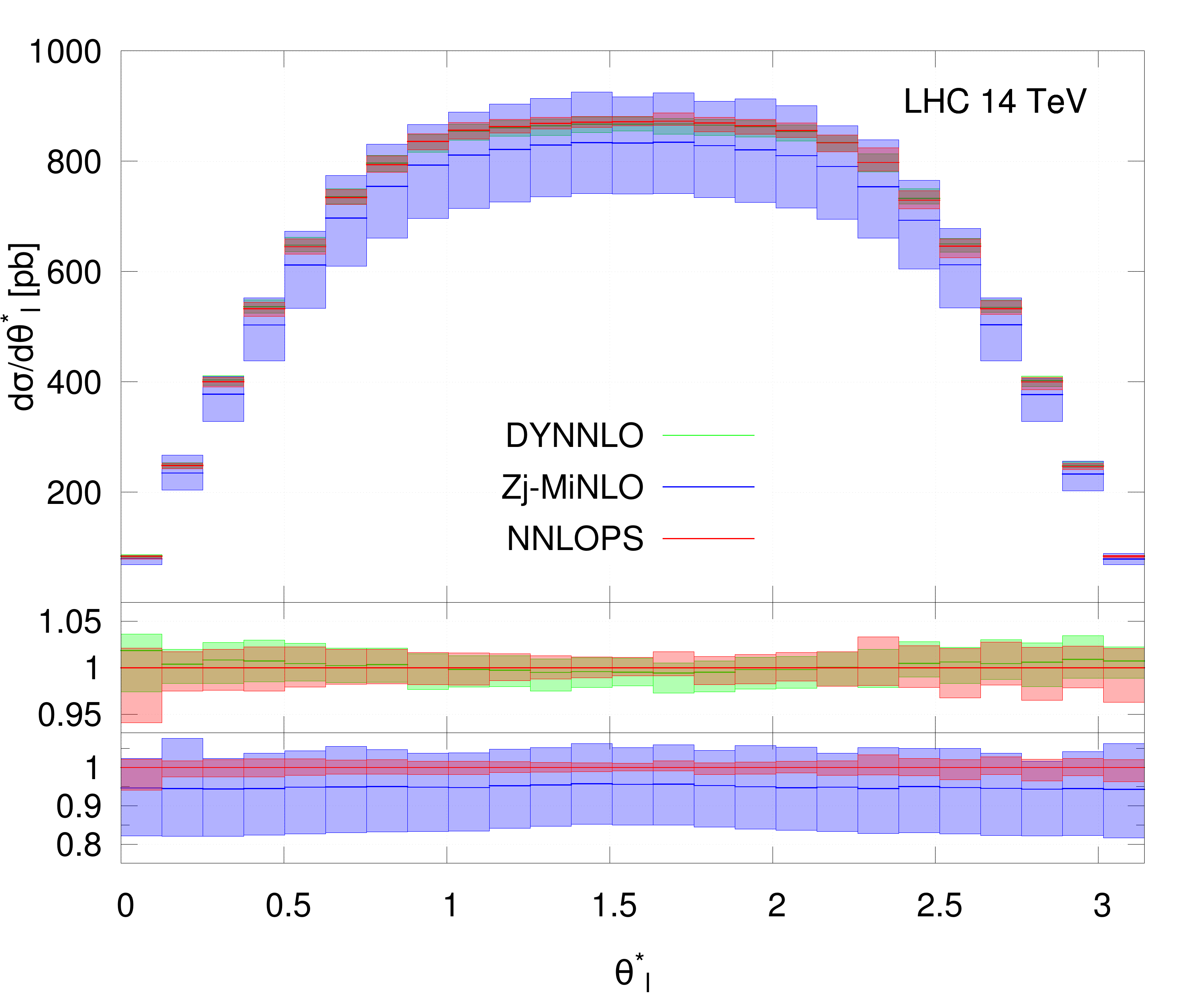}
    \par\end{centering}
    \caption{As for \cref{fig:val-yz} but for $m_\rm{ll}$ (left) and
      $\theta_l^*$ (right).}
    \label{fig:val-mll-theta}
\end{figure}

We proceed by examining the other two distributions that have been used in the
reweighting procedure, namely $a_\rm{m_\rm{ll}}$ and $\theta^*_l$. Instead of
showing $a_\rm{m_\rm{ll}}$ we plot in \cref{fig:val-mll-theta} the
invariant mass of the dilepton system $m_\rm{ll}$ which is directly related to
$a_\rm{m_\rm{ll}}$. We notice that the same features observed above hold:
\ZNNLOPSpPYTHIA{} agrees well with \DYNNLO{}, it tends to be about 5$-$10\%
higher than \ZNLOPSpPYTHIA{} and the uncertainty band is reduced by about a
factor 4. No sizeable difference in shape is observed in these two distributions,
when comparing \ZNLOPSpPYTHIA{} and \ZNNLOPSpPYTHIA{}.

\begin{figure}[!tbh]
  \begin{centering}
    \includegraphics[clip,width=0.49\textwidth]{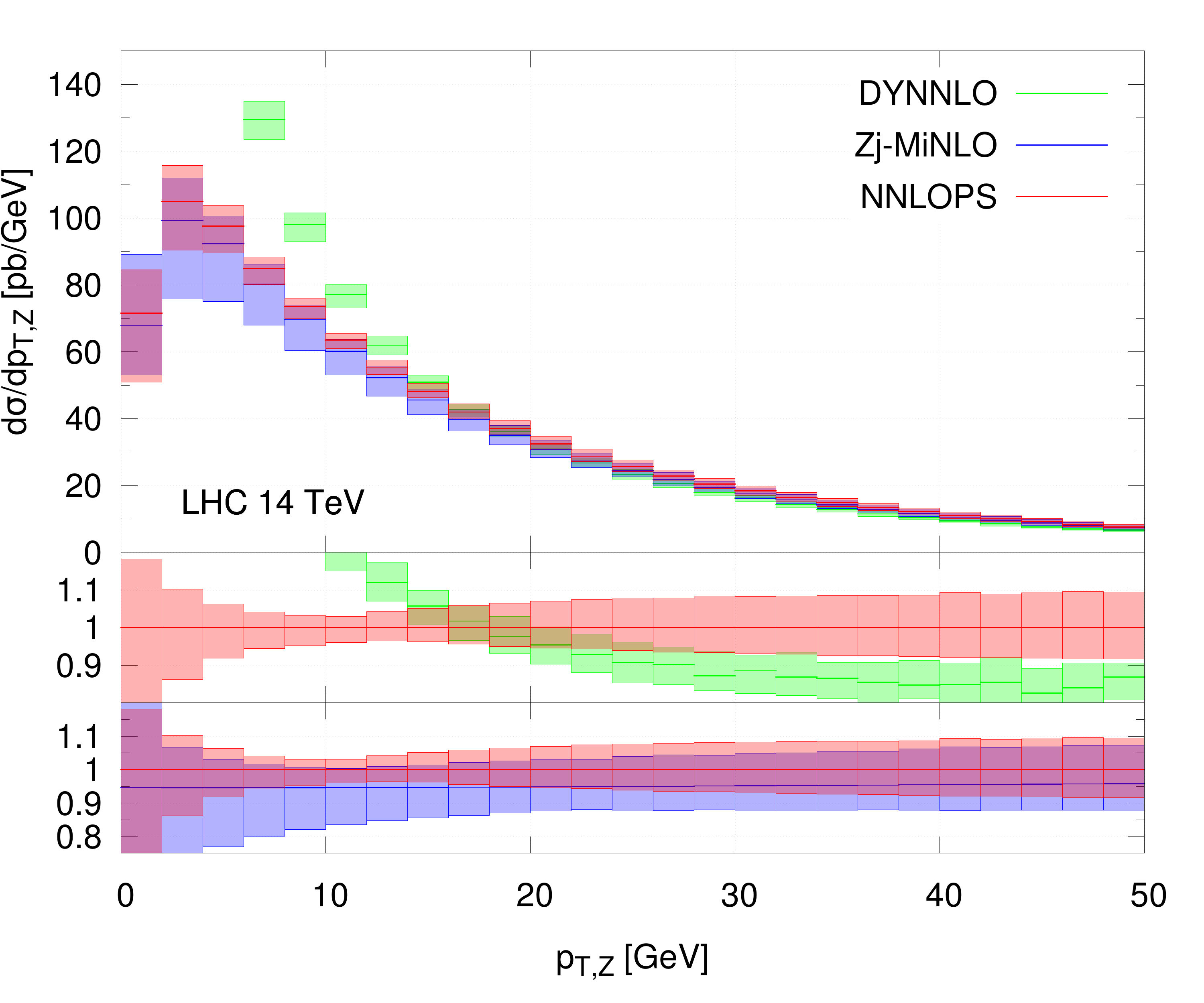}
    \hfill{}\includegraphics[clip,width=0.49\textwidth]{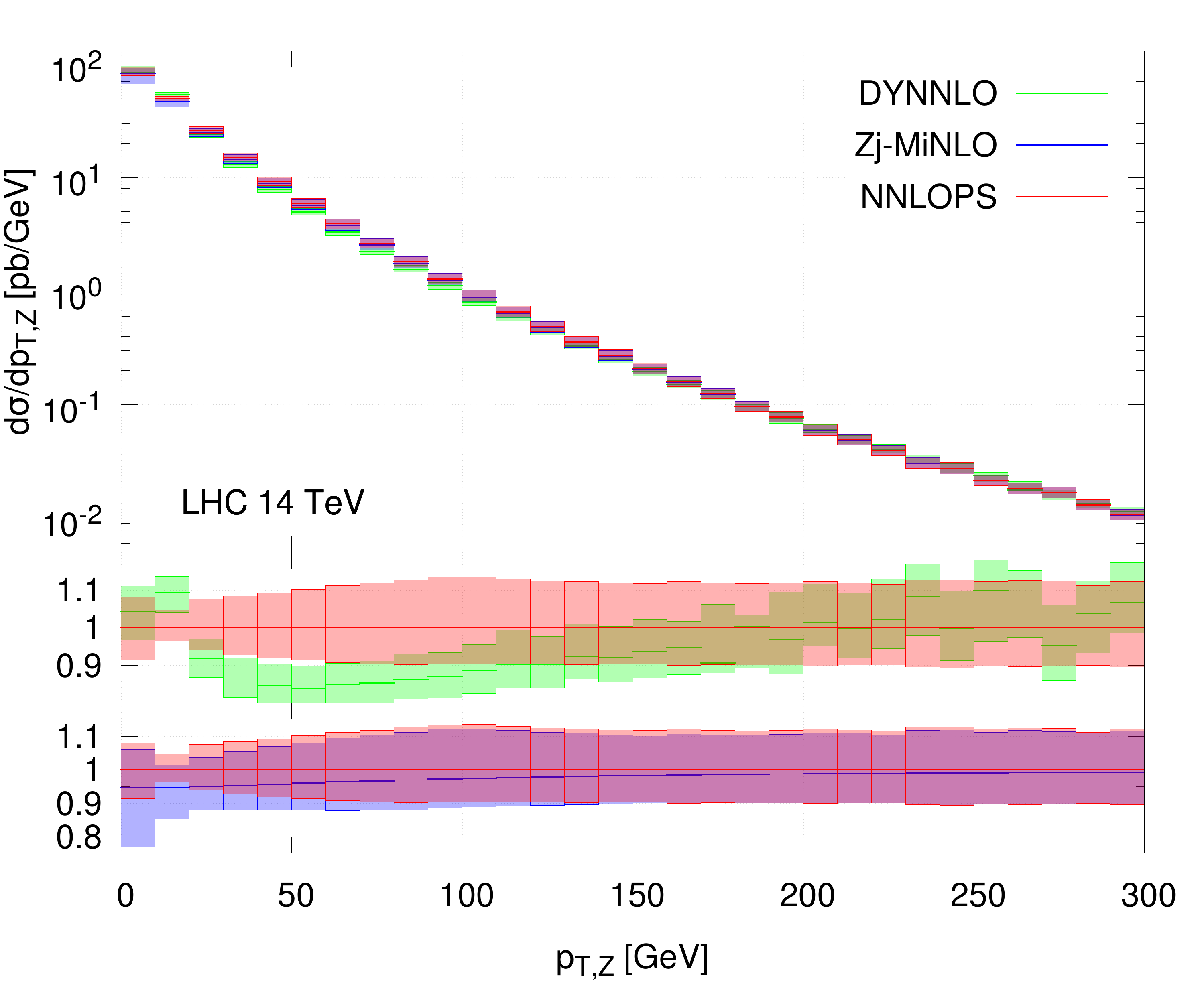}
    \par\end{centering}
    \caption{As for \cref{fig:val-yz} but for the $Z$-boson transverse
      momentum for two different ranges in $\ptz$.}
    \label{fig:val-ptz}
\end{figure}

We now show in \cref{fig:val-ptz} the $Z$ boson transverse momentum in two
different ranges. At finite values of $\ptz$ this quantity is described at NLO
accuracy only by all three codes, \DYNNLO{}, \ZNNLOPSpPYTHIA{} and
\ZNLOPSpPYTHIA{}.  In fact, at higher value of $\ptz$ the bands of
\ZNNLOPSpPYTHIA{} and \ZNLOPSpPYTHIA{} overlap and are very similar in size.

At small transverse momenta, while \DYNNLO{} diverges,
\ZNNLOPSpPYTHIA{} remains finite because of the Sudakov damping. The
difference in shape observed between \DYNNLO{} and \NNLOPS{} at finite
$\ptz$ has to do with the fact that in that region the fixed-order
calculation has to compensate for the divergent behaviour at small
$\ptz$.  We also note that the uncertainty band in \DYNNLO{} is far
too small when approaching the divergence at $\ptz = 0$.  The
uncertainty band of \ZNNLOPSpPYTHIA{} instead tends to increase at
very low transverse momenta, reflecting the fact that one is
approaching a non-perturbative region. One can however also note that
the uncertainty tends to shrink at about $\ptz = 10 \GeV$. We have
checked that this is not an artefact due to having used a 21-point
scale variation as opposed to a 49-point one.
We attribute this feature to the fact that the uncertainty band of the
fixed-order result shrinks in this region. This is true both for the
3-point and the 7-point scale variation in the fixed order, although
in the latter case this effect is slightly less pronounced.  We also
observe that our \ZNLOPSpPYTHIA{} result that uses the \MINLO{} scale
prescription does not show this feature. When we upgrade
\ZNLOPSpPYTHIA{} to NNLO accuracy, we necessarily inherit this feature
from the NNLO results we are using as input. 
It is also worth mentioning that this feature has been already
observed in several studies where an analytic resummation matched with
fixed-order results was performed for this
observable~\cite{Bozzi:2010xn,Becher:2010tm,Banfi:2012du}.

We end our discussion on neutral Drell-Yan production by looking
briefly into the effects of including non-perturbative contributions
in the \NNLOPS{} simulation, by turning on hadronisation, underlying
event and multiparton interaction (MPI). In particular, given the
small perturbative uncertainties found with $\ptz$, it is interesting
to see how much non-perturbative corrections affect the $5-15 \GeV$
region.

\begin{figure}[!tbh]
  \begin{centering}
    \includegraphics[clip,width=0.49\textwidth]{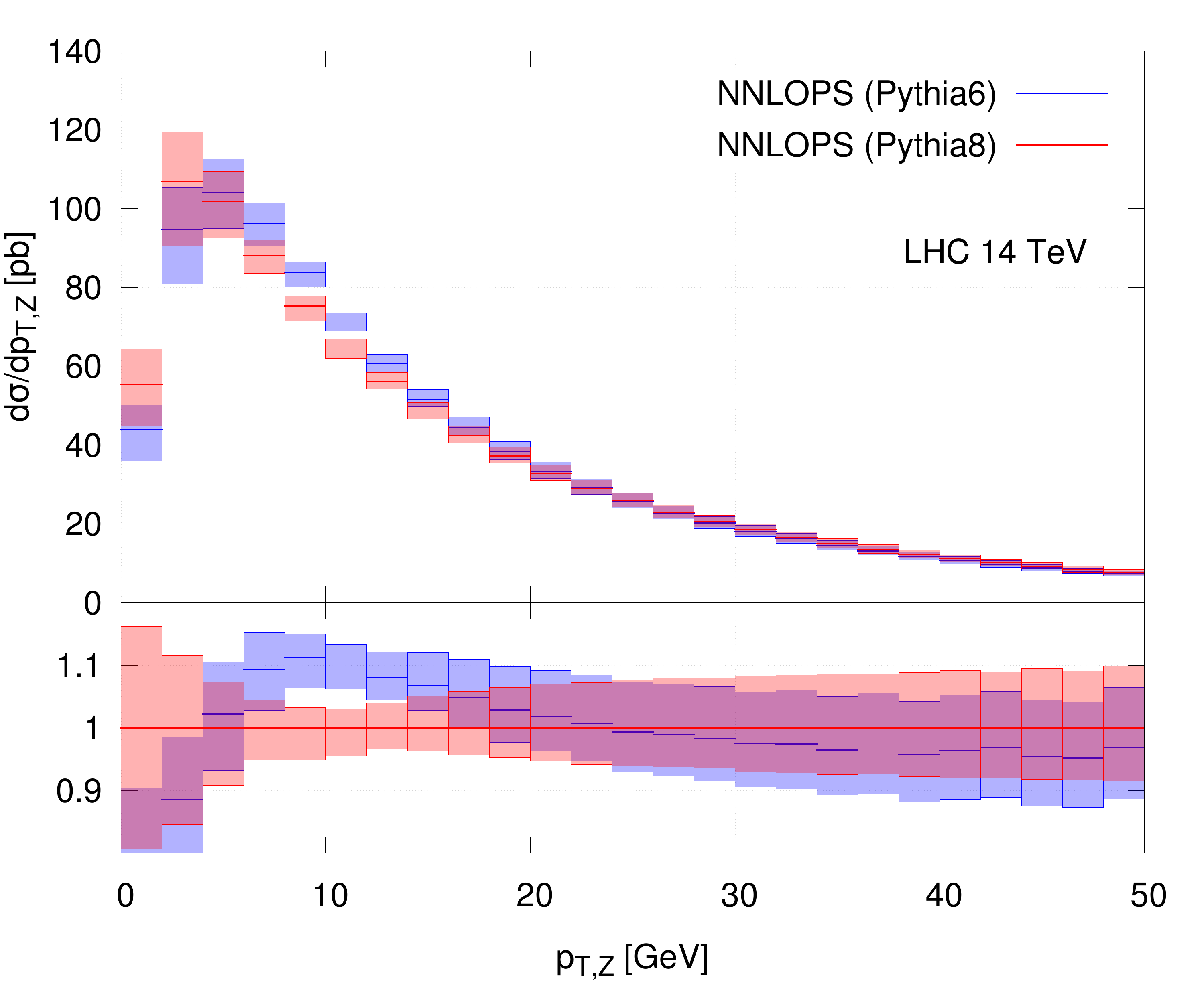}
    \hfill{}\includegraphics[clip,width=0.49\textwidth]{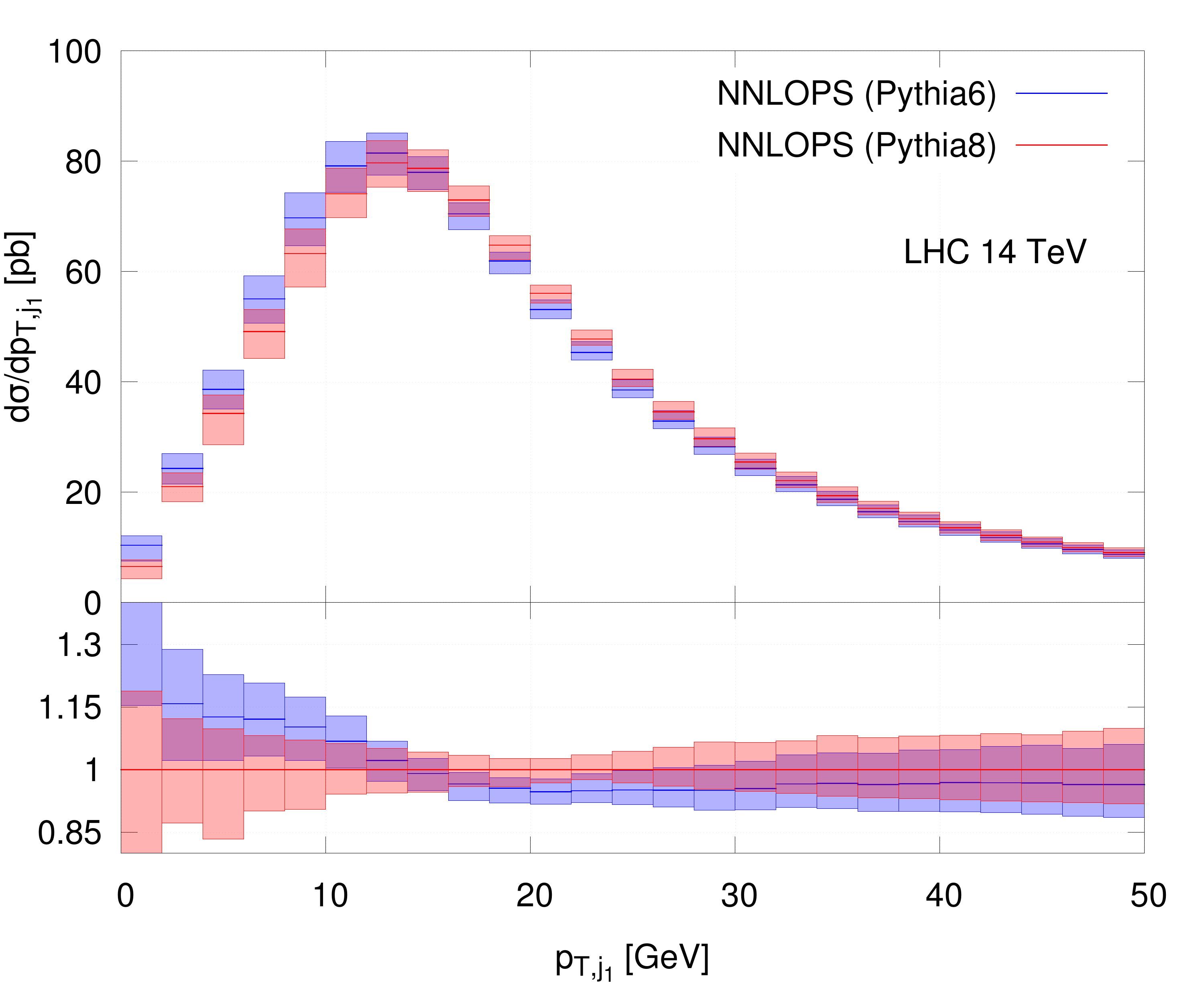}
    \par\end{centering}
  \caption{\NNLOPS{} predictions for the transverse momentum of the
    reconstructed $Z$ boson (left) and leading jet (right) obtained
    using \PYTHIA{6} (blue) and \PYTHIA{8} (red). Non-perturbative
    effects have been included here, by using the tunes mentioned in
    \cref{subsec:practical}.}
    \label{fig:par-ptz}
\end{figure} 
In \cref{fig:par-ptz} we show $\ptz$ (left) and the leading-jet transverse
momentum, defined according to the anti-$\kt$~\cite{Cacciari:2008gp} algorithm
with $R=0.7$ (right), after all non-perturbative stages are included, with
\PYTHIA{6} (blue) and \PYTHIA{8} (red).  We observe sizeable differences between
the two results, in particular for the $Z$ boson transverse momentum at $\ptz <
15 \GeV$. This is not surprising since this is a region dominated by soft
effects, hence the details of the modelling of non-perturbative effects are
expected to matter. For the jet-transverse momentum the difference between the
two shower models is slightly smaller. This can probably be attributed to the
fact that the Sudakov peak is at larger values of the transverse momentum,
compared with the $Z$ boson transverse momentum.

\begin{figure}[!tbh]
  \begin{centering}
    \includegraphics[clip,width=0.49\textwidth]{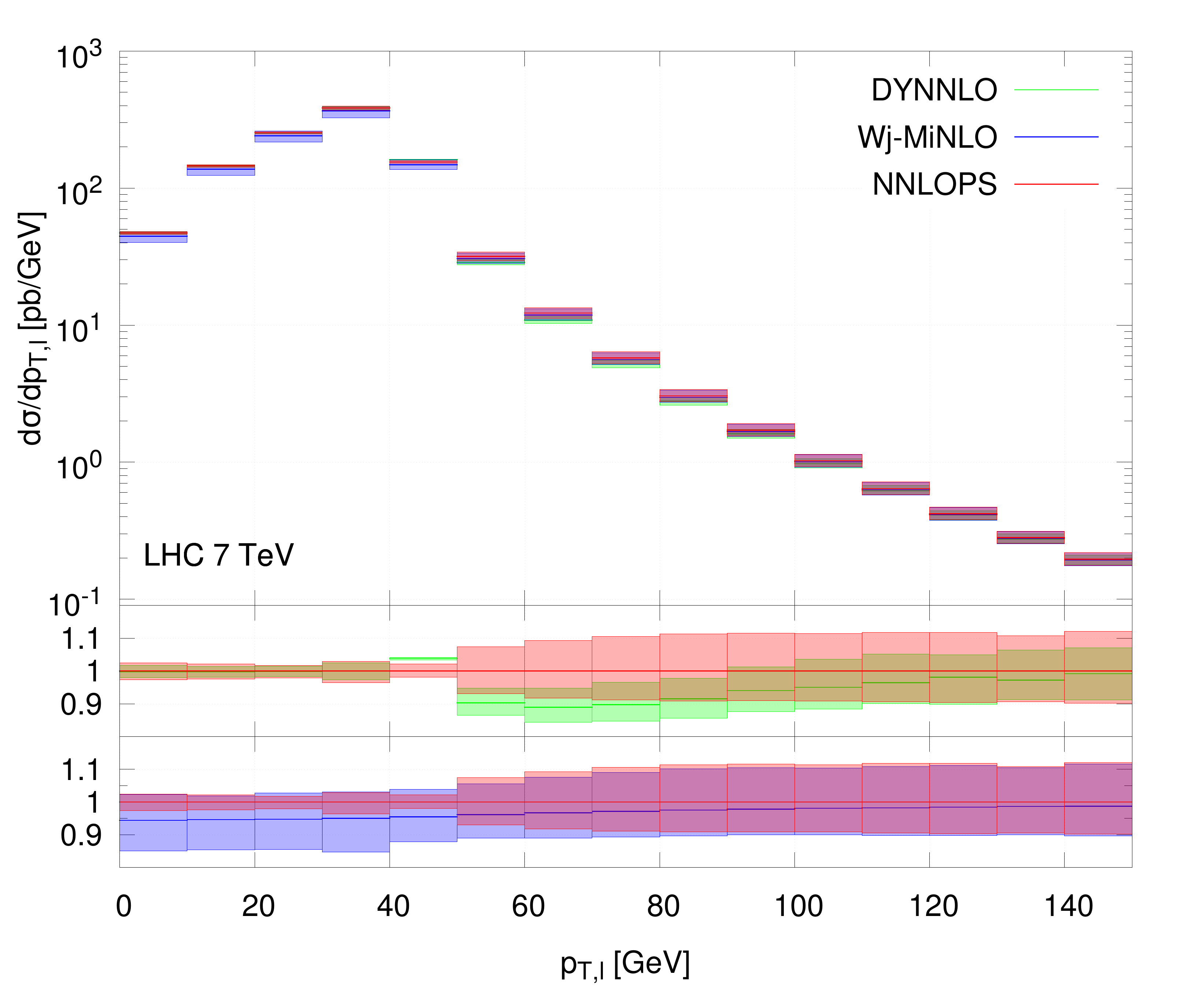}
    \hfill{}\includegraphics[clip,width=0.49\textwidth]{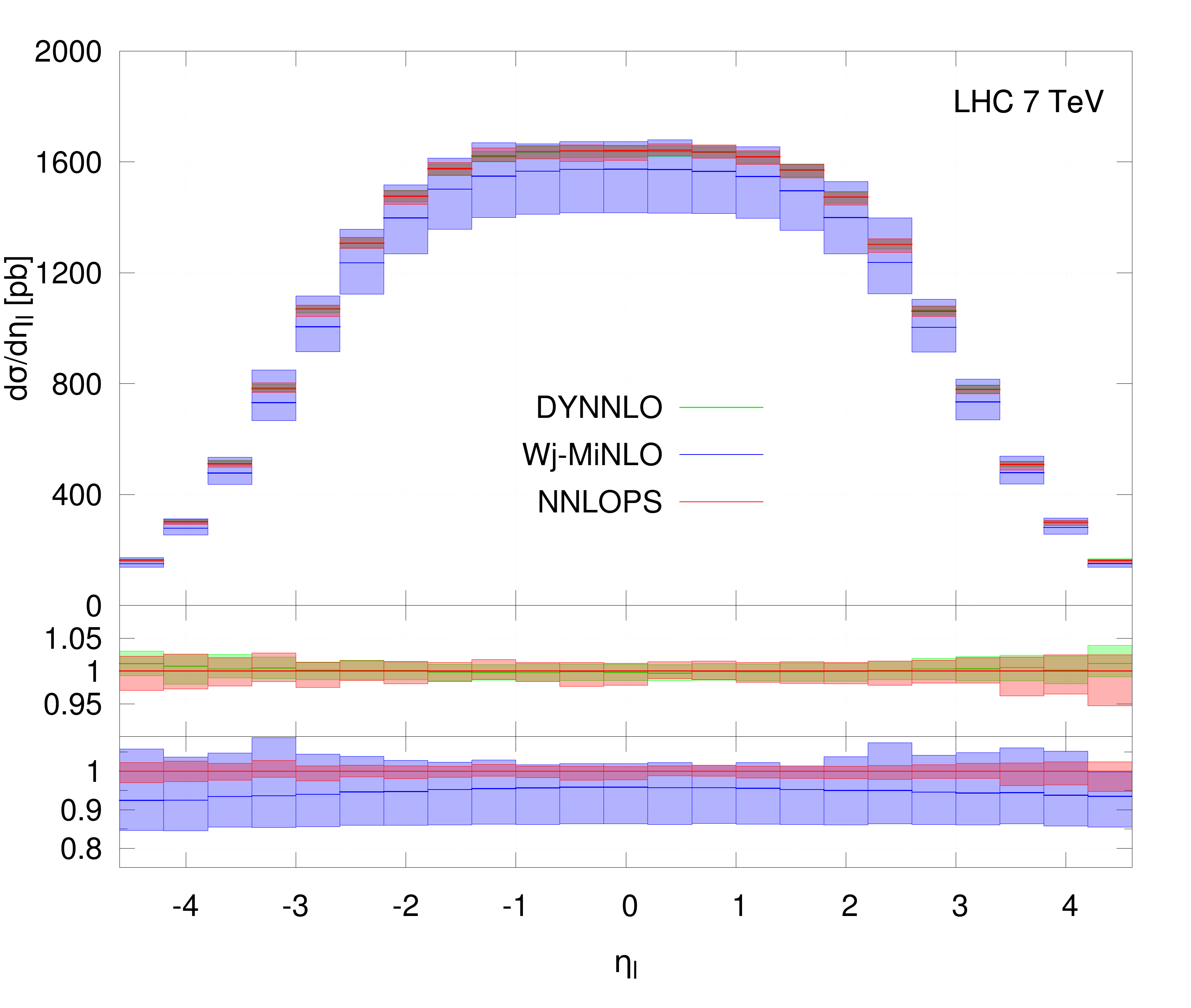}
    \par\end{centering}
    \caption{Comparison of the \NNLOPS{} (red), \ZJMINLO{} (blue) and
      \DYNNLO{} (green) 
      results for the transverse momentum (left) and rapidity distributions
      of the charged lepton (right) in $W$ production at the LHC running at
      7 TeV.  The \DYNNLO{} central scale is $\muf=\mur=\MW$, and its
      error band is the 3-point scale variation
      envelope. For
      \NNLOPS{} and \ZJMINLO{} the procedure to define the scale
      uncertainty is described in detail in
      \cref{subsec:Estimating-uncertainties}.  The lower panels show
      the ratio with respect to the \NNLOPS{} prediction obtained with
      its central scale choice.}
    \label{fig:val-lepplus}
\end{figure}

Finally, for illustrative purposes, we conclude this section by
showing in \cref{fig:val-lepplus} 
predictions for the transverse momentum (left) and pseudorapidity
(right) of the charged electron or positron in $W$ production at 7 TeV
(combining the $W^+$ and $W^-$ samples).
It is interesting to look at these leptonic observables since they don't
coincide with the quantities we are using to perform the NNLO
reweighting. Nevertheless, we should recover NNLO accuracy in the regions where
the lepton kinematics probes the fully-inclusive phase space: this is precisely
what we have found, as illustrated in \cref{fig:val-lepplus}. Here we include no
cuts on the final state other than requiring the transverse $W$ mass $\mtw =
\sqrt{2 (p_{{\scriptscriptstyle \mathrm{T,l}}}\, p_{{\scriptscriptstyle
      \mathrm{T}},miss} -\vec p_{{\scriptscriptstyle \mathrm{T,l}}} \,\cdot \vec
  p_{{\scriptscriptstyle \mathrm{T}},miss}) }$ to be larger than $40 \GeV$.  All
parameters and settings are the same as those used for the $Z$ production case
and \DYNNLO{} results have been obtained choosing $\mur=\muf=\MW$ as central
scale and performing a 3-point scale variation.

As expected, in the left panel of \cref{fig:val-lepplus} we observe a much
narrower uncertainty on the charged lepton transverse momentum for values of
$p_\rm{T, l}$ smaller than $\MW/2$, and a very good agreement with \DYNNLO{} in
this region, both in the absolute value of the cross-section as well as in the
size of the theoretical uncertainty band.
When $p_\rm{T, l}$ is larger than $\MW/2$ all distributions have NLO accuracy
only, since in this region the lepton kinematics requires non-vanishing values
for $\ptw$. We observe that in fact all uncertainty bands are larger in this
region. Our \NNLOPS{} result reproduces the \ZNLOPSpPYTHIA{} one well, while
there is some difference between \DYNNLO{} and \ZNLOPSpPYTHIA{}
predictions. This is expected since the scales used in the two calculations are
effectively different: \DYNNLO{} always uses the mass of the $W$ boson, whereas
in \MINLO{} the transverse momentum of the $W$ boson is used. Therefore when the
lepton is just slightly harder than $\MW/2$ we are probing phase-space regions
where the bulk of the cross-section typically has $0\lesssim \ptw \lesssim \MW$,
and hence \DYNNLO{} yields smaller cross-sections.

It is also worth mentioning that both the \NNLOPS{} and \ZNLOPSpPYTHIA{} plots
exhibit a smooth behaviour in proximity of the Jacobian peak $p_\rm{T, l}\simeq
\MW/2$, also when thinner bins (not shown) are used. This smooth behaviour is
due both to parton-shower effects and to the \MINLO{} Sudakov form factor.
On the contrary, the NNLO prediction displays the typical numerical instability
of fixed-order predictions due to the numerical cancellation between real and
virtual corrections close to the kinematical boundary. Furthermore, the NNLO
prediction has spuriously small uncertainties in this region.

Finally, in the right panel of \cref{fig:val-lepplus} we plot the rapidity
distribution of the charged lepton. Since in each bin of this distribution we
are fully inclusive with respect to QCD radiation, we observe the expected good
agreement with the NNLO prediction over the whole range, as well as a quite
narrow uncertainty band.

\subsubsection{Comparison to Analytic Resummations}
\label{subsec:resum}

The \MINLO{} method at the core of the results presented in this chapter works by
including NLL and (some) NNLL terms in the Sudakov form factors used to improve
the validity and accuracy of the underlying NLO computation.  Although
the formal logarithmic accuracy achieved by \MINLO{}-improved \POWHEG{}
simulations has not been addressed, it is interesting to compare \NNLOPS{}
predictions against results obtained with higher-order analytic resummation, for
observables where the latter are available. In this subsection we will show
results for $Z$ production at 7 TeV and focus on three quantities for which NNLL
resummation has been performed.

The classical observable to consider in Drell-Yan production to study the
effects of soft-collinear radiation is the transverse momentum of the
dilepton-pair system.
This observable has been extensively studied in the past and is now known to
NNLO+NNLL level~\cite{Bozzi:2010xn,Becher:2010tm}.
\begin{figure}[!tbh]
  \begin{centering}
    \includegraphics[clip,width=0.49\textwidth]{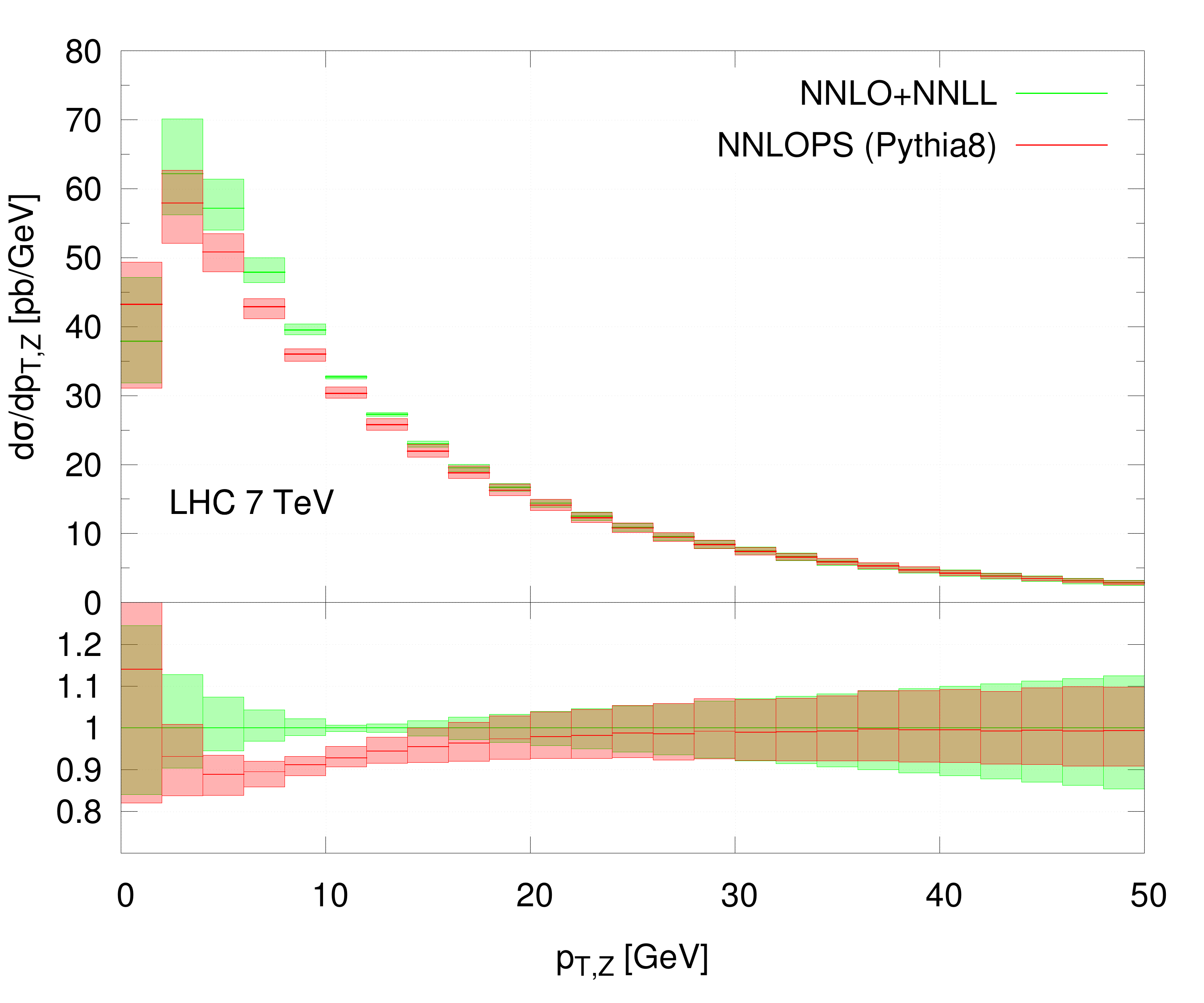}
    \includegraphics[clip,width=0.49\textwidth]{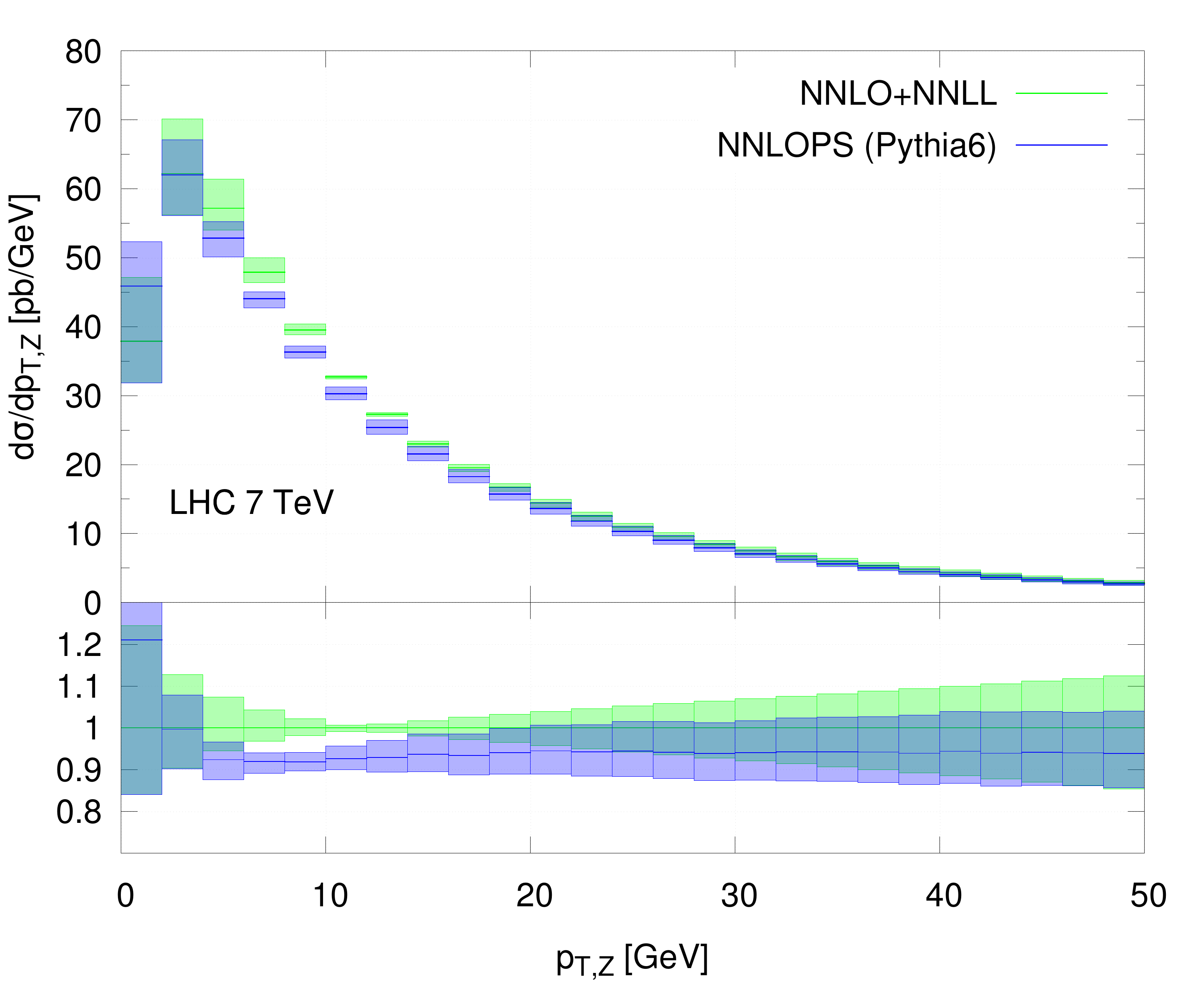}
    \par\end{centering}
    \caption{Comparison of \NNLOPS{} prediction and NNLO+NNLL
      resummation for $\ptz$ at the 7 TeV LHC. The \NNLOPS{}
      prediction is shown at parton level with parton showering
      performed with \PYTHIA{8} (left, red) and \PYTHIA{6} (right, blue).
      The resummed result is shown in green in both panels. The lower
      panels show the ratio to the NNLO+NNLL resummation.}
    \label{fig:res-ptZ}
\end{figure}
In \cref{fig:res-ptZ} we show a comparison between the NNLO+NNLL resummation
obtained using \DYQT{} and our \NNLOPS{} result obtained with \PYTHIA{8} (left)
and \PYTHIA{6} (right), switching off all non-perturbative effects
(\emph{i.e.}~hadronisation and MPI are switched off, primordial-$\kt$ is set to
zero). As usual, the uncertainty band for our results has been obtained as the
envelope of a 21-points scale variation. \DYQT{} uses as resummation scale
$m_{\textrm{ll}}$, and the associated band has been obtained varying $\mur$ and
$\muf$ among the usual 7 combinations. On top of this, for the central value of
$\mur=\muf$ we varied the resummation scale by a factor 2 up and down. This
gives 9 combinations, the envelope of which is used here to define the \DYQT{}
uncertainty.  The ratio plot shows a pattern quite similar to what was observed
in Figures 4 and 5 of \Bref{Hamilton:2013fea}, namely differences of up to
$\mathcal{O}(10-12\%)$ between analytic resummation and \NNLOPS{}, with a
slightly more marked difference in the very first bin. In the region $\ptz \sim
5-15 \GeV$, the uncertainty bands do not overlap, mainly because of the very
narrow uncertainty bands in both predictions, in particular in the NNLO+NNLL
result. In the case of Higgs production, instead, uncertainty bands are wider,
hence the predictions are more compatible.
Changing the $\beta$ parameter might improve this agreement, although we recall
that the \NNLOPS{} prediction does not have NNLL accuracy in this region.
By comparing the two \NNLOPS{} results shown in the two panels of
\cref{fig:res-ptZ}, we also observe that the spectra obtained with
\PYTHIA{8} are typically $\sim 5$ \% harder than those with \PYTHIA{6}, a
feature that was already noticeable in \cref{fig:par-ptz}, and which will be
present also in other distributions where \NNLOPS{} results are ``only'' NLO
accurate. Few percent differences between different NLO+PS results in these
kinematic regions can be due to subleading effects, such as differences in
details of the two parton-shower algorithms, as well as the use of different
tunes. 

Another interesting observable to consider is the $\phi^*$ distribution which is
a measure of angular correlations in Drell-Yan lepton
pairs~\cite{Banfi:2010cf}. This observable is defined as~\cite{Banfi:2012du}
\begin{equation} \phi^* = \tan \left(\frac{\pi - \Delta
\phi}{2}\right)\sin \theta^*\,,
\end{equation} where $\Delta \phi$ is the azimuthal angle between the
two leptons and $\theta^*$ is the scattering angle of the electron
with respect to the beam, as computed in the boosted frame where the $Z$
boson is at rest.
We note that ATLAS uses a slightly different definition of the angle $\theta^*$,
and defines it as
\begin{equation} \cos \theta^* = \mathrm{tanh}\left(\frac{y_{l^-}-y_{l^+}}{2}\right)\,.
\end{equation}
Since we will compare with ATLAS data in \cref{subsec:Wdata}, we will use the
latter definition throughout this work.
\begin{figure}[!tbh]
  \begin{centering}
    \includegraphics[clip,width=0.49\textwidth]{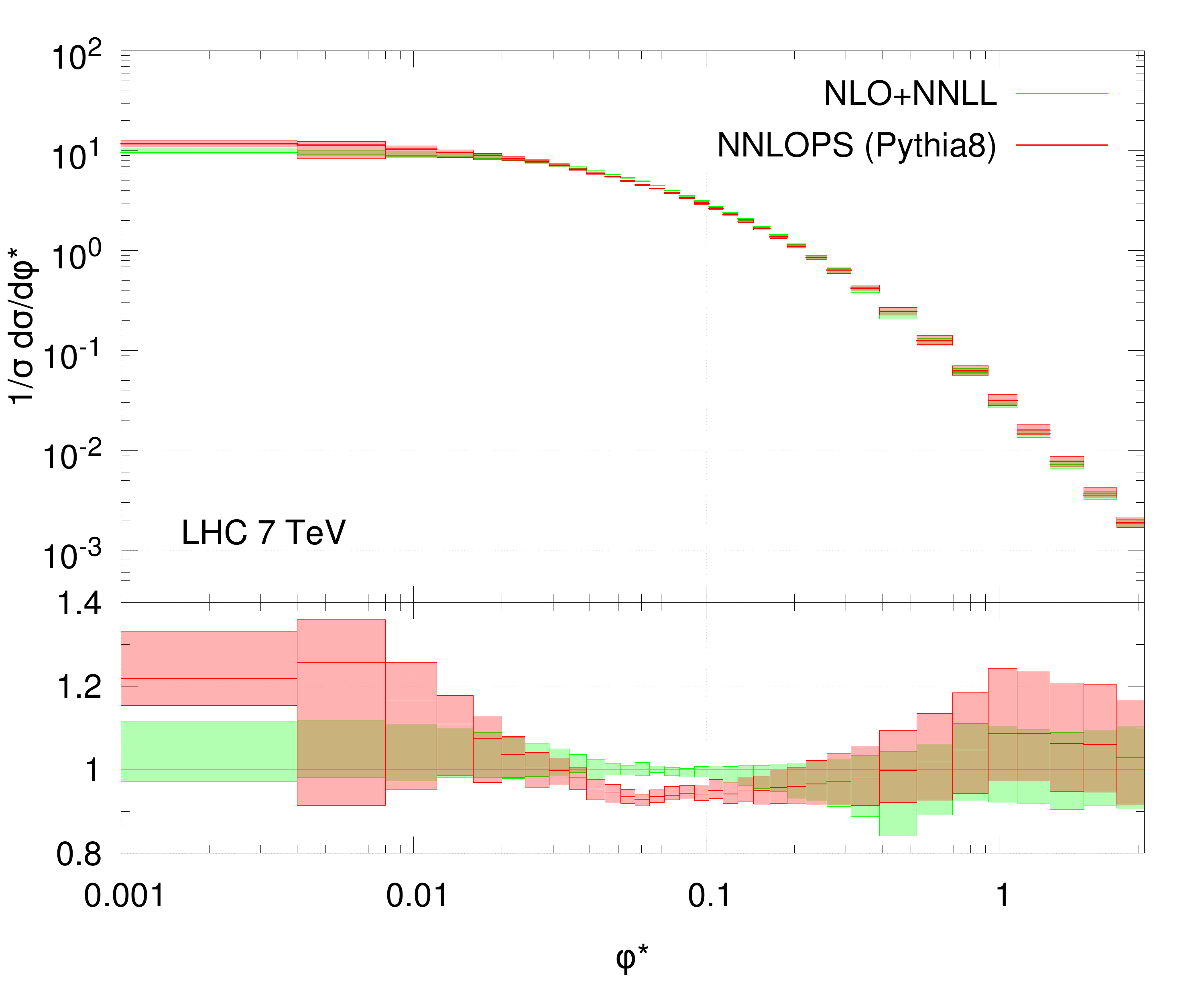}
    \includegraphics[clip,width=0.49\textwidth]{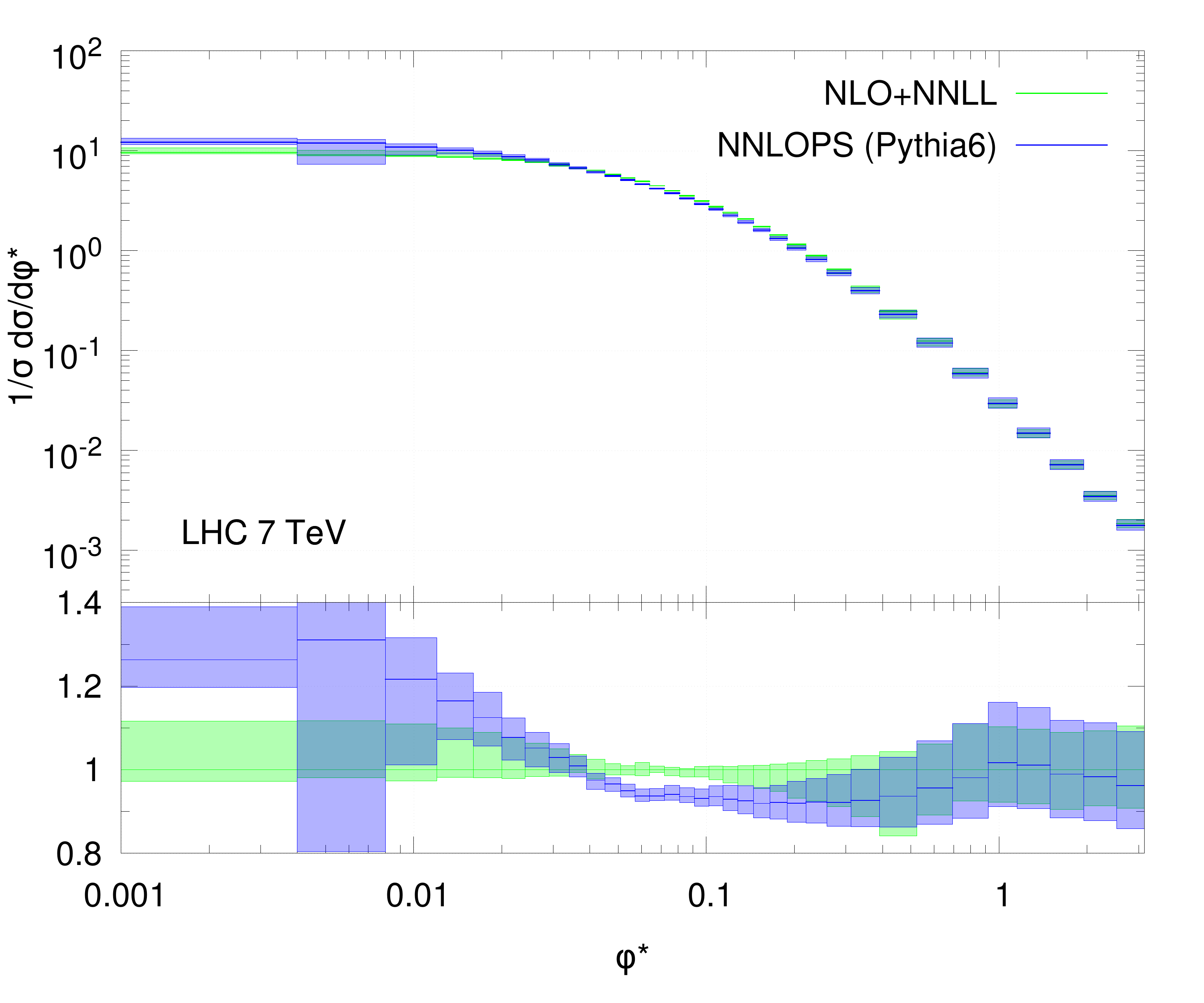}
    \par\end{centering}
    \caption{Comparison of \NNLOPS{} prediction and NLO+NNLL
      resummation for $\phi^*$ in $Z\to e^+e^-$ production at 7 TeV
      LHC. The \NNLOPS{} prediction is shown at parton level with
      parton showering performed with \PYTHIA{8} (left, red) and
      \PYTHIA{6} (right, blue).  The resummed result is shown in green
      in both panels.}
    \label{fig:res-phistar}
\end{figure}
In \cref{fig:res-phistar} we compare our \NNLOPS{} simulation with the
NLO+NNLL resummation of \Bref{Banfi:2012du}\footnote{We thank Andrea Banfi
  and Lee Tomlinson for providing us with their resummed results.}. From the
definition of $\phi^*$, it is clear that large values of $\phi^*$ correspond to
events where the $Z$ boson tends to be boosted, while for low values the $Z$
boson is almost at rest. We see that the two predictions agree reasonably well
for $\phi^* \gtrsim 0.2$, in particular when \PYTHIA{6} is used, while the
uncertainty bands to not overlap below that point.\footnote{We remark that the
  resummation of \Bref{Banfi:2012du} uses CTEQ6M parton distribution
  functions.}.  For high $\phi^*$, \PYTHIA{8} is slightly harder than
\PYTHIA{6}. Since for large values of $\phi^*$ the probed phase space regions
are dominated by large values of $\ptz$, this difference is expected, in view of
the discussion at the end of the previous paragraph.
We will show a comparison to data for this observable in
\cref{subsec:Zdata}, where we will also comment on the impact of
non-perturbative corrections.

Finally we consider the jet-veto efficiency which is defined as
\begin{equation}
\epsilon(p_\rm{T,veto}) \equiv \frac{1}{\sigma}\int_0^{p_{\rm
      T,veto}} d\ptjone \frac{d\sigma(\ptjone)}{d\ptjone}\,.  
\end{equation}
\begin{figure}[!tbh]
  \begin{centering}
    \includegraphics[clip,width=0.49\textwidth]{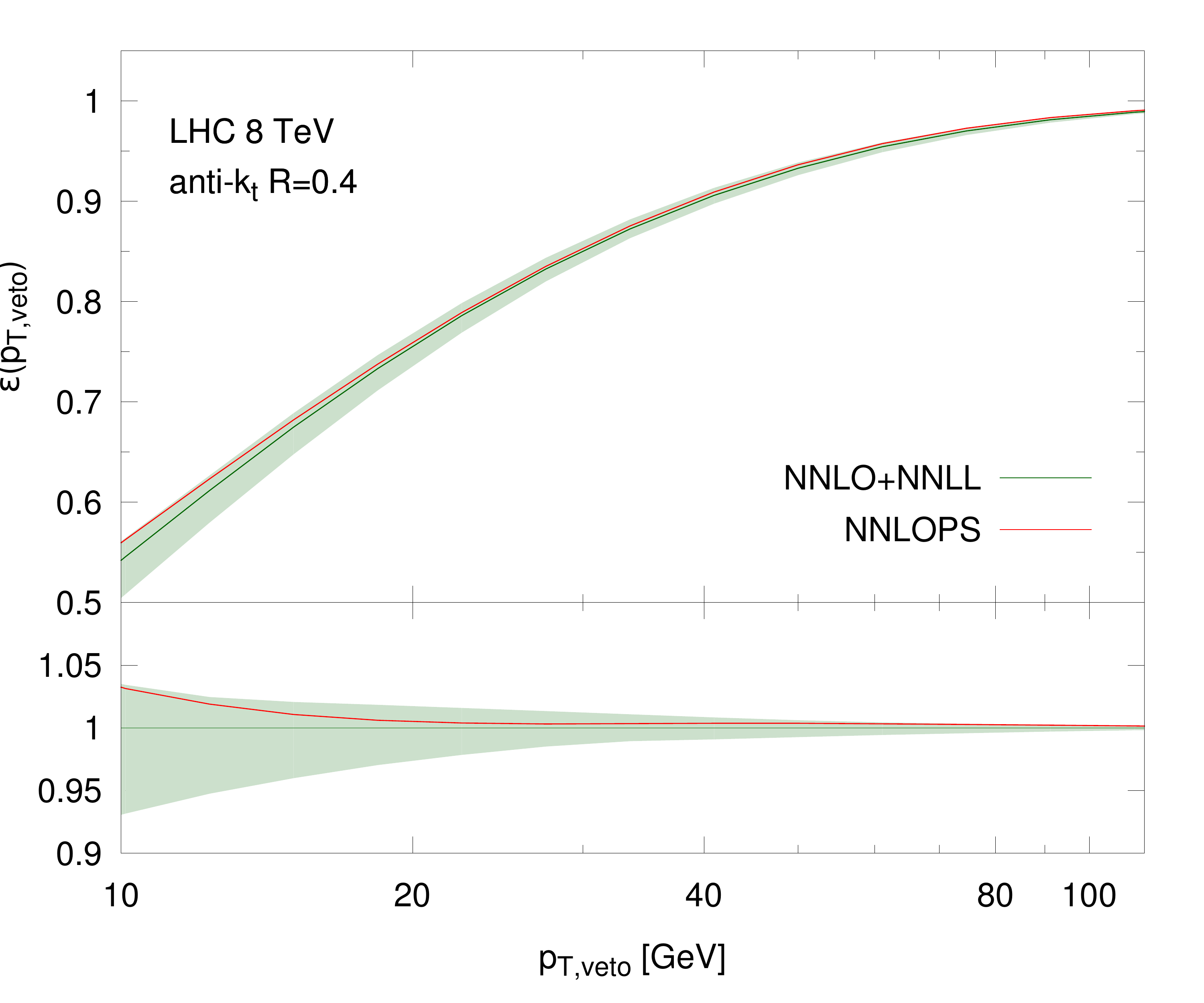}
    \includegraphics[clip,width=0.49\textwidth]{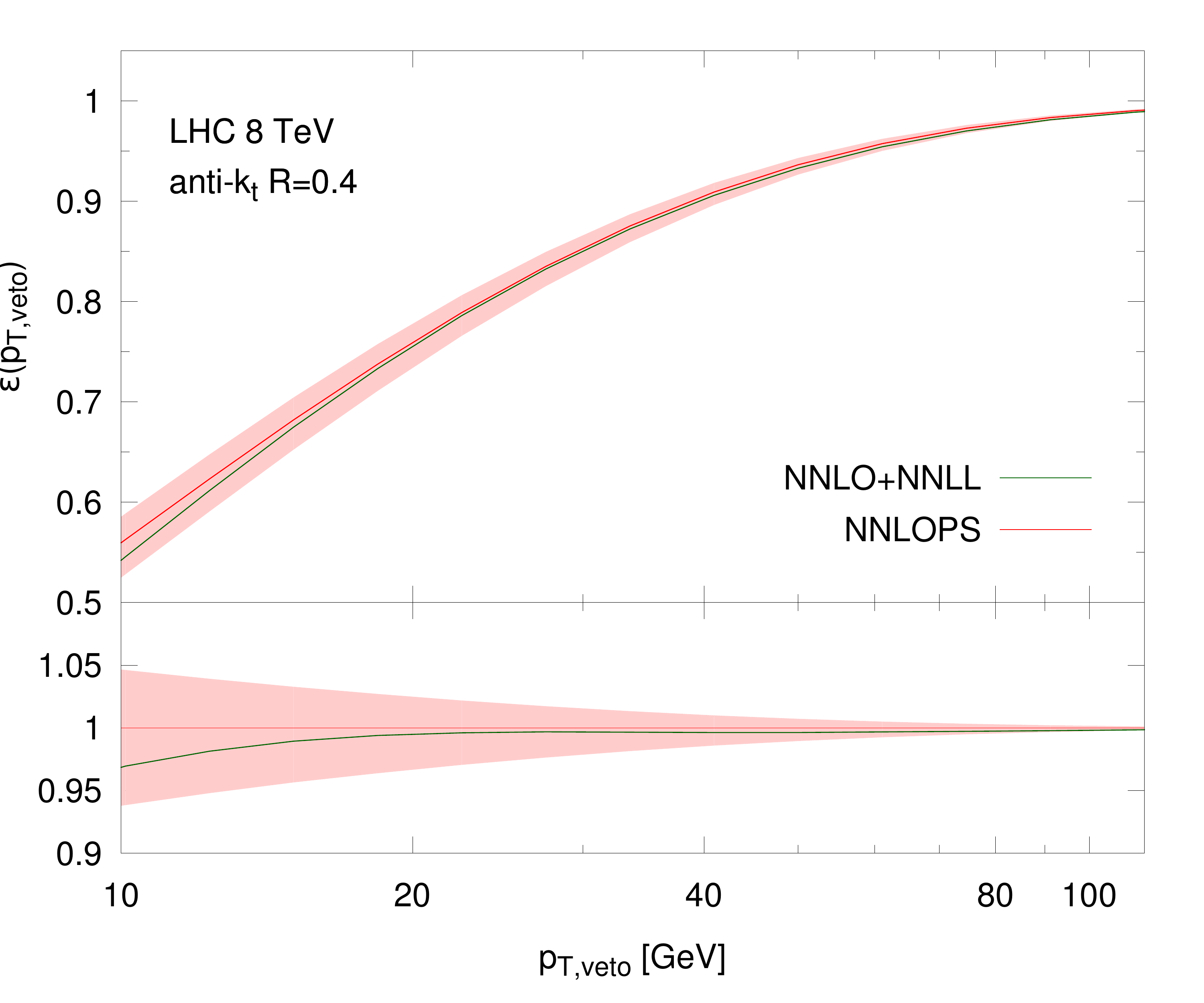}\\
    \includegraphics[clip,width=0.49\textwidth]{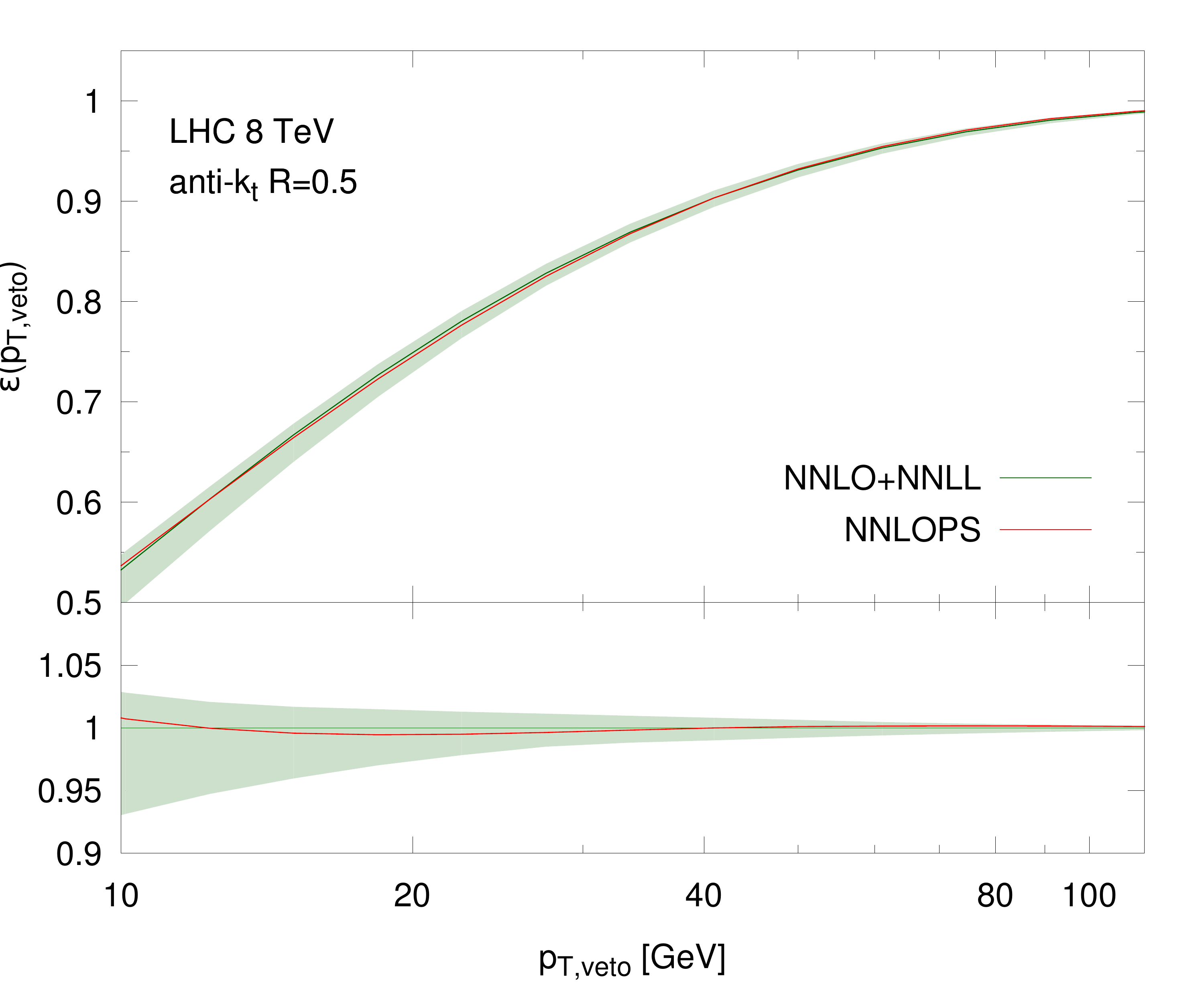}
    \includegraphics[clip,width=0.49\textwidth]{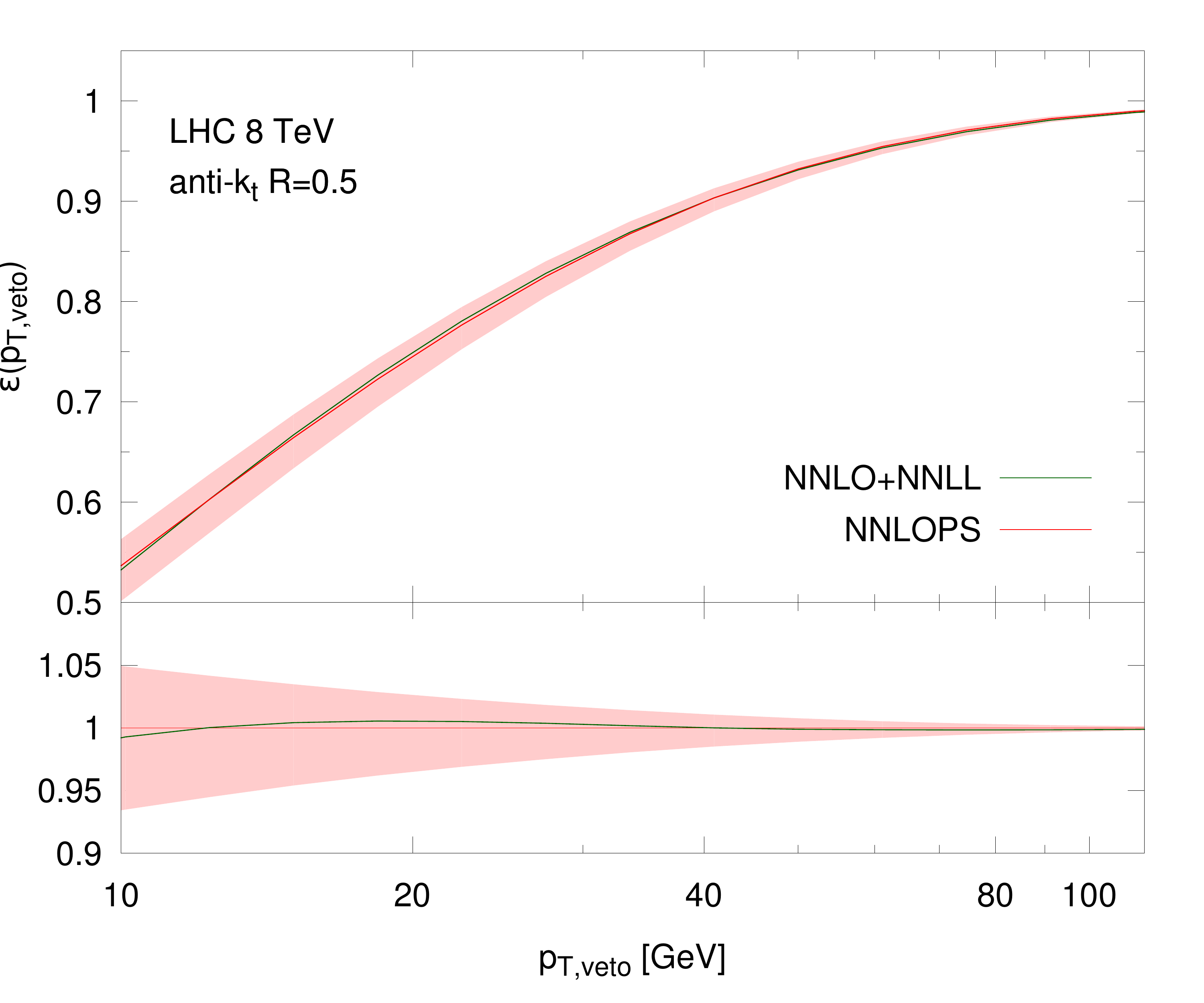}\\
    \includegraphics[clip,width=0.49\textwidth]{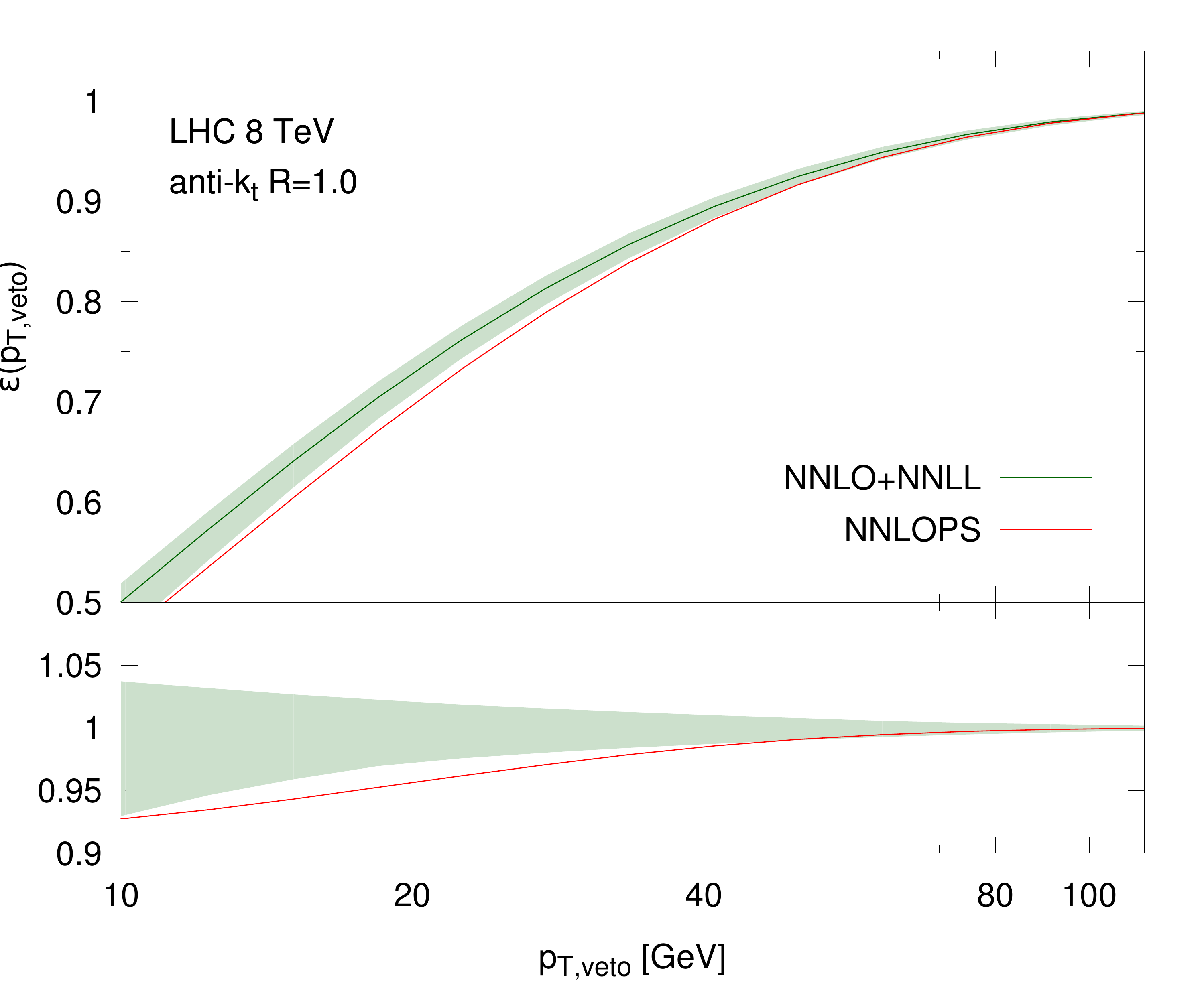}
    \includegraphics[clip,width=0.49\textwidth]{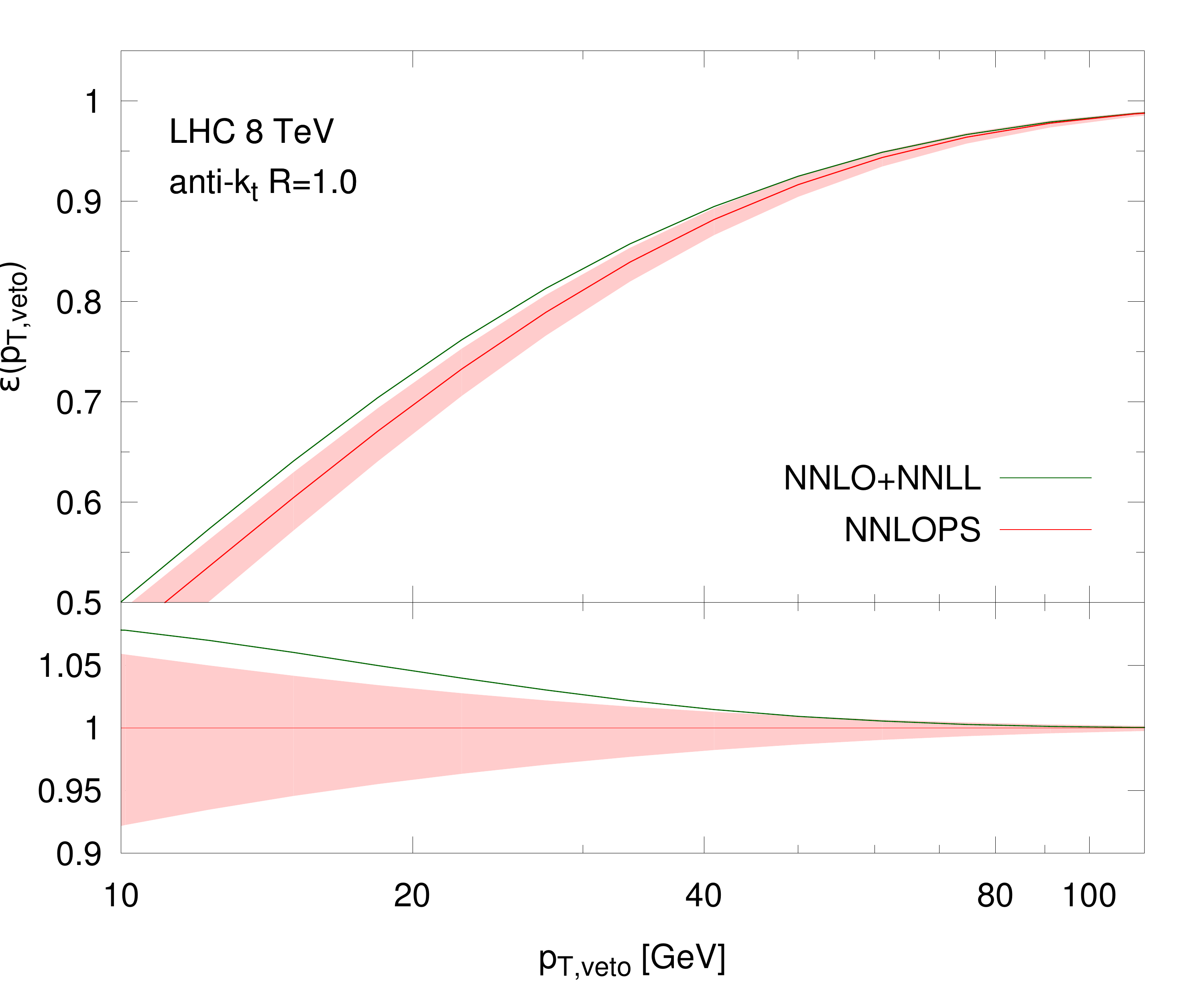}
    \par\end{centering}
    \caption{Comparison of \NNLOPS{} (red) prediction and NNLO+NNLL
      resummation (green) for the jet-veto efficiency for $Z$ production
      at 8 TeV LHC for three different values of the jet-radius. The
      \NNLOPS{} prediction is shown at parton level with parton
      showering performed with \PYTHIA{8}.}
    \label{fig:res-jetveto}
\end{figure}

Here results have been obtained for the LHC at 8 TeV, using the
anti-$k_t$ algorithm to construct jets.  We restrict ourselves to show
\NNLOPS{} results obtained with \PYTHIA{8}. Analytically resummed
results have been obtained using \JETVHETO{}~\cite{Banfi:2012jm}, and
have NNLL+NNLO accuracy. 
The \JETVHETO{} results have been obtained using its default setting, i.e. using
as a central value for the renormalisation, factorisation and resummation scale
$\MZ/2$\footnote{Since we use $\beta=1$ in the definition of $h(\pt)$ for the
  \NNLOPS{} results, it is interesting also to examine the \JETVHETO{} results
  using as a central value for renormalisation, factorisation and resummation
  scale $\MZ$. We have done so and find that there are minimal differences for
  $R=0.4$ and $R=0.5$. For $R=1.0$ we find a slightly worse agreement.}.

As is recommended in \Bref{Banfi:2012jm}, for the \JETVHETO{} results we
have obtained the uncertainty band as an envelope of eleven curves: we have
varied the renormalisation and factorisation scales independently giving rise to
the usual 7-scale choices, additionally, for central renormalisation and
factorisation scales we have varied the resummation scale up and down by a
factor 2 and looked at two different additional schemes to compute the
efficiency.

We observe a very good agreement between the two approaches for $R=0.4$ and
$0.5$, whereas for $R=1$ differences are more marked.  Few comments are in order
here: first, the pattern shown in the plots is consistent with what was already
observed in the Higgs case (Figure 7 of \Bref{Hamilton:2013fea}), namely
differences up to few percents, and good band overlap, for smaller values of
$R$, and larger differences for $R=1$. For very large values of $R$, the
leading-jet momentum will balance against the transverse momentum of the vector
boson. Given what we observed for $\ptz$, it is therefore no surprise that, when
$R=1$, we have $\mathcal{O}(3-5) \%$ differences with respect to the resummed
result for values of \mbox{$p_\rm{T,veto}\sim 25-30 \GeV$}, as used currently by
ATLAS and CMS in Higgs production.

\subsection{Comparison to Data}
\label{sec:compdata}
In this section we compare our predictions with a number of available data from
ATLAS, both for $Z$ and for $W$ production\footnote{Since the publication of
  this work a few more analyses have been released by both ATLAS and
  CMS~\cite{Aad:2015auj,Aad:2016izn,Aad:2016naf,CMS:2014jea,Khachatryan:2016yte}. They
  all show good agreement with the various theoretical predictions that they
  compare with.}.
\subsubsection{$Z$ Production}
\label{subsec:Zdata}
We show here a comparison to a number of measurements performed by ATLAS at 7
TeV~\cite{Aad:2011dm,Aad:2011gj,Aad:2012wfa}. ATLAS applied the following cuts:
they consider the leptonic decay of the $Z$ boson to electrons or muons and
require an electron (muon) and a positron (anti-muon) with $\pt > 20 \GeV$ and
rapidity $|y| < 2.4$. The invariant mass of the di-lepton pair should lie in the
window $66 \GeV < m_\rm{ll} < 116 \GeV$.

\begin{figure}[!t]
\begin{centering}
\includegraphics[clip,width=0.49\textwidth]{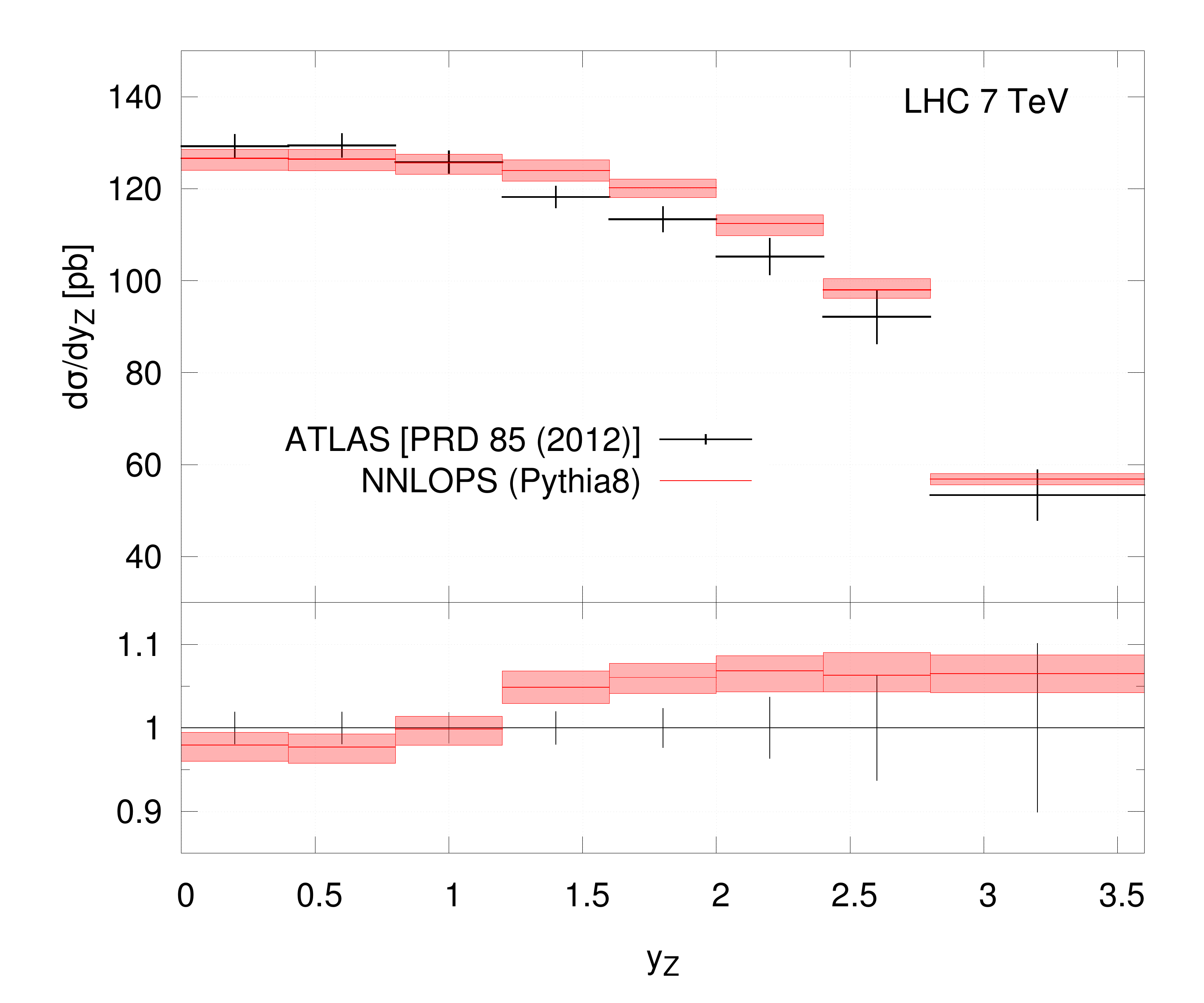}
\includegraphics[clip,width=0.49\textwidth]{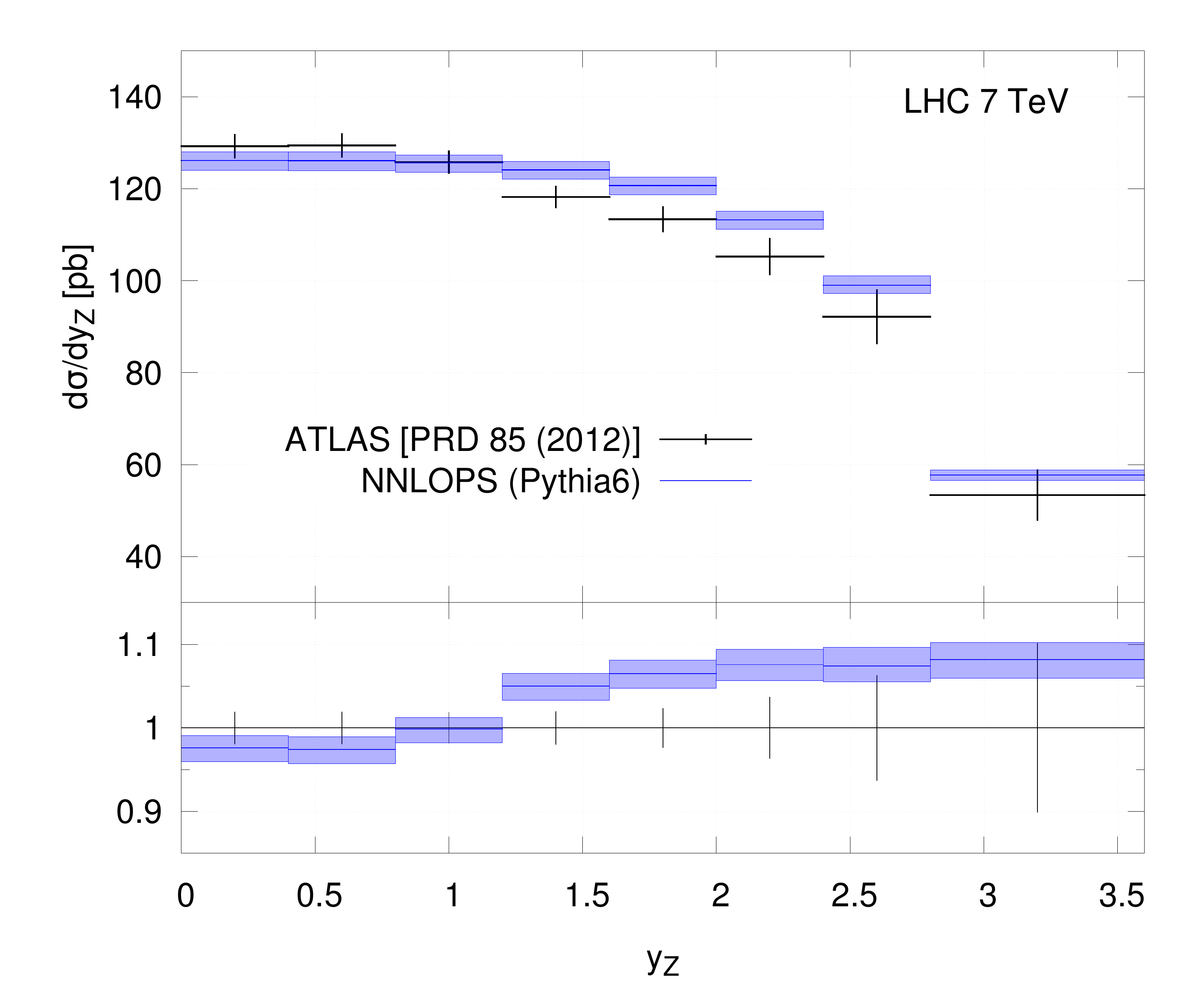}
\par\end{centering}
\caption{Comparison of \NNLOPS{} prediction obtained with \PYTHIA{8}
  (left) and \PYTHIA{6} (right) to data (black) from
  \Bref{Aad:2011dm} for the $Z$ boson rapidity distribution at 7
  TeV LHC.}
\label{fig:data-yz}
\end{figure}
We begin by showing in \cref{fig:data-yz} a comparison of \NNLOPS{}
results (with two versions of \PYTHIA{}) to data from \Bref{Aad:2011dm} for the
$Z$ boson rapidity distribution. As expected, our result displays a quite narrow
uncertainty band, due to having included NNLO corrections. Since this is a
fully inclusive observable, the absolute value of the cross-section and the size
of the uncertainty band will be driven by the NNLO reweighting: hence,
\PYTHIA{6} and \PYTHIA{8} results are almost indistinguishable, as expected.  We
also observe that we agree with data within the errors for central
rapidities. At high rapidity, however, there seems to be a tension between data
and our results. This discrepancy between data and pure NNLO was already
observed in the original ATLAS paper, although the NNLO results shown in
\Bref{Aad:2011dm} have a slightly larger uncertainty band since they also
contain PDF uncertainties.  We note that, at the moment, the dominant error is
coming from data. In a more recent analysis of $13 \TeV$ data~\cite{Aad:2016naf}
good agreement is found between data and \POWHEG{} events, which makes us expect
a similar agreement with our \NNLOPS{} results.

\begin{figure}[!tbh]
\begin{centering}
\includegraphics[clip,width=0.49\textwidth]{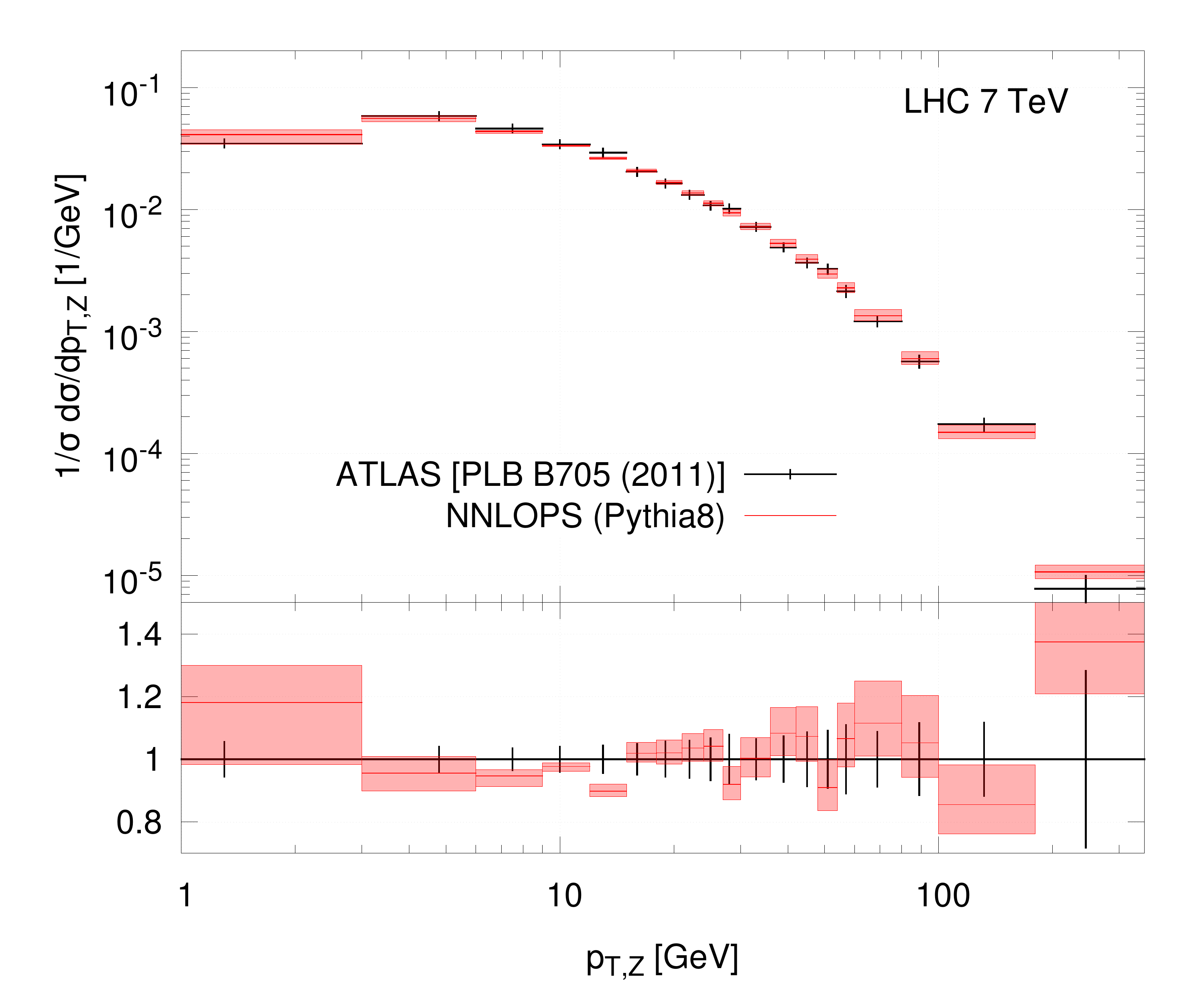}
\includegraphics[clip,width=0.49\textwidth]{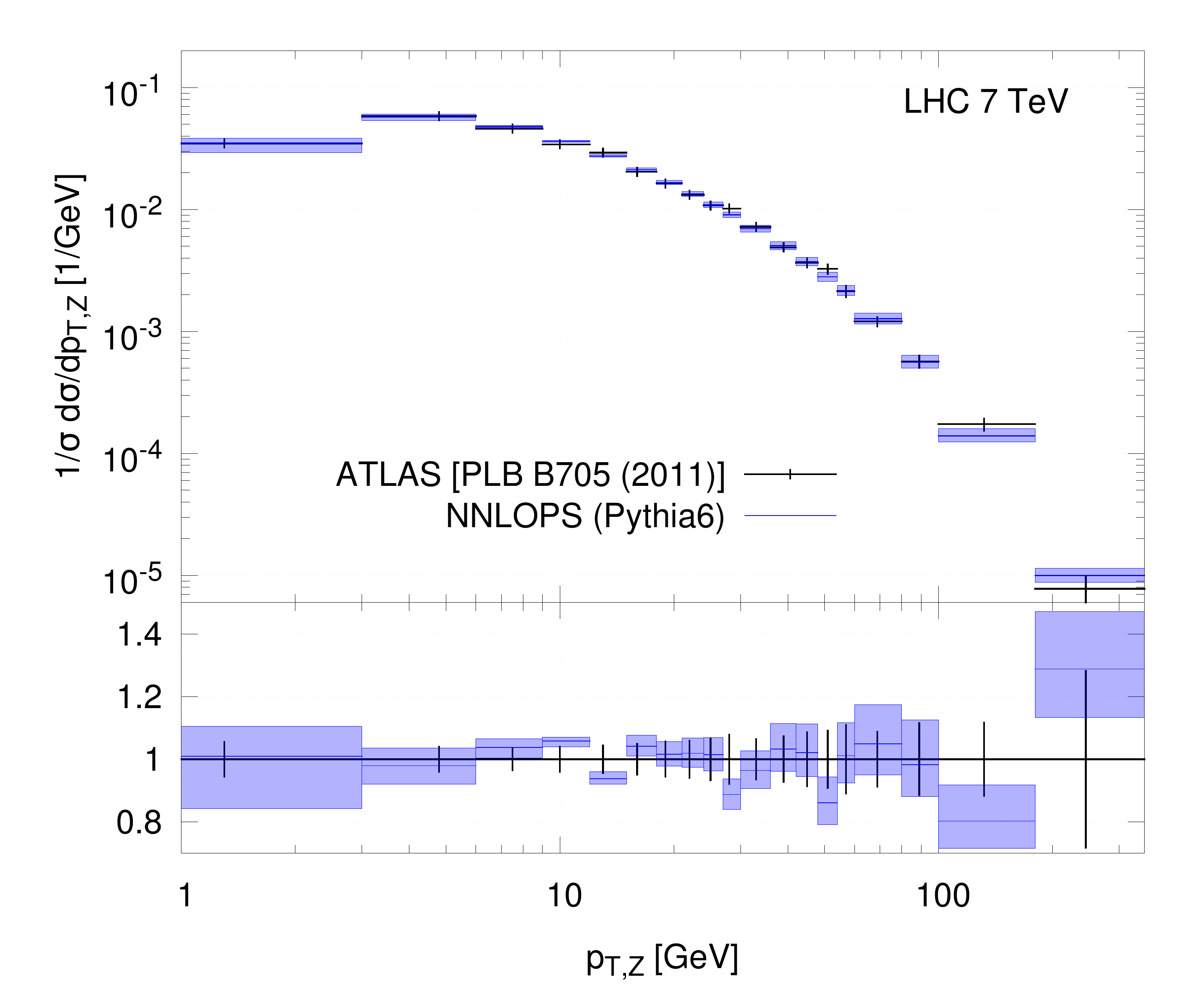}
\par\end{centering}
\caption{Comparison to data from \Bref{Aad:2011gj} for the $Z$ boson
  transverse distribution at 7 TeV LHC. Normalised data compared with \NNLOPS{}
  showered with \PYTHIA{8} (left plot, red) and \PYTHIA6{} (right plot,
  blue). Uncertainty bands for the theoretical predictions are obtained by first
  normalising all scale choices, as described in
  \cref{subsec:Estimating-uncertainties} and then taking the associated
  envelope of these normalised distributions.}
\label{fig:data-ptz}
\end{figure}
In \cref{fig:data-ptz} we now show the same comparison for the $Z$ boson
transverse momentum against data from \Bref{Aad:2011gj}. In the left panel
we use \PYTHIA{8} to shower events and in the right panel \PYTHIA{6}. We see
that there is very good agreement between our \NNLOPS{} prediction over the
whole $\pt$ range.  Only in the very first bin we observe a slight tension with
data in the case of \PYTHIA{8}, which can be due to a different modelling of
non-perturbative effects.  Given the fact that we use two different
parton-showers, with different tunes (and even different PDFs for the tunes), an
$\mathcal{O}(20\%)$ difference in the $1-3 \GeV$ region can be expected. It is
foreseeable that once a tuning of \PYTHIA{8} is done in conjunction with
\POWHEG{}, such differences will go away.
\begin{figure}[!tbh]
\begin{centering}
\includegraphics[clip,width=0.49\textwidth]{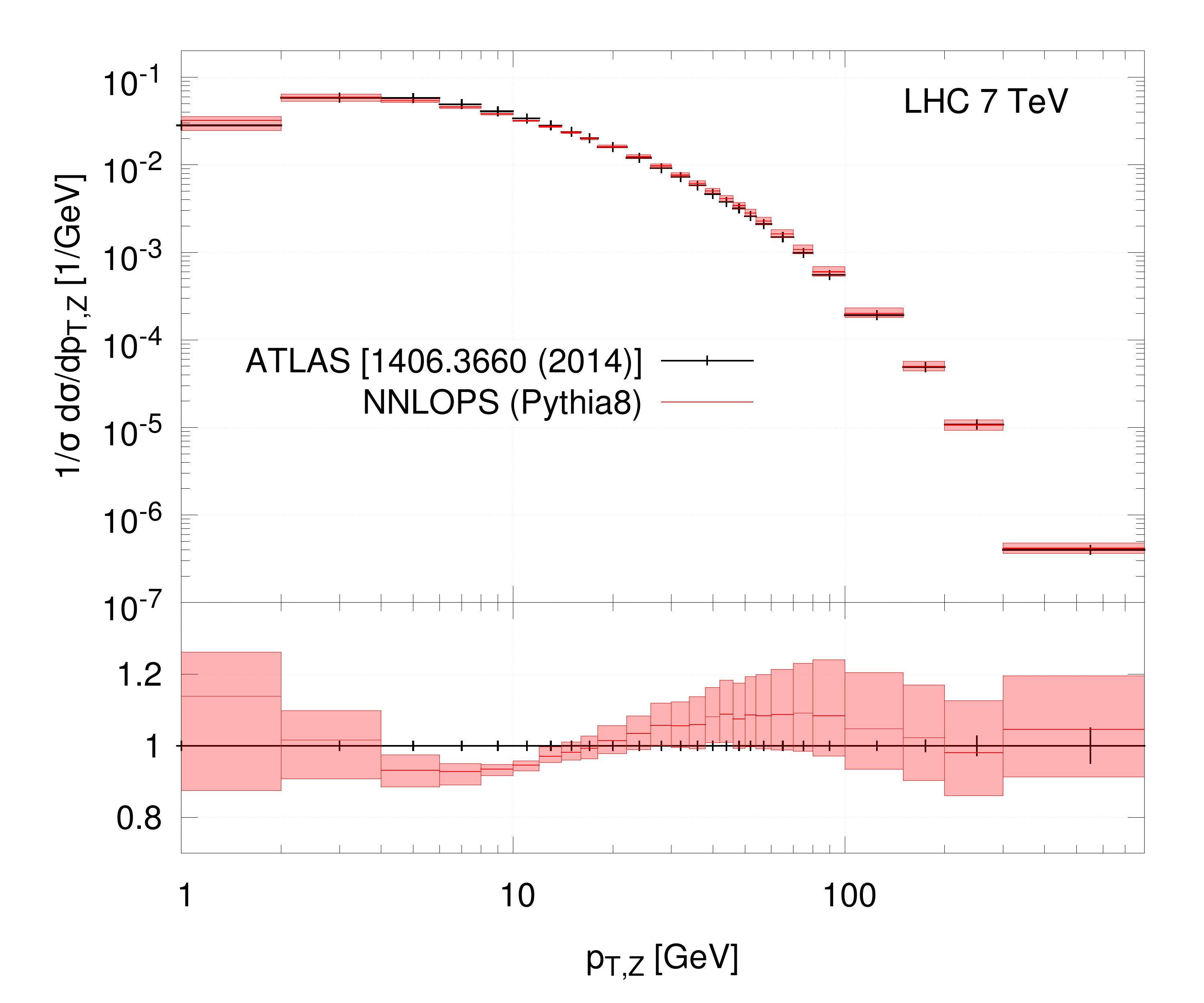}
\includegraphics[clip,width=0.49\textwidth]{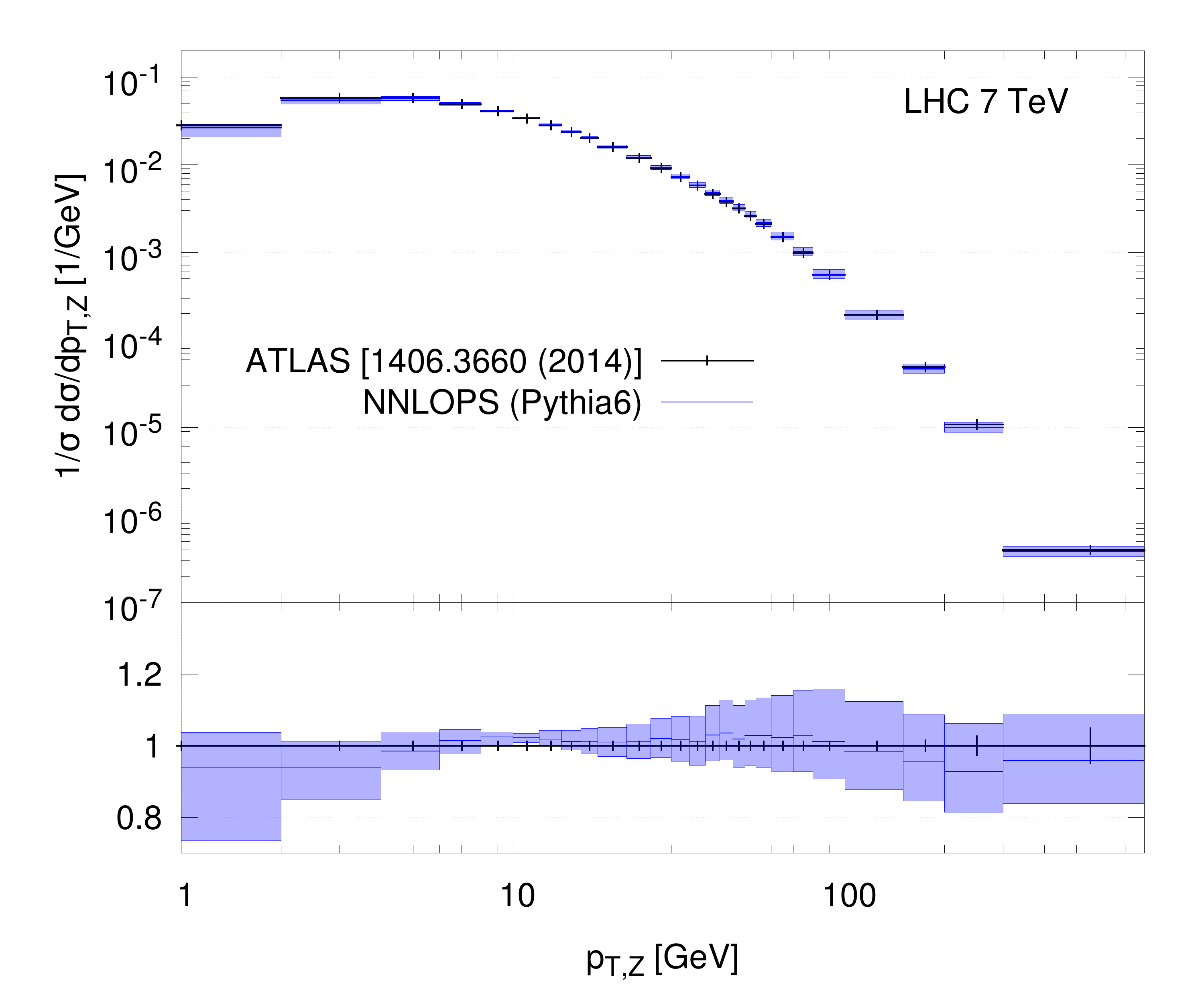}
\par\end{centering}
\caption{As in previous figure, but with more luminosity, thinner
  binning, and up to larger values of $\ptz$. Data are now from taken
  from \Bref{Aad:2014xaa}. }
\label{fig:data-ptz-new}
\end{figure}
\cref{fig:data-ptz-new} shows again $\ptz$, based now on $4.7\;\mathrm{fb}^{-1}$ of data
from ATLAS~\cite{Aad:2014xaa}. Due to the thinner binning, it is now possible to
appreciate clearly the differences between \PYTHIA{6} and \PYTHIA{8}: the
\NNLOPS{} result obtained with \PYTHIA{6} shows a remarkable agreement with data
across all the $\ptz$ range, whereas with \PYTHIA{8} we can observe differences
of up to 10\% for $ 5\lesssim \ptz \lesssim 100 \GeV$. What we observe from
these plots is consistent with \cref{fig:data-ptz}, the difference being that
here the improved precision in data allows us to conclude that for the setups
and tunes we are using, the best description is obtained when using \PYTHIA{6}.

\begin{figure}[!tbh]
\begin{centering}
\includegraphics[clip,width=0.49\textwidth]{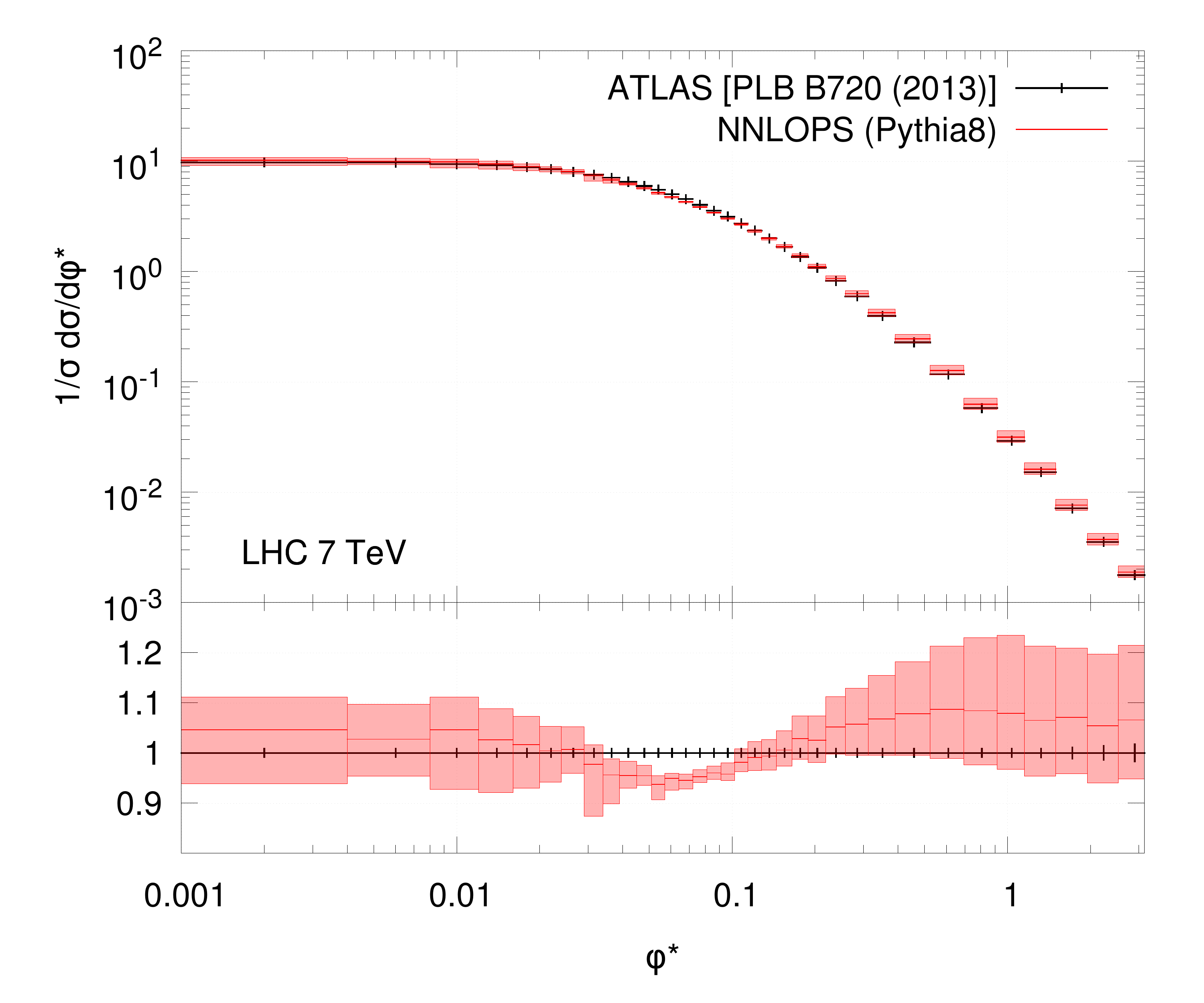}
\includegraphics[clip,width=0.49\textwidth]{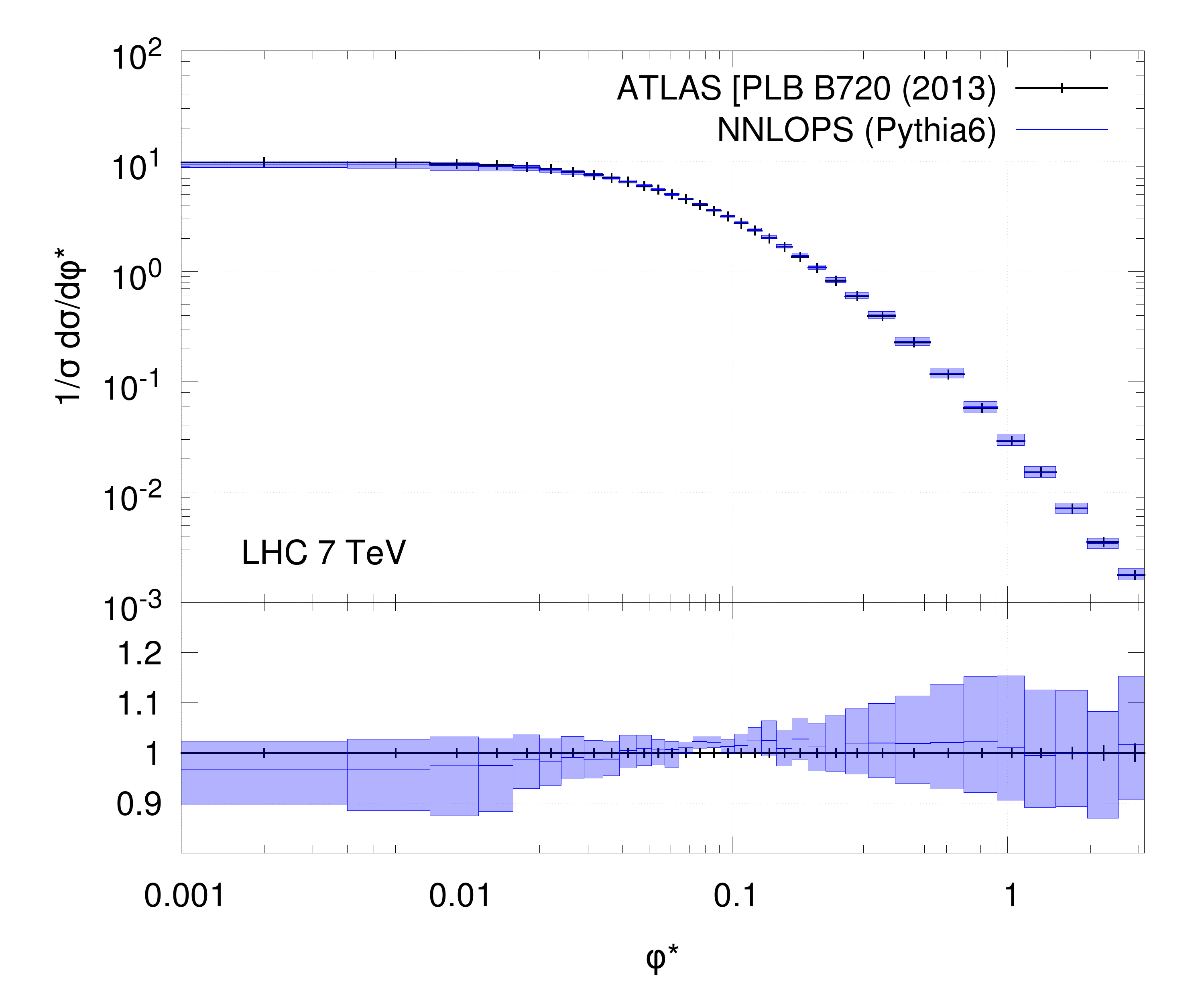}
\par\end{centering}
\caption{As in \cref{fig:data-ptz} for the $\phi^*$ distribution
  in $Z$ boson production. Data taken from \Bref{Aad:2012wfa}. }
\label{fig:data-phistar}
\end{figure}
Next, in \cref{fig:data-phistar}, we consider the comparison to data for the
$\phi^*$ distribution.  Although both our predictions are consistent with the
ATLAS measurement, it is clear that events showered with \PYTHIA{6} agree better
with data, whereas \NNLOPS{} showered with \PYTHIA{8} has a slightly different
shape, exhibiting a dip in the theory prediction compared with data, at around
$\phi^*=0.06$.  By comparing with our predictions before the inclusion of
non-perturbative effects, we have checked that these effects play a sizeable role
in the region below $\phi^*<0.1$, therefore it is not completely unexpected that
\PYTHIA{6} and \PYTHIA{8} give slightly different shapes. The fact that
non-perturbative effects introduce non-trivial changes on the shape of this
distribution was also noted in \Bref{Banfi:2011dx} (for predictions at the
Tevatron). We also observe the same pattern shown in \Bref{Banfi:2011dx}, namely
a moderate increase at $\phi^*\sim 0.05$ and a more pronounced decrease for very
low values of $\phi^*$ when non-perturbative effects are included.  Finally, it
is worth mentioning that the results we have obtained (especially with
\PYTHIA{6}) clearly show a better agreement with data than what was found in
\Bref{Aad:2011gj}, where different tunes were used, both for \PYTHIA{6} and
\PYTHIA{8}. It is difficult to draw a solid conclusion, since in
\Bref{Aad:2011gj} \POWHEG{}-\Z{} (as opposed to \ZJMINLO{}) was used, and events
were also reweighted using \noun{ResBos}~\cite{Balazs:1997xd}. Nevertheless it
seems clear that the best agreement are obtained when higher-order
perturbative corrections are included and modern tunes are used.

\subsubsection{$W$ Production} 
\label{subsec:Wdata}
In this section we compare our predictions to results of
\Brefs{Aad:2011fp,Aad:2013ueu}, and in particular we use the combined decay of
the $W$ to electrons and muons. The charged lepton is required to have $\pt >
20 \GeV$ and rapidity $|y| < 2.4$. The event must have a missing energy $p_{\rm
  T, miss} > 25 \GeV$ and the transverse mass of the $W$ boson defined as $\mtw =
\sqrt{2 (p_{{\scriptscriptstyle \mathrm{T}},l}\, p_{{\scriptscriptstyle
      \mathrm{T}},\mathrm{miss}} -\vec p_{{\scriptscriptstyle \mathrm{T}},l} \,\cdot \vec
  p_{{\scriptscriptstyle \mathrm{T}},\mathrm{miss}}) }$ must be above $40 \GeV$. As was the
case for the $Z$ studies, ATLAS only provides normalised distributions for the
standard observables we will show here.

\begin{figure}[!tbh]
\begin{centering}
\includegraphics[clip,width=0.49\textwidth]{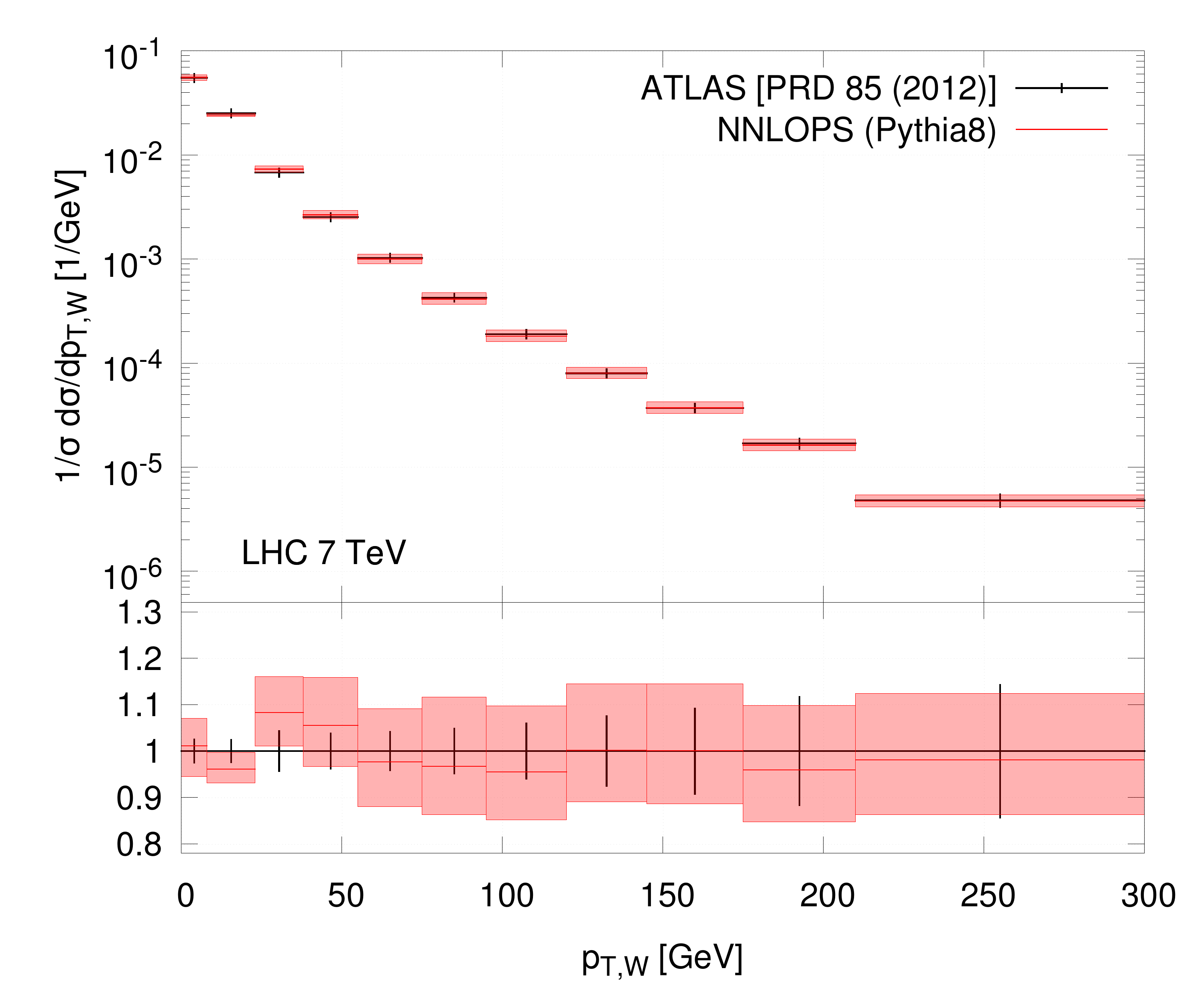}
\includegraphics[clip,width=0.49\textwidth]{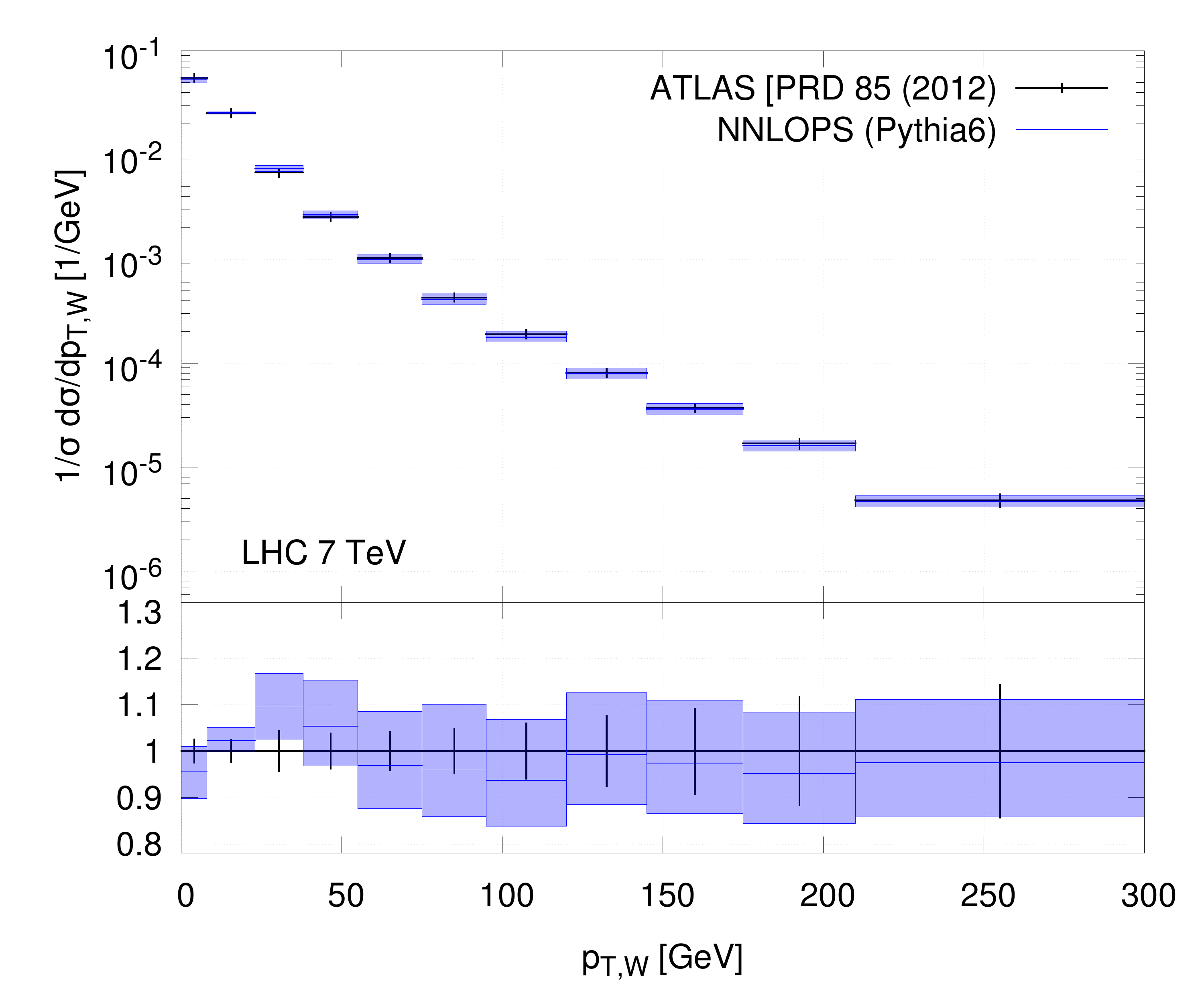}
\par\end{centering}
\caption{Comparison of \NNLOPS{} prediction (red) to data (black) from
  \Bref{Aad:2011fp} for the $W$ boson $\ptw$ distribution, using
  \PYTHIA{8} (left) and \PYTHIA{6} (right).}
\label{fig:data-ptw}
\end{figure}

We start by showing in \cref{fig:data-ptw} the transverse momentum of the
$W$ boson, $\ptw$. As expected, differences in the parton shower algorithm only
play a visible role in the small $\pt$ region, where minor differences between
\PYTHIA{6} and \PYTHIA{8} can be observed. For high $\ptw$, the two predictions
are consistent with each other, and agree quite well with data.

We also compared our predictions with the analysis performed by the ATLAS
collaboration in \Bref{Aad:2013ueu}. We show results for $\kt$-splitting
scales in $W+$jets events. These observables are defined as the smallest
distances found by the (inclusive) $\kt$-algorithm at each step in the
clustering sequence. The splitting scale $d_k$ is the smallest among all
distances found by the algorithm when going from $(k+1)$ to $k$
objects. Therefore $\sqrt{d_0}$ corresponds to the transverse momentum of the
leading jet, whereas $\sqrt{d_1}$ is the smallest distance among pseudo-jets
when clustering from $2$ to $1$ jet:
\begin{equation}
  \label{eq:d0}
  d_1=\min(d_{1B},d_{2B},d_{12})\,,
\end{equation}
where
\begin{eqnarray}
  d_{iB} & = & p^2_{\scriptscriptstyle \mathrm{T,i}}\,, \nonumber \\
  d_{ij} & = &\min(p^2_{\scriptscriptstyle \mathrm{T,i}},p^2_{\scriptscriptstyle \mathrm{T,j}})\frac{(\Delta R_{ij})^2}{R^2}\,,
\end{eqnarray}
are the usual distances used in the $\kt$-algorithm. Among other reasons, these
observables are interesting because they can be used as a probe of the details
of matching and merging schemes. Due to the underlying \ZJMINLO{} simulation,
our \NNLOPS{} prediction is NLO accurate for large values of $\ptjone$, and it
is at least LL accurate in describing the $1\to 0$ jet transition, which is
measured in the $d_0$ distribution. The second jet spectrum and the $2\to 1$ jet
transition (which is encapsulated in $d_1$) are instead described at LO+LL, due
to the underlying \POWHEG{} simulation. Since the definition of $d_1$ contains
$d_{12}$, this observable is a measure of the internal structure of the first
jet, and not only of the second jet transverse momentum.

\begin{figure}[!tbh]
\begin{centering}
\includegraphics[clip,width=0.49\textwidth]{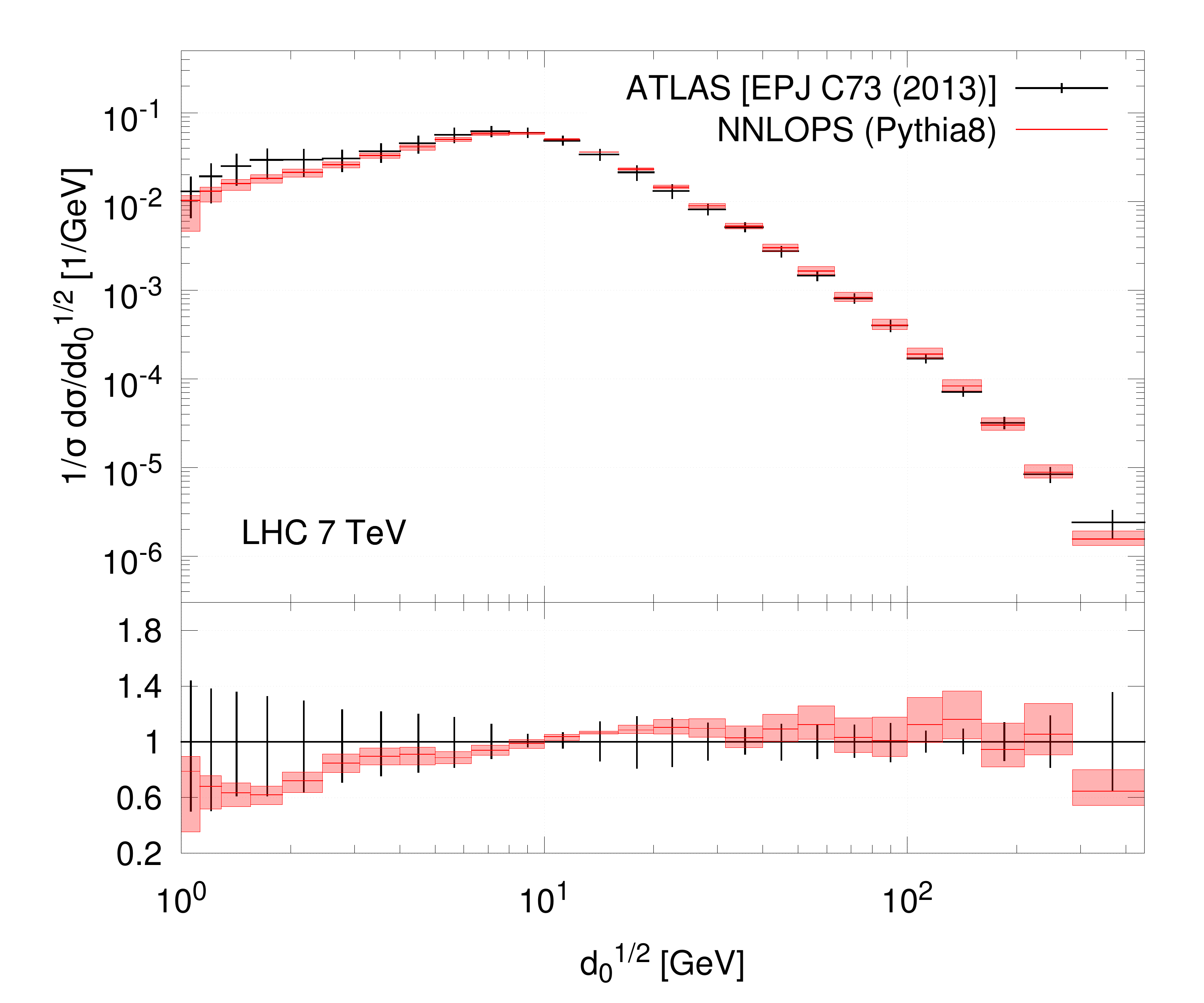}
\includegraphics[clip,width=0.49\textwidth]{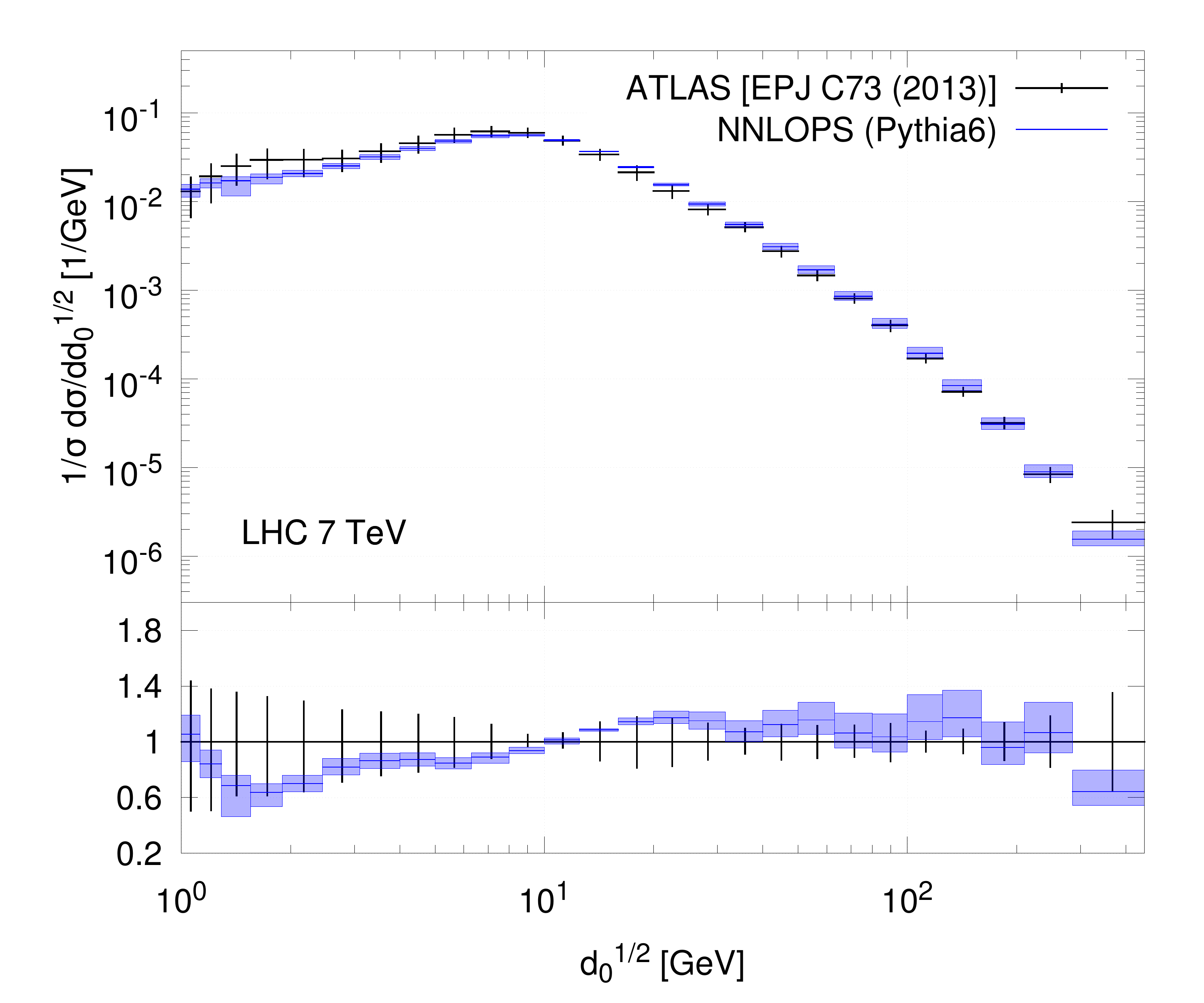}
\par\end{centering}
\caption{Comparison of \NNLOPS{} prediction (red) to 7 TeV LHC data (black) from
  \Bref{Aad:2013ueu} for the $W$ boson $\kt$ splitting scale
  $\sqrt{d_0}$ as defined in the text using \PYTHIA{8} (left) and
  \PYTHIA{6} (right).}
\label{fig:data-d0}
\end{figure}
\begin{figure}[!tbh]
\begin{centering}
\includegraphics[clip,width=0.49\textwidth]{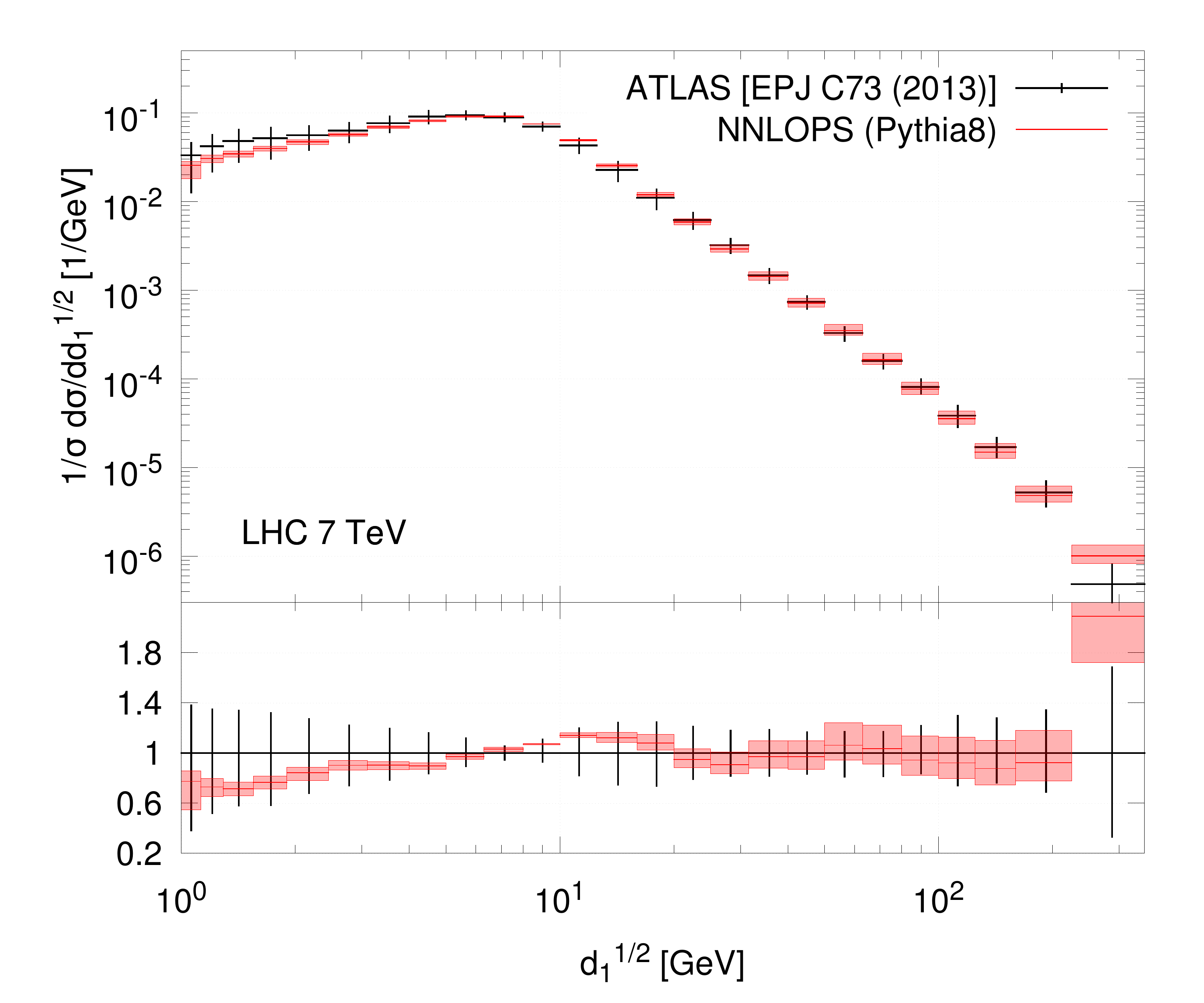}
\includegraphics[clip,width=0.49\textwidth]{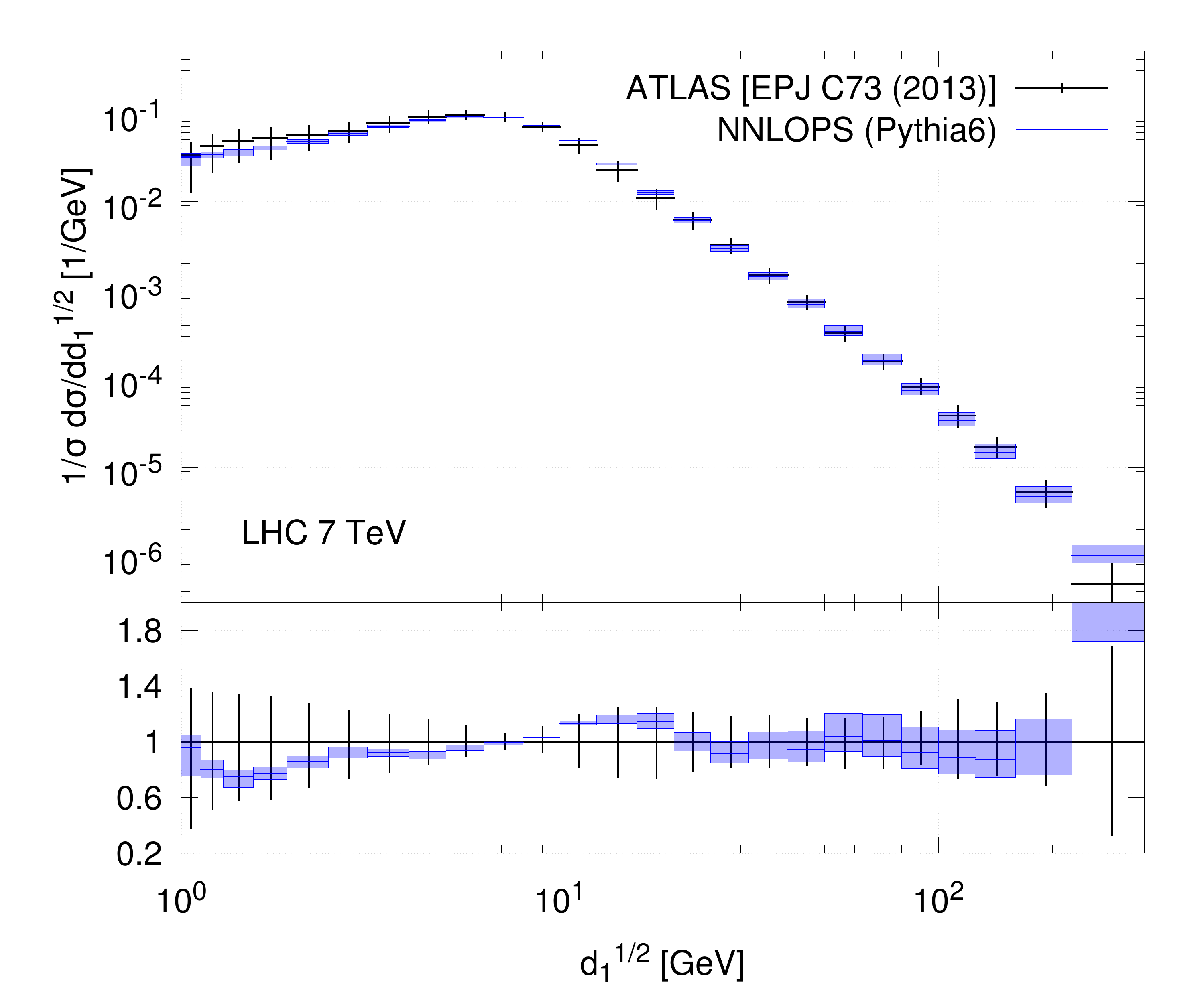}
\par\end{centering}
\caption{Comparison to 7 TeV LHC data from \Bref{Aad:2013ueu} for the $W$
  boson $\kt$ splitting scale $\sqrt{d_1}$ as defined in
  \cref{eq:d0} using \PYTHIA{8} (left) and \PYTHIA{6} (right).}
\label{fig:data-d1}
\end{figure}
In \cref{fig:data-d0,fig:data-d1} we show our \NNLOPS{}
predictions against ATLAS data, using as jet radius $R=0.6$. We find good
agreement, especially when $\sqrt{d_i}>10 \GeV$. Below this value, we are still
compatible with the experimental uncertainty bands, although we are
systematically lower than data. Once more, one should consider that the region
below $5-10 \GeV$ will be affected also by non-perturbative effects. For large
values of $d_i$ we are instead sensitive to the level of accuracy that we reach
in describing hard emissions. In this respect, it is no surprise that we have a
better agreement with data than the \POWHEG{} results shown in
\Bref{Aad:2013ueu}, where $d_1$ is poorly described since the second
emission is only described in the shower approximation. NLO corrections to the
$W+1$ jet region are included in the \NNLOPS{} simulation, and are very likely
the reason why we have a description of $d_0$ that is better than what was
observed in \Bref{Aad:2013ueu}.

\begin{figure}[!tbh]
\begin{centering}
\includegraphics[clip,width=0.49\textwidth]{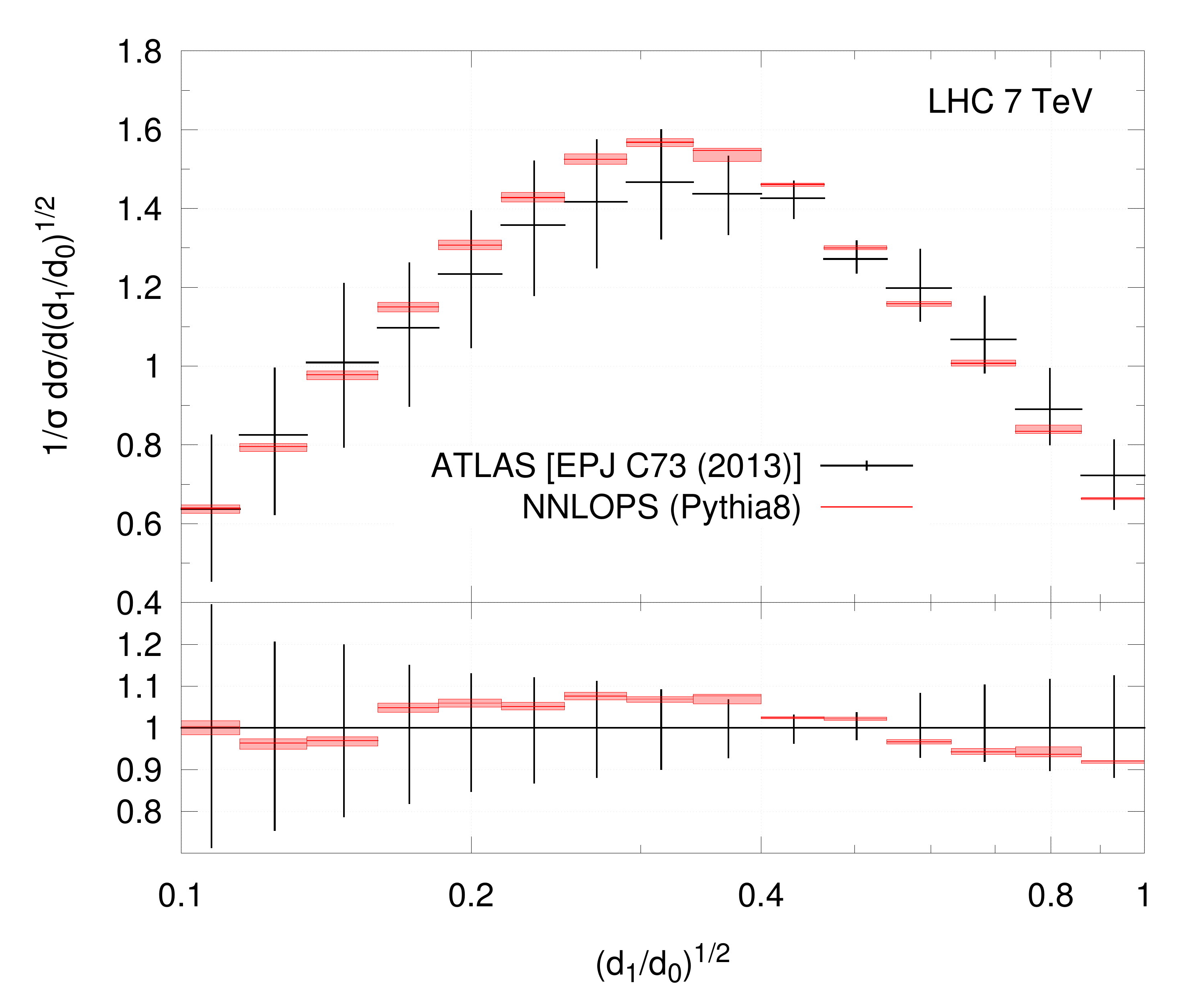}
\includegraphics[clip,width=0.49\textwidth]{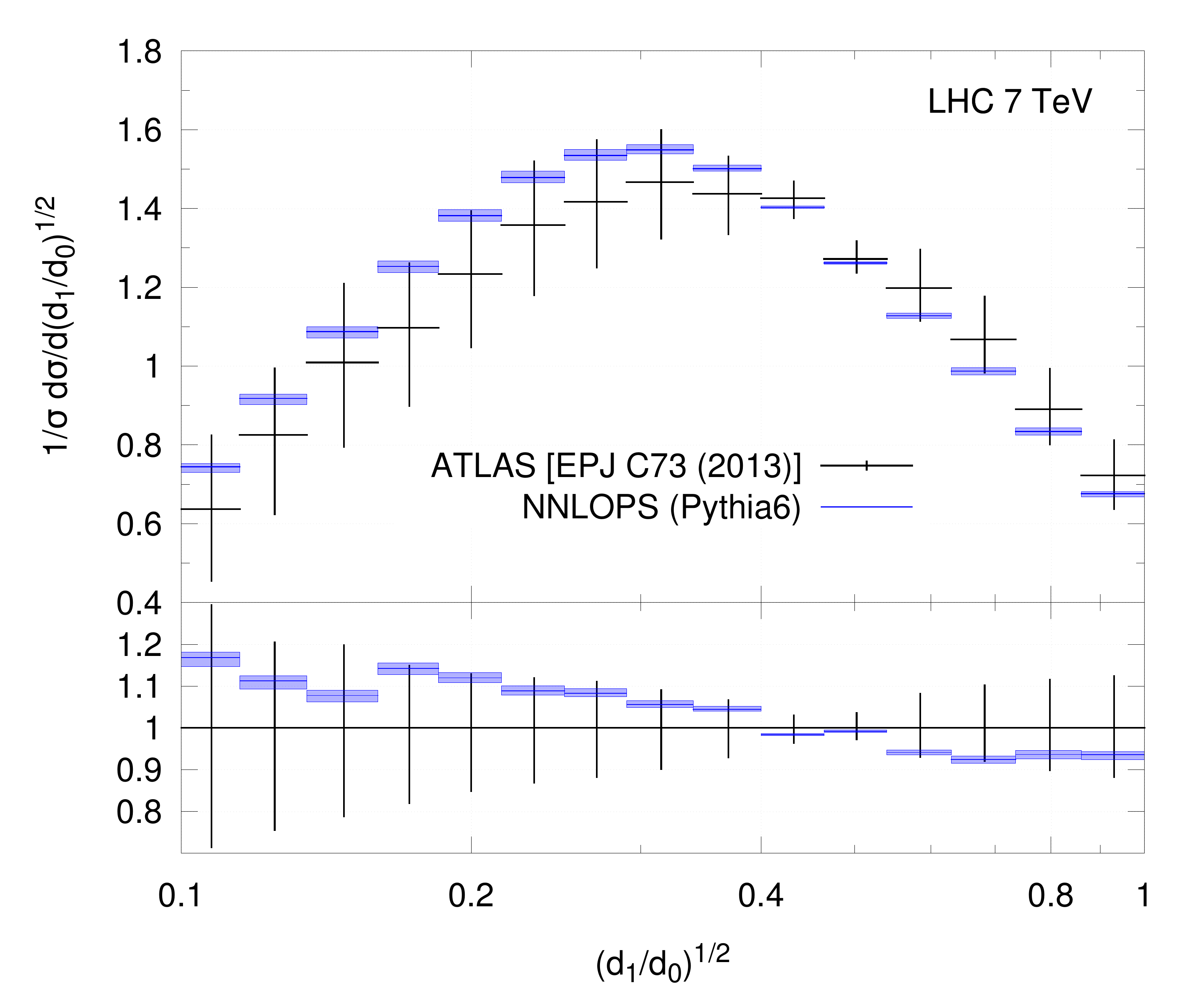}
\par\end{centering}
\caption{Comparison to 7 TeV LHC data from \Bref{Aad:2013ueu} for
  the $W$ boson ratio of the $\kt$ splitting scales $\sqrt{d_0}$ and
  $\sqrt{d_1}$ using \PYTHIA{8} (left) and \PYTHIA{6} (right).}
\label{fig:data-d1d0}
\end{figure}
Finally in \cref{fig:data-d1d0} we show the distribution for the ratio
$d_1/d_0$, for events with $\sqrt{d_0}>20 \GeV$.  Due to the ratio nature
of this quantity, a simultaneous over- or underestimation in predicting $d_1$
and $d_0$ should be partially compensated when plotting $d_1/d_0$. It is
therefore no surprise that the agreement with data is better than in
\cref{fig:data-d0,fig:data-d1}.

\subsubsection{$W$ and $Z$ Polarisation}
\label{subsec:polarisation}
Recently both ATLAS~\cite{ATLAS:2012au} and CMS~\cite{Chatrchyan:2011ig} have
published results on the polarisation of the $W$ boson at 7 TeV confirming the
Standard Model prediction that $W$ bosons are mostly left-handed in $pp$
collisions at large transverse momenta~\cite{Bern:2011ie}. Knowledge about the
$W$ boson polarisation is important, as it provides a discriminant in searches
for new physics.

We first very briefly review how to measure the polarisation in terms of angular
coefficients but refer the reader to the literature for a complete description
of the
topic~\cite{Bern:2011ie,Collins:1977iv,Lam:1980uc,Hagiwara:1984hi,Mirkes:1992hu,Mirkes:1994eb,Mirkes:1994dp,Hagiwara:2006qe}. Here
we will follow the derivation of~\cite{Bern:2011ie}. We then continue to compare
ATLAS data~\cite{ATLAS:2012au} to our \NNLOPS{} results for the $W$ boson
polarisation and present predictions for the angular coefficients for the $Z$
boson at 8 TeV.

The vector boson cross-section can be expanded in terms of $\cos{\theta^*}$ and
$\phi^*$ as
\begin{align}
  \frac{1}{\sigma}\frac{d\sigma}{d(\cos{\theta^*})d\phi^*} = &\frac{3}{16\pi}\Bigl[(1+\cos^2\theta^*)+A_0 \frac{1}{2}(1-3\cos^2{\theta^*})+A_1\sin 2\theta^*\cos\phi^* \notag \\
& + A_2\frac{1}{2}\sin^2\theta^*\cos 2\phi^* + A_3\sin\theta^*\cos\phi^* +A_4\cos\theta^* \notag \\ 
& + A_5\sin\theta^*\sin\phi^* + A_6\sin 2\theta^*\sin\phi^* + A_7\sin^2\theta^*\sin 2\phi^*\Bigr]\,, 
\label{eq:polexp}
\end{align}
where the polar angle $\theta^*$ and the azimuthal angle $\phi^*$ are defined in
some particular rest frame of the dilepton system. Here we will make use of two
different frames, the Collins-Soper frame\footnote{One defines in the laboratory
  frame the two beam directions by $\vec{b}_{+}=(0,0,1;1)$ and
  $\vec{b}_{-}=(0,0,-1;1)$. After boosting to the dilepton centre of mass frame,
  $\mathcal{O}^{'}$, one defines the z-axis as the bisector of $\vec{b}^{'}_{+}$
  and $-\vec{b}^{'}_{-}$ such that the z-axis points into the hemisphere of the
  $Z$ boson direction (in the lab frame). One then defines a q-axis lying in the
  plane spanned by the $\vec{b}^{'}_{+}$ and $\vec{b}^{'}_{-}$ vectors,
  orthogonal to the z-axis and pointing in the direction opposite to
  $\vec{b}^{'}_{+}+\vec{b}^{'}_{-}$.  $\theta^*$ is now defined with respect to
  the z-axis and $\phi^*$ with respect to the q-axis.}  \cite{Collins:1977iv}
for the $Z$ boson and the helicity frame for the $W$ boson defined as the
dilepton rest frame with the z-axis pointing along the direction of flight of
the $W$ boson in the lab frame. The cross-section can be differential in any
quantity that does not depend on the individual lepton kinematics.  The angular
coefficients can then be expressed in terms of expectation values defined as
\begin{align}
<f(\theta^*,\phi^*)> = \int_{-1}^{1}d(\cos\theta^*)\int_0^{2\pi}d\phi^*\frac{1}{\sigma}\frac{d\sigma}{d(\cos{\theta^*})d\phi^*}f(\theta^*,\phi^*)
\label{eq:expval}
\end{align} 
 by
\begin{align}
A_0&=4-<10\cos^2\theta^*>, &\quad A_1=&<5\sin 2\theta^*\cos\phi^*>, &\quad  A_2=&<10\sin^2\theta^*\cos 2\phi^*>\,, \notag \\
A_3&=<4\sin\theta^*\cos\phi^*>, &  A_4=&<4\cos\theta^*>,\phantom{\sin\theta^*} & A_5=&<5\sin^2\theta^*\sin 2\phi^*>\,, \notag \\
A_6&=<5\sin 2\theta^*\sin\phi^*>, & A_7=&<4\sin\theta^*\sin\phi^*>. &
\label{eq:angcoe}
\end{align}
For both $W$ and $Z$ production it is known that at $\mathcal{O}(\alpha_s)$
$A_5=A_6=A_7=0$. Here we have checked that with 20 million $Z$ events at 8 TeV,
and using only a cut on the invariant mass of $66 \GeV < m_{ll} < 116 \GeV$, the
coefficients do not deviate significantly from zero and we have therefore chosen
not to show them. It is also interesting to notice that as a consequence of the
spin-1 structure of the gluon, the Lam-Tung relation $A_0=A_2$\footnote{For
  $q\bar{q}$ initiated production the relation is exact to all orders. For $qg$
  initiated production it is violated at the NLO level~\cite{Bern:2011ie}. This
  is independent of the frame.} holds at LO~\cite{Lam:1978zr}. As we will see,
the deviations from this lowest-order result can become quite large,
$\mathcal{O}(20\%)$, in the $\ptz>10 \GeV$ region.  From the angular
coefficients we then define the left, $f_L$, right, $f_R$, and longitudinal,
$f_0$, polarisation fractions of the $W^\pm$ as
\begin{align}
f_L=\frac{1}{4}(2-A_0\mp A_4), \quad f_R=\frac{1}{4}(2-A_0\pm A_4), \quad f_0=\frac{1}{2}A_0\,.
\label{eq:polfra}
\end{align}
It is clear from these equations that the polarisation fractions are normalised
such that $f_L+f_R+f_0=1$ and it is therefore sufficient to show results for
$f_L-f_R$ and $f_0$. In this way we also separate the $A_0$ and $A_4$
dependence. In \cref{table:Wpol} we show our prediction for $f_L-f_R$ and
$f_0$ for combined $W^+$ and $W^-$ production at 7 TeV compared with ATLAS data
\cite{ATLAS:2012au} for two different $\pt$ regions. We find good agreement
between the data and predictions - three of them at the $1\sigma$ level and at
the $2\sigma$ level for $f_L-f_R$ in the low $\pt$ range. It should be noted
that the measurements are dominated by systematic uncertainties.
%\begin{center}
  \begin{table}[!tbh]
    \centering    
    \scalebox{0.8}{\begin{tabular} {l c c | c c}
          &  \multicolumn{2}{c}{$35\GeV < \ptw <50\GeV $} & \multicolumn{2}{c}{$\ptw >50\GeV $} \\
        \toprule
        & $f_L - f_R$ & $f_0$ &$f_L - f_R$ & $f_0$ \\
        \midrule
        Data &$0.238\pm 0.02 \pm 0.034$ & $0.219 \pm 0.033\pm 0.134$ & $0.252\pm 0.017\pm 0.034$&$0.127\pm 0.03\pm 0.108 $\\
        \NNLOPS{} &$0.317\pm0.002$ & $0.198\pm 0.004$&$0.289\pm 0.004$ &$0.214\pm 0.009$ \\
        \bottomrule
    \end{tabular}}
  \caption{A comparison between combined $W^+$ and $W^-$ at 7 TeV ATLAS data
    \cite{ATLAS:2012au} and our \NNLOPS{} prediction for $f_L-f_R$ and $f_0$ as
    defined in \cref{eq:polfra} and. For data the first uncertainty is
    statistical and the second one systematic. For the theoretical prediction
    the error is purely statistical. Except for the $\ptw$ cut the only cut
    imposed on the theoretical prediction is a transverse mass cut of $\mtw > 40
    \GeV$. The data and predictions are in good agreement.}
    \label{table:Wpol}
  \end{table}
%\end{center}

\begin{figure}[!tbh]
\begin{centering}
\includegraphics[clip,width=0.49\textwidth]{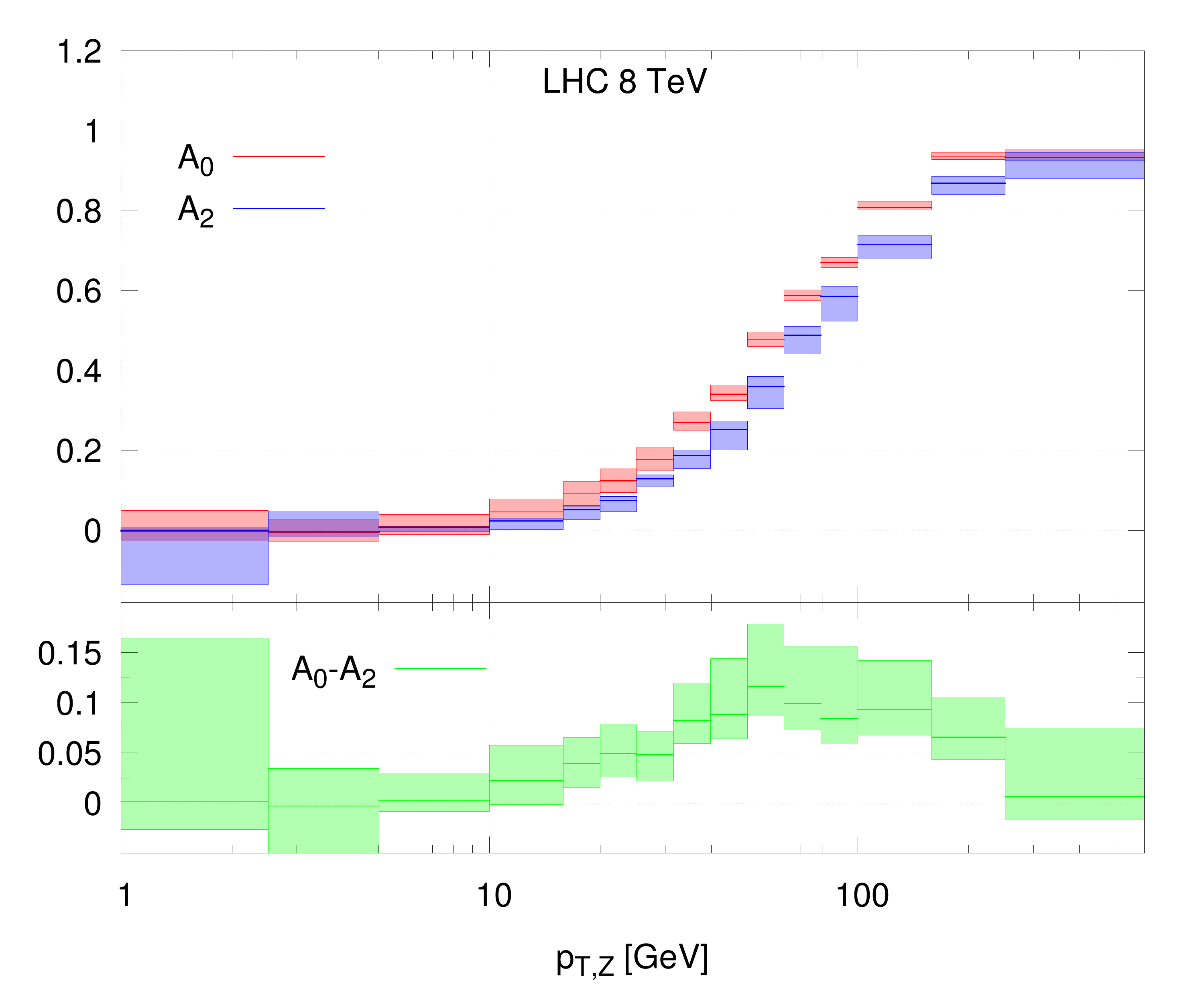}
\includegraphics[clip,width=0.49\textwidth]{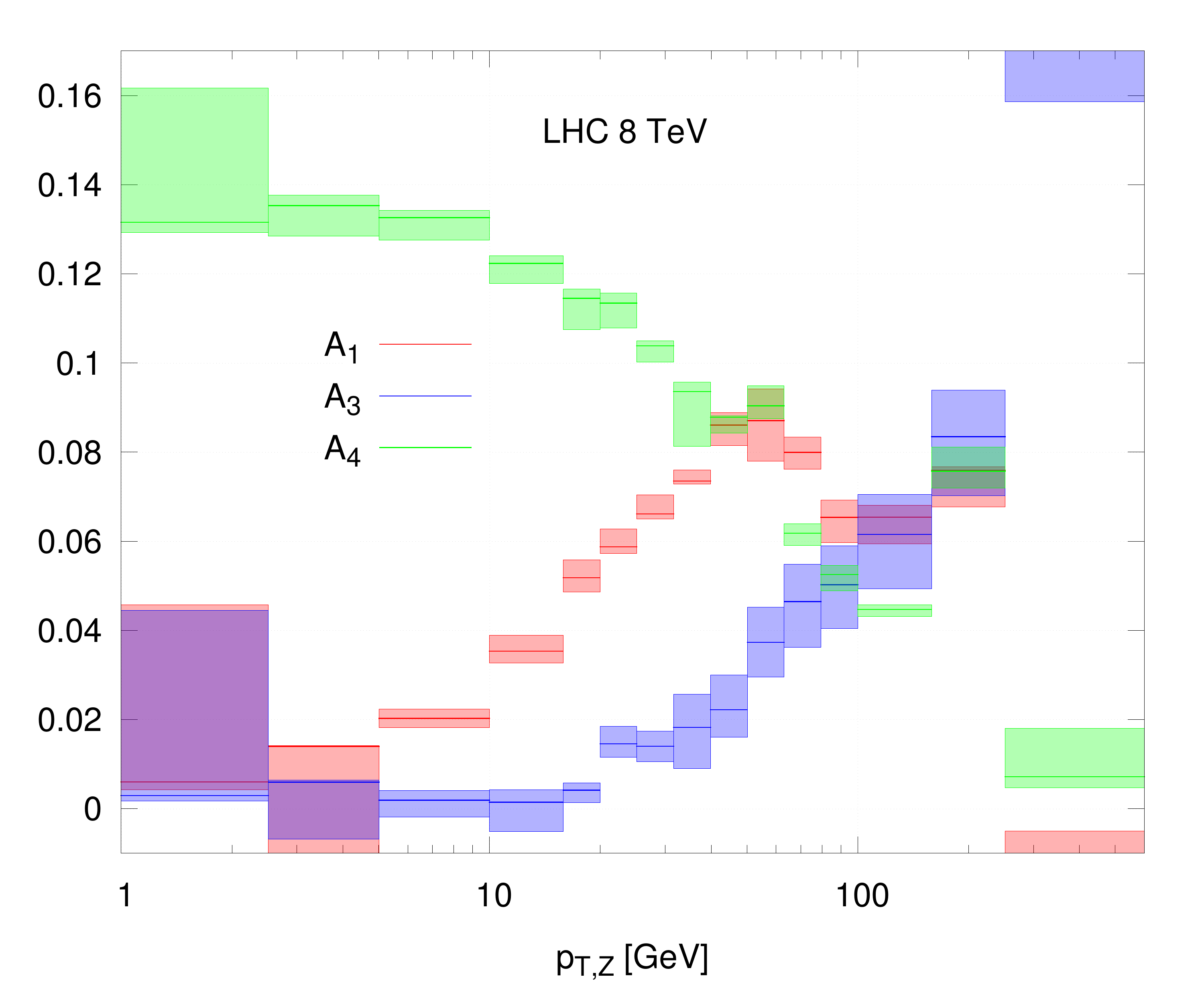}
\par\end{centering}
\caption{Predictions for angular coefficients \cref{eq:angcoe} in the
  Collins-Soper frame for $Z$ production at 8 TeV. In the left panel we show
  $A_0$ (red) and $A_2$ (blue) along with the difference $A_0-A_2$ (green) and
  in the right panel we plot $A_1$ (red), $A_3$ (blue) and $A_4$ (green). The
  only cut imposed is on the invariant mass of the dilepton system, $66 \GeV <
  m_{ll} < 116 \GeV$.}
\label{fig:angcoe}
\end{figure}

  When this work was carried out, there was no public measurement for the
  angular coefficients for the $Z$ boson in $pp$ collisions. Previously
  measurements of $A_0,A_2,A_3$ and $A_4$ in $p\bar{p}$ collisions had been
  published by the CDF collaboration~\cite{Aaltonen:2011nr} and very recently an
  analysis was published by ATLAS~\cite{Aad:2016izn} of the angular coefficients
  in $Z$-boson events at $8 \TeV$. This analysis finds significant deviations in
  $A_0 -A_2$ from theoretical predictions obtained with the \ZJMINLO{}
  generator. We here present our predictions for the angular coefficients as
  defined in \cref{eq:angcoe}, as a function of the boson transverse momentum at
  $8 \TeV$. We impose an invariant mass cut of $66 \GeV < m_{ll} < 116 \GeV$,
  but beyond that no other cuts are imposed. In the left panel of
  \cref{fig:angcoe} we plot $A_0$ (red) and $A_2$ (blue) along with the
  difference $A_0-A_2$, while in the right panel we plot $A_1,A_3$ and $A_4$.
For the two coefficients $A_0$ and $A_2$ we find that they are in very good
agreement for low transverse momentum, $\ptz < 10 \GeV$.  For higher
transverse momenta deviations start to appear peaking around $60 \GeV$.  For the
three other coefficients, we observe that the $\pt$ dependence is strongest for
larger transverse momenta. We can define polarisation fractions for the $Z$
boson analogously to those defined in \cref{eq:polfra}
by~\cite{Stirling:2012zt}
\begin{align}
f_L=\frac{1}{4}(2-A_0-\alpha A_4)\,, \quad f_R=\frac{1}{4}(2-A_0+\alpha A_4)\,, \quad f_0=\frac{1}{2}A_0\,,
\label{eq:polfraZ}
\end{align}
where $\alpha=\frac{c_L^2-c_R^2}{c_L^2+c_R^2}$, $c_L$ is the coupling to the
left-handed lepton and $c_R$ is the coupling to the right-handed lepton. As for
the $W$ boson these are normalised such that $f_L+f_R+f_0=1$ and we have in this
case that $f_L-f_R=-\frac{1}{2}\alpha A_4$. Using
$c_L=\sin^2\theta_W-\frac{1}{2}$ and $c_R=\sin^2\theta_W$ we get $\alpha\approx
0.032$. Comparing with \cref{fig:angcoe} we see that at low $\pt$ the
longitudinal polarisation of the $Z$ boson, $f_0$, is highly suppressed and it is
produced almost equally between the left- and right-handed polarisation. For
high $\pt$ the longitudinal polarisation starts to dominate.

\section{Conclusions} 
\label{sec:Conclusions}
In this chapter we reviewed the \MINLO{} procedure and described how a
\VJMINLO{} generator can be upgraded through a reweighting procedure to exhibit
\NNLO{}$^{(0)}$ accuracy. At the core of this method lies the fact that the
\ZJMINLO{} and \WJMINLO{} simulations achieve NLO accuracy also for fully
inclusive distributions, \emph{i.e.}~once the jet is integrated out. In the case
of Higgs production considered recently in \Bref{Hamilton:2013fea}, it was
enough to rescale the \NLOPS{} results to reproduce the NNLO Higgs rapidity
spectrum, thereby achieving \NNLOPS{} accuracy. In the present case instead, to
properly take into account the decay of the boson to leptons, \NNLOPS{} accuracy
is reached by performing a three-dimensional rescaling of the events generated
by \ZJMINLO{} and \WJMINLO{} using NNLO distributions as computed, for instance,
with \DYNNLO{}. This implies that the calculation is numerically more intensive.

We have validated our procedure by considering observables typically used to
study Drell-Yan processes. We have found extremely good agreement with NNLO
results for observables fully inclusive over QCD radiation, and all the features
expected in a computation matched with parton showers for more exclusive
observables. We have also compared our \NNLOPS{} predictions to state-of-the-art
analytic resummation for observables sensitive to soft-collinear QCD radiation,
which are not described accurately by a fixed-order NNLO calculation. Despite
the fact that the logarithmic accuracy of our simulation cannot be claimed to
reach the same precision as these analytic resummations, we have found
reasonably good agreement for the vector boson transverse momentum.  Slightly
more pronounced discrepancies were observed for $\phi^*$ and for the
jet-veto efficiency for large values of the jet radius $R$.

We have also compared with a number of available experimental results, not only
for fully inclusive observables but also for $\phi^*$, the transverse momentum
of the vector boson, as well as $\kt$ splitting scales. The successful outcome
of these comparisons is an indication that our computation can be used as a
state of the art prediction for future studies where a simultaneous inclusion of
NNLO corrections and parton shower effects are needed. We have also illustrated
how our generator can be used to study the polarisation of $W$ and $Z$ bosons
produced in Drell-Yan events.

We finally remark that we did not include in this work electroweak
corrections. These corrections are known exactly at one
loop~\cite{Dittmaier:2001ay,Dittmaier:2009cr} and can be included, at ${\cal
  O}(\alpha_{ew})$ either additively or multiplicative on top of the QCD
corrections included here. Version 3 of the program \FEWZ{}~\cite{Li:2012wna}
includes directly both EW and QCD corrections, however we could not use it to
generate the three dimensional distributions needed here.

%\part{Concluding Remarks}
\chapter{Final Remarks}\label{ch:final}
In this thesis we have presented a number of recent results in precision QCD,
spanning from fixed-order differential \NNLO{} and inclusive \NNNLO{}
calculations to the matching of \NLO{} and \NNLO{} calculations with parton
showers. Given the vast amount of data which the LHC is currently gathering,
these types of calculations will soon become crucial in determining the fate of
the Standard Model. In the absence of any new resonances or clear indications of
physics beyond the Standard Model, our ability to disentangle the truth from
data will be limited by the accuracy of our theoretical predictions. Several
complementary techniques exist to improve the reliability of these predictions
of which we have investigated a few here.

Except for the \POWHEG{} method presented in \cref{ch:powheg}, which is
extremely versatile, the methods we have studied here all have their
limitations. The \MINLO{} procedure can in principle be applied to any final
state, but the reweighting procedure that we introduced in \cref{ch:minlo}
becomes numerically unfeasible quickly as the number of external legs
grow. However, the success of the method indicates that we are on the right
track, and that we may expect to formulate a more general \NNLOPS{} method along
the lines of \POWHEG{} in the future. The calculation of \NNNLO{} accurate
inclusive VBF Higgs production presented in \cref{ch:incVBF}, was based on an
old observation that the VBF process factorises and can be described by the
proton structure functions. Although this point-of-view is not exact beyond
\NLO{}, it is accurate to the percent level, and most of the contributions
neglected by the approximation have been calculated fully and could in principle
be included. These are definitely needed as soon as one goes beyond \NNLO{}, as
the \NNNLO{} corrections and associated scale uncertainties are at the permille
level. Although permille corrections are not phenomenologically interesting
(yet), their smallness demonstrate the excellent convergence exhibited by
perturbative QCD. In addition to that, our knowledge of processes at \NNNLO{}
will in turn feed into the determination of PDFs, thereby improving our
modelling of the proton. Besides the inherent approximation of the structure
function approach, it has the further limitation of not being able to describe
the full final state due to the implicit integration over hadronic final
states. This limitation we eliminated in \cref{ch:vbfnnlo} by introducing the
``projection-to-Born'' method, effectively undoing the aforementioned
integration. The ``projection-to-Born'' method is very general in the sense that
it can be applied at any perturbative order. However, it only applies to
processes for which one can assign Born-like kinematics, and therefore it will
be interesting to see how applicable the method is in the future. At this
point it has already been used in the calculation of \NNLO{} t-channel single
top-quark production~\cite{Berger:2016oht} in addition to the calculation
presented here.

Although QCD has been around for almost half a century, and its phenomenology is
well established, we are still far from being done investigating the
consequences of the strong interactions in particle collisions. Even state of
the art calculations can at best claim an accuracy of a few percent when one
takes into account the effects of physics from PDFs, parton showering,
non-perturbative QCD, and EW effect. It is therefore not a bold claim to say
that the era of \emph{precision} QCD has only just begun, and that it will be
around for a very long time to come.

%% APPENDICES %% 
% Starts lettered appendices, adds a heading in table of contents, and adds a
%    page that just says "Appendices" to signal the end of your main text.
%\startappendices
% Add or remove any appendices you'd like here:
%\include{text/appendix-1}

%%%%% REFERENCES

% JEM: Quote for the top of references (just like a chapter quote if you're using them).  Comment to skip.
%\begin{savequote}[8cm]
%The first kind of intellectual and artistic personality belongs to the hedgehogs, the second to the foxes \dots
%  \qauthor{--- Sir Isaiah Berlin \cite{berlin_hedgehog_2013}}
%\end{savequote}

\setlength{\baselineskip}{0pt} % JEM: Single-space References
\bibliography{thesismaster}
%{\renewcommand*\MakeUppercase[1]{#1}%
%\printbibliography[heading=bibintoc,title={\bibtitle}]}

\end{document}